%% file: output.tex
\documentclass[a4paper, twoside, 12pt]{report}

\usepackage[english]{babel}
\usepackage[utf8]{inputenc}
\usepackage[T1]{fontenc}
\usepackage{lipsum}%nepotrebne
\usepackage{anyfontsize}%pre title
\usepackage{simpler-wick}%pre contractions

\usepackage[a4paper,top=3cm,bottom=2cm,left=3cm,right=3cm,marginparwidth=1.75cm]{geometry}
\usepackage{multirow,multicol} % for multiple columns in a page/table

\usepackage{amsmath,amssymb,amsthm}
\usepackage{physics,mathtools}
\usepackage{tikz,float}
\usepackage{dsfont}%for unit operator
\usepackage{subfigure}
\definecolor{tcd_blue}{RGB}{5, 105, 185}
\usepackage[numbers,sort&compress]{natbib}%makes[1-3] instead[1,2,3]
\usepackage[numbers]{natbib} % for citation. 
\usepackage{notoccite} % start citation ordering from text instead of table of contents. useful if there is citation in figure caption.
\usepackage{wasysym}%for the WR/FT diagram
\usetikzlibrary{matrix}%for the WR/FT diagram

\newcommand{\OH}{{\cal O}_H}
\newcommand{\OO}{{\cal O}}
\newcommand{\ra}{{\rightarrow}}

\def \nn {\nonumber}
\def \del {\partial}
\def\EE{\mathcal{E}}

\renewcommand{\cal}{\mathcal}%will add extra { and } around
\renewcommand{\bar}[1]{\overline{#1}}
\renewcommand{\tilde}[1]{\widetilde{#1}}
\renewcommand{\hat}[1]{\widehat{#1}}

\newcommand{\xp}{{x^+}}
\newcommand{\xpp}{{(x^+)}}
\newcommand{\xm}{{x^-}}

\def\qfr{\mathfrak{q}}
\def\wfr{\mathfrak{w}}
\newcommand{\ff}{{\mathfrak{f}}}
\newcommand{\Zt}{Z_{\rm scalar}} 
\newcommand{\Zv}{Z_{\rm shear}}  
\newcommand{\Zs}{Z_{\rm sound}}
\renewcommand{\aa}{\hat{a}}
\newcommand{\bb}{\hat{b}}
\newcommand{\cc}{\hat{c}}

\newcommand\tvev[1]{{\ensuremath{\left\langle{#1}\right\rangle}}}
\newcommand{\Zb}{\overline{Z}}
\newcommand{\be}{\begin{equation}}
\newcommand{\ee}{\end{equation}}
\def\et{{\widetilde E}}
\def\LL{{\cal L}}
\newcommand\lam{\lambda}
\newcommand\Lam{\Lambda}
\newcommand\om{\omega}
\newcommand\de{{\ensuremath{{\delta}}}}
\newcommand\De{{\ensuremath{{\Delta}}}}
\newcommand\ov{\over}
\newcommand\apr{{\ensuremath{{\alpha'}}}}
\def\le{\left}
\def\ri{\right}
\newcommand{\coo}[1]{{\mathcal{O}\left(#1\right)}}
\def\cL{{\cal L}}
\usetikzlibrary{patterns}
\makeatletter%make the "over" acceptable
\let\over\@@over
\makeatother

\usepackage{colortbl}%My color schemes for the (.) and [.]
\definecolor{ddred}{RGB}{198,31,31}
\definecolor{ddgreen}{RGB}{0,160,77}
\usepackage{hyperref}
\hypersetup{
    colorlinks=true,
    linkcolor=ddred,
    filecolor=blue,      
    urlcolor=ddred,
    citecolor=ddgreen}

\usepackage{titlesec}
\titleformat{\chapter}[hang]{\normalfont\huge\bfseries\color{tcd_blue}}{\thechapter}{1cm}{}{}

\title{Thermal Correlators and Black Holes:\\ \vspace{0.4cm}From Infinity to Singularity}
\author{Samuel Valach}
\newcommand{\cid}{21357714} % college id number, optional

\newcommand{\school}{School of Mathematics}
\newcommand{\department}{Trinity College Dublin, Dublin 2, Ireland}
\newcommand{\degree}{Doctor in Philosophy}
\newcommand{\supervisor}{Professor Andrei Parnachev}
\newcommand{\inputstatement}{A dissertation submitted in partial fulfillment of the degree of \\ \degree}

\begin{document}
\pagenumbering{roman}
\input{0_title/title} 

%% Declarations:
\input{0_formal/Declaration.tex}
\input{0_formal/Abstract.tex}
\input{0_formal/Dedication.tex}
\input{0_formal/Acknowledgement.tex}
\setcounter{page}{5} % set it such that the coverpage is page i. 
{\hypersetup{linkcolor=tcd_blue}%changes color in table of contents
\tableofcontents}

%% Imput Chapters:
\input{1_introduction/introduction.tex} % introduction to the project/thesis
\input{2_chapter/chapter_2.tex}
\input{3_chapter/chapter_3.tex}
\input{4_chapter/chapter_4.tex}
\input{5_chapter/chapter_5.tex}
\input{6_chapter/chapter_6.tex}
\input{7_conclusion/conclusion.tex}

\bibliographystyle{Z_bibs/JHEP}
\bibliography{Z_bibs/bibliography}
\addcontentsline{toc}{chapter}{Bibliography}

\input{A_appendices/appendix.tex}

\end{document}

%% file: 0_title/title.tex
\begin{titlepage}

\center % Center everything on the page

%% All the text parameters should be taken from the start of the main.tex file.
%% You should only alter stuff here if you want to change the layout

%----------------------------------------------------------------------------------------
%	LOGO SECTION
%----------------------------------------------------------------------------------------
%% Choose one of the following -- a colour or black-and-white logo

\includegraphics[width = 0.9\textwidth]{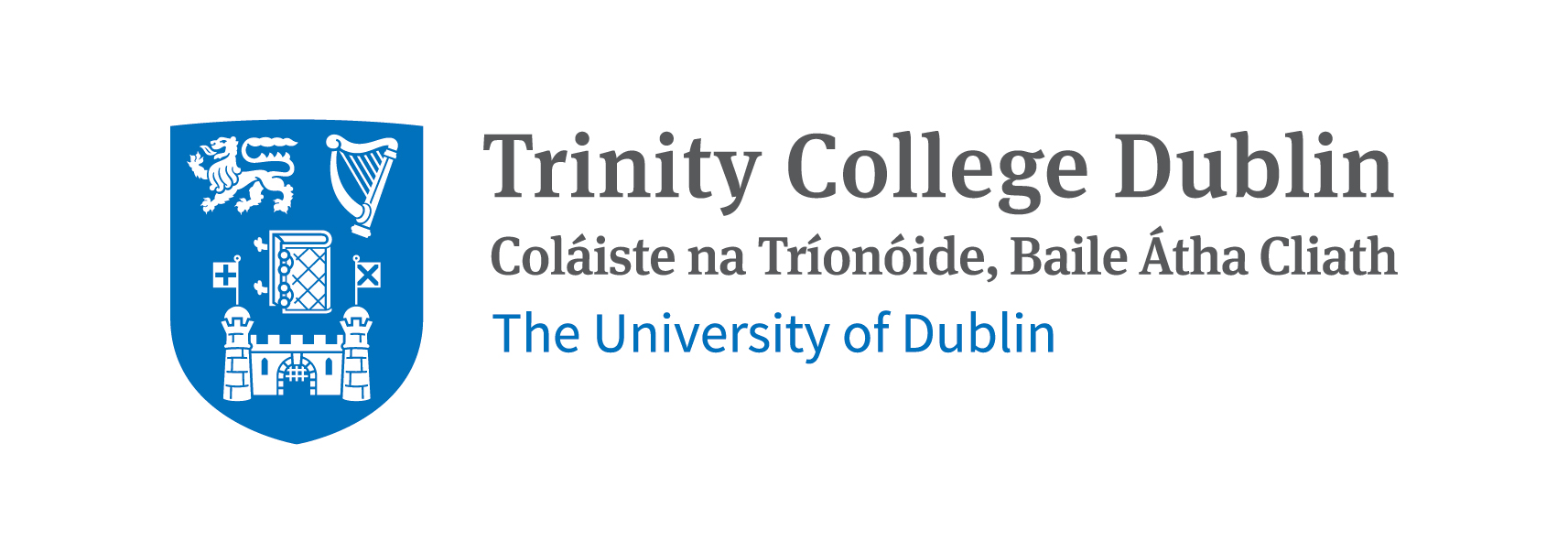}\\[2.0cm] 
%\includegraphics[width=12cm]{title/black-stacked-trinity.jpg}\\[1cm] 

%----------------------------------------------------------------------------------------
%	TITLE SECTION
%----------------------------------------------------------------------------------------
\makeatletter

{ \fontsize{24.2}{28.25}\selectfont \bfseries \color{tcd_blue} \@title}\\[1.5cm] % Title of your document

%----------------------------------------------------------------------------------------
%	AUTHOR SECTION
%----------------------------------------------------------------------------------------

\ifdefined\cid
{ \Large \bfseries \@author}\\[1.5cm] % Your name
%\cid \\[2cm] % Your Student ID
\else
\@author \\[1.5cm] % Your name
\fi

%----------------------------------------------------------------------------------------
%	Supervisor SECTION
%----------------------------------------------------------------------------------------

{ \large Supervisor: \supervisor}\\[1.0cm] % Their name
\ifdefined\cosupervisor
Co-supervisor: \cosupervisor\\[2cm] % Their name
\fi

%----------------------------------------------------------------------------------------
%	DATE SECTION
%----------------------------------------------------------------------------------------

{\large Submitted \\ June 1, 2025}\\[2cm] % Date, change the \today to a set date if you want to be precise.. Respectively: {\large Submitted \\ \inputdate}\\[2cm]

%----------------------------------------------------------------------------------------
%	TYPE OF THESIS SECTION
%----------------------------------------------------------------------------------------
\vfill
\inputstatement

\vfill % Fill the rest of the page with whitespace

\large \school\\[1.5ex] % Minor heading such as course title
\ifdefined\department
\large \department\\[1.5cm] % Minor heading such as course title
\fi

\end{titlepage}

%% file: 0_formal/Declaration.tex
\renewcommand{\abstractname}{Declaration and Co-authorship}
\begin{abstract}

I declare that this thesis has not been submitted as an exercise for a degree at this or any other university and it is entirely my own work. I agree to deposit this thesis in the open access institutional repository of the University, or allow the library to do so on my behalf, subject to Irish Copyright Legislation and Trinity College Library conditions of use and acknowledgement. I consent to the examiners retaining a copy of the thesis beyond the examining period, should they so wish. This thesis contains the results of four co-authored papers \cite{Karlsson:2022osn,Huang:2022vet,Esper:2023jeq,Ceplak:2024bja}. The paper \cite{Karlsson:2022osn} was co-authored with Robin Karlsson, Andrei Parnachev and Valentina Prilepina. The paper \cite{Huang:2022vet} was co-authored with Kuo-Wei Huang, Robin Karlsson and Andrei Parnachev. The paper \cite{Esper:2023jeq} was co-authored with Chantelle Esper, Kuo-Wei Huang, Robin Karlsson and Andrei Parnachev. The paper \cite{Ceplak:2024bja} was co-authored with Nejc Čeplak, Hong Liu and Andrei Parnachev.

\vspace{2.2cm}

\hspace{11.6cm}Samuel Valach

\hspace{12.1cm}June 2025
%Signed:~\rule{5cm}{0.3pt}\hfill Date:~\rule{5cm}{0.3pt}
\end{abstract}

%% file: 0_formal/Abstract.tex
\renewcommand{\abstractname}{Abstract}
\begin{abstract}
\vspace{1em}
This thesis investigates the relationship between thermal correlation functions in conformal field theories (CFTs) and the geometry of black holes within the AdS/CFT correspondence. Using a common technique -- a near-boundary expansion %dual to the boundary OPE 
-- we examine two main themes. 

First, we study deep geometric features of black holes from the CFT perspective. We demonstrate that information about the black hole singularity is encoded in the stress-tensor sector of the boundary correlator. More precisely, by resumming the contributions from the stress tensor and its composites, we discover branch-point singularities in the complex time plane. These singularities precisely correspond to the bulk geodesics that traverse the black hole interior, reflect off the singularity, and return to the boundary of the maximally extended spacetime. Furthermore, we explain the role of double-trace operators in the restoration of analyticity and KMS condition, and establish a map between individual bulk geodesics and OPE sectors of the thermal boundary correlator.

Second, we explore the universality of thermal stress-tensor correlators in CFTs with pure gravity duals. Through detailed holographic calculations in Einstein and Gauss-Bonnet gravity, we analyse the structure of two-point functions near the lightcone and identify a robust universal behaviour -- in this regime, the correlators are described by three universal functions. The bulk Lagrangian only affects the arguments of these functions through corrections to the cubic stress-tensor couplings and the thermal stress-tensor one-point function. We explain how this behaviour is connected to causality constraints, such as the averaged null energy condition (ANEC). In particular, we demonstrate that saturation of the ANEC leads to the correlator becoming independent of temperature.

Together, these results highlight the power of the near-boundary method as an effective tool for extracting physical data in holographic CFTs. Our results also suggest that thermal correlators -- especially their stress-tensor sector -- serve as sensitive probes of geometry beyond the black hole horizon and offer valuable insights into causality constraints and universal features of holographic theories.
\end{abstract}

%% file: 0_formal/Dedication.tex
\renewcommand{\abstractname}{\phantom{.}}
\begin{abstract}
\vspace{2.5em}
\centering{Dedicated to my brother Fridrich.}
\end{abstract}

%% file: 0_formal/Acknowledgement.tex
\renewcommand{\abstractname}{Acknowledgements}
\begin{abstract}
\vspace{1em}

First, I would like to thank my supervisor Andrei Parnachev for his invaluable advice and support throughout my PhD studies. I learned a lot about physics and about academia from him. I am grateful he was pushing me when I needed it and also when I didn't. Without him, I would not have accomplished what I have over the past four years. Thank you.

Much of the work during my PhD in Dublin was carried out in collaboration with my great friends Nejc \v{C}eplak, Robin Karlsson and Chantelle Esper, to whom I am deeply grateful. I would also like to thank Hong Liu, Kuo-Wei Huang and Valentina Prilepina for fruitful collaborations.

I would like to thank all the people at the School of Mathematics for creating a kind and welcoming environment, in particular Tristan McLoughlin, Manuela Kulaxizi and Nicolas Mascot. My work was partially supported by the Irish Research Council which I’m thankful for. I would also like to thank Jan Manschot and Arkady Tseytlin for being the examiners of this thesis.

The harsh weather in Dublin and absence of sunlight would be hard to withstand without all the great people I have met at Trinity College. I thank my fellow PhD students and postdocs Adolfo, Andrea, Aradhita, Chantelle, Chiara, Christina, Chris, Daniel, Elias, Erica, Georgio, Guy, Ilja, Ivan, Johannes, Koen, Luke, Liam, Liz, Luki, Marta, Matthias, Nancy, Nejc, Nikolaos, Riccardo, Robin, Rodrigo, Sean, Thibaut, Tomasso, Vera, Vincenzo, Yueke and Zhi-Zhen. 

I am especially thankful to my fiancée, Martina, who was always with me, and gave meaning to my everyday life. 

Finally, I am deeply grateful to my family for their unconditional support and for always being there for me. Without them, none of this would have been possible.

\end{abstract}

%% file: 1_introduction/introduction.tex
\chapter{Introduction}\label{ch1}
\pagenumbering{arabic}

The unification of general relativity and quantum mechanics remains one of the most ambitious and important goals in theoretical physics. Each theory, within its respective domain, has withstood rigorous experimental tests and driven advances in our understanding of nature. However, when confronted with regimes where both gravitational and quantum effects become equally important -- such as inside black holes or near the Big Bang -- the two frameworks appear to be mutually incompatible.

Efforts to construct a consistent quantum theory of gravity have given rise to a rich landscape of ideas, including string theory, loop quantum gravity, asymptotic safety, and more. Among these, the most conceptually profound and mathematically precise proposal for understanding quantum gravity in a non-perturbative regime has emerged through the lens of holography: the principle that the degrees of freedom of a quantum gravitational system are encoded on a lower-dimensional boundary.

An important hint toward this holographic nature of gravity arose from the study of black hole thermodynamics. The work of Bekenstein and Hawking \cite{Bekenstein:1973ur,Hawking:1976de,Hawking:1975vcx} revealed that black holes are not just classical solutions of general relativity, but thermodynamic objects with well-defined entropy and temperature. Remarkably, they found that the entropy \( S_{BH} \) of a neutral, non-rotating black hole is not proportional to its volume, as one might expect from conventional thermodynamics, but instead to the area \( A \) of its event horizon:
\begin{equation}\label{eq:SBH}
    S_{BH} = \frac{A c^3k_B}{4 G \hbar}\ . 
\end{equation}\vspace{-0.21cm}

This area scaling suggests that the fundamental degrees of freedom responsible for black hole entropy do not live in the bulk volume, but on the boundary surface -- the horizon itself. Such behaviour is radically different from that of ordinary quantum field theories, where entropy typically scales with volume. The understanding that gravitational entropy scales with area was one of the first concrete clues that a quantum theory of gravity might be holographic in nature.

This idea found a concrete realization in the Anti-de Sitter/Con\-formal Field Theory (AdS/CFT) correspondence \cite{Maldacena:1997re, Gubser:1998bc, Witten:1998qj}. In its original form, this correspondence posits a duality between Type IIB string theory on \( \text{AdS}_5 \times S^5 \) and \( \mathcal{N}=4 \) Super-Yang-Mills theory in four dimensions. More generally, it relates a gravitational theory in a \( (d+1) \)-dimensional asymptotically AdS spacetime to a \( d \)-dimensional conformal field theory living on the spacetime boundary. This duality has provided an invaluable non-perturbative tool for studying strongly coupled quantum field theories and has reshaped our understanding of quantum gravity.

To explore these ideas more systematically, it is often useful to focus on a class of conformal field theories that admit weakly-coupled semi-classical Einstein gravity duals. These so-called holographic CFTs provide a powerful and calculable framework for examining the duality. They are characterized by a large central charge and a large gap to higher-spin states, features which ensure the dual gravitational description remains under perturbative control. As such, they serve as an ideal laboratory for probing both boundary and bulk physics -- from transport and thermalization in the strongly coupled field theories, to the structure of black holes, including photon spheres, gravitational orbits, and even the nature of the black hole interior, see e.g.~\cite{Louko:2000tp,Kraus:2002iv,Fidkowski:2003nf,Festuccia:2005pi,Festuccia:2006sa,Amado:2008hw,Hartman:2013qma,LiuSuh13a,LiuSuh13b,Grinberg:2020fdj,Rodriguez-Gomez:2021pfh,Leutheusser:2021frk,deBoer:2022zps,David:2022nfn,Horowitz:2023ury,Parisini:2023nbd,Dodelson:2023nnr}. In this thesis, we will focus on this class of theories to investigate the interplay between gravitational dynamics and CFT correlation functions at finite temperature.

\section*{Thermal Correlators and Black Hole Geometry}

A particularly powerful feature of the AdS/CFT correspondence is that thermal states in the boundary CFT are dual to black hole geometries in the bulk \cite{Hawking:1982dh,Witten:1998zw}. This mapping allows for the use of semi-classical gravity to study quantum field theories at finite temperature. At the same time, it opens the possibility of exploring longstanding mysteries in black hole physics -- from thermodynamics to singularities and information loss -- through boundary field theory observables.

In the spirit of holography, the boundary of AdS plays a key role. According to the correspondence, the observables in the CFT -- $e.g.$ correlation functions of local operators -- encode information about the entire bulk spacetime. This thesis embraces this perspective and is driven by the following guiding question:
\vspace{0.45cm}

\emph{What can thermal correlators in the boundary theory reveal about the geometry of black holes in the bulk, and what can semi-classical analyses in black hole spacetimes reveal about the dynamics of strongly coupled thermal systems?}\vspace{0.45cm}

Thermal two-point functions in a conformal field theory are natural probes of both the field theory's dynamics and its dual gravitational background. Their structure reflects the response of the system to perturbations and encodes transport properties, equilibration timescales, and collective modes. In a holographic setting, these correlators are computed by solving field equations in black hole backgrounds and reading off the behaviour of the bulk fields near the AdS boundary. In this way, a precise dictionary between bulk physics and boundary data can be established.

Throughout this thesis we will be using a common computational technique: the \emph{near-boundary expansion} \cite{Fitzpatrick:2019zqz}. This method constructs solutions to the bulk field equations as an expansions near ``infinity'', $i.e.$ near the conformal boundary of AdS. By design, the expansion matches order by order with the Operator Product Expansion (OPE) -- the fundamental associative algebra of conformal field theories. This approach provides a powerful, effective, and remarkably universal tool for computing thermal correlators in theories with a holographic dual, and it offers a systematic way to explore the interplay between gravitational dynamics in the bulk and physical properties of the boundary theory. In Chapter \ref{ch2} we describe the necessary concepts for understanding this holographic method and explain the motivation and technical details of the expansion for the case of scalar fields. In Chapters \ref{ch3}-\ref{ch6} we apply this technique to investigate the two intertwined topics central to this thesis:\vspace{0.6cm}

\begin{enumerate}
    \item \textbf{Signatures of black hole singularities in thermal correlators:} Can field theory observables, computed entirely at the boundary, be sensitive to the spacetime region inside the black hole horizon? What signatures of the singularity, if any, can be detected in the analytic structure of CFT correlators?
    
    \item \textbf{The universality of stress-tensor correlators at finite temperature:} Are there features of the stress tensors that remain robust across large classes of gravitational duals? What constraints does the holographic framework impose on such correlators?
\end{enumerate}

As we will see, both of these deep physical questions can be effectively investigated within the same computational framework -- the near-boundary expansion -- that allows for the extraction of the information about the holographic CFT.

\section*{Theme I: Black Hole Singularity}

Black holes in general relativity are characterized by a horizon that cloaks a curvature singularity. While the horizon is a smooth and well-defined geometric object, the singularity represents a breakdown of classical physics: an endpoint where tidal forces become infinite and spacetime ceases to be well-defined. It is generally believed that, in a UV-complete theory of gravity, these singularities are resolved. However, the precise mechanism of this resolution remains to be understood. Holography provides an ideal framework to investigate this issue, as it allows us to apply a wide array of CFT techniques and incorporate quantum corrections in a controlled way. The first step in this direction is to answer the following question: What is the imprint of the singularity in the boundary CFT?

In Chapter \ref{ch3} of this thesis, which is based on \cite{Ceplak:2024bja}, we investigate this question by studying thermal scalar two-point functions in strongly coupled CFTs with gravitational duals described by minimally coupled scalar on a background of AdS-Schwarzschild black hole. We use geodesic approximations and the near-boundary method to solve the bulk equation of motion and analyse the behaviour of the correlators in various kinematic regimes.

Our analysis reveals that the information about the black hole singularity is encoded in the analytic behaviour of the stress-tensor sector of the dual CFT. More precisely, by resumming the corresponding conformal blocks we find non-analytic features -- \emph{bouncing singularities} -- which physically correspond to geodesics that penetrate the horizon, reflect off the singularity, and return to the boundary of the maximally extended spacetime. This connection is made precise through a detailed comparison between the bulk and boundary computations.

\section*{Theme II: Universality in Stress-Tensor Correlators}

The second major theme of this thesis concerns the structure and universality of thermal two-point functions of the stress-energy tensors in holographic CFTs. These correlators play a central role in the study of hydrodynamics, chaos, and thermalization (see $e.g.$ \cite{Kovtun:2003wp,Kovtun:2004de,Teaney:2009qa,Heinz:2013th,Buchel:2003tz,Iqbal:2008by,Kats:2007mq,Brigante:2007nu,Brigante:2008gz}), and they are directly related to bulk perturbations of the metric \cite{Policastro:2001yc,Policastro:2002se,Kovtun:2003wp,Kovtun:2004de}.

In Chapters \ref{ch4} through \ref{ch6}, which are based on \cite{Karlsson:2022osn,Huang:2022vet,Esper:2023jeq}, we systematically study stress-tensor correlators in theories where the holographic dual only contains the metric, $i.e.$ theories of pure gravity. Such theories are, in a sense, the simplest holographic models for thermal strongly-coupled matter on the boundary, as the only degrees of freedom in the CFT are the identity operator, the stress tensor and its composites. We start by examining the case of Einstein gravity, where we develop our computational tools and subsequently extend the calculations to higher-derivative Gauss-Bonnet gravity. We consider correlators in both position and momentum space, analyse their behaviour near the lightcone, and compute perturbative and non-perturbative contributions using a combination of analytic and numerical techniques.

One of the central findings of this part of the thesis is the emergence of \emph{near-lightcone universality}. In particular, we find that the behaviour of the correlators in this regime is governed by a universal structure that is largely insensitive to the details of the bulk action. Even when higher-curvature corrections are included, the structure of the correlator remains intact, suggesting a deep robustness rooted in higher dimensional ($d>2$) conformal field theories. To understand this universality, we decompose the correlators into conformal blocks and analyse the contributions of multi-stress tensor operators.
%We identify the precise structure of the individual cancellations that reflect the underlying constraints of the theory. 
Furthermore, we explore the implications of these results for energy conditions, such as the averaged null energy condition (ANEC), and their higher-spin generalizations.

\section{Structure of the Thesis}

The thesis is structured as follows:

\begin{itemize}
    \item \textbf{Chapter 2} introduces the computational tools used throughout the thesis. We review both sides of the AdS/CFT duality, present the near-boundary ansatz and demonstrate the extraction of the CFT data on a concrete example.

    \item \textbf{Chapter 3} is devoted to the study of scalar thermal correlators and their sensitivity to black hole singularities. We analyze both geodesic and OPE perspectives and demonstrate how bouncing singularities arise from specific contributions in the spectrum.

    \item \textbf{Chapter 4} initiates the analysis of stress-tensor thermal correlators in Einstein gravity. We compute two-point functions in various channels (scalar, shear, sound) and compare their structure to predictions from the conformal block decomposition, extracting important CFT data such as the anomalous dimensions of the double-stress tensors.

    \item \textbf{Chapter 5} extends this analysis to Gauss-Bonnet gravity. We investigate how higher-derivative corrections modify the correlators and study their implications for ANEC and its higher-order generalizations.

    \item \textbf{Chapter 6} focuses on the near-lightcone regime. We compute position- and momentum-space correlators, extract perturbative and nonperturbative behaviour, and explore universal features of the stress-tensor sector in holographic CFTs. 

    \item \textbf{Chapter 7} summarizes the main findings of the thesis, reflects on the implications and outlines directions for future research.
    
\end{itemize}

In Appendices \ref{a.cbd}-\ref{ap.eomsGB} we provide further results and technical details relevant for the Chapters \ref{ch2}-\ref{ch6}.

    \section{Conventions and Notation}

Throughout the thesis we will stick to the following conventions:

\begin{itemize}
    \item Signature in Minkowski space: $(-,+,+,+,\ldots)$
    \item Fourier transform: $f(\omega)\equiv\int_{-\infty}^\infty f(\tau)e^{-\tau\omega}\dd \tau$  and $f(\tau)\equiv\frac{1}{2\pi}\int_{-\infty}^\infty f(\omega) e^{\tau\omega}\dd \omega$
    \item Transformation diagram:
    \begin{center}
    \begin{tikzpicture}
    \matrix (m) [matrix of math nodes,row sep=4em,column sep=6em,minimum width=2em]
    {
     {\substack{\text{Minkowski,}\\\text{Momentum Space}}} & {\substack{\text{Minkowski,}\\\text{Position Space}}} \\
     {\substack{\text{Euclidean,}\\\text{Momentum Space}}} & {\substack{\text{Euclidean,}\\\text{Position Space}}} \\};
    \path[-stealth]
    (m-1-1) edge node [left] {$\substack{\omega\,\mapsto\,-i\omega\\A_{t}\,\mapsto\, iA_\tau}$} (m-2-1)
            edge node [above] {$\substack{\omega\,\mapsto\,+i\partial_{t}\\q\,\mapsto\,-i\partial_z}$} (m-1-2)
    (m-2-1) edge node [below] {$\substack{\omega\,\mapsto\,-i\partial_\tau\\q\,\mapsto\,-i\partial_z}$} (m-2-2)
    (m-1-2) edge node [right] {$\substack{t\,\mapsto\,-i\tau\\\partial_{t}\,\mapsto\, i\partial_\tau\\A_{t}\,\mapsto\, iA_\tau}$} (m-2-2);
    \end{tikzpicture}
    \end{center}
    \item $d$ denotes the spacetime dimension of the boundary CFT 
    \item $D=d+1$ denotes the dimension of the asymptotically AdS spacetime
\end{itemize}

%% file: 2_chapter/chapter_2.tex
\chapter{Holography Near the Boundary}\label{ch2}

In this chapter, we begin by introducing some foundational elements of conformal field theory (CFT), such as the Operator Product Expansion (OPE) and the conformal block decomposition, which will play a key role in our later analysis. For a more detailed review of CFTs, see $e.g.$ \cite{Simmons-Duffin:2016gjk,Penedones:2016voo} that inspired the first section of this chapter. 

In Section \ref{s2.AdSCFT} we briefly review the geometry of Anti-de Sitter (AdS) space and present the AdS/CFT correspondence in its bottom-up formulation, mostly following the notation of \cite{zaffaroni2000introduction} and \cite{hartman2015lectures}, and clarify what we mean by a holographic CFT. Following this, we motivate and introduce the near-boundary method \cite{Fitzpatrick:2019zqz} as a practical tool for computing correlators within the holographic framework in Section \ref{s2.nbe}. 

Finally, in Section \ref{s2.thrmlztion} we explain how to decompose the holographically calculated correlators in conformal blocks and extract important data characterizing the dual CFT. We conclude with a brief discussion of thermalization in this context, following \cite{Karlsson:2021duj}.

%%%%%%%%%%%%%%%%%%%%%%%%%%%%%%%%%%%%%%%%%%%%%

\section{Conformal Field Theories}\label{s2.cfts}

A conformal field theory is a quantum field theory invariant under conformal transformations. These are the transformations that rescale the metric by a positive real function
    \begin{equation}\label{e.confdefrov}
    \dd x^2\rightarrow(\dd x')^2=\Omega^2(x)\dd x^2
    \end{equation}
Infinitesimally we have $x'_\mu=x_\mu+v_\mu(x)$ and $\Omega(x)=1+\frac{\omega(x)}{2}$; substituting this into \eqref{e.confdefrov} yields
    \begin{equation}\label{e.macitarasoi}
        \partial_\mu v_\nu+\partial_\nu v_\mu=\omega(x)\eta_{\mu\nu}\ ,
    \end{equation}
where $\eta_{\mu\nu}$ is the flat metric in $d$ spacetime dimensions. Taking the trace on both sides of \eqref{e.macitarasoi} and dividing by $d$, we obtain the defining condition for infinitesimal conformal transformations
    \begin{equation}\label{e.makubara}
        \partial_\mu v_\nu+\partial_\nu v_\mu-\frac{2}{d}\left(\partial^\rho v_\rho\right)\eta_{\mu\nu}=0 \ .
    \end{equation}
Importantly, in $d=2$, this equation admits infinitely many independent solutions, leading to unique features of 1+1 dimensional CFTs. In this thesis we will only be interested in CFTs with $d>2$, where the equation \eqref{e.makubara} has finitely many solutions. These are
\begin{align}
    \qq{\emph{translations}} v_\mu(x)&=a_\mu\\
    \qq{\emph{rotations}} v_\mu(x)&=\rho_{\mu\nu}x^\nu\\
    \qq{\emph{dilatations}} v_\mu(x)&=\lambda x_\mu\\
    \qq{\emph{special conformal transformations}} v_\mu(x)&=b_\mu x^2-2x_\mu(b\cdot x)\ ,
\end{align}
where $a_\mu$, $\rho_{\mu\nu}$, $\lambda$ and $b_\mu$ are independent of $x$ and $\rho_{\mu\nu}=-\rho_{\nu\mu}$. Together they generate the conformal group $SO(d,2)$ (resp.\ $SO(d+1,1)$ in Euclidean signature). Abstractly, denote the corresponding algebra generators by $P_\mu$ (translations), $M_{\mu\nu}$ (rotations and boosts), $D$ (dilatations) and $K_\mu$ (special conformal transformations). 

In CFT we have a unitary action of the conformal group, $i.e.$ the generators $P_\mu$, $M_{\mu\nu}$, $D$ and $K_\mu$ are represented by hermitian operators which satisfy the commutation relations of the conformal algebra
    \begin{align}
    \comm{M_{\mu\nu}}{M_{\rho\sigma}}&=i\eta_{\mu\rho}M_{\eta\sigma}+\text{permutations}\\
    \comm{M_{\mu\nu}}{P_\rho}&=i(\eta_{\mu\rho}P_\nu-\eta_{\nu\rho}P_\mu)\\
    \comm{M_{\mu\nu}}{K_\rho}&=i(\eta_{\mu\rho}K_\nu-\eta_{\nu\rho}K_\mu)\\
    \comm{M_{\mu\nu}}{D}&=0\\
    \comm{K_\mu}{P_\nu}&=-2iM_{\mu\nu}-2i\eta_{\mu\nu}D\\
    \comm{D}{P_\mu}&=iP_\mu\\
    \comm{D}{K_\mu}&=-iK_\mu\ .
    \end{align}

For the classification of local operators in a CFT it is useful to introduce the notions of \emph{primary operators} and \emph{descendants}. A local operator $\mathcal{O}(x)$ is called primary if it satisfies the following transformation properties
    \begin{align}
    \comm{D}{\mathcal{O}(0)}&=\Delta\mathcal{O}(0)\\
    \comm{M_{\mu\nu}}{\mathcal{O}(0)}&=\mathcal{S}_{\mu\nu}\mathcal{O}(0)\\
    \comm{K_\mu}{\mathcal{O}(0)}&=0\ ,
    \end{align}
where $\Delta$ is called scaling dimension of operator $\mathcal{O}(x)$ and the precise form of the matrix $\mathcal{S}_{\mu\nu}$ depends on the spin $J$ of the local operator. Applying any number of generators $P_\mu$ on a primary $\mathcal{O}(x)$ one obtains an infinite tower of its descendants.

An important aspect of a CFT is that, in radial quantization, one can map local operators to states in the Hilbert space by
    \begin{equation}
    \mathcal{O}\longleftrightarrow\ket{\mathcal{O}}\equiv\mathcal{O}(0)\ket{0}\ ,    
    \end{equation}
where $\ket{0}$ is the vacuum. This relation is known as the \emph{operator-state correspondence}. In practice, it allows one to naturally introduce the notion of primary states and to easily transition between the language of states and operators in correlation functions.

Let us now focus on scalar primary operators. Conformal invariance is sufficiently constraining to completely fix all 1-point, 2-point and 3-point functions in the vacuum
    \begin{align}
    \expval{\mathcal{O}(x)}&=\begin{cases}0&\text{if}\quad \mathcal{O}(x)\neq\mathds{1}\\1&
    \text{if}\quad\mathcal{O}(x)=\mathds{1}\end{cases}\label{e.kukurickas}\\
    \expval{\mathcal{O}_1(x_1)\mathcal{O}_2(x_2)}&=N\frac{\delta_{\Delta_1\Delta_2}}{x_{12}^{2\Delta_1}}\\
    \expval{\mathcal{O}_1(x_1)\mathcal{O}_2(x_2)\mathcal{O}_3(x_3)}&=\frac{f_{123}}{x_{12}^{\Delta_1+\Delta_2-\Delta_3}x_{23}^{\Delta_2+\Delta_3-\Delta_1}x_{31}^{\Delta_3+\Delta_1-\Delta_2}}\ ,\label{e.thirdrelationss}
    \end{align}
where $x_{ij}\equiv x_i-x_j$, $N$ is a normalization constant and the coefficients $f_{ijk}$ are theory-dependent.

Similarly, one can determine the form of the 1, 2 and 3-point functions of spinning operators, in particular the 2-point function of the stress tensor $T_{\mu\nu}$
\begin{equation}\label{e.defifcentcah}
	\langle T_{\mu\nu}(x)T_{\rho\sigma}(0)\rangle = {C_T\over x^{2d}}\Big[{1\over 2}(I_{\mu\rho}I_{\nu\sigma}+{1\over 2}I_{\mu\sigma}I_{\nu\rho})-{1\over d}\delta_{\mu\nu}\delta_{\rho\sigma}\Big]\ ,
\end{equation}
where $I_{\mu\nu}=I_{\mu\nu}(x)=\delta_{\mu\nu}-\frac{2x_\mu x_\nu}{x^2}$ and $C_T$ is the central charge, which is related to the number of degrees of freedom and will play a crucial role in the notion of holographic CFTs.

Another central concept in conformal field theories is the \emph{operator product expansion} (OPE)
    \begin{equation}
    \mathcal{O}_i(x_1)\mathcal{O}_j(x_2)=\sum_{\mathcal{O}_k\in\text{primaries}}\lambda_{ijk}C_{ijk}(x_{12},\partial_2)\mathcal{O}_k(x_2)\ ,
    \end{equation}
where the sum runs over all primary operators, and all spin indices have been suppressed for simplicity. The OPE coefficients $\lambda_{ijk}$ are theory-dependent, while the function $C_{ijk}$ is completely fixed by conformal symmetry.\footnote{Note that by consistency, the OPE coefficients $\lambda_{ijk}$ are, up to a normalization factor, equal to $f_{ijk}$ in \eqref{e.thirdrelationss}.}

Let us now discuss two applications of the OPE: two-point function in a thermal state and the conformal block decomposition of a four-point function.

\subsubsection*{Application 1: Thermal 2-Point Function}

Studying quantum field theory at finite temperature $T$ is equivalent to placing the theory on a spacetime where the time direction is Euclidean and compactified into a circle of circumference $\beta=1/T$. This has important consequences for CFTs -- in particular, \eqref{e.kukurickas} no longer holds, and one can obtain non-trivial one-point functions for symmetric traceless tensors (see e.g.\ \cite{El-Showk:2011yvt,Iliesiu:2018fao})
    \begin{equation}\label{e.zajkos}
    \expval{\mathcal{O}^{\mu_1\ldots\mu_J}(x)}_\beta=\frac{b_{\mathcal{O}}}{\beta^\Delta}(e^{\mu_1}\ldots e^{\mu_J}-\text{traces})\ ,    
    \end{equation}
where $e^\mu$ is the unit vector in the Euclidean time direction and $b_{\mathcal{O}}$ is a theory-dependent coefficient.\footnote{In principle, this coefficient can be expressed in terms of the OPE data $\lambda_{ijk}$.} 

A thermal two-point function can be analysed using the OPE:
    \begin{equation}
    \big<\wick[offset=1.3em]{\mathcal{O}_1\c( x_1)\mathcal{O}_2\c(x_2)}\big>_\beta=\sum_{\OO_k\in\text{primaries}}\lambda_{12k}C_{12k}(x_{12},\partial_2)\expval{\OO_k(x_2)}_\beta\ .
    \end{equation}
For two identical scalar operators $\OO_1=\OO_2=\phi$ inserted at $x$ and 0, one can rewrite the above equation as
    \begin{equation}\label{e.marcipanec}
    \expval{\phi(x)\phi(0)}_\beta=\sum_{\OO\in\phi\times\phi}\lambda_{\phi\phi\OO}\abs{x}^{\Delta-2\Delta_\phi-J}x_{\mu_1}\ldots x_{\mu_J}\expval{\OO^{\mu_1\ldots\mu_J}(0)}_\beta\ ,
    \end{equation}
where $\Delta_\phi$ is the scaling dimension of $\phi$, while $\Delta$ and $J$ are the scaling dimension and spin of the exchanged operator $\OO$. The sum runs over all primary operators appearing in the OPE of two $\phi$ operators. Using Eq.\ \eqref{e.zajkos}, the expression in Eq.\ \eqref{e.marcipanec} can be rewritten as
    \begin{equation}\label{e.thermalblokies}
    \expval{\phi(x)\phi(0)}_\beta=\sum_{\OO\in\phi\times\phi}\lambda_{\phi\phi\OO}b_{\OO}\frac{J!}{2^J\left(\frac{d-2}{2}\right)_J}\text{C}_J^{\left(\frac{d-2}{2}\right)}\left(\frac{\tau}{\abs{x}}\right)\abs{x}^{\Delta-2\Delta_\phi}\ ,   
    \end{equation}
where $(a)_n\equiv\frac{\Gamma(a+n)}{\Gamma(a)}$ is the Pochhammer symbol and $\text{C}^{(a)}_J(y)$ are the Gegenbauer polynomials. Equation \eqref{e.thermalblokies} can be interpreted as an expansion in \emph{thermal conformal blocks}.\footnote{Note that Eq.\ \eqref{e.thermalblokies} is only valid for 2-point functions on $S_\beta^1\times\mathbb{R}^{d-1}$. In more general cases, such as $S_\beta\times S_R^{d-1}$, one can either attempt to generalize Eq.\ \eqref{e.thermalblokies}, or instead use a formalism based on vacuum 4-point functions. This perspective will be discussed in Sec.\ \ref{s2.thrmlztion}.}

\subsubsection*{Application 2: Vacuum 4-Point Functions}

Finally, let us examine the 4-point functions in the vacuum. For simplicity, consider the case where all four operators are identical scalar primaries $\phi$. Unlike the case of 1, 2 and 3-point functions, conformal symmetry can fix the four-point function only up to an arbitrary function:
    \begin{equation}\label{e.moznobudepoterba}
    \expval{\phi(x_1)\phi(x_2)\phi(x_3)\phi(x_3)}=\frac{g(u,v)}{x_{12}^{\Delta_\phi}x_{34}^{\Delta_\phi}}\ ,
    \end{equation}
where $u$ and $v$ are the conformally invariant cross-ratios
    \begin{equation}\label{e.crosratzinikz}
    u=\frac{x_{12}^2x_{34}^2}{x_{13}^2x_{24}^2}\qq{and}v=\frac{x_{23}^2x_{14}^2}{x_{13}^2x_{24}^2}\ .
    \end{equation}
Although the function $g(u,v)$ cannot be determined directly, it can be decomposed in a useful way using the OPE:
    \begin{equation}\label{e.odvodeieCBD}
    \begin{split}
    &\big<\wick[sep=0pt,offset=1.4em]{\c1\phi(x_1)\c1\phi(x_2)\c2\phi(x_3)\c2\phi(x_4)}\big>\\
    &\hspace{1.4cm}=\!\!\!\sum_{\OO,\OO'\in\phi\times\phi}\!\!\lambda_{\phi\phi\OO}\lambda_{\phi\phi\OO'}C_{12a}(x_{12},\partial_2)C_{34b}(x_{34},\partial_4)\big<\OO^a(x_2)\OO^{\prime b}(x_4)\big>\\
    &\hspace{1.4cm}=\!\!\!\sum_{\OO\in\phi\times\phi}\!\lambda^2_{\phi\phi\OO}C_{12a}(x_{12},\partial_2)C_{34b}(x_{34},\partial_4)\frac{I^{ab}(x_{24})}{x_{24}^{2\Delta}}=\frac{1}{x_{12}^{\Delta_\phi}x_{34}^{\Delta_\phi}}\!\sum_{\OO\in\phi\times\phi}\!\lambda_{\phi\phi\OO}^2g_{\Delta,J}(x_i).
    \end{split}
    \end{equation}
Here, we have assumed an orthonormalized set of exchanged operators $\OO$ and used the fact that the two-point function between spinning operators takes the form $\expval{\OO^a(x)\OO^{\prime b}(0)}=\Delta_{\OO\OO'}\frac{I^{ab}(x)}{x^{2\Delta}}$, where $I^{ab}(x)$ is a known kinetic structure completely fixed by conformal symmetry. The functions $g_{\Delta,J}(x_i)$ are called \emph{conformal blocks} and -- as is clear from their construction -- they are theory independent and in principle calculable.

By comparing Eqs.\ \eqref{e.moznobudepoterba} and \eqref{e.odvodeieCBD}, one obtains the decomposition of $g(u,v)$ into the functions $g_{\Delta,J}(x_i)$
    \begin{equation}\label{e.roscbd}
    g(u,v)=\sum_{\OO\in\phi\times\phi}\!\lambda_{\phi\phi\OO}^2g_{\Delta,J}(x_i)\ .    
    \end{equation}
Equation \eqref{e.roscbd} is known as the \emph{conformal block decomposition} and it serves as a key computational tool in many modern approaches to CFT.

In even dimensions one can find explicit expressions for the conformal blocks \cite{Dolan:2000ut,Dolan:2003hv}, for example in $d=4$ one has
    \begin{equation}\label{ecbdinsomeform}
    g_{\Delta,J}(u,v)=\frac{Z\bar{Z}}{Z-\bar{Z}}(k_{\Delta+J}(Z)k_{\Delta-J-2}(\bar{Z})-k_{\Delta-J-2}(Z)k_{\Delta+J}(\bar{Z}))\ ,    
    \end{equation}
where $k_\eta(\xi)=\xi^{\frac{\eta}{2}}\!\!\!\!\phantom{F}_2F_1(\tfrac{\eta}{2},\tfrac{\eta}{2},\eta,\xi)$ and $(Z,\bar{Z})$ are related to the cross ratios by
    \begin{equation}
     u=Z\bar{Z}\qq{and}v=(1-Z)(1-\bar{Z})\ .   
    \end{equation}
Note also that one can repeat the procedure \eqref{e.odvodeieCBD} for spinning operators \cite{Costa:2011mg,Costa:2011dw} -- we will return to this in the context of stress tensors in Chapter \ref{ch4}.

The importance of the conformal block decomposition lies in the fact that from a single 4-point function one can extract a vast amount of CFT data -- equivalent to computing a (infinite) tower of 3-point functions. 

Let us conclude with a final remark. Using the OPE one can systematically reduce any $n$-point function to 1, 2 and 3-point ones. In other words, the whole information contained in a CFT is encoded in the set of data $\{\Delta,\,\,J,\,\,\lambda_{ijk}\}$. The extraction of these coefficients will therefore be our primary goal.

%%%%%%%%%%%%%%%%%%%%%%%%%%%%%%%%%%%%%%%%%%%%%

\section{AdS/CFT}\label{s2.AdSCFT}

We now briefly turn to the gravitational side of the story, where the central role is played by Anti-de Sitter (AdS) spacetime -- the maximally symmetric solution of the Einstein equations in $D=d+1$ dimensions with negative cosmological constant $\Lambda<0$
    \begin{equation}
        \mathcal{R}_{\mu\nu}-\frac{1}{2}g_{\mu\nu}(\mathcal{R}-2\Lambda)=0\ ,
    \end{equation}
where $\mathcal{R}_{\mu\nu}$ is the Ricci tensor and $\mathcal{R}=\mathcal{R}_{\mu\nu}g^{\mu\nu}$ is the Ricci scalar. AdS spacetime can be realized geometrically as a quasi-sphere
    \begin{equation}\label{e.definingeqads}
        -X_0^2-X_{d+1}^2+\sum_{i=1}^dX_i^2=-R^2\ ,
    \end{equation}
embedded in the flat space $\mathbb{R}^{2,d}$, where $R$ is the constant radius of curvature of AdS spacetime, related to the cosmological constant via
    \begin{equation}
    \Lambda=-\frac{1}{2R^2}(D-2)(D-1)=-\frac{d(d-1)}{2R^2}\ .
    \end{equation}

Explicitly, the defining condition \eqref{e.definingeqads} can be solved by setting
    \begin{align}
        X_0&=R\cosh\hat{\rho}\cos\hat{t}\\
        X_{d+1}&=-R\cosh\hat{\rho}\sin\hat{t}\\
        X_{i}&=R\,\hat{x}_i\sinh\hat{\rho}\ ,
    \end{align}
where $i=1,2,\ldots,d$ and $\hat{x}_i$ are the coordinates on the unit $d$-sphere $S^d$, satisfying $\sum_{i=1}^d\hat{x}_i^2=1$. The coordinates $(\hat{\rho},\hat{t},\hat{x}_i)$ are known as the \emph{global coordinates}. It is important to mention that, a-priori, $\hat{t}\in[0,2\pi)$, however, from now on we will consider $\hat{t}\in\mathbb{R}$. In global coordinates, the AdS metric takes the form
    \begin{equation}
    \dd s^2=R^2(-\cosh^2\hat{\rho}\dd\hat{t}\,^2+\dd\hat{\rho}\,^2+\sinh^2\hat{\rho}\sum_{i=1}^d\dd\hat{x}_i^2)\ .
    \end{equation}

Another useful set of coordinates is defined by
    \begin{align}
    X_0&=\frac{1}{2r}(1+r^2(R^2+\vec{x}^2-t^2))\\
    X_i&=Rrx_i\\
    X_d&=\frac{1}{2r}(1-r^2(R^2-\vec{x}^2+t^2))\\
    X_{d+1}&=Rrt\ ,
    \end{align}
where $i=1,2,\ldots,d-1$ and we defined $\vec{x}\equiv(x_1,x_2,\ldots,x_{d-1})$. The coordinates $(r,t,\vec{x})$, known as the \emph{Poincaré coordinates}, will play a central role in most of the analyses in this thesis. In terms of $(r,t,\vec{x})$, the metric of $AdS_{d+1}$ takes the form
    \begin{equation}\label{e.poinckoordinaty}
    \dd s^2=R^2\left(\frac{\dd r^2}{r^2}+r^2(\dd x_\mu\dd x^\mu)\right)\ ,
    \end{equation}
where $\dd x_\mu\dd x^\mu$ is the Minkowski metric in $d$-dimensions. That is, in the Poincaré coordinates, the constant-$r$ slices are isomorphic to $\mathbb{R}^{1,d-1}$ plane, and we will refer to the slice at $r=\infty$ as the \emph{boundary of AdS}.\footnote{Strictly speaking, this is the conformal boundary of the AdS space.} It is important to note that, unlike global coordinates, Poincaré coordinates do not cover the entire AdS spacetime.

Finally, let us mention another useful set of coordinates, obtained by redefining the radial coordinate $r$:\footnote{In the presence of a black hole it is often convenient to redefine the radial coordinate $r$ using the location of the black hole horizon $r_+$ by $r\rightarrow u\equiv\frac{r_+^2}{r^2}$.}
    \begin{equation}
    r\rightarrow z\equiv\frac1r\quad\Longrightarrow\quad\dd s^2=\frac{R^2}{z^2}(\dd z^2+\dd x_\mu\dd x^\mu)\ .
    \end{equation}

    \subsection*{AdS/CFT correspondence}

The isometry group of AdS$_{d+1}$ is $SO(d,2)$, which is precisely the conformal group of a CFT$_d$. In a sense, a CFT in $d$ dimensions and a relativistic theory in (asymptotically) $AdS_{d+1}$ are two different realizations of the same $SO(d,2)$ symmetry. Motivated by this, \emph{define} the holographic\footnote{The word \emph{holographic} refers to the dimensional difference. AdS/CFT is a codimension 1 holographic duality.} correspondence between AdS and CFT \cite{Maldacena:1997re, Gubser:1998bc, Witten:1998qj} as the equivalence of partition functions
    \begin{equation}\label{e.adscftadscft}
    Z_{\rm AdS}\Big|_{\phi\rightarrow\hat{\phi}}\,=\,Z_{\rm CFT}[\,\hat{\phi}\,]\ .
    \end{equation}
Here, $Z_{\rm AdS}\big|_{\phi\rightarrow\hat{\phi}}$ denotes the partition function of a quantum gravity theory in $D=d+1$ dimensional (asymptotically) AdS spacetime, with all the fields $\phi$ in the theory taking the boundary values $\hat{\phi}$. By $Z_{\rm CFT}[\,\hat{\phi}\,]$ we mean the partition function of a CFT in $d$-dimensions with fields $\mathcal{O}$ sourced by $\hat{\phi}$, $i.e.$
    \begin{equation}\label{e.listttttt}
    Z_{\rm CFT}[\,\hat{\phi}\,]\equiv\expval{e^{\int\hat{\phi}\mathcal{O}}}_{\rm CFT}=e^{W[\,\hat{\phi}\,]}\ ,
    \end{equation}
where $W[\,\hat{\phi}\,]$ is the generating functional for connected correlators. Conceptually, one can view the AdS side as the \emph{bulk}, with the CFT living on its \emph{boundary}. 

Note that \eqref{e.adscftadscft} should be treated as a definition: fixing a theory on one side, Eq.\ \eqref{e.adscftadscft} defines the theory on the other side.\footnote{Relation \eqref{e.adscftadscft} is natural for a number of reasons. Besides the equivalence of the symmetries, it introduces key ingredients, such as the notion of a stress-tensor on the CFT side. This is because a stress tensor is (by definition) sourced by the boundary metric $\hat{g}_{\mu\nu}$ -- the boundary value of the bulk metric.} For example, given $Z_{\rm AdS}\big|_{\phi\rightarrow\hat{\phi}}$, one can compute any correlation function in the dual CFT via
    \begin{equation}
    \expval{\OO(x_1)\ldots\OO(x_n)}\propto\eval{\frac{\delta^nZ_{\rm CFT}[\,\hat{\phi}\,]}{\delta\hat{\phi}(x_1)\ldots\delta\hat{\phi}(x_n)}}_{\hat{\phi}=0}=\eval{\frac{\delta^nZ_{\rm AdS}\big|_{\phi\rightarrow\hat{\phi}}}{\delta\hat{\phi}(x_1)\ldots\delta\hat{\phi}(x_n)}}_{\hat{\phi}=0}.
    \end{equation}

Importantly, in practice one typically does not have access to the full quantum gravity path integral, but only to its semi-classical limit, where the bulk partition function is dominated by classical saddle points
    \begin{equation}\label{e.serdieckoo}
    Z_{\rm AdS}\Big|_{\phi\rightarrow\hat{\phi}}\approx e^{-S_{\rm AdS}(\phi\rightarrow\hat{\phi})}\ .
    \end{equation}
Here, $S_{\rm AdS}(\phi\rightarrow\hat{\phi})$ denotes the on-shell action evaluated on a solution $\phi$ of the equations of motion, satisfying the boundary condition $\phi\rightarrow\hat{\phi}$ as the boundary is approached.\footnote{Typically, the solution $\phi$ diverges near the boundary ($r\rightarrow\infty$). Thus, the condition $\phi\rightarrow\hat{\phi}$ should be understood as specifying the leading asymptotic behaviour, $e.g.$ $\phi(r,x_\mu)=r^{\Delta-4}\hat{\phi}(x_\mu)$, as we approach the conformal boundary.} Since the equations of motion are of second order, we need another boundary condition to have a uniquely determined solution. Path integrals are typically treated in the Eucludean signature, where the natural second condition is regularity in the bulk.

Combining Eqs.\ \eqref{e.listttttt} and \eqref{e.serdieckoo} we get the semi-classical AdS/CFT prescription
    \begin{equation}
    S_{\rm AdS}(\phi\rightarrow\hat{\phi})=-W[\,\hat{\phi}\,]\ .
    \end{equation}
In practice, if one is able to solve the equations of motion in the bulk, then any $n$-point function in the dual CFT is obtained by
    \begin{equation}
    \expval{\OO(x_1)\ldots\OO(x_n)}_c\propto\eval{\frac{\delta^nS_{\rm AdS}(\phi\rightarrow\hat{\phi})}{\delta\hat{\phi}^n}}_{\hat{\phi}=0}\ .
    \end{equation}

Let us also mention that by studying field theories on an (asymptotically) AdS background, one discovers a ``dictionary'' of relations connecting parameters on both sides of the AdS/CFT correspondence. For example, studying a minimally coupled scalar on an AdS background one finds that the scaling dimension of the dual operator $\OO(x_\mu)$ is related to the mass of the bulk scalar field $\phi(r,x_\mu)$ by
    \begin{equation}
    m^2=\Delta(\Delta-d)\ .
    \end{equation}

As we will show below, having a weakly coupled bulk theory leads to a strongly coupled CFT and vice versa, $i.e.$ AdS/CFT is a strong-weak duality. 

Another important aspect of the AdS/CFT dictionary is that
    \begin{equation}
    \qq{states on the boundary}\Longleftrightarrow\qq{geometries in the bulk}
    \end{equation}
In particular, studying the vacuum state in CFT corresponds to studying bulk theory on a pure AdS background. In other words, the bulk metric takes the form \eqref{e.poinckoordinaty} with small fluctuations $h_{\mu\nu}(r,x_\mu)$ around it.\footnote{Fluctuations $h_{\mu\nu}$ are universally present on the gravity side of the duality. Note that after quantization, they correspond to the gravitons.} In this thesis we want to study thermal states. As shown by the analysis of gravitational saddle points \cite{Hawking:1982dh,Witten:1998zw}, thermal states are dual to AdS black holes.\footnote{More precisely, there are two gravitational saddles that can contribute: the AdS black hole and the so-called thermal AdS. These correspond to two distinct thermodynamic phases, separated by the Hawking-Page phase transition \cite{Hawking:1982dh}, which occurs at a specific critical temperature. Throughout this thesis we assume the system is in the high-temperature phase -- far above the Hawking-Page transition -- where the dominant saddle is the AdS-Schwarzschild black hole.} This implies that, in order to study thermal correlators in a CFT, one must analyse the corresponding black hole geometry in the bulk.

\subsection*{Holographic CFTs}

Let us now briefly discuss the most famous example of the holographic duality \cite{Maldacena:1997re}:
    \begin{equation}
    \text{IIB Strings on}\,\,AdS_5\times S^5\quad\,\,\Longleftrightarrow\quad\,\,\mathcal{N}=4\,\,\,\text{SYM in}\,\,\,d=4
    \end{equation}
which has been studied and understood in great detail.\footnote{Note that in this concrete example we specify theories on both sides of the duality. The statement that one is the holographic dual of the other, in the sense of Eq.\ \eqref{e.adscftadscft}, is a non-trivial claim that one has to \emph{prove}.}

On the gravity side there are three adjustable scales: Planck scale $l_P$, string scale $l_s$ and AdS scale $l_{AdS}$,\footnote{For notational consistency we rename here the AdS scale $R$ to $l_{AdS}$.} which can be combined into two independent dimensionless ratios:
    \begin{equation}
    \frac{l_s}{l_{AdS}}\,\,\qq{and}\,\,\frac{l_P}{l_{AdS}}\ .    
    \end{equation}
The ratio $\frac{l_P}{l_{AdS}}$ controls quantum (loop) effects, $i.e.$ taking $\frac{l_P}{l_{AdS}}\ll1$ we get a semi-classical gravity. The ratio $\frac{l_s}{l_{AdS}}$ describes the stringy-corrections; more precisely, the effective theory takes the form
    \begin{equation}
    S_{\rm IIB}\propto\frac{1}{G_N}\int\sqrt{g}\,(\mathcal{R}+L_{\rm matter}+l_s^4\,\mathcal{R}^4+\ldots\,)\ .
    \end{equation}
The stringy states have masses of order $\mathcal{O}(1/l_s^2)$, so at energies below $\frac{1}{l_s^2}$ we only recover the classical Einstein gravity with coupled matter.

On the CFT side we have $\mathcal{N}=4$ supersymmetric Yang–Mills (SYM) theory in four dimensions with the gauge group $SU(N)$. Importantly, there are two dimensionless parameters in the theory: $N$, which roughly measures the number of degrees of freedom, and the Yang–Mills coupling constant $g_{\rm YM}$. It turns out it is useful to define their combination
    \begin{equation}
    \lambda\equiv g_{\rm YM}^2N\ ,
    \end{equation}
which is called 't Hooft coupling, and to describe this theory via $(\lambda,\frac{1}{N})$.

The holographic dictionary relates these dimensionless parameters by
    \begin{align}
    \lambda&\sim\left(\frac{l_{AdS}}{l_s}\right)^4\\
    N^2&\sim\left(\frac{l_{AdS}}{l_P}\right)^{d-1}\sim\frac{l_{AdS}^{d-1}}{G_N}\ .
    \end{align}
A crucial observation is that, in order to obtain semi-classical Einstein gravity in the bulk, both $\frac{l_s}{l_{AdS}}$ and $\frac{l_P}{l_{AdS}}$ must be small. This however corresponds to a regime $N\gg1$ and $\lambda\gg1$, $i.e.$ the dual CFT is strongly coupled and contains a large number of degrees of freedom. On the other hand, having a weakly-coupled theory on the CFT side, one has $\frac{l_s}{l_{AdS}}\gg1$, in which case the bulk theory departs significantly from the ordinary Einstein gravity. This illustrates that AdS/CFT is a \emph{strong/weak duality} and thus offers a potential window into physics not visible by standard perturbative methods.

Let us now discuss other examples of the holographic duality. It is important to note that, when one fixes a generic theory on one side and uses Eq.\ \eqref{e.adscftadscft} to define the dual, the resulting theory may be non-physical -- potentially violating unitarity, causality, or locality. It is therefore crucial to identify a class of theories for which both sides of the correspondence are ``well-behaved''. More precisely, we want to develop a notion of \emph{holographic CFTs} -- conformal field theories that admit weakly-coupled, semi-classical Einstein gravity duals.

In recent years there has been active discussion on the minimal set of conditions that define holographic CFTs (see $e.g.$ \cite{Heemskerk:2009pn,Camanho:2014apa}); the two crucial requirements are
    \begin{itemize}
        \item \textbf{$C_T\gg1$}\,:\,\, As in the case of $\mathcal{N}=4$ SYM, to have a semi-classical gravity dual we need a large number of degrees of freedom. In general CFTs, this is captured by the central charge $C_T$, defined in Eq.\ \eqref{e.defifcentcah}. Note that $\frac{1}{G_N}\sim N^2\sim C_T$.
        \item{$\Delta_{gap}\gg1$}\,:\,\, The dimension $\Delta_{gap}$ of the lightest (single-trace) primary operator with spin greater than 2 should be large. In the IIB Strings/$\mathcal{N}=4$ SYM example, this corresponds to a large gap between low-energy fields and the first stringy excitations. This condition is often associated with the CFT being strongly coupled.%This condition is often stated as a requirement to have a strongly coupled CFT.
    \end{itemize}
These conditions also drastically simplify the spectrum of the CFT, which is often well under control. OPE coefficients and scaling dimensions are naturally organized into expansions in $1/C_T$, and one can follow ``large-$C_T$ counting rules'' to determine the leading behaviour. For example, the leading contribution to a connected $n$-point function in a holographic CFT is of order $C_T^{-\frac{n-2}{2}}$.

Overall, holographic CFTs form a well-behaved class of theories that provide a valuable framework for studying both sides of the AdS/CFT correspondence.

%%%%%%%%%%%%%%%%%%%%%%%%%%%%%%%%%%%%%%%%%%%%%

\section{Near-Boundary Expansion}\label{s2.nbe}

In this thesis we focus on holographic CFTs with gravitational duals of the following schematic form:
    \begin{align}
    S_{\rm AdS}&\propto\int\dd^{d+1}x\sqrt{g}(\mathcal{L}_{\phi}+\mathcal{L}_{\rm grav})+S_{\rm bdry}\label{e.asinejakesetupek1}\\
    \mathcal{L}_\phi=\frac12(\partial&\phi)^2+\frac12m^2\phi^2\qq{and}\mathcal{L}_{\rm grav}=\mathcal{R}-2\Lambda+\ldots\label{e.asinejakesetupek2}
    \end{align}
where $S_{\rm bdry}$ is the boundary term necessary for a well-defined variational problem, the ellipsis indicates potential higher-curvature corrections, and we work in Euclidean signature. More concretely, in Chapters \ref{ch4}-\ref{ch6} we set $\mathcal{L}_\phi=0$ and study ``pure gravity models'' in Einstein and Gauss-Bonnet gravity,\footnote{Pure gravity models are conceptually simpler; however, one must deal with spinning correlators, since the only degrees of freedom are metric perturbations. This introduces several technical subtleties, which we study in detail in Chapters \ref{ch4}-\ref{ch6}.} while in this Chapter and Chapter \ref{ch3} we focus on a minimally coupled scalar and neglect higher-curvature corrections.

In these models, the spectrum of primary operators on the CFT side consists of the identity operator $\mathds{1}$, a scalar operator $\OO(x_\mu)$ dual to the bulk scalar field $\phi(r,x_\mu)$ and the stress tensor $T_{\mu\nu}(x_\mu)$, which is universally present in any CFT and is dual to the metric fluctuations in the bulk. In addition, there are infinitely many other primary operators composed of multiple $\OO(x_\mu)$ and $T_{\mu\nu}(x_\mu)$ -- the so called multi-trace operators.

Our goal is to investigate the thermal two-point function $\expval{\OO(x)\OO(0)}_\beta$ in these theories. For this correlation function there are only two relevant families of operators that can contribute:

\paragraph{Multi-Stress Tensors:} These are primary operators composed of $n$ stress tensors with any even number of derivatives and any number of index contractions. We denote them as $[T^n]_{\Delta,J}$, with $J=0,2,\ldots$ and $\Delta=nd+m$, where $m=0,2,\ldots$ corresponds to the number of derivatives. For $n>1$, the scaling dimensions $\Delta$ receive corrections -- called the anomalous dimensions -- that are suppressed in the large-$C_T$ limit. Nevertheless, these corrections will play a crucial role in Chapters \ref{ch4} and \ref{ch5}. By definition $T_{\mu\nu}\equiv[T^1]_{d,2}$ and the identity operator is included as $\mathds{1}\equiv[T^0]_{0,0}$. In the large volume limit, $i.e.$ $S_\beta^1\times S_R^{d-1}\rightarrow S_\beta^1\times\mathbb{R}^{d-1}$, the $n$-stress tensors with derivatives, as well as all the descendants, are suppressed by powers of $1/R$ \cite{Karlsson:2021duj}. Thus the only relevant operators are the multi-stress tensors
    \begin{equation}\label{e.stsrrr}
    [T^n]_J\equiv[T^n]_{dn,J}\qq{where}\Delta=dn\qq{and}J=0,2,4,\ldots,2n\ .
    \end{equation}
We refer to the set of all multi-stress tensors and their descendants as the \emph{stress-tensor sector}.

\paragraph{Double-Trace Scalars:} Another sector contributing to the correlator are the double-trace scalars defined by
    \begin{equation}
    [\OO\OO]_{n,l}\equiv\OO\partial_{\mu_1\ldots\mu_l}\partial^{2n}\OO\qq{where}\Delta=2\Delta_{\OO}+2n+l\qq{and}J=l\in2\mathbb{N}_0\ . 
    \end{equation}
We refer to the set of all double-trace scalars and their descendants as the \emph{double-trace sector}. Comparing the scaling dimensions of $[T^n]_J$ and $[\OO\OO]_{n,l}$, it is clear that for $\Delta_{\OO}\notin\mathbb{Z}$, the two sectors decouple, $i.e.$ their quantum numbers never coincide. The special case of integer $\Delta_{\OO}$ will be discussed in Appendix \ref{AppendixA}.

Knowledge of the relevant CFT spectrum can serve as a guiding principle in developing an effective bulk ansatz for computing $\expval{\OO(x)\OO(0)}_\beta$ from gravity \cite{Fitzpatrick:2019zqz}, an approach we will use in various forms throughout this thesis. We begin with the general setup \eqref{e.asinejakesetupek1} and \eqref{e.asinejakesetupek2} in $d\in2\mathbb{N}$, assuming a planar AdS black hole ($i.e.$, a black brane),\footnote{Throughout this thesis we restrict our attention to the large-volume limit where the AdS scale satisfies $R\gg\beta$.}
    \begin{equation}\label{e.genbhlkafp}
    \dd s^2=r^2f(r)\dd\tau^2+\frac{\dd r^2}{r^2f(r)}+r^2\dd\vec{x}\,^2\ ,   
    \end{equation}
where we set the AdS scale to 1 for simplicity. The function $f(r)$ depends on the specific form of $\mathcal{L}_{\rm grav}$ in \eqref{e.asinejakesetupek2}; for example, in Einstein gravity, $f(r)=1-\frac{\mu}{r^d}$, where $\mu$ is a parameter related to the black hole mass.

To compute the boundary correlator $\expval{\OO(x)\OO(0)}_\beta$, we have to solve the equation of motion for the scalar field in the black hole background \eqref{e.genbhlkafp}. In practice it is useful to formulate this problem in terms of the \emph{bulk-to-boundary propagator}, defined as
    \begin{equation}
    \Phi(r,x_1-x_2)\equiv\expval{\phi(r,x_1)\mathcal{O}(x_2)}\ .
    \end{equation}
Setting $x_1=(\tau,\vec{x})$ and $x_2=0$, $\Phi(r,\tau,\vec{x})$ satisfies the Klein–Gordon equation in the background \eqref{e.genbhlkafp}:
    \begin{equation}\label{e.whatweneedtosolverly}
    (-\Box+m^2)\Phi(r,\tau,\vec{x})=0
    \end{equation}
subject to the boundary condition
    \begin{equation}
    \lim_{r\rightarrow\infty}\Phi(r,\tau,\vec{x})\propto r^{\Delta_{\OO}-d}\delta(\tau)\delta^{(d-1)}(\vec{x})
    \end{equation}
along with regularity in the bulk interior.

In this framework, the thermal boundary two-point function is given by:
    \begin{equation}
    \expval{\OO(\tau,\vec{x})\OO(0)}_\beta=\lim_{r\rightarrow\infty}r^{\Delta_{\OO}}\Phi(r,\tau,\vec{x})\ .
    \end{equation}
Note that a similar analysis, although technically more involved, can also be carried out for spinning operators. We will discuss this explicitly for the case of the stress-tensor two-point function in Chapters \ref{ch4}-\ref{ch6}.

In summary, our task is to solve Eq.\ \eqref{e.whatweneedtosolverly} in the black hole background \eqref{e.genbhlkafp}. For this we use the near-boundary ansatz developed by Fitzpatrick and Huang in \cite{Fitzpatrick:2019zqz}. The logic behind it can be summarized in the following steps:
\begin{enumerate}
    \item The structure of the CFT should be reflected in the structure of $\Phi$; in particular, the solution should decompose as
        \begin{equation}
        \Phi(r,\tau,\vec{x})=\Phi_T(r,\tau,\vec{x})+\Phi_{[\OO\OO]}(r,\tau,\vec{x})\ ,
        \end{equation}
    where $\Phi_T(r,\tau,\vec{x})$ captures the contribution from the stress-tensor sector ($i.e.$  the exchange of $[T^n]_J$ operators in the dual theory), while $\Phi_{[\OO\OO]}(r,\tau,\vec{x})$ encodes the double-trace sector ($i.e.$ the exchange of $[\OO\OO]_{n,l}$ operators). When $\Delta_{\OO}\notin\mathbb{Z}$, the two contributions are completely decoupled and can be studied independently. Let us therefore only focus on the stress-tensor part $\Phi_T(r,\tau,\vec{x})$.
    \item In the small temperature limit, the dominant contribution to $\Phi_T(r,\tau,\vec{x})$ arises from the exchange of the identity operator in the dual CFT. In the bulk this corresponds to the vacuum bulk-to-boundary propagator, which is known explicitly \cite{Witten:1998qj}
        \begin{equation}
        \Phi_{AdS}(r,\tau,\vec{x})=\left[\frac{r}{1+r^2(\tau^2+\vec{x}\,^2)}\right]^{\Delta_{\OO}}.
        \end{equation}
    We therefore write the finite-temperature result as:
        \begin{equation}\label{e.step1}
        \Phi_T(r,\tau,\vec{x})=\Phi_{AdS}(r,\tau,\vec{x})\psi(r,\tau,\vec{x})\ .
        \end{equation}
    \item We are interested in the OPE regime on the boundary, $i.e.$ the regime where the two scalar operators are close to each other compared to the other scales in the theory. In the bulk, the physical intuition is that the closer the operators are to each other, the less they should depend on the events in the bulk interior. In other words, it seems natural to expect that the OPE expansion on the boundary is dual to a $1/r$ expansion in the bulk. We are thus led to assume:
        \begin{equation}\label{e.step2}
        \psi(r,\tau,\vec{x})=\sum_{n=0}^\infty\frac{\psi_n(\tau,\vec{x})}{r^{dn}}\ .
        \end{equation}
    Here, symmetry considerations -- specifically the form of the function $f(r)$ -- suggest that only terms of the form $r^{-dn}$ appear in the expansion.
    \item In the case of AdS$_3$/CFT$_2$, where the geometry is described by the BTZ black hole, the bulk-to-boundary propagator is known in closed form \cite{Keski-Vakkuri:1998gmz}. Studying its structure provides insight into the general form of the solution in higher dimensions. In particular, the bulk-to-boundary propagator is naturally expressed in a specific set of coordinates $(r,w,\rho)$ defined by
        \begin{equation}
        \rho^2=r^2\vec{x}\,^2\qq{and}w^2=1+r^2(\tau^2+\vec{x}\,^2)\ ,
        \end{equation}
    and one finds that $\psi_n(w,\rho)$ is a polynomial in $\rho^2$.
    \item Consistency with the thermal block expansion \eqref{e.thermalblokies} ensures that the sum over $\rho^{2m}$ in $\psi_n(w,\rho)$ truncates at finite order for any $n$. One finds
        \begin{equation}\label{e.step3}
        \psi_n(w,\rho)=\sum_{m=0}^n\psi_{nm}(w)\rho^{2m}\ .
        \end{equation}
    \item Finally, the functions $\psi_{nm}(w)$ are determined by substituting Eqs.\ \eqref{e.step1}-\eqref{e.step3} into the equation of motion \eqref{e.whatweneedtosolverly}, expressed in coordinates $r$, $w$ and $\rho$ and requiring the correlator to be ``well-behaved for any $r$''. More concretely, any terms in $\psi_{nm}(w)$ that would cause divergences at finite $r$, or in the $r\rightarrow\infty$ limit (aside from the expected $\delta$-function), must be discarded. This procedure uniquely fixes $\psi_{nm}(w)$ as finite-order polynomials in $w^2$.
\end{enumerate}

Combining steps 1-6, we get the complete ansatz for the stress-tensor sector of the correlator:
    \begin{equation}\label{e.thegeneralansatzdominguoile}
    \Phi_T(r,w,\rho)=\left(\frac{r}{w^2}\right)^{\Delta_{\OO}}\left(1+\sum_{n=1}^\infty\sum_{m=0}^n\sum_{k=k_{\rm min}}^{k_{\rm max}}a_{mk}^n\frac{\rho^{2m}w^{2k}}{r^{dn}}\right)\ ,
    \end{equation}
where, in the case of Einstein gravity, one finds:
    \begin{equation}
    k_{\rm min}=-n\qq{and}k_{\rm max}=\frac{dn}{2}-m\ .
    \end{equation}
This ansatz is highly effective for computing $\Phi_T$ to high orders in $n$, and thus provides a straightforward way to study the stress-tensor sector of the correlator.

Several comments are in order. Note that the ansatz \eqref{e.thegeneralansatzdominguoile} can be generalized to the case of spherical black holes \cite{Fitzpatrick:2019zqz}, integer scaling dimensions $\Delta_{\OO}$, or for external operators with spin. Importantly, note that one might follow the same procedure for the case of the double-trace scalars, $i.e.$ to develop an ansatz for $\Phi_{[\OO\OO]}$. However, unlike in the stress-tensor case, step 6 alone is not enough to fully determine the $w$-dependence of $\Phi_{[\OO\OO]}$. The fact that the near-boundary ansatz can \emph{not} fully fix the whole solution $\Phi(t,\tau,\vec{x})$ should not be surprising, since only one boundary condition has been imposed. In order to fully determine the bulk-to-boundary propagator, it is necessary to impose (regularity) condition deep in the bulk interior which our expansion is insensitive to. Surprisingly, despite this limitation, the stress-tensor sector can still be recovered—reflecting the universal nature of the stress tensor and its composites.

We conclude by taking the boundary limit of $\Phi_T$, which yields the stress-tensor part of $\expval{\OO(x)\OO(0)}_\beta$
    \begin{equation}\label{e.reftothisalot}
    G_T(\tau,\vec{x})\equiv\expval{\OO(x)\OO(0)}_\beta\Big|_{\{[T^n]_J\}}\!\!=\lim_{r\rightarrow\infty}r^{\Delta_{\OO}}\Phi_T(r,w=\sqrt{1+r^2(\tau^2+\vec{x}^2)},\rho=r\abs{\vec{x}})\ .
    \end{equation}
In the next section we decompose this result and extract the corresponding CFT data.

%%%%%%%%%%%%%%%%%%%%%%%%%%%%%%%%%%%%%%%%%%%%%

\section{Thermalization}\label{s2.thrmlztion}

Our aim is to study the physics of holographic CFTs, which is encoded in the spectrum, the OPE coefficients $\lambda_{\mathcal{O}\mathcal{O}'\mathcal{O}''}$ and the thermal one-point coefficients $b_{\OO}$. We therefore take the holographically computed result \eqref{e.reftothisalot}, decompose it into its fundamental constituents and extract the relevant dynamical data. More concretely, the decomposition is given in terms of thermal conformal blocks \eqref{e.thermalblokies}, corresponding to the exchange of the operators $[T^n]_J$ \eqref{e.stsrrr}.

From the construction of the ansatz \eqref{e.thegeneralansatzdominguoile}, the term of order $1/r^{nd}$ in the bulk is mapped to the thermal blocks corresponding to the exchange of $n$-stress tensors. Note that for $n>1$ there are $n+1$ distinct $n$-stress tensors, distinguished by their spin. The CFT data we get from this decomposition are combinations of the OPE coefficients $\lambda_{\OO\OO[T^n]_J}$ and the thermal one-point coefficients $b_{[T^n]_J}$.

Before writing any explicit results, let us introduce an alternative perspective. Instead of decomposing in thermal conformal blocks \eqref{e.thermalblokies}, we may use a more robust framework of vacuum four-point conformal blocks \eqref{e.odvodeieCBD}. This approach is equivalent because, at the level of individual blocks, a thermal block associated with an operator $\OO'$ is -- up to a constant prefactor -- identical to a (T-channel) flat-space four-point conformal block for the same exchanged operator $\OO'$
    \begin{equation}\label{e.magicrelation}
    \expval{\OO(x)\OO(0)}_\beta\bigg|_{\OO'}\,\,\propto\,\,
    \begin{tikzpicture}[baseline=(current bounding box.center), scale=0.65]
  % Nodes
  \node (Otop) at (0,1) {$\mathcal{O}$};
  \node (Obot) at (0,-1) {$\mathcal{O}$};
  \node (Btop) at (4,1) {$\Psi$};
  \node (Bbot) at (4,-1) {$\Psi$};

  % Middle node position (just for lines)
  \coordinate (left) at (1.2,0);
  \coordinate (right) at (2.8,0);
  \node at (2,0.5) {$\mathcal{O}'$};

  % Lines/arrows
  \draw[-] (Otop) -- (left);
  \draw[-] (Obot) -- (left);
  \draw[-] (right) -- (Btop);
  \draw[-] (right) -- (Bbot);
  \draw (left) -- (right); % horizontal line, no arrow

\end{tikzpicture}
    \end{equation}
where $\Psi$ denotes a scalar operator. This equivalence follows from the fact that (T-channel) conformal blocks, in the large-volume limit, reduce to Gegenbauer polynomials \cite{Karlsson:2021duj}.

One can thus use the language of four-point conformal blocks in vacuum to derive the kinematics of the individual contributions in the decomposition of $\expval{\OO(x)\OO(0)}_\beta$. Although this may appear as an unnecessary complication -- ultimately leading to equivalent kinematics -- we will adopt this description throughout the thesis, as it provides a more robust and flexible framework that can be easily generalized to more complicated settings ($e.g.$ finite $R$).

Note that in some cases, one can make an even stronger claim than \eqref{e.magicrelation}: instead of block-by-block equivalence, there may be complete equivalence between the correlators of a subset $\{\tilde{\OO}_i\}$ of operators in the thermal state and those in a scalar state $\ket{\Psi}$:
    \begin{equation}
    \expval{\tilde{\OO}_i}{\Psi}=\big<\tilde{\OO}_i\big>_\beta\qq{for}\forall i\ .
    \end{equation}
In such cases, we say that the operators $\{\tilde{\OO}_i\}$ \emph{thermalize} in the state $\ket{\Psi}$.

A concrete example is the thermalization of the stress-tensor sector $\{[T^n]_J,n=0,1,2,\ldots\}$ in heavy states $\ket{\Psi}=\ket{\OO_H}$ with scaling dimension $\Delta_H\sim C_T$ \cite{Karlsson:2021duj}, $i.e.$
    \begin{equation}\label{e.gensttrmlztion}
    \expval{[T^n]_J}{\OO_H}=\expval{[T^n]_J}_\beta\ ,
    \end{equation}
which holds to leading order in $C_T^{-1}$. Note that the equation for $n=1$:\footnote{This can be done without loss of generality. The non-trivial content of thermalization lies in the claim that this relation extends consistently to all $n$ and $J$ in \eqref{e.gensttrmlztion}.}
    \begin{equation}
    \expval{T_{\mu\nu}}{\OO_H}=\expval{T_{\mu\nu}}_\beta
    \end{equation}
defines the relation between the thermal one-point coefficient $b_{T_{\mu\nu}}$ and the OPE coefficient $\lambda_{HHT}$, which is fixed by a Ward identity as a function of $\Delta_H$ and the spacetime dimension $d$. Focusing on $d=4$, this relation can be conveniently written as
    \begin{equation}
    \frac{b_{T_{\mu\nu}}}{\beta^4}=-\frac{\mu C_T S_4}{40}\ ,
    \end{equation}
where $S_d=\frac{2\pi^{d/2}}{\Gamma(d/2)}$ and we have introduced a parameter $\mu\propto \Delta_H/C_T$ that is mapped to the black hole mass parameter $\mu$ in the bulk. More details on thermalization will be provided in Chapter \ref{ch4}.

To summarize, we can use four-point conformal blocks to decompose the holographically computed correlator \eqref{e.reftothisalot} and instead of the thermal data, express everything in terms of the parameter $\mu$. Indeed, from the bulk perspective, the data is naturally expressed via $\mu$; each term at order $1/r^{nd}$ is proportional to $\mu^n$. We therefore use $\mu$ as a counting parameter: it determines the order of the $1/r$ expansion in the bulk and counts the number of
stress tensors that are being exchanged in the conformal block decomposition on the boundary.

Explicitly, decomposing $G_T(\tau,\vec{x})$ \eqref{e.reftothisalot} into T-channel conformal blocks \eqref{ecbdinsomeform} corresponding to the $[T^n]_J$ exchanges,\footnote{More details on the conformal blocks can be found in Appendix \ref{a.cbd}.}
    \begin{align}
    G_T(\tau&,\vec{x})=\frac{1}{(Z\bar{Z})^\Delta}\sum_{\Delta',J'}c_{\Delta',J'}g_{\Delta',J'}(Z,\bar{Z})\\
    \qq{where}&Z\stackrel{R\rightarrow\infty}{=}\tau+i\abs{\vec{x}}\qq{and}\bar{Z}\stackrel{R\rightarrow\infty}{=}\tau-i\abs{\vec{x}}\ ,
    \end{align}
one finds the dynamical data $c_{\Delta',J'}$ as
{\allowdisplaybreaks{\small{
    \begin{align}
    T_{\mu\nu}&:\quad\,\,c_{4,2}=\frac{\Delta}{120}\mu\label{e.dalekyskokik}\\
    T_{\mu\nu}T^{\mu\nu}&:\quad\,\,c_{8,0}=\frac{\Delta  \left(\frac{\Delta ^4}{28800}-\frac{\Delta ^3}{4480}+\frac{\Delta
   ^2}{2016}-\frac{\Delta }{2520}+\frac{1}{4200}\right)}{(\Delta -4) (\Delta -3)
   (\Delta -2)}\mu^2\label{eaestedakn}\\
    T_{\mu\alpha}T^{\alpha}_\nu&:\quad\,\,c_{8,2}=\frac{\Delta  \left(\frac{\Delta ^3}{28800}-\frac{23 \Delta ^2}{201600}+\frac{11 \Delta
   }{100800}+\frac{1}{16800}\right)}{(\Delta -3) (\Delta -2)}\mu ^2\\
    T_{\mu\nu}T_{\rho\lambda}&:\quad\,\,c_{8,4}=\frac{\Delta  \left(\frac{\Delta ^2}{28800}+\frac{\Delta }{33600}+\frac{1}{50400}\right)}{\Delta -2}\mu ^2\\
    \vdots\hspace{0.5cm}&\hspace{1.62cm}\vdots\nonumber
    \end{align}}}}%
where $\Delta$ is the scaling dimension of the external scalar operator $\OO$.

The resulting coefficients agree with all available results obtained through other methods -- for example, those derived from bootstrap and related techniques
\cite{Karlsson:2019dbd,Li:2019zba,Li:2020dqm,Karlsson:2020ghx,Dodelson:2022eiz},
as well as with coefficients obtained by solving the Fourier-transformed equation order by order in the OPE
\cite{Dodelson:2022yvn,Dodelson:2023vrw,Esper:2023jeq,Parisini:2023nbd}\footnote{See e.g. \cite{Katz:2014rla,Manenti:2019wxs} for additional examples of the OPE analysis of finite-temperature holographic correlators.}.
Other non-trivial checks include comparison with the Regge limit holographic data
\cite{Kulaxizi:2019tkd,Karlsson:2019qfi,Karlsson:2019txu,Parnachev:2020zbr}
and with geodesic calculations in asymptotically AdS spacetimes \cite{Parnachev:2020fna,Rodriguez-Gomez:2021pfh}.

To conclude, we now have all the technical apparatus necessary for the calculations presented in the main body of this thesis. In the next four chapters we will apply these tools to explore various interesting physical questions.

%% file: 3_chapter/chapter_3.tex
\chapter{Black Hole Singularity from OPE}\label{ch3}

This chapter is devoted to the study of scalar thermal correlators and their sensitivity to physics beyond the black hole horizon. It has long been known that black hole singularities leave certain signatures in boundary thermal two-point functions, associated with null bulk geodesics bouncing off the singularity (bouncing geodesics) \cite{Fidkowski:2003nf,Festuccia:2005pi}. In this chapter, we clarify how black hole singularities manifest in the dual CFT. We decompose the boundary correlator using the OPE, focusing on contributions from the identity, the stress tensor, and its composites. We show that this part of the correlator develops singularities precisely at the points connected by bulk bouncing geodesics. Black hole singularities are thus encoded in the analytic structure of boundary correlators determined by multiple stress-tensor exchanges. Furthermore, we show that in the large-conformal-dimension limit, the sum of multi-stress-tensor contributions develops a branch-point singularity, as predicted by the geodesic analysis. Finally, we argue that the appearance of complexified geodesics, is tied to the contributions of double-trace operators in the boundary CFT.

\section{Motivation}
\label{sec.introduction}
 
{The AdS/CFT duality}~\cite{Maldacena:1997re, Gubser:1998bc, Witten:1998qj} provides a powerful laboratory for  understanding quantum gravity in the bulk and conformal field theory (CFT) on the boundary.
{In particular,} it relates  black holes in AdS$_{d+1}$ to the boundary CFT$_d$ at finite 
temperature~\cite{Witten:1998zw,Mal01}.  {CFT observables} can hence be used to probe the
interior of black holes and possibly the black hole singularity, see e.g.~\cite{Louko:2000tp,Kraus:2002iv,Fidkowski:2003nf,Festuccia:2005pi,Festuccia:2006sa,Amado:2008hw,Hartman:2013qma,LiuSuh13a,LiuSuh13b,Grinberg:2020fdj,Rodriguez-Gomez:2021pfh,Leutheusser:2021frk,deBoer:2022zps,David:2022nfn,Horowitz:2023ury,Parisini:2023nbd,Dodelson:2023nnr}. 

In~\cite{Fidkowski:2003nf,Festuccia:2005pi}, for $d \geq 3$, signatures of black hole singularities were identified in the boundary thermal two-point functions\footnote{For simplicity, the spatial separation of the operators has been set to zero.}
\be \label{Tcor}
G (t) = \tvev{\phi(t,\vec 0)\phi(0,\vec 0)}_\beta
\ee
where  $\phi$ is a scalar CFT operator 
of conformal dimension $\Delta$ and
$\beta$ is the inverse temperature.
In the limit of large $\Delta$,  $G(t)$ can be computed using 
bulk geodesics connecting the two boundary points where the operators are inserted.
In~\cite{Fidkowski:2003nf} it was found that a specific analytically continued $G(t)$, which we denote as $\hat G(t)$,  exhibits singularities of the form 
\be \label{Tcor1}
\hat G (t) \propto {1 \over (t-t_c^\pm)^{2 \De}}, \quad t \to t_c^\pm = \pm  {\beta e^{\mp {i \pi\over d}} \over 2 \sin {\pi\over d}}
 \equiv \pm {\tilde \beta \over 2}- i {\beta \over 2}  , \quad \De \to \infty  \ .
\ee
On the gravity side, considering complex time of the form $t=t_L-i{\beta}/{2}$, with $t_L \in \mathbb{R}$, corresponds to analysing geodesics in the two-sided eternal AdS-Schwarzschild black hole. The behavior~\eqref{Tcor1} arises because the spacelike geodesics which connect the two asymptotic regions of the black hole,  approach a null geodesic bouncing off the future (past) black hole singularity
as $t \to t_c^+$ ($t \to t_c^-$), see right of Figure~\ref{fig:Penrose}.
We will refer to the singular behavior~\eqref{Tcor1} as {the \textit{bouncing singularities}.}
 
The correlator  $G(t)$  can only have singularities  at $t =0$ and $t=- i \beta$. 
To obtain $\hat G (t)$,
which exhibits bouncing singularities,
one observes~\cite{Fidkowski:2003nf} that%
\footnote{To be more precise, $\cL(t)$ is obtained by an analytic continuation from the analogous expression using the Euclidean correlator $G(\tau)$, with $t = i\tau$. 
This analytic continuation is subtle and as a consequence the limit in \eqref{eq:cllim} is not always well defined.}
\be
\label{eq:cllim}
\LL (t) =  -\lim_{\Delta \to \infty} \frac{1}{\Delta} \log G(t), 
\ee
develops a branch point singularity at $t = - i {\beta \ov 2}$.
Analytically continuing $\LL(t)$ through a branch cut emanating from this branch point to the second sheet gives $\hat \LL (t)$.
The latter is given by the proper length of the bouncing geodesic and can be used to define
\be\label{ghas}
\hat G(t) \equiv e^{-\De \hat \LL (t)}\,,
\ee
which exhibits~\eqref{Tcor1}. 
In contrast,  for $t=t_L-i\beta/2$, 
$G(t)$ is given by a sum of two complex geodesics which are are regular for all $t_L$.

In~\cite{Festuccia:2005pi} it was found that the same bouncing geodesics have direct signatures in the Fourier transform $G (\om)$ of~\eqref{Tcor},\footnote{For notational simplicity, we use the same notation $G(\om)$ for the Fourier transform of $G(t)$, distinguishing them only by the arguments. Similarly, below we will use the same notation $G (\tau)$ for the Euclidean analytic continuation of $G(t)$.}  albeit for large imaginary frequencies, 
\be \label{Tcor2}
G (\om)  \propto \om^{2 \De} e^{i t_c^\pm \om} , \quad \om \to \pm i \infty  , \quad \De \to \infty \ .
\ee

The results~\eqref{Tcor1} and~\eqref{Tcor2} raised a number of questions whose boundary understanding has been elusive: 

\begin{enumerate}
\item What is the CFT origin of the singular behavior in~\eqref{Tcor1}? 
\item
What is CFT origin of the branch point singularity of $\LL (t)$
at $t = -{i \beta} /{2}$ in the large $\De$ limit?
\item
What is the boundary interpretation of the pair of complex geodesics
which dominate the correlator at  $t =t_L- i{\beta}/{ 2} $
in the large $\De$ limit?
\item What is the CFT origin of~\eqref{Tcor2}?
\item  The behavior~\eqref{Tcor1} and~\eqref{Tcor2} applies only to $d \geq 3$,
as for $d=2$ (the BTZ black hole) the black hole singularities are  orbifold singularities rather than curvature singularities.
What is the boundary origin of this difference? 
\end{enumerate}

In this chapter we  address these questions by investigating the behavior of thermal correlators of scalar operators by performing the Operator Product Expansion (OPE) in
the boundary theory.  We will restrict our discussion to the  boundary theory on flat space.

The OPE of two  operators $\phi$ separated in Euclidean time $\tau = i t$ 
can be written schematically as 
\be \label{eyh}
\phi (\tau,0) \phi (0,0) = \sum_n C_n \tau^{\De_n - 2 \De} \OO_n (0), 
\ee
where $n$ collectively labels all operators $\OO_n$ with $\De_n$ being their conformal dimension.
The Euclidean analytic continuation of $G(t)$ can be written in terms of OPE as 
\be \label{eun}
G (\tau)  \equiv G (t=-i \tau)  = {1 \ov \tau^{2 \De}} \sum_n C_n v_n \le({\tau \ov \beta} \ri)^{\De_n} , \quad \tvev{\OO_n}_\beta = v_n \beta^{-\De_n}  \ .
\ee
Note that the sums in (\ref{eyh}), (\ref{eun}) may have operators of the same conformal dimension which only differ by their spin.
In this case, each such operator contributes a separate term in (\ref{eyh}) and (\ref{eun}).

As mentioned in Chapter \ref{ch2}, in holographic theories that are dual to Einstein gravity in the bulk, the correlator~\eqref{eun} has a particularly simple structure
\begin{align}\label{lte}
    G(\tau)  = G_T(\tau) + G_{[\phi\phi]}(\tau)\, .
\end{align}
$G_T(\tau)$ is the contribution from the stress-tensor sector \cite{Karlsson:2019dbd}, $i.e.$ operators $[T^n]_J$, while $G_{[\phi\phi]}(\tau)$ is the contribution from the double-trace scalars $\phi (\partial^2)^n \partial_{i_1} \cdots \partial_{i_l} \phi$.
It will also be convenient to define 
\be
\label{eq:llt}
\LL_T (\tau) =  -\lim_{\Delta \to \infty} \frac{1}{\Delta} \log G_T(\tau)
\ee
and similarly $\LL_{[\phi\phi]}(\tau)$. 

Using the near-boundary ansatz \cite{Fitzpatrick:2019zqz} described in Section \ref{s2.nbe}, one can compute the OPE coefficients of multi-stress tensors, hence getting $G_T (\tau)$ explicitly as a series expansion in $\tau$. By studying the OPE data of $n$-stress tensor exchanges in the regime of large $n$,  we are able to resum the full stress-tensor sector for finite $\De$ and show that $G_T (\tau)$ contains a singularity precisely of the form~\eqref{Tcor1},\footnote{Since~\eqref{Tcor1} only applies to the large $\De$ limit, the exponents are also consistent.}
\be \label{Tcor3}
G_T (\tau) \propto {1 \ov (\tau - \tau_c^\pm)^{2 \De-{d \ov 2}}} , \quad \tau \to \tau_c^\pm = {\beta \ov 2} \pm i {\tilde \beta \ov 2}  = i t_c \ .
\ee
We want to emphasise that despite the similar form to \eqref{Tcor1}, this singular behaviour is obtained in the opposite limit, by first fixing $\Delta$ at a finite value and then finding the divergent behaviour of the stress-tensor sector near the singularity.
The result \eqref{Tcor3}, combined with general structure of the double-trace contribution $G_{[\phi \phi]} (\tau)$, as well as the geodesic analysis, gives a boundary picture that sheds light on various questions mentioned  earlier. 
Here we highlight the main elements: 

\begin{enumerate}
\item 
Black hole singularities are encoded in the analytic structure of the stress tensor sector of thermal correlation functions. In particular, the bouncing singularity is present at  finite $\De$ and can be accessed by analytically continuing the stress tensor sector contribution $G_T (\tau)$ to complex values of $\tau$ without the need of going to a different sheet, which is needed to obtain 
$\hat \LL(t)$. 

Heuristically, we may interpret the black hole geometry as being obtained from the empty AdS by ``condensing'' multiple gravitons, which are roughly dual to multiple stress tensors on the boundary. It thus makes intuitive sense that the black hole singularities reflect the analytic behavior of the stress tensor sector. The stress-tensor sector of thermal correlators thus possesses a large degree of universality and can serve as a direct probe of the black hole structure.

\item We expect that the OPE of $G(\tau)$ is uniformly convergent for all $\De$ only for $|\tau| < {\beta}/{2}$. 
This is the regime where we can take the $\De\to\infty$ limit inside the OPE and in
doing so neglect the contributions of double-trace operators. 
This is why for $|\tau| < {\beta}/{2}$  the stress-tensor sector OPE fully reproduces the geodesic length~\cite{Fitzpatrick:2019zqz}.
At  leading order in the large $\De$ limit, 
\begin{align}\label{gsb}
    \lim_{\Delta\to\infty}G (\tau) =
    \lim_{\Delta \to \infty}  \begin{cases}
   G_T(\tau) &  \tau < \frac{\beta}{2}\,, \cr
   G_{[\phi\phi]}(\tau) & \tau > \frac{\beta}{2}\,, \cr
   G_T (\tau) + G_{[\phi\phi]}(\tau)  & \tau = \frac{\beta}{2}  + i t_L, \; t_L \in \mathbb{R} \,.
   \end{cases}   
\end{align}
The branch point singularity observed in $\LL (t)$ at $t = -{i \beta \ov 2}$ in the geodesic analysis can  be understood from the large $\De$ limit of $G (\tau)$. The appearance of new geodesic saddles at  this point is the  consequence of the ``sudden'' turn-on of the double trace contribution. In particular, the two terms in the last line of~\eqref{gsb} can be identified respectively with the contributions of the two complex geodesics in the gravity analysis. 

In addition, for $\tau >\beta/2$, the double-trace contribution ensures that the full correlator satisfies the Kubo-Martin-Schwinger (KMS) condition \cite{Kubo:1957mj, Martin:1959jp}, which states that the correlator is periodic in real $\tau$ with period $\beta$.\footnote{Any finite-temperature correlator must satisfy the KMS condition. For the derivation and more details see Appendix \ref{s.kms2d}} As such, double traces are inherently linked with the periodicity in the temporal circle.

Finally, the full thermal correlator cannot contain singularities of the type \eqref{Tcor3}. As such, the double-trace sector must also contain a singularity at the same values as the stress-tensor sector, but with opposite sign, so that the full correlator is regular.

\item We observe numerically that as $\De$ is increased from a finite value to infinity, $\LL_T(\tau)$ computed using the OPE transitions between  a function which is regular at $\tau=\beta/2$ and has a singularity  at \eqref{Tcor3} and a function whose radius of convergence is $\beta/2$, where it develops a branch cut.
\item For $d=2$, $G(\tau)$ is known exactly from conformal symmetry, and equals the corresponding Virasoro vacuum block of the heavy-heavy-light-light correlator~\cite{Fitzpatrick:2015foa}.
All contributions come from the Virasoro descendants of the identity, which are the multi-stress operators,\footnote{We discuss this in detail in Appendix~\ref{s.kms2d}.} 
with no double-trace contributions for any $\tau$
\be
G (\tau)  = G_T (\tau), \quad d=2 \ .
\ee
Since $G(\tau)$ cannot have the bouncing singularities~\eqref{Tcor3}, neither can $G_T (\tau)$. This is related to the corresponding bulk geometry being regular. 

When the CFT is put on a circle, the double-trace contributions are needed for  thermal correlators to be periodic on the spatial circle. On the other hand, the corresponding bulk BTZ geometry develops an orbifold singularity. This suggests that the appearance of the black hole orbifold singularity is intrinsically linked with  non-trivial double-trace contributions. This is in stark contrast with the situation in $d \geq 3$, where the analytic behavior of multiple stress-tensor exchanges appears to  reflect bulk curvature singularities. 

\item The behavior~\eqref{Tcor3} does not directly say anything regarding the momentum space behavior~\eqref{Tcor2}. 
It is known that~\eqref{Tcor2} does survive to finite $\De$~\cite{festucciathesis}. It is thus tempting to speculate a relation 
\be 
G(i \om_E)  \sim \int_{-\infty}^\infty d \tau \, e^{ i \om_E \tau} G_T (\tau) , \quad \om_E \to \pm \infty ,
\ee
although with current available information, it is not possible to be more precise.
\end{enumerate}

In Section~\ref{sec:gff} we show that \eqref{gsb} also applies in 
the generalized free field (GFF) case
where the thermal correlator is given by the sum over  thermal images of the vacuum correlator, 
\be
\label{eq:GFF}
  G(\tau)^{(GFF)} = \langle \phi(\tau) \phi(0) \rangle_\beta^{(GFF)} = \sum_{m\in \mathbb{Z}} {1\over (\tau + m\, \beta)^{2\Delta}  }\,,
\ee
The sum over images ensures that the correlator satisfies the KMS 
condition \cite{Kubo:1957mj,Martin:1959jp}, $G(\tau)=G(\beta-\tau)$.
Note that in \eqref{eq:GFF} the $m=0$ term is the contribution of the identity, while all other terms correspond to multi-trace contributions \cite{Iliesiu:2018fao,Alday:2020eua}\footnote{See also \cite{El-Showk:2011yvt,Fitzpatrick:2014vua,Komargodski:2012ek} for related earlier developments and
\cite{Parisini:2023nbd,Marchetto:2023xap} for  examples of recent work on  manifestations of KMS conditions in CFT.}.
However, there are also differences -- unlike in the holographic case,  $\LL(\tau)^{(GFF)}$ does not develop a 
branch point at $\tau=\beta/2$.

The fact that double-trace operators do not contribute to holographic correlators for $\tau<\beta/2$  in the large-$\Delta$ limit was discussed in \cite{Fitzpatrick:2019zqz}
(see also \cite{Rodriguez-Gomez:2021pfh}).
Another situation  where the boundary correlator only receives contributions from the stress-tensor sector appeared in~\cite{Parnachev:2020fna}, where a particular near-lightcone limit was considered. 
In this limit, the  correlators receive contributions only from the leading twist multi-stress tensor operators and can be related to  spacelike one-sided geodesics.

The rest of this chapter is organised as follows. In Section~\ref{sec:SC} we review the relation between spacelike geodesics in eternal Schwarzschild-AdS black holes and thermal correlation functions. For definiteness, we focus on $d=4$. In particular, we analyse the singularity associated with the bouncing geodesic. In Section~\ref{s.bhope} we then analyse the OPE coefficients associated with $n$-stress-tensor exchanges in $d=4$. By analysing the large-$n$ behavior, we find~\eqref{Tcor3}. In Section~\ref{sec:inter}, we use the OPE analysis of Section~\ref{s.bhope} to give a boundary interpretation of the results of Section~\ref{sec:SC}.  In particular, we argue that the bouncing singularities originate from the singular behaviour found at finite $\Delta$ in the OPE analysis of the stress-tensor sector.
We discuss the results in other dimensions and the generalization to the boundary CFT on a sphere,  possible resolutions of the black hole singularities from $\apr$ and $G_N$ corrections, and various future perspectives in Section~\ref{sec:Disc}.

Some more technical details are presented in the appendices. In Appendix~\ref{a.cbd} we discuss the general structure of thermal two-point functions. In Appendix~\ref{app:eomcoefz} we state the partial differential equation that we solve to determine the holographic OPE data, while in Appendix~\ref{a.sumz} we discuss the validity of approximating the OPE series expansion by an integral. We also present some additional arguments to support the main claims of this chapter. The analysis of the thermal correlator in two-dimensions is performed in detail in Appendix~\ref{s.kms2d}. In Appendix~\ref{app:DeltaTau} and Appendix~\ref{app:XCorr} we analyse the subleading terms in the correlator near the bouncing singularity. In the main part of the chapter, we focus only on the $d=4$ case, but we show in Appendix~\ref{a.ope6d} that the stress-tensor sector has a singularity at the complexified time $t_c$ which corresponds to the bouncing geodesic  in other dimensions as well (we explicitly checked $d=6$ and $d=8$). Finally, in Appendix~\ref{app:LTanal} we discuss in detail the lowest-twist contributions to the OPE.

%%%%%%%%%%%%%%%%%%%%%%%%%%%%%%%%%%%%%%%%%%%%%%%%%%%%%
%%%%%%%%%%%%%%%%%%%%%%%%%%%%%%%%%%%%%%%%%%%%%%%%%%%%%

\section{Black hole singularity from geodesics}
\label{sec:SC}

The main object of interest in this chapter are thermal two-point correlation functions of identical operators
\begin{equation}\label{eq:DefCorrFun}
        G(t,\vec{x})=\expval{\phi(t,\vec{x})\phi(0,\vec{0})}_\beta\,,
\end{equation}
where $\beta = T^{-1}$ denotes the inverse temperature. In field theories with  holographic duals  these correlators can be calculated using the Green's function of the scalar field propagating in the
asymptotically  Anti-de Sitter black hole. When the mass, $m$, associated with the bulk scalar field, or equivalently,  the conformal dimension  of the dual field theory operator, $\Delta$,  is large, then the correlation function can be approximated by summing over classical saddles of the relevant path integral \cite{Louko:2000tp}
\begin{align}
\label{eq:Saddles}
	G(t, \vec{x}) \sim \sum_{\rm saddles}e^{-\Delta \,L}\,.
\end{align}
These saddles correspond to the geodesics in the black-hole background that connect the boundary points at which the operators are inserted and $L$ denotes the regularised proper length of the geodesic.

We will focus on the case where the bulk spacetime is a black brane in five dimensions%
\footnote{Examples in higher  dimensions are discussed in Appendix~\ref{a.ope6d}.}
with the metric 
\begin{equation}
    \label{eq:d=4BlackBranemet}
     ds^2~=~-r^2\,f(r)\,dt^2+\frac{dr^2}{r^2\,f(r)}+r^2\,d\vec{x}\,^2\,,
\end{equation}
where $\vec{x}=(x,y,z)$, so that  $d\vec{x}\,^2$  denotes the flat metric on $\mathbb{R}^3$, and 
\begin{align}
\label{eq:d=4fFunction}
    f(r) \equiv 1 -\frac{\mu}{r^4}\,.
\end{align}
Note that in most of this chapter we set the radius of AdS to unity. Near the spacetime boundary, $r\to \infty$, the metric \eqref{eq:d=4BlackBranemet} reduces to that of the Poincar\'e patch of  AdS$_5$. As such, the conformal boundary is just four-dimensional Minkowski space $\mathbb{R}^{1,3}$. The parameter $\mu$ is related to the inverse temperature $\beta$ through $\mu=(\pi/\beta)^4$. For the remainder of this section, we set $\mu =1$, which means that $\beta = T^{-1}= \pi$ and the location of the black-hole horizon is given by $r_0 = 1$. The curvature singularity is at $r=0$.

\begin{figure}[t]
    \centering
    \includegraphics[scale=1]{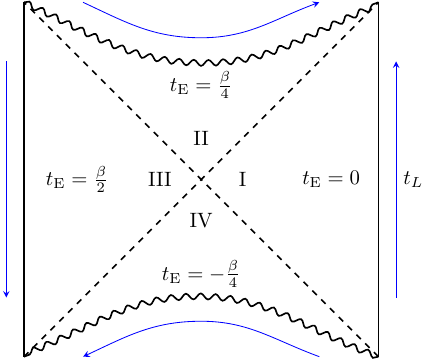}
    \caption{The Penrose diagram for the Lorentzian section of the maximally extended black-hole geometry in AdS$_5$. The spacetime separates into four regions with different constant values for the imaginary part of the time-coordinate.
    The blue arrows depict the direction of Lorentzian time, $t_L$, in each region.}
    \label{fig:Penrose0}
\end{figure}

We are interested in the maximally extended spacetime which can be described by using complexified Schwarzschild coordinates. The time coordinate $t$ then has a real and imaginary part, which we denote as\footnote{Note that we use a different sign convention for the imaginary part compared to \cite{Fidkowski:2003nf, Festuccia:2005pi}.}
\begin{align}
\label{eq:ComplexTime}
    t = t_{\rm L} - i\, t_{
    \rm E}\,, \qquad t_{\rm L} \in \mathbb{R}\,,\quad  0\leq t_{
    \rm E} < \beta\,. 
\end{align}
The Lorentzian section of this spacetime can be divided into four wedges in which the imaginary part takes  different constant values, summarised in 
Figure~\ref{fig:Penrose0}. 
Regions I and III describe two spacelike separated regions outside the  horizons. We choose the imaginary part of the time coordinate in Regions I and III to be given by $t_{ \rm E} = 0$ and $t_{ \rm E} = {\beta}/2$ respectively. Region II describes the interior of the black hole with $t_{ \rm E} = {\beta}/4$, while Region IV is the white hole region, where  $t_{ \rm E} = - {\beta}/4$. In essence, crossing a horizon corresponds to shifting the imaginary part of the time coordinate by  $\beta/4$.

\begin{figure}[t]
    \centering
    \includegraphics[width=1\textwidth]{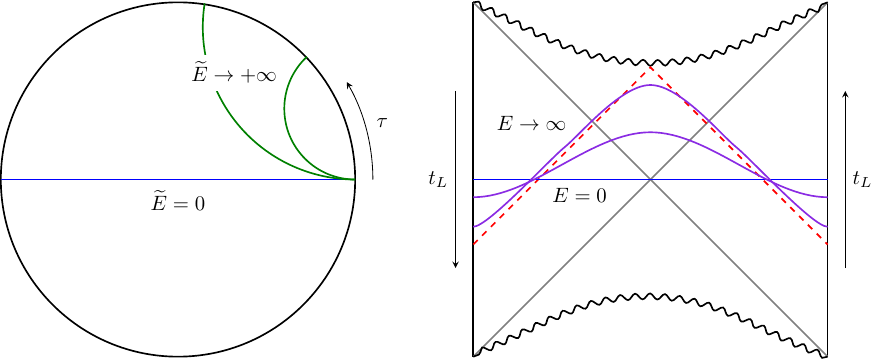}
    \caption{Diagrams for the Euclidean section (left) and the Lorentzian section (right) of the complexified black-hole spacetime.
    In both, we plot spacelike geodesics: The $\et =  E = 0$  geodesic (blue) can be drawn in both sections.
    On the left, the green curves depict geodesics with increasing  $\et$, which correspond to taking the limit $\tau \to 0$. 
    On the right, in purple, we plot real geodesics that probe the black-hole interior. As $E = - i \et \to \infty$, the geodesics become light-like signalling singular behaviour -- these are the Bouncing geodesics \cite{Fidkowski:2003nf, Festuccia:2005pi}.}
    \label{fig:Penrose}
\end{figure}

In what follows we  review the analysis of the geodesic approximation to 
the correlator $G(t_L - i\beta/2, \vec{x})$, which can be interpreted as a two-sided correlator with the operators in \eqref{eq:DefCorrFun} being inserted at different asymptotic regions of the complexified spacetime \eqref{eq:d=4BlackBranemet}. As shown in \cite{Fidkowski:2003nf, Festuccia:2005pi}, real spacelike geodesics that connect the two asymptotic regions probe the interior of the black hole, see Figure~\ref{fig:Penrose}. As they probe deeper into the interior, the geodesics  become more and more light-like with their proper length vanishing. This singular behavior is incompatible with the general properties of thermal correlation functions, which shows that such ``bouncing geodesics'' cannot contribute to the path integral.
However, in the next section we show that the  diverging behaviour related to the bouncing geodesics is still encoded in the stress-tensor sector of the OPE.

\subsection{Geodesics in the Euclidean section}

We begin by analysing geodesics in the Euclidean section of the black-brane spacetime
\begin{align}
\label{eq:EuclidMetric}
     ds^2~=~r^2\,f(r)\,d\tau^2~+~\frac{dr^2}{r^2\,f(r)}~+~r^2\,d\vec{x}\,^2\,,
\end{align}
where $f(r)$ is defined in (\refeq{eq:d=4fFunction}) and
$\tau \sim \tau + \beta$ is periodically identified.%
\footnote{The coordinate $\tau$ is related to the usual time coordinate through Wick rotation, $\tau = i\,t = t_{\rm E}+ i\,t_{\rm L}$, which explains our choice of sign for the imaginary part in \eqref{eq:ComplexTime}. In this subsection, we take $\tau \in \mathbb{R}$, while we will extend it to full complex space in subsequent sections.}
We can use the symmetries of the metric to reduce the problem to the motion in a one-dimensional effective potential. In general, we can introduce the energy, $\et$, associated with the time-translation invariance, and linear momenta $P_i$, related to the $\mathbb{R}^3$ isometry. In the main text we limit ourselves to the case with $x=0$, so that all linear momenta are set to zero.\footnote{The case with non-vanishing spatial separation is considered in Appendix~\ref{app:XCorr}.}

Let the geodesic be parameterised by an affine parameter $s$ and denote the derivative with respect to this parameter with a dot, for example $\dot{\tau}(s)$.
The energy of a geodesic is given by
\begin{align}
    \et = r^2\,f(r)\,\dot \tau\,,
\end{align}
defined in such a way that $\dot \tau >0$ for $\et>0$.
The condition that the geodesic is everywhere spacelike can be rearranged into 
\begin{align}
\label{eq:TurnEuclid}
    \dot r^2 = r^2\,f(r) - \et^2\,.
\end{align}
Geodesics in the Euclidean section are pictured on the left of Figure~\ref{fig:Penrose}.
They start at $r= \infty$ and probe the space up to a minimal value, $r_t$, which we call the turning point, before returning to the asymptotic boundary.
The turning point is given by the largest real root at which \eqref{eq:TurnEuclid} vanishes
\begin{align}
\label{eq:EuclideanTurning}
    r_t^2 = \frac12\left(\et^2+ \sqrt{\et^4+4}\right)\,.
\end{align}

The time difference between the endpoints of a geodesic  is given by \begin{align}
\label{eq:TauEuclid}
    \tau &\equiv \tau_f-\tau_i = 2 \int_{r_t}^{\infty}\frac{\et\,dr}{r^2\,f(r)\,\sqrt{r^2\,f(r)- \et^2}}\nonumber\\
    &= \frac12 \log\left(\frac{\et^2 +2\,\et+2}{\sqrt{\et^4+4}}\right)- \frac{i}{2}\log\left(\frac{\et^2 +2\,i\,\et-2}{\sqrt{\et^4+4}}\right)\,.
\end{align}
Let us focus on the behaviour at high energies, $\et \to \infty$. 
In this limit, the turning point goes toward the asymptotic boundary -- such geodesics probe only the asymptotic region of space.
To see how the time difference scales with the energy, we expand \eqref{eq:TauEuclid}
\begin{align}
\label{eq:EofTauExp}
    \tau = \frac{2}{\et}-  \frac{8}{5\,\et^5}+ \coo{\et^{-9}}\,, \quad \Longrightarrow \quad \et = \frac{2}{\tau }-\frac{\tau ^3}{10}+ \coo{\tau^{7}}\,,
\end{align}
meaning that as the energy is increased, the time difference goes to 0.

The regularised proper length of the geodesic is given by
\begin{align}
\label{eq:ProperLengthEuclid}
    L &= 2\lim_{r_{\rm max}\to \infty}\left[\int_{r_t}^{r_{\rm max}}\frac{dr}{\sqrt{r^2\,f(r) -\et^2}}- \log r_{\rm max}\right]= \frac12\,\log\left(\frac{16}{\et^4+4}\right)\,,
\end{align}
where $r_{\rm max}$ is the UV cut-off length \cite{Festuccia:2005pi}. By expressing the energy as a function of $\tau$ we find a logarithmic divergence at the origin
\begin{align}
\label{eq:OneSidedProperLength}
    L = 2\,\log\tau - \frac{\pi^4}{40}\left(\frac{\tau}{\beta}\right)^4 - \frac{11\,\pi^8}{14400}\left(\frac{\tau}{\beta}\right)^8+ \coo{\tau^{12}}\,,
\end{align}
where we have reinstated the appropriate units, using $\beta/\pi = 1$. When $\Delta \gg 1$, the sum over geodesics gives an approximation to the thermal correlation function.
And by exponentiating \eqref{eq:OneSidedProperLength} we get the contribution from the spacelike geodesic connecting two points on the Euclidean time circle
\begin{align}
    \label{eq:NearCoincidentCorrelator}
   e^{-\Delta\,L} =\frac{1}{\tau^{2\Delta}}\left[1+
    \frac{\Delta\,  \pi^4}{40}\,\left(\frac{\tau}{\beta}\right)^4
    +\frac{\left(9 \Delta ^2+22 \Delta \right) \pi
   ^8}{28800}\,\left(\frac{\tau}{\beta}\right)^8+
   \coo{\tau^{12}}\right]\,.
\end{align}
In the next section we will show that this result is completely reproduced by only the stress-energy sector in the large-$\Delta$ limit.

\subsection{Contributions from different saddles} \label{sec:geo}

The expressions~\eqref{eq:TauEuclid} and~\eqref{eq:ProperLengthEuclid} can be analytically continued to complex $\widetilde E$ to obtain geodesics in the real section of Schwarzschild spacetime. In particular, to obtain two-sided correlation functions, we need to consider 
\be \label{tLd}
\tau = {\pi \ov 2} + i t_L , \quad t_L \in \mathbb{R} \ .
\ee 
For this purpose, expanding~\eqref{eq:TauEuclid} around $\widetilde E =0_+$, we find 
\be \label{deo}
\tau - {\pi \ov 2} =  - {\widetilde E^3 \ov 6} + O(\widetilde E^5), \qquad L = {\log[2]} - {\widetilde E^4 \ov 8} + O(\widetilde E^8)\,,
\ee
which implies that $L (\tau)$ has a branch point singularity of the form $(\tau-{\pi \ov 2})^{4 \ov 3}$ at $\tau = {\pi \ov 2}$. 
Solving $\widetilde E$ in terms of $\tau$ we find three branches of solutions 
\be 
\widetilde E_0 = (6r)^{1 \ov 3}\,  e^{i \theta \ov 3} 
 , \quad \widetilde E_1 = \widetilde E_0 \,e^{-{2 \pi i \ov 3}} , \quad 
\widetilde E_2 = \widetilde E_0\, e^{-{4 \pi i \ov 3}}, \quad {\pi \ov 2} - \tau \equiv r \,e^{i \theta}  \ .
\label{eq:branches}
\ee
For ${\pi \ov 2} - \tau  > 0$ ($i.e.$ $\theta =0$), $\widetilde E_0$ branch should be chosen. To analytically continue  $L (\tau)$ to
$\tau$ given by~\eqref{tLd}, we decrease $\theta$ from $0$ to $-{\pi \ov 2}$ (for definiteness take $t_L > 0$). A careful analysis~\cite{Fidkowski:2003nf}\footnote{See Section 3.4 of the paper. While the discussion in Section 3.4  was phrased as a model, it can be justified using Section 4.4 of~\cite{Festuccia:2005pi}.}
shows that in the large $\De$ limit:
\begin{enumerate}
    \item For  $\theta  \in (-{\pi \ov 8},0]$, $\widetilde E_0$ is the only saddle contributing to $G(\tau)$. 
    \item For $\theta \in (- {\pi \ov 2}, -{\pi \ov 8} ]$, $\widetilde E_0$ is the dominant saddle, but now $\widetilde E_1$ also contributes as a subdominant saddle, $i.e.$  
\be \label{enq}
G (\tau) \sim e^{- \De L (\widetilde E_0)} + e^{- \De L (\widetilde E_1)} \ .
\ee
    \item  At $\theta = -{\pi \ov 2}$, $i.e.$ for two-sided geodesics, the two terms in~\eqref{enq}  have the same norm and thus contribute equally. Both $\widetilde E_0$ and $\widetilde E_1$ are complex, corresponding to complex geodesics in the black hole spacetime.\footnote{See \cite{Chapman:2022mqd,Aalsma:2022eru} for recent discussion of complexified geodesics in the de Sitter context.}
    \item  For  $\theta \in [-{7 \ov 8}\pi, -{\pi \ov 2})$, we still have~\eqref{enq}, but now $\widetilde E_0$ contribution is subdominant. 
    \item For $\theta \in [-\pi, -{7 \ov 8}\pi)$ only $\widetilde E_1$ contributes. In particular, for $\theta = -\pi$, $i.e.$ $\tau > {\beta}/{2}$, we have $\widetilde E_1$ becomes real, negative, and 
\be 
-{1 \ov \De} \log G (\tau) =   L (\widetilde E_1 (\tau)) =  L (\widetilde E_0 (\beta-\tau)) =-{1 \ov \De} \log G (\beta-\tau), \quad \tau \in \le({\beta \ov 2}, \beta \ri)  \ .
\ee 
This shows that $G(\tau)$ in the large $\Delta$ limit satisfies the KMS condition. 

\item The saddle corresponding to $\widetilde E_2$ does not lie on the steepest descent contour of  
the integral to obtain $G(\tau)$~\cite{Fidkowski:2003nf,Festuccia:2005pi}, and 
never contributes to the correlator.

\end{enumerate}

\subsection{Bouncing geodesics}
\label{ssec:bounce}

The $\widetilde E_2$ branch in \eqref{eq:branches}, which never contributes to the correlator, corresponds to the two-sided real geodesics presented in the right plot of Figure~\ref{fig:Penrose}. More explicitly, for $\theta = - {\pi \ov 2}$ we have 
\begin{equation}
    \widetilde E_2  = i\,E\,,\qquad E \in \mathbb{R}_+ \, ,
\end{equation}
The turning point is then given by%
\footnote{The analytic structure of the turning point as a function of the energy,  $r_t(\et)$,  is discussed in \cite{Festuccia:2005pi}. This function has branch cuts in the complex plane that can be associated to quasinormal modes of the black hole. Then one can define $r_t(\et)$ in the full complex $\et$ plane by starting from \eqref{eq:TurnEuclid} for $\et \in \mathbb{R}$ and analytically continue through the origin. Physically this can be thought of as following the geodesics in the Euclidean section as the real part of the energy is decreased from  infinity to 0. At this point the geodesic crosses the cap or equivalently traverses the double-sided Lorentzian section (see the two geodesics in blue in Figure~\ref{fig:Penrose}). One then increases the imaginary part of the energy, $E$, causing the geodesics to probe the region behind the horizon.}
\begin{align}
\label{eq:P=0Turning}
    r_t^2 = \frac12\left(\sqrt{E ^4+4}-E ^2\right)\,.
\end{align}
One can see that the turning point is inside the horizon for $E ^2>0$. In fact, as $E \to \infty$, the turning point approaches the origin as $r_t \sim 1/E$, meaning that in this regime the geodesics probe the region near the singularity, see right of Figure~\ref{fig:Penrose}.

The time difference and the proper length of the two-sided geodesics corresponding to $\widetilde E_2$ can be obtained by taking $\et = i\,E$ in \eqref{eq:TauEuclid} and \eqref{eq:ProperLengthEuclid}
\begin{subequations}
\label{eq:LorentzianIntegrals}
    \begin{align}
    \label{eq:deltatRes1}
         \tau &=  \frac{\pi}{2}\left(1+i\right) +\frac{1}{2}\log\left(\frac{E ^2 -2i\,E  -2}{\sqrt{4+ E ^4}}\right)- \frac{i}{2}\log\left(\frac{E ^2 +2\,E  +2}{\sqrt{4+ E ^4}}\right)\,,\\
        \label{eq:cLRes1}
        \hat \LL & \equiv L (\widetilde E_2)  = \frac12\log{\left(\frac{16}{E ^4+4}\right)}\,.
    \end{align}
\end{subequations}
which matches the known results \cite{Fidkowski:2003nf, Festuccia:2005pi}. We have also identified $\hat \LL$, which was introduced around~\eqref{ghas}, as the geodesic distance associated with $\tilde E_2$. 
We keep the constant time shift explicit---the real part corresponds to the shift of $\tau = \beta/2$ which comes from the spacelike geodesic crossing two horizons as it goes from the I to the III patch in the Lorentzian section of the spacetime.

Denote the  time shift observed above as
\begin{align}
\label{eq:CriticalTauDef}
    \tau_c \equiv \frac{\pi}{2}(1+i) = \frac{\pi}{\sqrt{2}}\,e^{\frac{i\,\pi}{4}}\,,
\end{align}
and expanding \eqref{eq:deltatRes1} in large $E $ gives
\begin{align}
\label{eq:LargeEdeltat}
\tau=  \tau_c - \frac{2\,i}{E } + \frac{8\,i}{5\,E ^5}  + \coo{E ^{-9}}\,,
\end{align}
meaning that as $E \to \infty$, and  the spacelike geodesics become increasingly null-like, $\tau \to \tau_c$. Let
\begin{align}
    \label{eq:Deltatdef}
    \delta \tau \equiv \tau_c-\tau\,,
\end{align}
and perturbatively invert \eqref{eq:deltatRes1} to express $E $ as a function of $\delta \tau$
\begin{align}
    \label{eq:Eoft}
    E= \frac{2i}{\delta \tau} - \frac{i}{10}\,(\delta \tau)^3+ \coo{(\delta t)^{7}}\,.
\end{align}
One can then insert this into the expression for proper length and again find a logarithmic divergence, only now as $\tau \to \tau_c$
\begin{align}
    \label{eq:PropLenDeltat}
    \hat \cL = 2 \log\delta \tau + \frac{\pi^4}{160}\,\left(\frac{\delta \tau}{\tau_c}\right)^4+\coo{\delta \tau^{8}}\,,
\end{align}
where we have reinstated the units using $\pi^4 = \beta^4 = - 4\,\tau_c^4$. We note that the first correction to the logarithmic divergence appears at order $(\delta \tau)^4$.
We then find 
\begin{align}
\label{eq:GeoCOrrSmallDeltaTauExp}
   \hat G \equiv   e^{-\Delta \,\hat \cL}
    = \frac{1}{(\delta \tau)^{2\Delta}}\Bigg[1 -\frac{\Delta\,\pi^4}{160}\,\left(\frac{\delta \tau}{\tau_c}\right)^4 + \coo{\delta \tau^{8}}\Bigg]\,,
\end{align}
which exhibits the singular behavior~\eqref{Tcor1}.

%%%%%%%%%%%%%%%%%%%%%%%%%%
%%%%%%%%%%%%%%%%%%%%%%%%%%

\section{Black hole singularity from OPE}
\label{s.bhope}

In this section we holographically extract the OPE coefficients of the $n$-stress tensor contributions to the thermal correlator.
We then find the large-$n$ behaviour of these coefficients, which allows us to resum the stress-tensor sector of the correlator near the radius of convergence of the OPE.
In $d=4$, for finite $\Delta$ we find singularities in the complex $\tau$-plane, located at $\tau_c=\frac{\beta}{\sqrt{2}}e^{i\frac{\pi}{4}+ik\frac{\pi}{2}}$ for $k\in\mathbb{Z}$.
These correspond to the singularities associated with bouncing geodesics discussed in the previous section and are direct signatures of the black hole singularity. 
We further show that these bouncing singularities disappear in the large $\Delta$ limit, where we recover the branch point singularity at $\tau = \beta/2$. This is consistent with the geodesic analysis performed in the previous section.

\subsection{Stress-tensor sector of the correlation function}
\label{ss.holo}

Consider the scalar two-point function at finite temperature $T = \beta^{-1}$
\begin{equation}\label{eq:DefCorrFunagain}
        G(\tau,\vec x)=\expval{\phi(\tau,\vec x)\phi(0,0)}_\beta\,.
\end{equation}
Let the CFT be holographically dual to a  planar AdS-Schwarzschild black hole in $(d+1)$-dimensions and the scalar operators be dual to minimally coupled scalar fields in the bulk.
We again use the black hole metric in the Euclidean signature \eqref{eq:EuclidMetric}
    \begin{equation}\label{e.bhg}
         ds^2~=~r^2\,f(r)\,d\tau^2~+~\frac{dr^2}{r^2\,f(r)}~+~r^2\,d\vec{x}^2\,,
    \end{equation}
with  $\vec{x}=(x,y,z)$,  and $f(r)=1-\frac{\mu}{r^d}$.
In this section we keep the parameter \mbox{$\mu=(4\pi/d\,\beta)^d$} explicit. 
The equation of motion for the minimally coupled scalar
    \begin{equation}\label{e.weworig}
        (\Box-m^2)\phi=0\, ,\qquad m^2=\Delta(\Delta-d)
    \end{equation}
can be solved by a near-boundary expansion \cite{Fitzpatrick:2019zqz}.
We first perform a coordinate transformation $(\tau,\vec{x},r)\rightarrow(w,\rho,r)$
    \begin{equation}
    \rho^2=r^2\,\vec{x}^2\,,\qquad w^2=1+r^2(\tau^2+\vec{x}^2)    \,,
    \end{equation}
and write
    \begin{equation}\label{e.anz1}
        \phi(w,\rho,r)=\left(\frac{r}{w^2}\right)^\Delta \psi(w,\rho,r)\ ,
    \end{equation}
where $(r/w^2)^\Delta$ is the solution in the pure AdS space. 
The equation of motion \eqref{e.weworig} then reduces to a differential equation for $\psi(w,\rho,r)$, whose explicit form in arbitrary dimension $d$ is given in Appendix~\ref{app:eomcoefz}.
Importantly, using the standard AdS/CFT dictionary, the boundary correlator can be obtained by taking the limit
    \begin{equation}\label{eq:LimGen}
        G(\tau,\vec{x})
        =\frac{1}{(\tau^2+\vec{x}\,^2)^\Delta}\lim_{r\rightarrow\infty}\psi\,.
    \end{equation}
Determining the function $\psi$ using the bulk equation of motion determines the boundary correlation function.

As we described in Chapter \ref{ch2}, $\psi$ can be written as a sum of two contributions
    \begin{equation}
        \psi=\psi_{T}+\psi_{[\phi\phi]}\,,
    \end{equation}
where $\psi_{T}$ denotes the stress-tensor sector, which also includes the contribution dual to the identity operator, while $\psi_{[\phi\phi]}$ denotes the double-trace contributions. We restrict ourselves to non-integer conformal dimensions, for which the two sectors decouple, and focus on the stress-tensor sector. Its near-boundary expansion is given by \eqref{e.thegeneralansatzdominguoile}, $i.e.$,
    \begin{equation}
    \label{e.theansatz}
        \psi_{T}=1+\sum_{i=1}^\infty\sum_{j=0}^i\sum_{k=-i}^{\frac{d\,i}{2}-j}a_{j,k}^i\frac{\rho^{2j}w^{2k}}{r^{d\,i}}\ ,
    \end{equation}
where 1 corresponds to the contribution of the identity operator. By inserting this ansatz   into the equation of motion of the bulk scalar and expanding to arbitrary order in $1/r$, one is able to determine the coefficients $a_{j,k}^i$. Most importantly, through the dictionary \eqref{eq:LimGen}, this large-$r$ expansion on the bulk side systematically maps to the OPE of the boundary correlator, which for $\vec{x}=0$ reads%
\footnote{See Appendix \ref{a.cbd} for more details.}
    \begin{equation}\label{e.resumL}
G_T(\tau)=\frac{1}{\tau^{2\Delta}}\sum_{n=0}^\infty\Lambda_n\left(\frac{\tau}{\beta}\right)^{d\,n}\,,
    \end{equation}
    where the subscript in $G_T(\tau)$ denotes that this is only the 
stress-tensor sector of the full correlator, $G(\tau)$.%
\footnote{With an abuse of nomenclature, we will still refer to $G_T(\tau)$ as the correlator. The subscript should serve as a reminder that it is just the stress-tensor sector contribution.} Through this expression we are able to determine the stress-tensor contributions to $\Lambda_n$ using the near-boundary expansion of $\psi_{T}$.

In what follows, we focus on the correlation functions where the operators are inserted at the same spatial points, with correlators at $\vec{x} \neq 0$ considered in Appendix~\ref{app:XCorr}.

\subsection{OPE coefficients and the KMS condition}
\label{ss:NCLim}

\begin{figure}
    \centering
    \includegraphics[width=\textwidth]{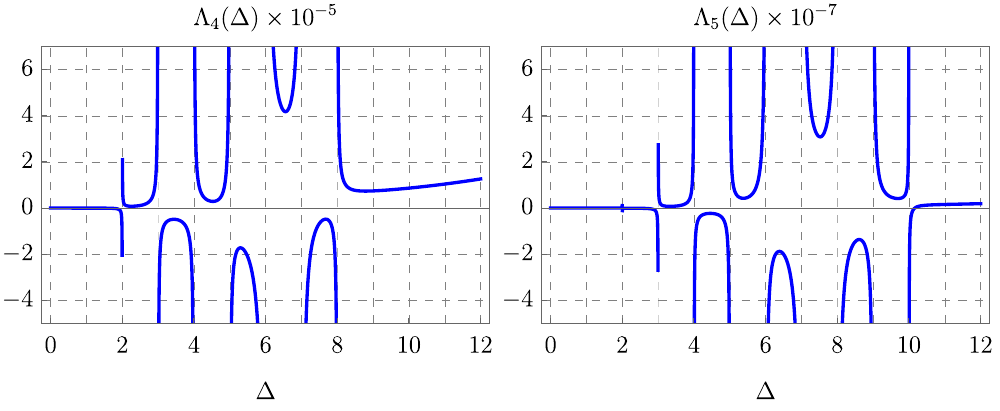}
    \vspace{-0.5cm}
    \caption{Plots for $\Lambda_4$ and $\Lambda_5$ as functions of $\Delta$. 
    These OPE coefficients have poles at $\Delta=2,3,\ldots,2n$ and are regular for $\Delta >2n$.}
    \label{fig:l4l5}
\end{figure}

Let us now set $d=4$. For small values of $n$, one can efficiently calculate the $\Lambda_n$ as explicit functions of $\Delta$. After that, calculating $\Lambda_n$ for general $\Delta$ becomes too time-demanding, so we first fix $\Delta$ to a certain value and only  then calculate $\Lambda_n$. Namely, as $n$ grows, these coefficients become more complicated functions of $\Delta$. For example, the first few terms in the small $\tau$ expansion of the correlator are given by
{\small{
\begin{equation}
\label{eq:NCCorr}
    G_T(\tau) \approx \frac{1}{\tau^{2\Delta}}\!\left[1 + \frac{\pi^4\,\Delta}{40}\left(\frac{\tau}{\beta}\right)^4\!\!+ \frac{\pi ^8 \,\Delta  \left(63 \Delta ^4-413 \Delta ^3+672 \Delta ^2-88 \Delta
   +144\right)}{201600 (\Delta -4) (\Delta -3) (\Delta -2)}\left(\frac{\tau}{\beta}\right)^8\!+ \ldots\right].
\end{equation}}}%
The expressions for $\Lambda_n$ with higher $n$ follow the same pattern as $\Lambda_2$ above (see Figure~\ref{fig:l4l5} for $\Lambda_4$ and $\Lambda_5$ as functions of $\Delta$) and 
can be schematically written as 
\begin{align}
\label{eq:LambdanStructure}
    \Lambda_n \sim \frac{h_n(\Delta)}{\prod_{k=2}^{2n}(\Delta-k)}\,,
\end{align}
where $h_n(\Delta)$ is a polynomial function. 

\begin{figure}[t]
    \centering
    \includegraphics[width=\textwidth]{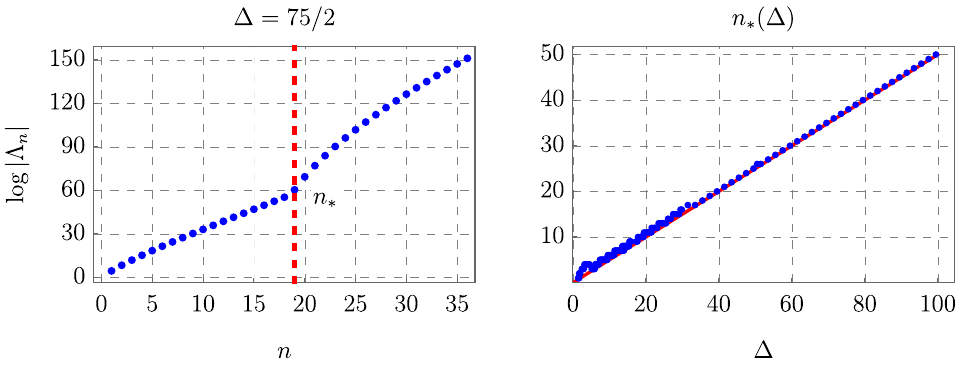}
    \vspace{-0.5cm}
    \caption{On the left we plot the values of $\log|\Lambda_n|$ for $\Delta = \frac{75}2$. We see that at $n_* = 19$, there is a change of behaviour of the OPE coefficients. 
    On the right, we plot the behaviour of $n_*$ as a function of $\Delta$ and find that $n_* = \Delta/2$ (red). }
    \label{fig:nstar}
\end{figure}

Fixing $n$ and varying $\Delta$, we can distinguish two regimes: $\Delta \leq 2n$, where we find poles at $\Delta=2,3,\ldots,2n$, and $\Delta >2n$, where the OPE coefficients have no poles in $\Delta$.
For fixed $\Delta$, the OPE coefficients in these two regimes behave differently, which can be seen in Figure~\ref{fig:nstar}. We see a significant change in the behaviour  at a particular value $n=n_*$.
Analysing the dependence of this cross-over point on the conformal dimension, we find that $n_* = \Delta/2$,
\footnote{In practice, we have taken $n_*$ to be the lowest integer greater than $\Delta/2$, which is why all points lie just above the $\Delta/2$ line on the right plot in Figure~\ref{fig:nstar}.}
which is exactly where we find the last pole in $\Lambda_n$.

That the cross-over happens at this precise value of $n$ should not be surprising. Namely, recall from \eqref{e.resumL} that the conformal dimension of the $T^n\equiv[T^n]_J$ multi-stress tensor exchange in $d$ dimensions is $\Delta_{T^n} = n\,d$, while those for double-trace operators have $\De_m = 2 \De + 2m$, $m=0, 1, 2, \ldots$. They can mix when 
\mbox{$\De_{T^n} = \De_m$}, which are exactly the locations of the poles of $\De$  in $\Lam_n$ for a fixed $n$, and are 
only possible for $n \geq n_*$ with 
\begin{align}
    \label{eq:cross-over}
    n_* \equiv \frac{2 \Delta}{d} \stackrel{d=4}{=} {\De \ov 2} \,,\quad \Longleftrightarrow \quad \Delta_{T^n}^* \equiv n_* \,d = 2\,\Delta\, .
\end{align}
  This explains the different behavior of $\Lam_n$ for $n > n_*$ and $n < n_*$ observed in the left plot of Figure~\ref{fig:nstar}.
We can equivalently say that the cross-over point happens exactly at the point where double-trace operators start contributing;  for $n < n_*$, the series expansion for $G (\tau)$ has only contributions from $G_T (\tau)$.

\begin{figure}[t]
    \centering
   \includegraphics[scale=0.8]{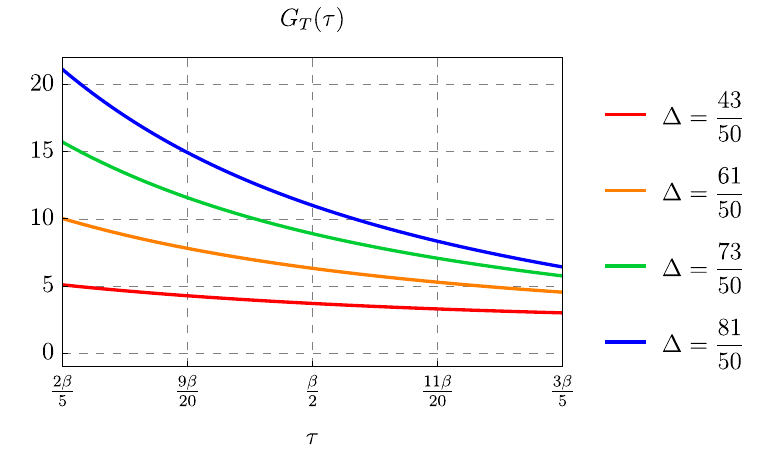}
    \caption{The value of stress-tensor contribution to the thermal correlator for different values of $\Delta$ near $\tau= \beta/2$. 
    We see that the stress-tensor sector is not symmetric around $\tau= \beta/2$ and thus does not satisfy the KMS condition, $G_T(\tau) = G_T(\beta-\tau)$.
 }
    \label{fig:noKMS}
\end{figure}

In our analysis, we have no control over the double-trace contributions. 
Therefore, for finite $\Delta$, $G_T(\tau)$ does not reproduce the full thermal correlation function. A simple check of the importance of the double-trace exchanges comes from the failure of the stress-tensor sector to satisfy the KMS condition.  In Figure~\ref{fig:noKMS}, we  plot numerically $G_T (\tau)$ for various $\De$, and see that $G_T(\tau)$ is not symmetric around $\tau= \beta/2$ and thus $G_T(\tau)\neq G_T(\beta-\tau)$. The stress-tensor sector contribution by itself does not satisfy the KMS condition and contains no knowledge about the periodicity of the~(Euclidean) time circle. Hence one role of the double-trace sector is to ensure that the full correlator satisfies the KMS condition.

\subsection{Asymptotic analysis of OPE coefficients for finite \texorpdfstring{$\De$}{Delta} and bouncing  singularities}

Let us now focus on the analysis of the coefficients $\Lambda_n$ for large values of $n$ and finite $\Delta$. We are thus interested in the regime where $n>n_*$ and both the stress-tensor and double-trace contributions are appearing in the full correlator. As already mentioned above, calculating $\Lambda_n$ as explicit functions  of $\Delta$ is currently out of our computational reach. Instead we first fix $\Delta$ to a finite number: For each value of $\Delta$ considered, we calculated $n \approx 50$ coefficients, which on a standard desktop computer takes around 10 days per $\Delta$.

\begin{figure}[t]
\includegraphics[scale=0.9]{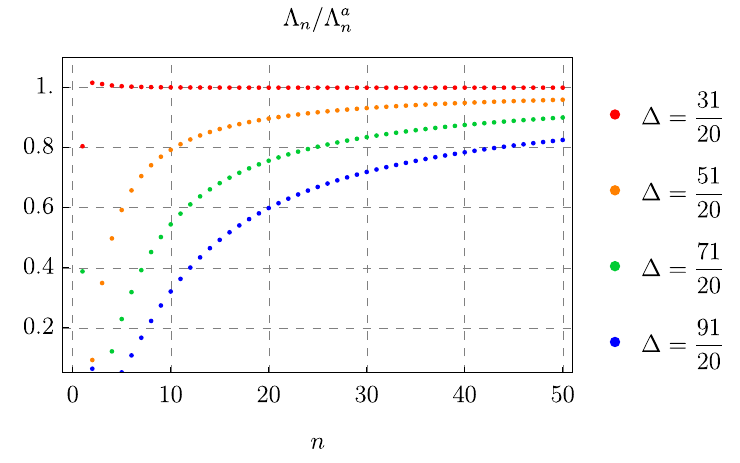}
	\centering
    \caption{Ratio of the explicit results for $\Lambda_n$ to the leading large-$n$ prediction for different values of $\Delta$ in $d=4$.}
    \label{fig:d4Ratio}
\end{figure}

We find that for large values of $n$, the  $\Lambda_n$ can be approximated by
(see Figure~\ref{fig:d4Ratio})
\begin{align}
\label{eq:Lambdand=4Ansatz}
   \Lambda_n^a = c(\Delta)\, \frac{n^{2\Delta -3}}{\left(\frac{1}{\sqrt{2}}\right)^{4n}\,e^{i\,\pi\,n}}\,.
\end{align}
One can be more precise and include $1/n$ corrections to \eqref{eq:Lambdand=4Ansatz}.  We analyse such terms and how we obtained the form for $\Lambda_n^a$ in detail in Appendix~\ref{app:DeltaTau}. The $1/n$ corrections become especially important at large $\Delta$, which can already be seen in Figure~\ref{fig:d4Ratio}, where the approach to the asymptotic form is slower for higher $\Delta$.

\begin{figure}[t]
    \centering
    \includegraphics[scale=0.8]{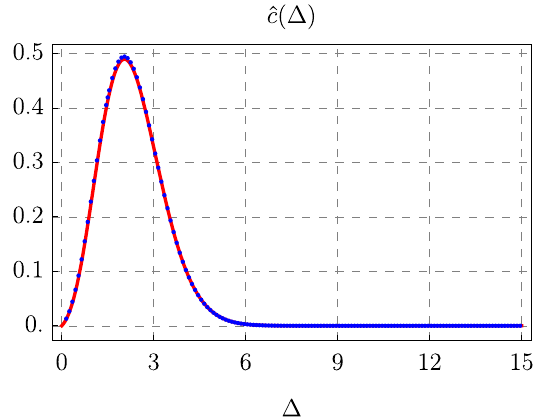}
    \caption{Numerical data for $\hat c(\Delta)$ (blue) compared with the function given in \eqref{eq:cdeltafun} (red).}
    \label{fig:c(Delta)}
\end{figure}

The coefficients $\Lambda_n^a$ contain a non-trivial function $c(\Delta)$, which has poles at $\Delta =2,3,\ldots$.%
\footnote{As mentioned earlier, at integer values of $\Delta$ the double trace operators can mix with the stress tensor sector. The addition of these double traces should cancel the divergence and make the correlation function finite \cite{Fitzpatrick:2019zqz}.}
This diverging behaviour can be isolated through
\begin{align}
\label{eq:cdeltafun}
    c(\Delta) \equiv \frac{\pi\,\Delta\,\left(\Delta-1\right)}{\sin\left(\pi\,\Delta\right)}\,\hat c(\Delta)\,, \qquad \qq{for}\Delta >0\,,
\end{align}
where the function $\hat c(\Delta)$ is free of poles.
It turns out (see Figure~\ref{fig:c(Delta)}), that this residual function can be approximated by\footnote{That is, for $\Delta\gtrsim 5/4$  and up to $\Delta \approx 15$, the ratio between the ``numerical'' values and  \eqref{eq:cdeltafun} is equal to 1 up to 2\%.} 
\begin{align}
\label{eq:chatfun}
    \hat c(\Delta)  \approx  \frac{\Delta}{\Gamma\left(2\,\Delta+ \frac{3}2\right)}\,\frac{4^{2\Delta}}{20}\,.
\end{align}

\begin{figure}[t]
\centering
\includegraphics[scale=1]{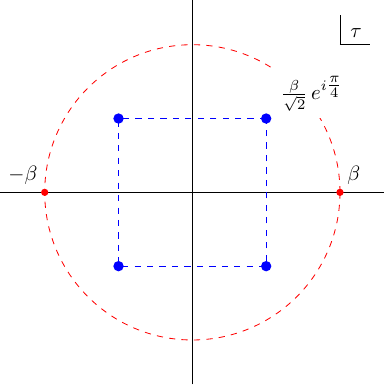}
\caption{The poles of the stress-tensor sector of the thermal correlator in four dimensions in the complex $\tau$ plane. The poles are located inside the circle of radius $\beta$.}
\label{fig:Polesd4}
\end{figure}

We now insert the asymptotic form of $\Lambda_n$ into the OPE of the correlator \eqref{e.resumL} and approximate the sum with an integral%
\footnote{We discuss the validity of this approximation in Appendix \ref{a.sumz}.}
\begin{align}
\label{e.sin1}
        G_T(\tau)&\approx\frac{1}{\tau^{2\Delta}}\int_0^\infty\Lambda_n^a\left(\frac{\tau}{\beta}\right)^{4n}d n
        %
        %
        %&
        =\frac{c(\Delta)\,\Gamma(2\Delta-2)}{\tau^{2\Delta}}\, \left[-\log\left(\frac{\tau^4}{\left(\frac{\beta}{\sqrt2}\right)^4\,e^{i\pi}}\right)\right]^{-(2\Delta-2)}\!.
\end{align}
While this form of the correlator is not valid for all values of $\tau$, it gives us information about its behaviour near singular points: Namely, the \eqref{e.sin1} will diverge whenever the argument of the logarithm is equal to 1, which is precisely at 
\begin{equation}\label{e.poles}
       \tau =  \tau_c\equiv\frac{\beta}{\sqrt{2}}e^{i\frac{\pi}{4}+ik\frac{\pi}{2}}\qq{for}k\in\mathbb{Z}\,.
\end{equation}
Let $\delta \tau \equiv \tau_c - \tau$. Near the critical values, the correlator takes the form
\begin{align}
\label{eq:CorrDivergenced4}
    G_T(\tau \approx \tau_c) \sim \frac{c(\Delta)\,\Gamma(2\Delta-2)}{4^{(2\Delta-2)}}\,\frac{1} {\tau_c^{2}}\frac{1}{\delta\tau^{2\Delta-2}}\,,
\end{align}
which is precisely of the form of the bouncing singularities~\eqref{Tcor1} (see Figure~\ref{fig:Polesd4}).
The four-fold rotational symmetry of the singularities~\eqref{e.poles} originates from the 
fact that multiplying by a fourth root of unity $\tau \to e^{i\frac{k\pi}{2}} \tau$
leaves all terms inside the sum in the OPE~\eqref{e.resumL} invariant.

At first sight, one can now take the large-$\Delta$ limit of the correlator near the poles. 
 To make a valid comparison, we first define
\begin{align}
\label{eq:LogOfCorr}
    \overline{\cL}_{T} \equiv -\frac{1}{\Delta}\,\log G_T(\tau)\,,
\end{align}
which, using \eqref{eq:CorrDivergenced4}, gives
\begin{align}
\label{eq:LeadingLogRes}
    \overline\cL_T \simeq -\frac{1}{\Delta}\log\left[\frac{c(\Delta)\,\Gamma(2\Delta-2)}{4^{(2\Delta-2)}}\,\frac{1} {\tau_c^{2}}\right]+ \frac{2\Delta-2}{\Delta}\log\delta\tau\,.
\end{align}
If then one inserts the expression \eqref{eq:cdeltafun} (including \eqref{eq:chatfun}) for $c(\Delta)$ then in the large-$\Delta$ limit all $\Delta$ dependence disappears%
\footnote{At integer values of $\Delta$, the prefactor $c(\Delta)$ diverges (see \eqref{eq:cdeltafun}). However, in these cases, one finds $\log(\Delta)/\Delta$ behaviour which vanishes as $\Delta\to \infty$.} and
\begin{align}
\label{eq:ProperLengthNearLim}
    \lim_{\Delta \to \infty}\overline \cL_T \simeq  2\log\delta\tau \,.
\end{align}
This appears to give the precise form of divergence expected from the bouncing geodesic \eqref{eq:PropLenDeltat}.

However, the order of limits for~\eqref{eq:ProperLengthNearLim} is different from that of~\eqref{eq:PropLenDeltat}.
To obtain (\refeq{eq:ProperLengthNearLim}), we first focused on the limit $\tau\to\tau_c$ and then took the large $\De$ limit. 
On the other hand, the bulk geodesic analysis corresponds to taking the large $\De$ limit first. 
That the order of limits is important can be seen from the structure of the $\Lambda_n$ as functions of $\Delta$, 
as indicated in Figure~\ref{fig:l4l5} and~\ref{fig:nstar}. To obtain~\eqref{eq:CorrDivergenced4}, 
we first fixed $\Delta$ and analyzed the asymptotic large $n$ behavior of $\Lam_n$, which is in the regime $n>n_* = \Delta/2$. 
However, the geodesic analysis corresponds to taking first $\Delta\to \infty$, where the cross-over point also goes to infinity, $n_* \to \infty$, and thus  we are always in the regime $n < n_*$.

\subsection{Asymptotic behavior in the large \texorpdfstring{$\De$}{Delta} limit} \label{s.abintldl}

To properly analyse the large $\Delta$ limit and make contact with the geodesic results we first need to take $\Delta \to \infty$ and only then determine the OPE coefficients $\Lambda_n$. We show in this subsection that the 
Euclidean geodesic results~\eqref{eq:TauEuclid} and~\eqref{eq:ProperLengthEuclid} are indeed recovered.

We begin by expanding the logarithm of \eqref{eq:NCCorr} in small $\tau$
\begin{align}
\label{eq:LOS}
    \overline{\cL}_T = 2\,\log\tau - \frac{\pi^4}{40}\left(\frac{\tau}{\beta}\right)^4 - \frac{\left(77 \Delta ^3-483 \Delta ^2+712 \Delta +72\right)\,\pi ^8  }{100800\,(\Delta -4) (\Delta -3) (\Delta -2)}\,\left(\frac{\tau}{\beta}\right)^8 + \ldots
\end{align}
and compare this with the proper length of the geodesic in Euclidean space \eqref{eq:OneSidedProperLength}. 
We see that logarithmic and the $\tau^4$ terms match,%
\footnote{These terms are related to the exchange of identity and a single stress-tensor operator in the OPE and are fixed by symmetry.}
while the $\tau^8$ (and all higher order) terms differ.
This should not be surprising: The geodesic result can be meaningfully compared to the correlation function only in the  $\Delta\to \infty$ limit, in which case
\begin{align}\label{ldeB}
   \cL_T\equiv  \lim_{\Delta\to \infty}\overline{\cL}_T = 2\,\log\tau - \frac{\pi^4}{40}\left(\frac{\tau}{\beta}\right)^4 - \frac{11\,\pi^8}{14400}\,\left(\frac{\tau}{\beta}\right)^8 + \ldots\,,
\end{align}
is equal to \eqref{eq:OneSidedProperLength}. We checked the agreement explicitly up to order $(\tau/\beta)^{20}$.
This importantly shows that in the limit where the conformal dimension of the probe field is large, the stress tensor sector completely reproduces the geodesic result. 

When $\Delta \to \infty$, we can also analyse the asymptotic behaviour of the expansion coefficients in $\cL_T$. 
In particular, inserting \eqref{e.resumL} into \eqref{eq:LogOfCorr}, we obtain
\begin{align}
    \overline{\cL}_T = 2\log\tau - \sum_{n=1}^{\infty} \frac{\widetilde \Lambda_n}{\Delta}\left(\frac{\tau}{\beta}\right)^{4n}\,,
\end{align}
where $\widetilde \Lambda_n$ are appropriate combinations of $\Lambda_n$ and we have used the fact that $\Lambda_0=1$.
One can then take the limit $\Delta \to \infty$ as 
\begin{align}
\label{eq:cltexp}
    \cL_T= \lim_{\Delta \to \infty}\overline{\cL}_T = 2 \log \tau - \sum_{n=1}^{\infty} L_n\,\left(\frac{\tau}{\beta}\right)^{4n}\,,
\end{align}
where we have assumed that one can take the limit inside the series and thus
\begin{align}
    L_n \equiv \lim_{\Delta \to \infty}\frac{\widetilde \Lambda_n}{\Delta}\,.
\end{align}\vspace{-0.7cm}
\begin{figure}[H]
    \centering
    \includegraphics[scale=0.8]{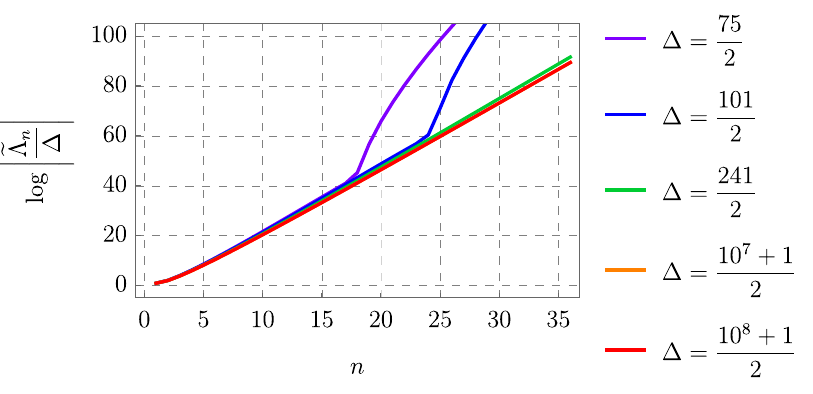}
    \vspace{-0.5cm}
    \caption{The values of $\log\big|\widetilde\Lambda_n/\Delta\big|$ for different $\Delta$. The discrete data is joined for a clearer presentation. When $\Delta$ is small, we observe a change of behaviour at $n = n_* = \Delta/2$. For large $\Delta$, this cross-over point disappears and the data converges to a fixed curve, analysed in \eqref{eq:Ansatz1}.}
    \label{fig:plogL}
\end{figure}

\begin{figure}[t]
    \centering
    \includegraphics[scale=0.9]{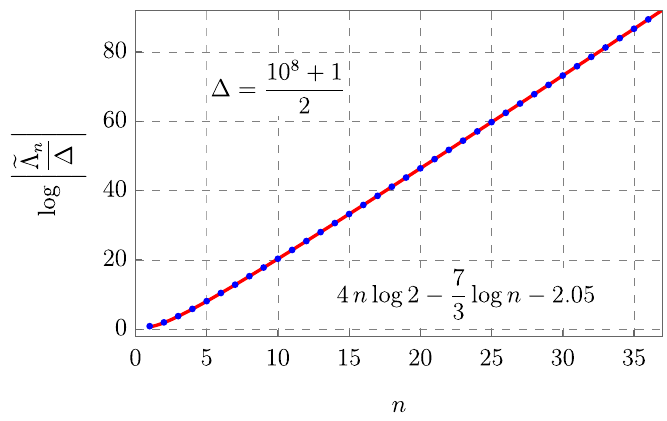}
    \vspace{-0.2cm}
    \caption{Comparing the holographic data for $\Delta = (10^8+1)/2$ to the ansatz \eqref{eq:Ansatz1} and the numerical data given by \eqref{eq:NumericalData}.}
    \label{fig:plogfit}
\end{figure}

The first few $L_n$ are given implicitly in \eqref{ldeB}. For higher values of $n$, we again analyse the data for large but fixed  $\Delta$. For some  $\Delta$, we plot the values of $\log\big|\widetilde\Lambda_n/\Delta\big|$ in Figure~\ref{fig:plogL}. For small enough  $\Delta$, we again observe the cross-over at $n=n_*$ to the asymptotic behaviour analysed in the previous subsection. For large $\Delta$, not only does this cross over get pushed to infinity, but the curve stabilises to an asymptotic value, which we assume is given by
\begin{align}
\frac{\widetilde \Lambda_n}{\Delta}  \xrightarrow[\Delta\to\infty]{} L_n + \coo{\frac{1}{\Delta}}\,.
\end{align}
In practice, we analyse  $\Delta = (10^8+1)/2$ and estimate $L_n$ up to $1/\Delta$ corrections. We find that at large values of $n$, one can make an ansatz 
\begin{align}
\label{eq:Ansatz1}
    L_n = c\,n^b\,a^n\left(1+ \coo{\frac1n}\right)\,,
\end{align}
where $a$, $b$, and $c$ are all constants. 
Then for $n\gg 1$
\begin{align}
    \log L_n = \log c + b \,\log n+ n\,\log a + \coo{\frac1n}\,,
\end{align}
which can be directly fitted to the $\widetilde \Lambda_n/\Delta$ data for $\Delta = (10^8+1)/2$ and we find% 
\footnote{We used the data from $n=10$ to $n=36$ and used the ansatz \eqref{eq:Ansatz1} with up to $1/n^{12}$ corrections.}
(see also Figure~\ref{fig:plogfit})
\begin{align}
\label{eq:NumericalData}
   & \log a = 4\,\log 2 \pm 10^{-6}\,,&&  b = -\frac73  \pm 10^{-4} && \log c = -2.050 \pm 10^{-3}\,,
\end{align}
In what follows we neglect the uncertainties which can be traced back to both $1/\Delta$ and $1/n$ effects. 
We insert these values into the expansion \eqref{eq:cltexp} and approximate the sum by an integral%
\footnote{We use the generalized exponential integral \begin{align*}
    E_n(x) \equiv \int_1^{\infty}t^{-n}\,e^{-t\,x}\,dt\,.
\end{align*}}
{\allowdisplaybreaks{
\begin{align}
    \cL_T &\approx 2 \log\tau - c\,\sum_{n=1}^{\infty}n^{-\frac{7}{3}}\,2^{4n}\,\left(\frac{\tau}{\beta}\right)^{4n}\approx 2\log\tau -c \int_1^\infty\,dn\,n^{-\frac73}\,\left(\frac{2\,\tau}{\beta}\right)^{4n}\nonumber\\*
   &= 2\log \tau - c\,E_{\frac73}\left(-\log\left[\left(\frac{2\tau}{\beta}\right)^4\right]\right)\,.
    \label{eq:ExponentialIntegal}
\end{align}}}
This function has a branch point at all points at which the argument of the logarithm is equal to 1, which is 
\begin{align}
    \tau_{b} = \frac{\beta}{2}\,e^{\frac{i\pi\,k}{2}}\,,\qquad k \in \mathbb{Z}\,, 
\end{align}
which suggests four singularities in the complex-$\tau$ plane.
Focusing on $\tau =\beta/2$, we expand \eqref{eq:ExponentialIntegal} around this point and find
\begin{align}
\label{eq:sing_bovertwo}
    \cL_T \approx  - c\,\Gamma\left(-\frac{4}{3}\right)\,\left(\frac{8}{\beta}\right)^{\frac{4}{3}}\,\left[\frac{\beta}{2}-\tau\right]^{\frac43} + \ldots \approx - 1.364 \le({\pi \ov 2}-\tau \ri)^{4 \ov 3} + \ldots \,,
\end{align}
where we kept only the leading non-analytic behavior at $\tau = \beta/2$, and in the second equality inserted the value of $c$ and taken $\beta = \pi$. The leading non-analytic behavior at  $\tau = \beta/2$ of the Euclidean geodesic results~\eqref{eq:TauEuclid} and~\eqref{eq:ProperLengthEuclid} can be readily found from~\eqref{deo}, which gives 
\be 
L = - {6^{4 \ov 3} \ov 8} \le({\pi \ov 2}-\tau \ri)^{4 \ov 3} + \ldots \approx  - 1.363  \le({\pi \ov 2}-\tau \ri)^{4 \ov 3} + \ldots  \ .
\ee
We see that the asymptotic analysis reproduces the location, the exponent, and the prefactor of the branch point structure observed from bulk geodesics. 

One can also study the large $\Delta$ limit of the logarithm of the OPE coefficients. 
One can perform the same analysis as above for a large value of $\Delta$ and obtain
\begin{align}
\log\Lambda_n \sim n \log \Delta - \log \Gamma(n+1)  + n\,\log\left(\frac{\pi^4}{40}\right)+\coo{\frac{1}{\Delta}}\,,
\end{align}
where the numerical value of $\log\left(\frac{\pi^4}{40}\right) \approx 0.89004$ gets reproduced by fitting up to $\approx 10^{-6}$ for $\Delta = (10^8+1)/2$.
Exponentiating the above result yields
\begin{align}
    \Lambda_n = \frac{1}{n!}\,\left(\frac{\Delta\,\pi^4}{40}\right)^n\left(1+ \coo{\frac{1}{\Delta}}\right)\,.
    \label{eq:StarOPE}
\end{align}
Inserting this into the OPE and assuming that one can trust these OPE coefficients up to $n \to \infty$ yields
\begin{align}
    \label{eq:expres}
     G_T(\tau) 
     \approx  \frac{1}{\tau^{2\Delta}}\sum_{n=0}^{\infty}\,\frac{1}{n!}\left(\frac{\pi^4\,\Delta\,\tau^4}{40\,\beta^4}\right)^n 
     \approx \frac{1}{\tau^{2\Delta}}\exp\left[\frac{\pi^4\,\Delta\,\tau^4}{40\,\beta^4}\right]\,,
\end{align}
where we neglected $1/\Delta$ corrections. This result matches  \cite{Rodriguez-Gomez:2021pfh}, where starting from \eqref{eq:NearCoincidentCorrelator}, a double-scaling limit is taken with  $\Delta \to \infty$ while $\tau/\beta \to 0$ such that $\Delta\tau^4/\beta^4$ is held fixed.
The significance of this double-scaling limit is that it can be obtained from the exponentiation of the single stress tensor exchange.
Indeed, we see that the argument of the exponent is precisely the first non-logarithmic term in the proper length in \eqref{eq:OneSidedProperLength}.
Therefore, taking the large $\Delta$ limit at the level of the OPE coefficients neglects all the higher order terms in \eqref{ldeB}, since the information about these terms is encoded in the $1/\Delta$ corrections in \eqref{eq:StarOPE}, which are discarded.
Since these higher order terms are crucial for the analytic structure of the proper length, we are unable to see the branch cut when taking the $\Delta\to\infty$ limit at the level of the correlation function.
As seen above, one needs to analyse the proper length in order to see the branch cut.

\subsection{Summary} \label{sec:sum}

Here we summarize the main results obtained in this section: 

\begin{enumerate}
\item For a finite $\De$, as indicated by Figure~\ref{fig:noKMS}, $G_T (\tau)$ does not satisfy the KMS condition on its own. 
 
The double-trace contribution is thus also needed such that the KMS condition is satisfied by the full correlator. 

\item \label{item:2}
For a finite $\De$, $G_T (\tau)$ has singular behavior~\eqref{eq:CorrDivergenced4} at~\eqref{e.poles}. 
 
Full thermal correlation functions cannot have such singular behavior, and thus~\eqref{eq:CorrDivergenced4} should be cancelled by $G_{[\phi\phi]} (\tau)$. That is, $G_{[\phi\phi]} (\tau)$ must have the same singular behavior, but with an opposite sign. 

\item\label{item:3}
 In the $\De \to \infty$ limit, the small $\tau$ expansion of $\LL_T (\tau)$ (defined in~\eqref{ldeB}) precisely recovers the small $\tau$ expansion of $L (\tau)$ obtained from gravity. 

While we have only checked the agreement between $\LL_T (\tau)$ and $L (\tau)$ in this limit up to $\tau^{20}$, we will assume below that the agreement in fact holds to all orders in the small $\tau$ expansion. 

\item There is a crossover behavior in $\Lambda_n$
at $n_*=\Delta/2$. As a result, in the large $\De$ limit, the singularity of $G_T (\tau)$ at $\tau_c$ obtained at a finite $\De$ is not directly seen, instead we find a branch point singularity in $\LL_T (\tau)$ at $\tau = \beta/{ 2}$.

A possible scenario is that as $\De$, and therefore $n_*$, increases the size of the region around $\tau=\tau_c$ where 
$G_T(\tau)$ diverges shrinks and finally vanishes when $\Delta\to \infty$.
At the same time, as $\De$ increases, the behavior of $\LL_T (\tau)$ near $\tau=\beta/2$ gets closer and closer to the singular behavior observed in the geodesic analysis \eqref{eq:sing_bovertwo}.
This would be another signature of the
noncommutativity of limits $\tau \to \tau_c$ and $\De\to\infty$.

\end{enumerate}

%%%%%%%%%%%%%%%%%%%%%%%%%%%%%%%%%%%%%%%%%%%%%%%%%%%%%
%%%%%%%%%%%%%%%%%%%%%%%%%%%%%%%%%%%%%%%%%%%%%%%%%%%%%

\section{Boundary interpretation of the gravity results}\label{sec:inter}

We now use the results obtained from the OPE analysis of Section~\ref{s.bhope} to interpret the gravity results reviewed in Section~\ref{sec:SC}. While the singular behavior~\eqref{eq:CorrDivergenced4} has the same form  as that in~\eqref{Tcor1}, 
the connection is not immediate, as~\eqref{eq:CorrDivergenced4} and~\eqref{Tcor1} appear in different quantities and in different regimes (one in $G_T$ at finite $\De$, while the other in $\hat G$ at infinite $\De$).  
What is more, as discussed at the end of last section, the singular behavior~\eqref{eq:CorrDivergenced4} disappears in the large $\Delta$ limit, which obscures its connection to the bouncing singularities and  black hole singularities. 
 
In this section, we argue that the singular behavior~\eqref{eq:CorrDivergenced4} can indeed be identified as the boundary origin 
of~\eqref{Tcor1} and thus the black hole singularities. 
Below we will restore $\beta$, and when we say ``for all $\tau$'' it always refers to ${\rm Re} \, \tau \in (0, \beta)$.

\subsection{Role of the double-trace contributions in the large \texorpdfstring{$\De$}{Delta} limit}\label{s.dt}

We first discuss the role of the double-trace contributions in the large $\De$ limit. In Section~\ref{s.bhope} we found that in the $\De \to \infty$ limit, the small $\tau$ expansion of $\LL_T (\tau)$  precisely recovers that of $L (\tau)$ obtained from 
the geodesic analysis ($i.e.$ item~\ref{item:3} of Section~\ref{sec:sum}). 
Furthermore, we have shown that in the large $\Delta$ limit, $\LL_T (\tau)$  has a branch point singularity of the form $({\beta \ov 2}-\tau)^{4 \ov 3}$ at $\tau ={\beta}/{2}$. 
That is, for large $\De$, $\LL_T (\tau)$ (and $L(\tau)$) has a radius of convergence $\beta/2$, and 
the double-trace contribution $G_{[\phi \phi]} (\tau)$ can be neglected in the large $\De$ limit for $\tau < {\beta}/{2}$. 

That $G_{[\phi \phi]} (\tau)$ can be neglected for small $\tau$ has a simple explanation from the OPE structure.  The double-trace contribution $G_{[\phi\phi]} (\tau)$ has a small $\tau$ OPE of the form 
\begin{align}\label{dtope}
G_{[\phi\phi]}(\tau) = \frac{1}{\beta^{2\Delta}}\,\sum_{k=0}^{\infty}\,D_k\left(\frac{\tau}{\beta}\right)^{2\,k}
\end{align}
where $D_k$ are some constants. The appearance of even powers in~\eqref{dtope} comes from the fact that the double-trace operators $\phi (\partial^2)^n \partial_{i_1} \cdots \partial_{i_l} \phi$
have dimensions $2 \De + 2n +l$ and only the ones with even $l$ can have nonzero thermal expectation values in~\eqref{eun}. 

Comparing~\eqref{dtope} with the analogous expression
for the stress tensor sector~\eqref{e.resumL} gives 
\begin{align}
    \frac{G_{[\phi\phi]}(\tau)}{G_T(\tau)} \sim \left(\frac{\tau}{\beta}\right)^{2\Delta}\,\left(1 + \coo{\tau^2}\right), 
\end{align}
with the full $G (\tau)$ having the structure 
\be 
G(\tau) = {1 \ov \tau^{2 \De}} \le[\sum_{n=0}^\infty\Lambda_n\left(\frac{\tau}{\beta}\right)^{d\,n}\ 
+\ \sum_{k=0}^{\infty}\,D_k\left(\frac{\tau}{\beta}\right)^{2\,k + 2 \De} \ri] \ .
\ee
We may  conclude from the above equation that, in the large $\De$ limit, the double-trace contribution is always suppressed for all $\tau$. This, however, assumes that we can take the large $\De$ limit {\it inside} the sum, which 
requires  $G(\tau)$ to be uniformly convergent for all $\Delta$ for all $\tau$. This is incorrect, since as mentioned earlier, $\LL_T (\tau)$ develops a branch point singularity at $\tau = {\beta}/ { 2}$ in the large $\De$ limit, and only has the radius of convergence equal to $\beta \ov 2$.\footnote{Even if the series $G_T (\tau)$ remains convergent in the large $\De$ limit for all $\tau$, it is possible that the full series $G (\tau)$ may not be uniformly convergent for all $\De$ for all $\tau$. We will see an example of this in Section~\ref{sec:gff}.}
Thus for $|\tau| \geq {\beta}/ { 2}$, we need to sum  the series for $G (\tau)$ first for finite $\De$ and then take the large $\Delta$ limit.

For $|\tau| \geq {\beta}/ { 2}$, we expect $G_{[\phi\phi]}(\tau)$ will be needed. After all, as mentioned earlier, the double-trace contributions are needed for the full correlator to satisfy the KMS condition. Now comparing with the discussion of Section~\ref{sec:geo}, we can identify 
\be \label{ll0}
\LL_T (\tau) = L (\widetilde E_0 (\tau))
\ee
where $L$ on the RHS is the geodesic distance obtained from gravity. From~\eqref{enq}, we can further identify 
\be \label{ll00}
\LL_{[\phi\phi]} (\tau) =  L (\widetilde E_1 (\tau))
\ee
which becomes important for $\theta \leq -{\pi \ov 2}$, with $\theta$ defined as ${\beta \ov 2} - \tau = r e^{i \theta}$. 
Furthermore, the relative dominance between $\widetilde E_0$ and $\widetilde E_1$ saddles discussed there 
can now be written in terms of the boundary language as 
\begin{align}
\label{eq:RealCorr}
    \lim_{\Delta\to\infty}G (\tau) =
    \lim_{\Delta \to \infty}  \begin{cases}
   G_T(\tau) &  0< \tau < \frac{\beta}{2} \cr
   G_{[\phi\phi]}(\tau) & \frac{\beta}{2} < \tau < \beta \cr
   G_T (\tau) + G_{[\phi\phi]}(\tau)  & \tau = \frac{\beta}{2}  + i t_L, \; t_L \in \mathbb{R} 
   \end{cases}     \ .
\end{align}
In particular, $G_T (\tau)$ and $G_{[\phi\phi]}(\tau)$ correspond respectively  to the contributions from the two complex geodesics 
on the gravity side. 

One can ask why the uniform convergence stops at $\tau = \beta/2$. If we consider a theory with time-reversal symmetry, $\tau \to - \tau$, that satisfies the KMS condition, $G(\tau) = G(\beta-\tau) $, then the branch point at which the OPE sum and the $\Delta \to \infty$ limit cannot be exchanged  can be in principle at any $0\leq \beta^*\leq \beta/2$. It would be interesting to  explore this direction further. 

\subsection{Boundary interpretation of the bouncing singularities}\label{s.biotbs}

We now turn to the boundary interpretation of the bouncing singularities~\eqref{Tcor1}.

The identification~\eqref{ll0} is consistent with our numerical observation that the singular behavior~\eqref{eq:CorrDivergenced4}  is not directly seen in the large $\De$ limit, as $\widetilde E_0$ corresponds to a complex geodesic for $\tau = {\beta}/ { 2} + i t_L$, and is regular at $\tau_c = {\beta \ov 2} (1+i)$. 
The same statement applies to $G_{[\phi \phi]} (\tau)$. As we discussed in item~\eqref{item:2} of Section~\ref{sec:sum},
at a finite $\De$, $G_{[\phi \phi]} (\tau)$ should also have a bouncing singularity at $\tau_c$, but the identification~\eqref{ll00} implies that the singularity is no longer there in $\LL_{[\phi \phi]} (\tau)$. 
Nevertheless, we would like to argue that the bouncing geodesic singularities~\eqref{Tcor1} do originate from~\eqref{eq:CorrDivergenced4}. 

More explicitly, after obtaining $G_T (\tau)$ from the OPE of multiple stress tensors, we can expand $\cL_T (\tau)$ near $\tau = {\beta}/ { 2}$ as 
\be \label{loq1}
\LL_T (\tau) \sim  \le({\beta \ov 2} - \tau\ri)^{4 \ov 3} + \ldots \equiv f_T (y)\,, \quad y \equiv \le({\beta \ov 2} - \tau\ri)^{1 \ov 3}  \,,
\ee
where we neglect the constant prefactor and terms that are regular at $\tau = \beta/2$. 
Since on the gravity side $\tilde E_1 (\tau)$ and $\tilde E_2 (\tau)$ are related to $\tilde E_0 (\tau)$ 
by  phase multiplications~\eqref{eq:branches}, we can then write 
\be \label{loq2}
\LL_{[\phi \phi]} (\tau) = f_T \le(e^{-{2 \pi i \ov 3}} y \ri), \quad \hat \LL (\tau)  = f_T \le(e^{-{4 \pi i \ov 3}} y \ri) \ .
\ee
That is,  through the branch point singularity developed by $\LL_T (\tau)$ at $\tau = {\beta}/ { 2}$, both $\LL_{[\phi \phi]} (\tau)$ coming from $G_{[\phi \phi]} (\tau)$, and $\hat \LL (\tau)$~\eqref{eq:PropLenDeltat} corresponding to the bouncing geodesic are fully determined from $\LL_T (\tau)$.

While the singular behavior~\eqref{eq:CorrDivergenced4} of 
$G_T (\tau)$ and $G_{[\phi \phi]} (\tau)$ have seemingly both disappeared in themselves in the large $\De$ limit, the bouncing singularities are not lost, but are transferred to $\hat \LL (\tau)$.  They are just not as manifest. 
This phenomenon is quite remarkable. As emphasized in Section~\ref{s.bhope}, the bouncing singularities of $G_T (\tau)$ at finite $\De$ are controlled by the asymptotic large $n$ behavior in the regime $n > n_*$, while  $\LL_T (\tau)$ is given by $\Lam_n$  in the regime $n< n_*$ after first taking $n_* \to \infty$. 
Nevertheless, $\LL_T (\tau)$ does contain the information regarding the bouncing singularities, albeit indirectly.

At finite $\De$, both $G_T (\tau)$ and $G_{[\phi \phi]} (\tau)$ have bouncing singularities, then why do we say that they originate from the stress tensor sector, not from the double-trace sector? We already saw that in the large $\De$ limit, both 
$\LL_{[\phi \phi]} (\tau)$ and $\hat \LL (\tau)$ are determined from $\LL_T (\tau)$. This is not an accident; in fact, $G_{[\phi \phi]} (\tau)$ is also determined by $G_T (\tau)$ for any $\De$. 

To make the point clear, recall the OPE computation of a vacuum four-point function 
\be\label{o4}
\vev{VVWW} = \sum_{\OO} C_{VV\OO} C_{WW\OO} \vev{\OO\OO}  ,
\ee
where the four-point function is fully determined from the OPE data and vacuum two-point functions. 
In contrast, in the OPE computation~\eqref{eun} of $G(\tau)$, the right hand side of the equation involves the thermal expectation values of double-trace operators 
\be 
\tvev{\phi (\partial^2)^n \partial_{i_1} \cdots \partial_{i_l} \phi}_\beta
\ee
whose values require knowledge of thermal two-point functions of $\phi$, which are just $G(\tau)= \left\langle\phi(\tau)\phi(0)\right\rangle_\beta$ (and derivatives thereof) evaluated at a fixed point. These two-point functions are unknown but necessary in order to determine the OPE expansion. In other words, both sides of~\eqref{eun} depend on $G(\tau)$, and it should be viewed as an equation {\it to solve} for $G(\tau)$, rather than as an expression to calculate the left hand side as is the case for~\eqref{o4}.  The inputs that we need to solve~\eqref{eun} are $G_T (\tau)$, the OPE coefficients 
of $\phi (\partial^2)^n \partial_{i_1} \cdots \partial_{i_l} \phi$, as well as imposing that $G(\tau)$  obeys the KMS condition and be analytic in ${\rm Re} \, \tau \in (0, \beta)$.  Thus we can say $G(\tau)$ and $G_{[\phi \phi]} (\tau)$ are determined from $G_T (\tau)$. 

Finally, we comment that $G_T (\tau)$ is closely connected to the bulk geometry, and thus enjoys some level of universality. 
The OPE coefficients for $k$-stress tensor exchanges correspond in the bulk to couplings of the bulk dual $\Phi$ of $\phi$ 
to multiple gravitons, $i.e.$ couplings of the form $\Phi \Phi h^k$, where $h$ schematically denotes the bulk graviton. The thermal one-point function $v_n$ for $k$ stress tensors is given schematically by $(\langle T_{\mu \nu} \rangle_\beta)^k$, $i.e.$ $k$-th power of the boundary stress tensor in the black hole geometry.

\subsection{A speculation on the momentum space behavior and bouncing singularities} 

Now suppose the behavior~\eqref{Tcor2} is also present at finite $\De$. Then given~\eqref{eq:CorrDivergenced4}, 
it is tempting to speculate that 
\be \label{ejp}
G(i \om_E) \sim \int_{-\infty}^\infty d \tau \, e^{i \om_E \tau} \, G_T (\tau)  \ .
\ee
We use $\sim$ in the above equation to indicate that while (under some assumptions) the right hand side gives the correct qualitative behavior 
for $G (i \om_E)$ in the limit $\om_E \to \pm \infty$, at the moment we do not have enough information to specify a precise relation. As $\om_E \to +\infty$, we assume the behavior of $G_T (\tau)$ at infinity is such that we can close the contour in the upper half complex $\tau$-plane. Then the integral will mainly receive contributions from the neighborhood of $\tau_c^+ = {\beta \ov 2} + i {\widetilde \beta \ov 2}$, leading to 
\be 
G(i \om_E) \sim e^{i \om_E \tau_c} \om_E^{2 \De-3} \propto  (i \om_E)^{2 \De-3} e^{i (i \om_E) t_c^+} , \quad \om_E \to +\infty \ .
\ee
Similarly, for $\om_E \to - \infty$, we close the contour in the lower half complex $\tau$-plane, with the main contribution coming from integration around $\tau_c^- = {\beta \ov 2} - i {\widetilde \beta \ov 2}$, leading to 
\be 
G(i \om_E) \sim  (i \om_E)^{2 \De-3} e^{i (i \om_E) t_c^-} , \quad \om_E \to -\infty \ .
\ee

\subsection{Double-trace contributions in the GFF example} \label{sec:gff}

Now consider the following example 
\be
\label{eq:GFF1}
G(\tau)^{(GFF)}=   \langle \phi(\tau) \phi(0) \rangle_\beta^{(GFF)} = \sum_{m\in \mathbb{Z}} {1\over (\tau + m\, \beta)^{2\Delta}  }\,,
\ee
which is obtained by taking the vacuum Euclidean two-point function of $\phi$ and adding images in the $\tau$ direction such that it has periodicity $\beta$. The $m=0$ term is the contribution of the identity while all other terms correspond to multi-trace contributions~\cite{Iliesiu:2018fao,Alday:2020eua}, as expanding $m \neq 0$ terms in $\tau$ we find a power series of the form~\eqref{dtope}.  
It will also be convenient to define
\be\label{eq:Lgff}
   \LL(\tau)^{(GFF)} = -\lim_{\De\to\infty} {1\over\De} \log G(\tau)^{(GFF)}
\ee
It is easy to compute $\LL(\tau)^{(GFF)} $: it is given by
\be
   \LL(\tau)^{(GFF)} = 2 \log \tau, \; \tau<{\beta\over 2}, \qquad 
    \LL(\tau)^{(GFF)} = 2 \log (\beta-\tau), \; \tau>{\beta\over 2}
\ee
and has a cusp at $\tau=\beta/2$.

We will see momentarily that the series in $\tau$ 
which defines ${1\over\De} \log G(\tau)^{(GFF)}$ is uniformly convergent for all $\De$ only for $\tau < {\beta}/ { 2}$.
Moreover, the GFF example satisfies ~\eqref{eq:RealCorr},
which also applies in holography.
It will be convenient to write
\be
\label{eq:logg}
  -{1\over\De} \log G(\tau)^{(GFF)} = 2 \log \tau - {1\over\De} 
          \log \left(1 +\sum_{m\neq0} {\tau^{2\De}\over (\tau+m \beta)^{2\De}} \right)
\ee
where we have separated the $m=0$ term (the identity operator contribution). 
We can now expand the second term for small $\tau$, 
\begin{equation}
  -{1\over\De} \log G(\tau)^{(GFF)}
    = 2 \log \tau+ \lim_{n \to \infty} \LL^{(n)}(\tau),\qquad
   \LL^{(n)}(\tau)=  \sum_{k=0}^{n}c_k(\Delta)\,\le({\tau \ov \beta} \ri)^{2\Delta +2 k}  
\end{equation}
where we have written the infinite series as a limit. 
Taking the large $\De$ limit inside the series amounts to the order of limits 
\begin{equation}\label{wbq}
    \lim_{n \to \infty}\,\lim_{\Delta\to \infty}\LL^{(n)}(\tau) =0,
    \qquad \tau <\beta\,,
\end{equation}
in which case  all
terms vanish.
However,  the limits $n\to \infty$ and $\Delta \to \infty$ may not be exchangeable.
For the other order of limits, we should take the large $\De$ limit directly in~\eqref{eq:logg}, which gives 
\begin{equation}\label{ejhw}
   \lim_{\Delta\to \infty}\,\lim_{n\to \infty}  \LL^{(n)}(\tau) 
    =\begin{cases}
    0, & \tau<\frac{\beta}{2} \\
    -2 \log \tau + 2 \log (\beta-\tau) , &  \tau>\frac{\beta}{2}
  \end{cases} \ .
\end{equation}
By comparing~\eqref{wbq} and~\eqref{ejhw}  we see that the limits can only be exchanged when $\tau<\beta/2$. %
From the Moore-Osgood theorem, it follows that the limit $n\to \infty$~(the OPE) cannot be uniformly convergent in $\Delta$ for $\tau >\beta/2$.
This example nicely illustrates the role of double traces: when $\tau$ is small, only the stress-tensor sector (identity) contributes in the large-$\Delta$ limit and the double traces can be safely ignored. 
When $\tau >\beta/2$, the double trace contribution dominates. 

Furthermore, for $\tau = {\beta \ov 2} + i t_L, t_L \in \mathbb{R}$ 
at large $\De$ terms with $m=0$ and $m=-1$ both contribute
\be 
G^{(GFF)}\left(\tau={\beta \ov 2} + i t_L\right) \approx \frac{1}{\tau^{2\Delta}} + \frac{1}{(\tau- \beta )^{2\Delta}}  \ .
\ee
We thus see the GFF example~\eqref{eq:GFF1} has exactly the same behavior as~\eqref{eq:RealCorr} except that $G_T (\tau)$ only has contribution from the identity and thus is always trivially convergent for all $\tau$ even in the large $\De$ limit. 

%%%%%%%%%%%%%%%%%%%%%%%%%%%%%%%%%%%%%%%%%%%%%%%%%%%%%
%%%%%%%%%%%%%%%%%%%%%%%%%%%%%%%%%%%%%%%%%%%%%%%%%%%%%

\section{Discussion and Implications}\label{sec:Disc} 

We showed that the stress tensor sector of the thermal two-point function of  scalar operators exhibits  bouncing singularities at finite $\De$.  We  argued that this singular behavior matches the one observed in~\cite{Fidkowski:2003nf} from the bulk geodesic analysis, and can be interpreted as a boundary reflection of  black hole singularities. Since contributions from multi-stress tensor exchanges do not depend on the specific details of the operators, they encode the universal features of  black hole singularities as probed by generic bulk fields. 
Furthermore, we expect that two-point functions of general ``light'' operators in generic ``heavy'' states  exhibit  thermal 
behavior at  leading order in the  $1/N$ expansion. Hence our conclusion can also be used to explain the universality of black hole singularities in single-sided black holes formed from gravitational collapse. 

Our results also help elucidate the role of double-trace operator contributions, and in particular connect them to geodesic results in the limit of  large conformal dimensions. 

Below we first discuss  results in other dimensions, then give a boundary interpretation of the gravity results for the boundary on a sphere,  and finally offer some future perspectives. 

\subsection{Other spacetime dimensions}

In this chapter, we have focused on $d=4$ for definiteness. Here we discuss the results in other dimensions. 

For $d=2$, which is discussed in detail in Appendix~\ref{s.kms2d}, Euclidean function $G(\tau, x)$~(including the spatial dependence) is known exactly from conformal symmetry, and can be shown to solely come from  the Virasoro descendants of the identity, which are the multi-stress operators. The stress-tensor sector  satisfies the KMS condition by itself, and does not have any unphysical singularity. 
The corresponding bulk geometry is described by the AdS Rindler, and there is no singularity behind the Rindler horizon.  
 
The $d=4$ discussion of the main text can be straightforwardly generalized to $d=6$ and $d=8$. The geodesic analysis of~\cite{Fidkowski:2003nf} yields the bouncing singularities for general $d$ at 
\begin{equation}\label{tauc0}
   \tau_c  = {\beta \over 2} \pm i {\beta \over 2}  \ {\cos {\pi\over d} \over \sin {\pi\over d}} = \pm i {\beta e^{\mp {i \pi\over d}} \over 2 \sin {\pi\over d}}\ . 
\end{equation}
As discussed in Appendix~\ref{a.ope6d},  these singularities are exactly reproduced from the asymptotic analysis of the stress tensor OPE in $d=6$ and $d=8$, with the behavior near the singularities 
\be 
G_T (\tau) \propto {1 \ov (\de \tau)^{2 \De -{d \ov 2}}} , \quad \de \tau = \tau_c - \tau  \ .
\ee
The $-{d \ov 2}$ term in the exponent $2 \De - {d \ov 2}$ is somewhat curious and it would be interesting to understand its meaning further.

Note that for $d=6$, $|\tau_c| = \beta$, while for $d=8$, $|\tau_c| > \beta$, which again highlights that the stress tensor sector contribution $G_T (\tau)$ does not obey the KMS condition. It is also curious to note 
that for both $d=6$ and $d=8$ there are additional singularities other than~\eqref{tauc0} (or their reflection in the left $\tau$ plane) at 
\begin{eqnarray}
&& d=6: \qquad \tau =  \beta , \\
&& d=8: \quad \tau =  {\beta \ov 2}  \frac{e^{\pm i\frac{\pi}{8}}}{\sin\frac{\pi}{8}}\,,
\end{eqnarray}
These singularities have ${\rm Re} \, \tau > {\beta}/ { 2}$. 
It would be interesting to see whether they play a role similar to the
bouncing singularities.  

\subsection{Boundary theory on a sphere} 

Consider the boundary CFT on a sphere $S^{d-1}$ of radius $R$ for $d >2$. 
The decomposition~\eqref{lte} still applies but the structure of each term becomes significantly more complicated. For example, now the descendants of the multiple stress tensor operators can also contribute, and there is a new dimensionless parameter ${\beta \ov R}$.
Nevertheless, we expect that the boundary theory interpretation we gave in this chapter should still apply qualitatively. 
That is, there should be singularities in the stress tensor sector contribution $G_T (\tau)$ at a general finite $\De$, which in the large $\De$ limit, ``become'' the bouncing singularities seen in the geodesic analysis. Below we give a boundary interpretation of the gravity analysis given in Section 3.3 of~\cite{Fidkowski:2003nf}. 

\begin{figure}
\centering
\includegraphics[scale=1]{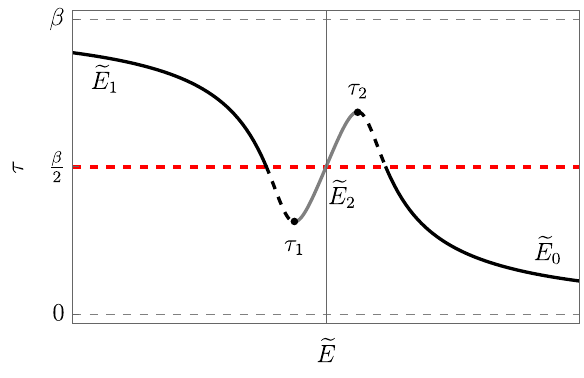}
\caption{$\tau$ as function of $\widetilde E$ obtained from the Euclidean geodesic analysis for the boundary theory on a sphere.
}
\label{fig:sphere}
\end{figure}

Figure~\ref{fig:sphere} gives Euclidean time separation $\tau$ as a function of $\widetilde E$ obtained from bulk Euclidean geodesics. In contrast with~\eqref{deo}, at $\tau = {\beta}/ { 2}$, there are three different real solutions of $\widetilde E$, labelled as $\widetilde E_0, \widetilde E_1, \widetilde E_2$  in the figure.  
Contribution from the $\widetilde E_0$  branch, which is the unique real solution for sufficiently small $\tau$, should again be identified with the large $\De$ limit of $\LL_T$. Contribution from the $ \widetilde E_1$ branch should be identified with the double trace contributions $\LL_{[\phi \phi]}$, while the middle branch $\widetilde E_2$ (in gray) does not contribute  and gives rise to the bouncing geodesics when extended to complex $\tau = {\beta \ov 2} + i t_L$. $\widetilde E_0, \widetilde E_1$ again correspond to complex geodesics for such complex $\tau$. The geodesic saddle $\widetilde E_1$ corresponding to the double-trace piece starts appearing at some value $\tau_1 < {\beta}/ { 2}$ (denoted with a dot in Figure~\ref{fig:sphere}), but is subdominant for $\tau < {\beta}/ { 2}$. Its contribution becomes the same as that from $\widetilde E_0$ at $\tau = {\beta}/ { 2}$. For $\tau > {\beta}/ { 2}$,  $\widetilde E_1$ ($i.e.$ the double-trace piece) dominates. We see that the structure is identical to~\eqref{eq:RealCorr}. 

In the large $\De$ limit, $\LL_T (\tau)$ is now analytic at $\beta/2$, but develops a square root branch point singularity at some value $\tau_2  > {\beta}/ { 2}$ (see Figure~\ref{fig:sphere}). 
Near $\tau = \tau_2$, the geodesic distance  $\hat \LL (\tau)$ corresponding to $\widetilde E_2$  branch can be obtained from $\LL_T$ as follows, 
\be \label{eya}
\LL_T (\tau) = f_T (y), \quad  y \equiv (\tau_2 - \tau)^{1 \ov 2}, \quad \hat \LL (\tau) = f_T (-y) \ .
\ee
Analytically continuing $\hat \LL (\tau)$ to complex $\tau = {\beta \ov 2} + i t_L$ should yield the bouncing singularities. 
In this case,  the double-trace piece $\LL_{[\phi \phi]}$ is still determined from $\LL_T (\tau)$ through a more involved procedure. After finding $\hat \LL (\tau)$ from $\LL_T (\tau)$ at $\tau_2$ using~\eqref{eya}, we  continue $\hat \LL (\tau)$ to $\tau_1$, where $\hat \LL (\tau)$ should encounter another square root branch point singularity.  At $\tau_1$ we can obtain $\LL_{[\phi\phi]} (\tau)$ from $\hat \LL (\tau)$ using a parallel procedure as~\eqref{eya}. 

\subsection{Future perspectives} 

Here we comment on some immediate extensions of our results and other future directions:
\vspace{1cm}
\begin{enumerate}
\item {\bf Subleading corrections and spatial dependence.} 

One can try to go a step further and analyze subleading corrections and spatial dependence to the leading singular behavior we found in $G_T (\tau)$ -- we discuss corrections in $\delta\tau$ in Appendix~\ref{app:DeltaTau} and the corrections in small $x$ in Appendix~\ref{app:XCorr}.
Our current analysis is not enough for drawing further conclusions.  
We also analyzed  the contributions of lowest-twist operators in the stress-tensor sector. We summarise our early analysis in Appendix~\ref{app:LTanal}.
Interestingly, we  find that the leading twist OPE coefficients grow quicker than the  $\Lambda_n$ coefficients 
(which contain contributions from the $[T_{\mu\nu}]^n$ operators of all twist)
at large $n$ and furthermore do not exactly reproduce the locations of the bouncing singularities. 
This shows that the leading-twist exchanges are not the dominant exchanges in the stress-tensor sector and that the subleading twists are crucial to reproduce the singular behavior. 

\item {\bf Leading twist OPE.} 

It is not surprising that the resummation of the leading twist contributions 
does not lead to the correct bouncing singularities.
The kinematical regime where the leading twist contributions are dominant
is the near-lightcone regime, where $x^-\to0$ with $x^- (x^+)^3/\beta^4$ fixed.
Clearly, this regime is very different from the $\vec x=0$ kinematics discussed in 
this chapter.
The near-lightcone limit has its advantages -- the OPE coefficients are universal \cite{Fitzpatrick:2019zqz}
and, furthermore, can be computed using CFT bootstrap \cite{Karlsson:2019dbd}.
In addition, the geodesic analysis in this limit takes a simple form \cite{Parnachev:2020fna}.
One may hope that generalization of the results of this chapter to the near-lightcone regime
may provide a useful way of computing analytic observables which can probe   black hole singularities.

\item {\bf Nature of singularities of BTZ black holes?}

For $d=2$, when the spatial direction is a circle, the bulk system is described by a  BTZ black hole, which has an orbifold singularity. The singularity is invisible to the kind of geodesic analysis reviewed in Section~\ref{sec:SC}, and thus to the kind of boundary correlation functions discussed in this chapter. The singularity is, however, visible from correlation function of the form 
\be \label{owm}
\tvev{\Phi (X) \phi (0,0)}_\beta
\ee
when the bulk point $X$ is taken to the black hole singularity~\cite{HamKab06b} (here $\Phi (X)$ is the bulk field dual to boundary operator $\phi$). $\Phi (X)$ can be expressed in terms of boundary operators using bulk reconstruction, and thus~\eqref{owm} can be written in terms of boundary correlation functions. It would be interesting to see whether it is possible to understand the nature of the BTZ singularities in terms of features of the stress tensor or double-trace contributions to the thermal correlation functions. 

\item {\bf OPE structure for the boundary CFT on a sphere.} 

It would be desirable to use the stress tensor sector for the CFT on a sphere 
for $ d> 2$ to check explicitly whether the bouncing singularities can be recovered
and whether the radius of convergence of the stress-tensor sector
at large $\De$ is larger than $\beta/2$.
For this purpose one can use the results of~\cite{Fitzpatrick:2019zqz}
for a spherical black hole geometry and repeat our analysis.

\item {\bf Effects on the black hole singularities from $\apr$ or $G_N$ corrections.}

Connecting   black hole singularities to the singularities of the stress tensor sector of boundary thermal correlators opens 
up new avenues for understanding effects on the black hole singularities from $\apr$ or $G_N$ corrections, and their possible 
resolutions at finite $\apr$ or $G_N$.%
\footnote{An alternative approach addressing black hole singularities in higher derivative gravity can be found for example in \cite{Koshelev:2024wfk} and references therein.}

We will use the example of ${\cal N} =4$ super-Yang-Mills theory with gauge group $SU(N)$ as an illustration, for which $\apr$ and $G_N$ translates into 
$1/\sqrt{\lam}$ and $1/N^2$, where $\lam$ is the 't Hooft coupling. 

Consider first a finite, but large $\lam$ in the large $N$ limit, so that we can still talk about 
bulk gravity in terms of spacetime geometry. In this case, there are two important changes:
(i) The coefficients $\Lam_n$ in~\eqref{e.resumL} become $\lam$-dependent \cite{Fitzpatrick:2020yjb}. (ii) 
At finite $\lam$, single-trace operators other than the stress tensor, which can include light operators dual to other supergravity fields and operators corresponding to stringy modes, may develop nonzero thermal expectation values, and can 
contribute to the exchanges in~\eqref{eun}. More explicitly, suppose the 
thermal expectation value of a single-trace operator $O$ is nonzero at finite $\lam$, and $O$ appears in the OPE 
expansion of $\phi (\tau)  \phi (0)$, then in~\eqref{lte} we should add contributions from the exchanges of $O^n$ for $n \in \mathbb{N}$. Denote such a contribution as $G_O (\tau)$, and we should sum over all possible such $G_O (\tau)$. 

(i) arises from $\lam$-dependence of the OPE coefficients as well as the $\lam$-dependence of the thermal expectation value of the stress tensor. These corrections could change the location of the bouncing singularities we saw in $G_T (\tau)$, and can in principle be studied perturbatively in $1/\sqrt{\lam}$ expansion. 
(ii) will make the connection of the singularities in $G_T$ to spacetime singularities discussed in Section~\ref{sec:inter}  more intricate. For example, in the large $\De$ limit,  both $G_T (\tau)$ and $G_O (\tau)$ may contribute in the small $\tau$ regime, with the contribution from $G_O$ interpreted as smearing of geodesics from interacting with stringy excitations. 

At finite $N$ ($i.e.$ finite $G_N$ in the bulk), equation~\eqref{lte} no longer applies. Neither the notion of multiple stress tensor nor double-trace operators make exact sense. But in the large $N$ limit, it should be possible to study perturbative $1/N$ corrections to $G_T (\tau)$, and could lead to insights into the nature of the black hole singularities under perturbative $G_N$ corrections. 

\item {\bf Black hole in the presence of matter fields.} 

In this chapter we considered the simplest Schwarzschild black hole. It is interesting to consider solutions with matter fields turned on which can deform the black hole interior as well as the region near the black hole singularities (see e.g.~\cite{Frenkel:2020ysx} for an example). On the boundary side, this corresponds to turning on marginal or relevant deformations. It would be interesting to see whether the stress energy sector 
still captures the bouncing singularities. 

Similarly, one can consider a spacetime corresponding to a spherically symmetric star. The spacetime would still be described by the Schwarzschild metric outside the star's radius, which is greater than the horizon radius of a black hole with equal mass, while the star's interior would depend on the matter field configuration.
It would be interesting to see how our results are altered in this case, especially because such a star solution is free of curvature singularities.

Since the stress-tensor sector is determined solely from the asymptotic expansion of the bulk equations of motion and the star and black hole spacetimes are asymptotically identical, the stress-tensor sector contributions to the OPE would be identical as well. 
As such, one can speculate that all the details about the matter content and the star's interior will be contained within the double-trace sector.

However, a hidden working assumption used in this chapter is that the near-boundary expansion provides a reliable solution to the Klein-Gordon equation in the region $r\in(0,\infty)$. For a Schwarzschild-AdS solution this is justified since the spacetime is analytic everywhere apart from the singularity at $r = 0$. For solutions with matter fields, this is no longer the case, since the Schwarzschild metric ceases to be the valid description at a non-zero radius $R_0 > 0$ (radius of the star), where a discontinuity in the metric appears. One might thus question the validity of the method in this case. It would be interesting to resolve this issue in detail.

\item {\bf Possible boundary origin of the BKL behavior.} 

In the presence of perturbations, there can be intricate chaotic dynamics in the approach to the black hole singularities~\cite{BKL70,Mis69}, called the BKL  or mixmaster behavior. Recently, such chaotic behavior has been obtained inside a four dimensional asymptotically AdS planar black hole~\cite{DeCHat23}. The connection between the black hole singularities and multiple stress tensor exchange on the boundary should be helpful to understand the boundary origin of the chaotic behavior.
\end{enumerate}

%% file: 4_chapter/chapter_4.tex
\chapter{Thermal \texorpdfstring{$TT$}{TT} Correlators}\label{ch4}

This chapter initiates the analysis of stress-tensor thermal correlators, focusing on the CFT dual to pure Einstein gravity. In a sense, this provides a minimal model for strongly interacting matter, as the only degrees of freedom are the stress tensor and its composites. We compute two-point functions in various channels -- scalar, shear, and sound -- and decompose them in the conformal blocks, extracting important CFT data such as the anomalous dimensions of the double-stress tensors and combinations of OPE coefficients and thermal one-point functions.

\section{Motivation}

Hydrodynamics describes low-energy excitations in matter at finite temperature and density \cite{Landau1987}.
A lot of interest was attracted to the hydrodynamics of conformal field theories at strong coupling and large central charge $C_T$,
which  admit a dual
gravitational (holographic) description \cite{Maldacena:1997re,Gubser:1998bc,Witten:1998qj}.
Transport coefficients  can be extracted from the two-point functions of the stress tensor ($TT$-correlators) at finite temperature
and holography maps these correlators to two-point functions of metric perturbations in a black hole background
 \cite{Policastro:2001yc,Policastro:2002se,Kovtun:2003wp,Kovtun:2004de}.

The holographic value of the shear viscosity is much closer to the experimentally observed values for quark-gluon plasma
than perturbative calculations  (see $e.g.$ \cite{Teaney:2009qa,Heinz:2013th}  for  reviews).
The ratio of the shear viscosity to the entropy density was shown to be universal,
$\eta/s =\hbar/4\pi k_B$,  in all theories
with  Einstein gravity duals \cite{Kovtun:2003wp,Kovtun:2004de,Buchel:2003tz,Iqbal:2008by}.
However, the addition of higher derivative terms to the gravity Lagrangian 
changes this  value  \cite{Kats:2007mq,Brigante:2007nu,Brigante:2008gz}. What does this imply for the hydrodynamics of strongly interacting
field theories?

In a way, gravity provides a minimal model for strongly interacting matter, where the only degrees of freedom 
are  the stress tensor and its composites, multi-stress tensors  -- they are encoded by the fluctuations of the metric in the dual theory.
From a CFT point of view, such a minimal model is defined by the OPE coefficients and the spectrum of anomalous dimensions of
multi-stress tensor operators. 
Consider the OPE coefficients which determine the three-point functions of the stress tensor, 
which are specified by the three parameters  in $d>3$ dimensions.
They change as the bulk couplings in front of the gravitational higher derivative terms are varied.\footnote{
Note that we expect consistent holographic models with generic graviton three-point couplings to also contain
higher spin fields \cite{Camanho:2014apa}. }
Presumably these OPE coefficients do not completely determine the theory, but is it possible that some sector of the theory 
is universal?

We can make progress in answering this question by decomposing the $TT$ correlator using the OPE expansion. In a minimal theory
the operators that appear are multi-stress tensor operators\footnote{
In this chapter we consider Einstein gravity as a holographic model -- it is believed to be a consistent 
truncation. In other words, in the dual CFT language, couplings to other operators and corrections to the OPE
coefficients are suppressed by the (large) gap in the spectrum of the conformal dimensions of higher spin operators -
see $e.g.$ \cite{Meltzer:2017rtf,Afkhami-Jeddi:2018own} for  recent discussions.}
 and one can in principle deduce the conformal data working order-by-order
in the temperature $T=\beta^{-1}$  [in $d$ spacetime dimensions $k$-stress tensors naturally contribute terms $\OO(\beta^{-dk})$]. 
This can be effectively done using the near-boundary expansion \cite{Fitzpatrick:2019zqz} that we explained in Chapter \ref{ch2} and used in Chapter \ref{ch3} for the scalar  two-point function.\footnote{In \cite{Karlsson:2019dbd} an alternative way of computing the stress tensor sector of the scalar correlator using conformal bootstrap
	and an ansatz, motivated by \cite{Kulaxizi:2019tkd}, was proposed. 
	The procedure of \cite{Karlsson:2019dbd}  allows one to compute the OPE coefficients with the leading twist multi-stress tensors.
	The result has many similarities to the Virasoro HHLL vacuum block (see $e.g.$ \cite{Fitzpatrick:2014vua,Fitzpatrick:2015zha,Kulaxizi:2018dxo,Karlsson:2021mgg})  but at the moment
	the full resummed correlator in $d>2$ is only known in the $\Delta_L \ra \infty$ limit \cite{Parnachev:2020fna}.
	(see \cite{,Karlsson:2019qfi,Li:2019tpf,Huang:2019fog,Fitzpatrick:2019efk,Li:2019zba,Karlsson:2019txu,Huang:2020ycs,Karlsson:2020ghx,Li:2020dqm,Fitzpatrick:2020yjb,Parnachev:2020zbr,Karlsson:2021duj,Rodriguez-Gomez:2021pfh,Huang:2021hye,Rodriguez-Gomez:2021mkk,Krishna:2021fus,Huang:2022vcs,Dodelson:2022eiz} for related work).} Note that, for scalar two-point function, two classes of contributions were relevant: multi-stress tensors that are captured by our method, and double-trace scalars, that are left undetermined in the near-boundary expansion.

What happens when a thermal state (or, in the dual language, a black hole) is probed by the stress tensor operators?
In this chapter we attempt to decompose this correlator by
generalizing the near-boundary analysis \cite{Fitzpatrick:2019zqz} to the case of external operator being the stress tensor.
Here we consider the contributions of the identity operator, the stress tensor and the double stress tensors
to the correlator.
One immediate technical complication that we face is that the external operator with which we probe the system, namely the
stress tensor, has integer conformal dimension.
In   \cite{Fitzpatrick:2019zqz}  it was observed that some OPE coefficients have poles for integer values of the conformal dimension of the external scalar operator.
This feature is related to mixing of double stress and double trace operators.
The OPE coefficients for both series have poles which cancel,
leaving behind logarithmic terms.
One can also observe that the coefficients of these terms cannot be fixed by the near-boundary analysis \cite{Fitzpatrick:2019zqz,Li:2019tpf}.

In the case of the stress tensor the double-trace operators made out of the external operator $T_{\mu\nu}$ {\it are also}  double-stress tensor operators.
One may wonder if their OPE coefficients can be determined from the near boundary analysis.
The answer turns out to be no.
Another important difference from  \cite{Fitzpatrick:2019zqz}  is related to the leading behaviour of the OPE coefficients of two
stress tensors and a double stress tensor.
This OPE coefficient scales like one (for unit-normalized operators), as opposed to $\OO(1/C_T)$ in the scalar case, and gives rise to the disconnected
part of  the correlator.
This implies that the connected part of the $TT$ correlator contains information about conformal data which is subleading in the
$1/C_T$ expansion.
This leads to some complications\footnote{
In particular, it has been observed  in the $d=2$ case that thermalization of multi stress-tensor operators happens only
to leading order in the $1/C_T$ expansion (see \cite{Basu:2017kzo,He:2017txy,Datta:2019jeo} for recent discussions).}
 but in the end, we succeed at extracting the leading $1/C_T$ contributions to the anomalous dimensions of the double trace operators
and to the leading lightcone behavior of the $TT$ correlators.
Other conformal data at this order remains undetermined -- it should be thought of as an analog of the double trace operator data in the 
external scalar case.

Let us mention another technical difficulty that we need to confront in the case of external stress tensors.
In \cite{Fitzpatrick:2019zqz}  the symmetry implies that the two-point function depends on the time $t$, the spatial radial coordinate $\rho$ and 
the AdS radial coordinate $r$.
This is no longer true in the stress tensor case, due to the presence of distinct polarizations.
We handle this by computing stress tensor correlators integrated over two parallel $xy$-planes separated in the transverse spatial 
direction, which we denote by $z$. 
There are three independent choices of polarization, distinguished by the  transformation properties with respect
to rotations of the plane of integration.
A suitable modification of the ansatz used in  \cite{Fitzpatrick:2019zqz} allows us to solve the stress tensor problem.
However, integrating over the $xy$-plane leads to some divergent contributions and to additional logarithmic terms.
Fortunately, this does not affect our ability to extract the anomalous dimensions.

The rest of this chapter is organized as follows. In Section \ref{Section:Bulk}, we consider metric perturbations on top of a planar AdS-Schwarzschild black hole and compute the stress tensor two-point function in a near-boundary expansion (OPE limit in the dual CFT). In Section \ref{CFTside4d}, we perform the OPE expansion of the stress tensor thermal two-point functions in $d=4$ and by comparison to the bulk calculations in the previous section, we read off the anomalous dimensions of double-stress tensor operators with spin $J=0,2,4$
and determine the near-lightcone behavior of the correlators. We conclude with a discussion in Section \ref{sec:disc}. 
In Appendix \ref{AppendixA}, we treat the simpler example of scalar perturbations in the bulk as a toy model for the metric perturbations, 
focusing  on the subtleties that arise for external operators with integer dimensions.
In addition, we consider scalar correlators integrated over the $xy$-plane and 
show  how the correct OPE data is recovered in this case. 
Appendix \ref{AppendixB} lists some of the results that are too lengthy to present in Section \ref{Section:Bulk}.
 In Appendix \ref{app:SpinningBlocks}, we present conventions and details on the spinning conformal correlators
 relevant for the decomposition of  thermal stress tensor two-point functions.

\section{Holographic calculation of thermal  \texorpdfstring{$TT$}{TT} correlators}\label{Section:Bulk}

Recently some OPE coefficients of scalars and multi-stress tensors  were calculated in 
the context of holographic models \cite{Fitzpatrick:2019zqz,Li:2019tpf}. 
As explained in Chapter \ref{ch2}, this was accomplished by making a  comparison 
between the CFT conformal block decomposition of vacuum four-point function with two heavy external scalars (so-called HHLL correlators) on the CFT side 
and a near-boundary expansion of
the bulk-to-boundary propagator in the AdS-Schwarzschild background on the bulk side.

Our goal in this chapter is to use an analogous approach to extract the CFT data\footnote{By the CFT data we mean products of the OPE coefficients and thermal one-point functions and anomalous dimensions of the double-trace stress tensors. This will be explained in greater detail in the next section.} for the stress tensor two-point function in a thermal state dual to the AdS-Schwarzschild black hole, in this section we will focus on the bulk part of this calculation. In practice we will consider the integrated version of the correlator
\begin{equation}\label{eq:defHTTH}
	G_{\mu\nu,\rho\sigma}(t,z):=\int_{\mathbb{R}^2} dx dy\langle T_{\mu\nu}(x^\alpha)T_{\rho\sigma}(0)\rangle_{\beta}.
\end{equation}

To compute the $TT$ correlator, it is necessary to consider the linearized Einstein equations in the black hole background.
For technical reasons, we take the large volume limit, where all conformal descendants decouple; this corresponds to considering the planar asymptotically AdS  black hole. 
The corresponding system of PDEs is technically difficult to solve because different polarizations mix with each other. 
To make the problem tractable, we integrate the correlator over two spatial directions in \eqref{eq:defHTTH}.
The resulting fluctuation equations simplify to  three independent PDEs for the three different polarizations.
We   show explicitly that  an  ansatz of  \cite{Fitzpatrick:2019zqz,Li:2019tpf}, suitably modified to fit our needs, 
successfully solves these equations.

As a warm-up exercise, we consider the scalar case, discussed in \cite{Fitzpatrick:2019zqz,Li:2019tpf}, but now integrate 
over the $xy$-plane.
The details of this calculation are described in Appendix \ref{AppendixA}, but the summary is as follows.
For non-integer values $\Delta_L$  of the conformal dimensions of the external scalar operator
all coefficients in the ansatz are fixed, order-by-order,
by imposing the scalar field equations of motion in the bulk.
Matching to the conformal block expansion then yields the OPE coefficients of scalars and multi-stress tensors,
which reproduce the results of \cite{Fitzpatrick:2019zqz}.
Note that the integrals are only convergent for large $\Delta_L$, but their analytic continuation to small  $\Delta_L$ yields the correct results.

For integer $\Delta_L$ there is mixing between multi-stress and multi-trace operators, which results in logarithmic terms \cite{Fitzpatrick:2019zqz}.
This mixing is reflected in the appearance of the $\log r$ terms in the bulk ansatz \cite{Li:2019tpf}; a closely related fact is 
that not all coefficients in the ansatz are now determined by the bulk equations of motion.
For example, for $\Delta_L=4$  there is one undetermined parameter at $\OO(\mu^2)$; it corresponds to
an undetermined factor in a double-trace OPE coefficient.

As explained in Appendix \ref{AppendixA}, the addition of spatial integration leads to an additional undetermined coefficient in the ansatz. 
This  coefficient is, roughly speaking, related to the volume of the $xy$-plane we are integrating over.
In practice, we use dimensional regularization, so instead of the volume, a $1/\epsilon$ pole appears in the expression for
this undetermined coefficient.
The other undetermined coefficient is related to the logarithmic term, just as in the non-integrated case.
In summary, we conclude that in the  scalar case, the spatial integration does not affect our ability to read off the OPE data.

In this section we perform the bulk calculations for the case where the external operator is the stress tensor. 
In other words, we compute the OPE expansion for the thermal $TT$ correlator in holographic CFTs.
This section is organized as follows. First we consider metric perturbations around a planar AdS-Schwarzschild black hole. Then we integrate out two out of five space-time directions and, following \cite{Kovtun:2005ev,Kovtun:2006pf}, we utilize the resulting $O(2)$ symmetry together with the bulk gauge freedom to reformulate the problem in terms of the 
three gauge invariant combinations of the gravitational fluctuations in the AdS-Schwarzschild background.
The resulting PDEs can then be solved one by one using the ansatz \cite{Fitzpatrick:2019zqz,Li:2019tpf}, naturally adapted to the integrated case. Finally, using the holographic dictionary, we derive the stress tensor two-point function in a thermal state for various polarizations. In Section \ref{CFTside4d} we compare these results with the CFT conformal block decomposition and extract conformal data.  

\subsection{Linearized Einstein equations}

We consider the Einstein-Hilbert action with a cosmological constant\footnote{Throughout Chapters \ref{ch4}–\ref{ch6}, we will exclusively use Euclidean time. To simplify notation, we denote it by $t$.}
\begin{equation}\label{EHAction}
	S=\frac{1}{16\pi G_5}\int\dd^5x\sqrt{g}(\mathcal{R}-2\Lambda),
\end{equation}
where $G_5$ is the five-dimensional gravitational constant, $\mathcal{R}$ is the Ricci scalar and $\Lambda$ is the cosmological constant. Decomposing the metric as the background part plus a small perturbation $h_{\mu\nu}$, one obtains the linearized Einstein equations
in the form
\begin{equation}\label{LEE}
	\mathcal{R}^{(1)}_{\mu\nu}+dh_{\mu\nu}=0,
\end{equation}
where $\mathcal{R}_{\mu\nu}^{(1)}$ is the linearized Ricci tensor and $d$ is the dimension of the conformal boundary, $i.e.$ $d=4$ in our case. 

We will be interested in the planar AdS-Schwarzschild black hole as the background spacetime,
\begin{equation}\label{bhmetrics}
	\dd s^2 = r^2f(r)\dd t^2+r^2\dd \vec{x}\,^2+\frac{1}{r^2f(r)}\dd r^2,
\end{equation}
where $\vec{x}=(x,y,z)$ and $f(r)=1-\frac{\mu}{r^4}$.
Here and in the rest of the chapter we set the radius of the AdS space to unity.

By solving the linearized Einstein equations (\ref{LEE}) with the appropriate boundary conditions, we obtain the metric perturbation $h_{\mu\nu}$ and, in principle, the holographic dictionary then precisely determines the correlators in the four-dimensional CFT on the boundary. However, due to the complicated form of these equations, this is difficult to do in practice.

To make this problem tractable, we integrate the bulk-to-boundary propagator over the $xy$-plane. This will simplify the equations of motion to three independent PDEs, which we will be able to solve using the ansatz \cite{Fitzpatrick:2019zqz,Li:2019tpf}.
As a result, the corresponding CFT correlators, which  we obtain via holographic dictionary, will be integrated over 
the $xy$ directions. This will be  studied in Section \ref{CFTside4d} from the  CFT point of view.

\subsection{Polarizations and gauge invariants}

Our aim is to solve the linearized Einstein equations \eqref{LEE} in the background \eqref{bhmetrics}, with the solution integrated over two spatial directions, which we choose to be $x$ and $y$.

Upon  integration, the (linearized) gravitational action will exhibit an $O(2)$ rotational symmetry. This property allows us to divide the components $h_{\mu\nu}$ into three representations (referred to as channels in this context) which can be studied separately:
\begin{align}
	\text{Sound channel}&:\qquad h_{tt},\,h_{tz},\,h_{zz},\,h_{rr},\,h_{tr},\,h_{zr},\,h_{xx}+h_{yy}\\
	\text{Shear channel}&:\qquad h_{tx},\,h_{ty},\,h_{zx},\,h_{zy},\,h_{rx},\,h_{ry}\\
	\text{Scalar channel}&:\qquad h_{\alpha\beta}-\delta_{\alpha\beta}(h_{xx}+h_{yy})/2.
\end{align}
The sound channel has spin 0, shear channel has spin 1 and the scalar channel (whose equations of motion will be identical 
to that of the scalar) has spin 2 under $O(2)$.

In every channel, we can define a quantity $Z_i$ \cite{Kovtun:2005ev,Kovtun:2006pf}, that is invariant under the gauge transformations $h_{\mu\nu}\rightarrow h_{\mu\nu}-\nabla_\mu\xi_\nu-\nabla_\nu\xi_\mu$ of the gravitational bulk theory. In the position space these are
\begin{align}
	Z_1&=\partial_zH_{tx}-\partial_tH_{xz}\label{vectorinvdef}\\
	Z_2&=2f\partial_z^2H_{tt}-4\partial_t\partial_zH_{tz}+2\partial_t^2H_{zz}-\left((f+\frac{r}{2}f')\partial_z^2+\partial_t^2\right)(H_{xx}+H_{yy})\label{soundinvdef}\\
	Z_3&=H_{xy}\label{tensorinvdef},
\end{align}
where $H_{tt}=h_{tt}/fr^2$, $H_{ti}=h_{ti}/r^2$ and $H_{ij}=h_{ij}/r^2$ for $i,j\in\{x,y,z\}$, $f=f(r)$ is the function appearing in the black hole metric and the prime denotes the derivative with respect to $r$. 
As is conventional, we refer to  $Z_1$, $Z_2$  and  $Z_3$ as  the shear channel invariant, the sound channel invariant and the scalar channel
invariant, respectively.

We can now choose a particular channel, take the linearized Einstein equations \eqref{LEE} and assume the metric perturbation 
to be of the form $h_{\mu\nu}=h_{\mu\nu}(t,z,r)$. Combining the resulting equations, we get PDEs for the invariants. 
It will be useful to define the following quantities:
\begin{align}
	c_1&\coloneqq (3\mu^2-8\mu r^4+5r^8)/r^5\\
	c_2&\coloneqq 2\mu(r^4-\mu)/r^5\\
	c_3&\coloneqq (\mu-r^4)^2/r^4\\
	c_4&\coloneqq 16\mu^2(r^4-\mu)/(3r^{10})\\
	c_5&\coloneqq 1+\mu(\mu-4r^4)/(3r^8)\\
	c_6&\coloneqq 2-4\mu/(3r^4)\\
	c_7&\coloneqq (\mu^2-6\mu r^4+5r^8)/r^5\\
	c_8&\coloneqq (r^4-\mu)(9\mu^2-16\mu r^4+15r^8)/(3r^9)\\
	c_9&\coloneqq -(\mu-3r^4)(\mu-r^4)^2/(3r^8).
\end{align}
The equations of motion for the invariants are then given by\footnote{These are the equations one finds by Wick rotating and Fourier transforming the corresponding ODEs presented in \cite{Kovtun:2005ev}.}
\begin{align}
	0&=(\partial_t^2+f\partial_z^2)^2Z_1+\left(c_1(\partial_t^2+f\partial_z^2)+c_2(\partial_t^2-f\partial_z^2)\right)Z_1'+c_3(\partial_t^2+f\partial_z^2)Z_1''\label{es1}\\
	0&=(c_4\partial_z^2+c_5\partial_z^4+c_6\partial_t^2\partial_z^2+\partial_t^4)Z_2+(c_7\partial_t^2+c_8\partial_z^2)Z_2'+(c_3\partial_t^2+c_9\partial_z^2)Z_2''\label{es2}\\
	0&=(\partial_t^2+f\partial_z^2)Z_3+c_7Z_3'+c_3Z_3''.\label{es3}
\end{align}

\subsection{Ansatz and the vacuum propagators}\label{susucica}

In order to solve (\ref{es1})-(\ref{es3}) we need to find the bulk-to-boundary propagators $\mathcal{Z}_i$, which are related to the invariants by
\begin{equation}\label{prevodis}
	Z_i(t,z,r)=\int\dd t'\dd z'\mathcal{Z}_i(t-t',z-z',r)\hat{Z}_i(t',z'),
\end{equation}
where $\hat{Z}_i$ is related to the boundary value (up to derivatives) of $Z_i$ as will be explained below. To solve the equations of motion we use the ansatz \cite{Fitzpatrick:2019zqz, Li:2019tpf} introduced for the case of a scalar field in a black hole background, suitably modified for our integrated case. Concretely, we want to solve the bulk equations of motion in the regime:\footnote{Note that in the original non-integrated case one has $|\vec{x}|$ instead of $z$.}
\begin{equation}\label{limitoss}
	r\rightarrow\infty\qquad\text{with}\qquad rt,\, rz\,\,\,\text{fixed}
\end{equation}
Let us remind that the intuition behind this limit is the expectation that the bulk solution becomes sensitive only to the near-boundary region as the CFT operators approach each other.

To realize \eqref{limitoss} in practice, we introduce new coordinates defined by
\begin{align}
	\rho&\coloneqq rz\\
	w^2&\coloneqq 1+r^2t^2+r^2z^2\,.
\end{align}
In these coordinates the limit is $r\rightarrow\infty$ with $w$ and $\rho$ held fixed.

According to \cite{Fitzpatrick:2019zqz}, one expects the solution to be of the form of the product of the pure AdS propagator and an expansion in $1/r$, where at each order we have a polynomial $\sum_i\alpha_i(w)\rho^i$. Substituting this into the equations of motion, we can find analytical solutions for all $\alpha_i(w)$. Imposing regularity in the bulk and demanding the proper boundary behaviour\footnote{By the proper boundary behaviour we mean that the boundary limit of the bulk solution should reproduce the form of the boundary correlators expected from the boundary CFT.}, we determine the integration constants and find the coefficients $\alpha_i(w)$ as polynomials in $w$.

If there are logarithmic terms\footnote{Logarithmic terms appear, for example, in the case of a scalar field with integer conformal dimension $\Delta_L$ or in the presence of anomalous dimensions as in the case of the stress tensor thermal two-point function. They can  also be produced upon integration. We comment more on the origin of these terms in appendix \ref{AppendixA}.}
$\mathcal{Z}_i$ takes the form \cite{Li:2019tpf}
\begin{equation}\label{thetheansatz}
	\mathcal{Z}_i=\mathcal{Z}_i^{AdS}\left(1+\frac{1}{r^4}\left(G^{4,1}_i+G^{4,2}_i\log r\right)+\frac{1}{r^8}\left(G^{8,1}_i+G^{8,2}_i\log r\right)+\ldots\right),    
\end{equation}
where $\mathcal{Z}_i^{AdS}$ is the vacuum bulk-to-boundary propagator for the invariant $Z_i$ and $G^{4,j}_i$, $G^{8,j}_i$, $\ldots$, $j\in\{1,2\}$, $i\in\{1,2,3\}$ are given by\footnote{We use the bounds of the sums as they were derived for the case of a scalar field \cite{Fitzpatrick:2019zqz,Li:2019tpf}. As we will see, this will be valid also for the stress tensor case in the scalar and shear channels. In the sound channel we will need a slight modification of the ansatz.} (we suppress the channel index for simplicity)
\begin{align}
	&G^{4,j}=\sum_{m=0}^2\sum_{n=-2}^{4-m}(a^{4,j}_{n,m}+b^{4,j}_{n,m}\log w)w^n\rho^m\\
	&G^{8,j}=\sum_{m=0}^6\sum_{n=-6}^{8-m}(a^{8,j}_{n,m}+b^{8,j}_{n,m}\log w)w^n\rho^m\\
	&\phantom{wwwwww}\vdots\nonumber
\end{align}
Here $G^{4,j}$ corresponds to the stress tensor contribution ($\propto\mu^1$) and $G^{8,j}$ corresponds to the double-stress tensor contributions. We expect to find $b^{4,1}=0$ and $G^{4,2}=0$ in all three channels. 

The vacuum propagator $\mathcal{Z}_i^{AdS}$ for the $i$-th channel can be determined using the AdS bulk-to-boundary propagators for the various components of the metric perturbation.
Let us describe this calculation in more detail. The AdS propagator for $H_{\mu\nu}$ was computed in \cite{Liu:1998bu} and in the five dimensional bulk case can be expressed as 
\begin{equation}
	\mathfrak{G}_{\mu\nu,\rho\sigma}=\frac{10r^4}{\pi^2(1+r^2(t^2+\vec{x}^2))^4}J_{\mu\alpha}J_{\nu\beta}P_{\alpha\beta,\rho\sigma},
\end{equation}
where $J_{\mu\nu}$ and $P_{\mu\nu,\rho\sigma}$ are given by
\begin{align}
	J_{\mu\nu}&=\delta_{\mu\nu}-\frac{2x_\mu x_\nu}{\frac{1}{r^2}+t^2+\vec{x}^2}\\
	P_{\mu\nu,\rho\sigma}&=\frac12(\delta_{\mu\rho}\delta_{\nu\sigma}+\delta_{\nu\rho}\delta_{\mu\sigma})-\frac14\delta_{\mu\nu}\delta_{\rho\sigma}.
\end{align}
Integrating out the $x$ and $y$ directions, we get
\begin{equation}\label{vacHsgreen}
	\mathcal{G}_{\mu\nu,\rho\sigma}(t,z,r)\coloneqq\int_{\mathbb{R}^2}\dd x\dd y\,\mathfrak{G}_{\mu\nu,\rho\sigma}(t,x,y,z,r),
\end{equation}
from which we find the (integrated) AdS solution for $H_{\mu\nu}$ 
\begin{equation}\label{vacHs}
	H_{\mu\nu}(t,z,r)=\int\dd t'\dd z'\mathcal{G}_{\mu\nu\rho\sigma}(t-t',z-z',r)\hat{H}_{\rho\sigma}(t',z'),
\end{equation}
where $\hat{H}_{\mu\nu}$ are the sources, $i.e.$ the values of the bulk solution on the conformal boundary.

Substituting \eqref{vacHs} into the definitions of the invariants (\ref{vectorinvdef})-(\ref{tensorinvdef}), one can
accordingly read off the AdS bulk-to-boundary propagators $\mathcal{Z}^{AdS}_i$.

Here we list the resulting expressions for some particular choices of the sources:
{\setlength{\tabcolsep}{19pt}
	\renewcommand{\arraystretch}{2.1}%2.2
	\begin{center}
		\begin{tabular}{c|c c}
			Sources     &  $(t,z,r)$-result & $(w,\rho,r)$-result\\
			\hline\hline
			$\hat{H}_{xy}$    &   $\mathcal{Z}_3^{AdS}=\frac{2r^2}{\pi\left(r^2\left(t^2+z^2\right)+1\right)^3}$    &$=\frac{2r^2}{\pi w^6}$\\[1ex]
			\hline
			$\hat{H}_{tx}$    &   $\mathcal{Z}_1^{AdS}=-\frac{12 r^4 z}{\pi\left(r^2\left(t^2+z^2\right)+1\right)^4}$     &$=-\frac{12r^3\rho}{\pi w^8}$\\
			$\hat{H}_{xz}$    &   $\mathcal{Z}_1^{AdS}=\frac{12 r^4 t}{\pi  \left(r^2 \left(t^2+z^2\right)+1\right)^4}$      &$=\frac{12r^3\sqrt{w^2-1-\rho^2}}{\pi w^8}$\\[1ex]
			\hline
			$\hat{H}_{tz}$    &   $\mathcal{Z}_2^{AdS}=-\frac{384 r^6 t z}{\pi  \left(r^2 \left(t^2+z^2\right)+1\right)^5}$  &$=-\frac{384r^4\rho\sqrt{w^2-1-\rho^2}}{\pi w^{10}}$\\
			$\hat{H}_{tt}$    &   $\mathcal{Z}_2^{AdS}=-\frac{24 \left(r^6 \left(t^2-7 z^2\right)+r^4\right)}{\pi  \left(r^2 \left(t^2+z^2\right)+1\right)^5}$  &$=-\frac{24r^4(w^2-8\rho^2)}{\pi w^{10}}$\\
			$\hat{H}_{xx}$   &   $\mathcal{Z}_2^{AdS}=\frac{24 r^4-72 r^6 \left(t^2+z^2\right)}{\pi  \left(r^2 \left(t^2+z^2\right)+1\right)^5}$  &$=\frac{24r^4(4-3w^2)}{\pi w^{10}}$\\  
			$\hat{H}_{zz}$    &   $\mathcal{Z}_2^{AdS}=\frac{24 r^4 \left(r^2 \left(7 t^2-z^2\right)-1\right)}{\pi  \left(r^2 \left(t^2+z^2\right)+1\right)^5}$  &$=\frac{24r^4(7w^2-8(1+\rho^2))}{\pi w^{10}}$
		\end{tabular}
\end{center}}

At this point, we have all the pieces needed for the ansatz (\ref{thetheansatz}). 
Inserting it into equations (\ref{es1})-(\ref{es3}), we can determine the coefficients $a_{n,m}^{k,j}$ and $b_{n,m}^{k,j}$. 
We next proceed to discuss the results channel-by-channel.

\subsection{Scalar channel}\label{sekskalal}

We begin by considering the scalar channel where the equation of motion (\ref{es3}) has the simplest form. In this chapter we confine our attention to the contributions due to the identity operator ($\mu^0$), the stress tensor ($\mu^1$) and double-stress tensors ($\mu^2$). We are therefore interested in finding $G^{4,1}_i$, $G^{4,2}_i$, $G^{8,1}_i$ and $G^{8,2}_i$ in the ansatz (\ref{thetheansatz}). 

In the scalar channel, we may either turn on the source $\hat{H}_{xy}\neq0$ or $\hat{H}_{xx}=-\hat{H}_{yy}\neq0$. Since these differ only by an $O(2)$ rotation, the corresponding bulk solutions, as well as the form of the action will be identical. 
For this reason, we will restrict our attention to the case where $\hat{H}_{xy}\neq0$. Hence, the invariant $Z_3$ 
is given by
\begin{equation}\label{formofZ3}
	Z_3(t,z,r)=\int\dd t'\dd z'\mathcal{Z}_3^{(xy)}(t-t',z-z',r)\hat{H}_{xy},
\end{equation}
where $\mathcal{Z}_3^{(xy)}$ is the bulk-to-boundary propagator\footnote{The superscript index in the parenthesis specifies the choice of the non-zero sources.}.

Transforming  (\ref{es3}) into the $(w,\rho,r)$-coordinates with $\mathcal{Z}_3^{(xy)}$ given by (\ref{thetheansatz}), we 
find the solution at $\OO(\mu)$,
\begin{equation}
	\mathcal{Z}_3^{(xy)}\Big|_{\mu^1}=\frac{\mu \left(w^6+w^4+6 w^2-2 \rho ^2 \left(w^4+2 w^2+3\right)-12\right)}{5 \pi  r^2 w^8}.
\end{equation}
As expected, there are no log terms in this case. At $\OO(\mu^2)$ we find 
\begin{equation}
	\begin{split}
		\mathcal{Z}_3^{(xy)}\Big|_{\mu^2}=&\frac{\mu^2}{4200 \pi  r^6 w^{10}} \Big[120 w^{10} \left(-4 \rho ^2+5 w^2-6\right) (\log (w)-\log (r))+655 w^8\\
		\phantom{\frac11}&+448 w^6+3136 w^4-12656 w^2+56 \rho ^4 \left(10 w^8+20 w^6+35 w^4+44w^2+36\right)
		\\&-4 \rho ^2 \left(750 w^{10}+40 w^8+345 w^6+476 w^4+448 w^2-2016\right)+8064\Big]\\
		&+\frac{2}{\pi  r^6}\left[\left(1-6 \rho ^2\right) a^{8,1(xy)}_{6,0}+a^{8,1(xy)}_{8,0} \left(w^2-8 \rho ^2\right)\right],
	\end{split}
\end{equation}
where the coefficients $a^{8,1(xy)}_{6,0}$ and $a^{8,1(xy)}_{8,0}$ are not fixed by the near-boundary analysis. 
We also see the presence of log terms which are due to the $xy$-integration and the anomalous dimensions of the double-stress tensors.

\subsubsection{$G_{xy,xy}$}\label{holoprvy}

We now use the holographic dictionary to determine the thermal correlator $G_{xy,xy}$. The action for the scalar invariant $Z_3$ (and $Z_1$ and $Z_2$ below) can be obtained by Fourier transforming and Wick rotating the result obtained in \cite{Kovtun:2005ev}:
\begin{equation}\label{S3ss}
	S_3=\frac{\pi^2C_T}{160}\lim_{r\rightarrow\infty}\int\dd t\dd zr^5\left(1-\frac{\mu}{r^4}\right)\partial_rZ_3(t,z,r)Z_3(t,z,r).
\end{equation}
The invariant $Z_3(t,z,r)$ is fully determined by the bulk-to-boundary propagator $\mathcal{Z}_3^{(xy)}$ via 
eq. \eqref{formofZ3}. 
To compute the action \eqref{S3ss}, we expand $\mathcal{Z}_3^{(xy)}$ near $r=\infty$ as
\begin{equation}
	\mathcal{Z}_3^{(xy)}(t,z,r)=\delta^{(2)}(t,z)+\frac{1}{r^4}\zeta_{3}^{(xy)}(t,z)+\ldots,
\end{equation}
where the ellipses represent subleading contact terms of $\mathcal{O}(\frac{1}{r^2})$ of the schematic form $\partial^n\delta/r^n$ as well as contributions analytic in $(t,z)$ that are $\mathcal{O}(\frac{1}{r^6})$. As we will see, in the scalar channel $G_{xy,xy}\propto\zeta_3^{(xy)}$.

To proceed, we substitute the bulk-to-boundary propagator into the action (\ref{S3ss}):
\begin{align}
	S_3=\frac{\pi^2C_T}{160}\lim_{r\rightarrow\infty}\!\int&\dd^2x\dd^2x'\dd^2x''(r^5\!-\!\mu r)\partial_r\mathcal{Z}_3^{(xy)}(x\!-\!x',r)\mathcal{Z}_3^{(xy)}(x\!-\!x'',r)\hat{H}_{xy}(x')\hat{H}_{xy}(x'')\nonumber\\
	=&-\frac{\pi^2C_T}{40}\int\dd^2x\dd^2x'\zeta^{(xy)}_3(x-x')\hat{H}_{xy}(x)\hat{H}_{xy}(x')\,,
\end{align}
where in the second line we have integrated the delta function. 
We have  used an abbreviated notation $x=\{t,z\}$, $x'=\{t',z'\}$ and $x''=\{t'',z''\}$ and omitted contact terms
 (see $e.g.$ \cite{Skenderis:2002wp} for a review on holographic renormalization and the treatment of contact terms).

We can now compute the CFT correlator,
\begin{equation}
	G^{bulk}_{xy,xy}=\expval{T_{xy}(t,z)T_{xy}(0,0)}_\beta=-\frac{\delta^2S_3}{\delta\hat{H}_{xy}(t,z)\delta\hat{H}_{xy}(0,0)}=\frac{\pi^2C_T}{20}\zeta^{(xy)}_3(t,z)
\end{equation}
Inserting the explicit bulk solution, we obtain the following results order-by-order in $\mu$:
\begin{align}
	G_{xy,xy}^{(bulk)}\Big|_{\mu^0}=&\frac{\pi C_T}{10(t^2+z^2)^3}\label{eq:Bulkxyxy0}\\
	G_{xy,xy}^{(bulk)}\Big|_{\mu^1}=&\frac{\pi \mu C_T(t^2-z^2)}{100(t^2+z^2)^2}\label{eq:Bulkxyxy1}\\
	G_{xy,xy}^{(bulk)}\Big|_{\mu^2}=&\frac{\pi \mu^2C_T}{4200} \left(3 \left(5 t^2+z^2\right) \log \left(t^2+z^2\right)-\frac{2 \left(75 t^2 z^2+61 z^4\right)}{t^2+z^2}\right)\nonumber\\
	&+\frac{1}{10} \pi  C_T \left(a^{8,1(xy)}_{8,0} \left(t^2-7 z^2\right)-6 z^2 a^{8,1(xy)}_{6,0}\right)\label{eq:Bulkxyxy2}.
\end{align}
We will compare them with the CFT calculations in the next section.

\subsection{Shear channel}

We can repeat the above procedure  to solve the shear channel bulk equation (\ref{es1}) for the sources $\hat{H}_{tx}$ and $\hat{H}_{xz}$
and express the results in terms of  $w$, $\rho$ and $r$.
The explicit expressions are listed in appendix \ref{AppendixB}.
We will now use them to determine $G_{tx,tx}$ and $G_{xz,xz}$ using the AdS/CFT dictionary;
these calculations are summarized below.

\subsubsection{$G_{tx,tx}$ and $G_{xz,xz}$}

The action for the shear channel invariant is given by\footnote{Note the presence of the inverse operator $(\partial_t^2+\partial_z^2)^{-1}$ which is a Fourier transform of $(\omega^2+q^2)^{-1}$  that appears in the action derived in  \cite{Kovtun:2005ev}.} \cite{Kovtun:2005ev}
\begin{equation}\label{shearshearaction}
	\begin{split}
		S_1=&\frac{\pi^2C_T}{160}\lim_{r\rightarrow\infty}\int\dd t\dd z\frac{\left(1-\frac{\mu}{r^4}\right)r^5}{\partial_t^2+\partial_z^2\left(1-\frac{\mu}{r^4}\right)}\partial_rZ_1(t,z,r)Z_1(t,z,r)\\
		=&\frac{\pi^2C_T}{160}\lim_{r\rightarrow\infty}\int\dd t\dd z\left(\frac{r^5}{\partial_t^2+\partial_z^2}+\mathcal{O}(r^2)\right)\partial_rZ_1(t,z,r)Z_1(t,z,r).
	\end{split}
\end{equation}

We begin by turning on  the source $\hat{H}_{tx}$ and follow the same approach as in  subsection \ref{holoprvy}. The shear 
channel invariant is then given by
\begin{equation}\label{formofZ1}
	Z_1(t,z,r)=\int\dd t'\dd z'\mathcal{Z}_1^{(tx)}(t-t',z-z',r)\hat{H}_{tx},
\end{equation}
where $\mathcal{Z}_1^{(tx)}$ is the bulk-to-boundary propagator corresponding to our choice of  source.

The near-boundary expansion of $\mathcal{Z}_1^{(tx)}$ reads
\begin{equation}\label{expansiono}
	\mathcal{Z}_1^{(tx)}=\partial_z\delta^{(2)}(t,z)+\frac{1}{r^4}\zeta_1^{(tx)}+\frac{\log{r}}{r^4}\zeta_{1,log}^{(tx)}+\ldots,
\end{equation}
where the ellipses correspond to contact terms which are $\mathcal{O}\left(\frac{1}{r^2}\right)$ and non-contact terms which are $\mathcal{O}\left(\frac{1}{r^6}\right)$. Here, however, we encounter $\log r$ terms in the expansion,
\begin{equation}
	\zeta_{1,log}^{(tx)}=-\frac{z\left(840 a^{8,2(tx)}_{8,0}+41 \mu^2\right)}{70 \pi }.
\label{zetazero}
\end{equation}
This term diverges as $r\rightarrow \infty$, unless the value of the coefficient $a^{8,2(tx)}_{8,0}$ is fixed to be
\begin{equation}
	a^{8,2(tx)}_{8,0}=-\frac{41}{840}\mu^2.
\end{equation}

Using the expansion \eqref{expansiono} in the action (\ref{shearshearaction}) and proceeding as in the scalar channel case, we obtain
\begin{equation}
	G_{tx,tx}^{bulk}=\frac{\pi^2C_T}{20}\frac{\partial_z}{\partial_t^2+\partial_z^2}\zeta_1^{(tx)}
\end{equation}
Thus, we arrive at
\begin{align}
	G_{tx,tx}^{(bulk)}\Big|_{\mu^0}=&-\frac{1}{\partial_t^2+\partial_z^2}\frac{3 \pi  C_T \left(t^2-7 z^2\right)}{5 \left(t^2+z^2\right)^5}\label{eq:Bulktxtx0}\\
	G_{tx,tx}^{(bulk)}\Big|_{\mu^1}=&\frac{1}{\partial_t^2+\partial_z^2}\frac{3 \pi  \mu C_T \left(t^4-6 t^2 z^2+z^4\right)}{200 \left(t^2+z^2\right)^4}\label{eq:Bulktxtx1}\\
	G_{tx,tx}^{(bulk)}\Big|_{\mu^2}=&-\frac{1}{\partial_t^2+\partial_z^2}\Bigg[\frac{\pi  \mu^2 C_T}{8400} \left(\frac{2 \left(669 t^4 z^2+804 t^2 z^4+271 z^6\right)}{\left(t^2+z^2\right)^3}+123 \log \left(t^2+z^2\right)\right)\nonumber\\
	&+\frac{3}{5} \pi  a^{8,1(tx)}_{8,0} C_T\Bigg].\label{eq:Bulktxtx2}
\end{align}
Here we keep the inverse operator $(\partial_t^2+\partial_z^2)^{-1}$ explicit, as in the later comparison we will 
act on the corresponding CFT expressions with the operator  $\partial_t^2+\partial_z^2$.
The  correlator $G^{(bulk)}_{xz,xz}$ can be computed in a similar way and the result is presented  order-by-order in $\mu$ in Appendix \ref{AppendixB}.
%If we instead repeat the identical procedure, but with the source $\hat{H}_{xz}$, we find that $a^{8,2(xz)}_{8,0}=\frac{\mu^2}{24}$ in order to cancel the divergent $\log r$ term. The resulting correlator $G^{(bulk)}_{xz,xz}$ is listed order-by-order in $\mu$ in appendix \ref{AppendixB}.

\subsection{Sound channel}

We now consider the sound channel.
Closer inspection reveals that in the sound channel the form of the ansatz must be modified due to a  technical issue present for the diagonal sources. We first explain how it arises and how to treat it and then proceed with the computation of the holographic
$TT$  correlators.

\subsubsection{Modified ansatz}

We find that for the source $\hat{H}_{tz}$, we are able to extract the corresponding
results in the sound channel using the same ansatz as in the scalar and shear channels. 
However, we observe that if we turn on any of the diagonal sources $\hat{H}_{tt}$, $\hat{H}_{zz}$, $\hat{H}_{xx}$ or $\hat{H}_{yy}$, 
then the ansatz of the form (\ref{thetheansatz}) is no longer valid. 

The reason for this stems from the structure of the vacuum solution $\mathcal{Z}_2^{AdS}$ in these cases. 
Let us take $\hat{H}_{tt}\neq0$ as an example. In this case the AdS propagator 
is  $-(24r^4(w^2-8\rho^2))/(\pi w^{10})$.
From  (\ref{thetheansatz}) it is clear that the ansatz will only be valid if the actual solution of the bulk equations is 
proportional to $(w^2-8\rho^2)$ to all orders in $\mu$. This condition is too restrictive and, as one can show directly, is not satisfied in the case of  (\ref{es2}).

To solve this issue for the diagonal terms, we separate the identity contribution\footnote{
The form of this ansatz is deduced from the structure of the expected CFT results.
We don't explicitly quote the corresponding equations in the chapter, but they are the diagonal analogues of 
Eqs.\  \eqref{eq:TxyxyIntEinstein}, \eqref{eq:TtxtxIntEinstein}, \eqref{eq:TtztzIntEinstein} at $\OO(\mu)$ and \eqref{eq:xyxyCFT}, \eqref{eq:txtxCFT} and \eqref{eq:tztzCFT} at $\OO(\mu^2)$.
}:
\begin{equation}\label{thetheansatzmod}
	\mathcal{Z}_i^{diag}=\mathcal{Z}_i^{AdS}+\left(G^{4,1}_i+G^{4,2}_i\log r\right)+\frac{1}{r^4}\left(G^{8,1}_i+G^{8,2}_i\log r\right)+\ldots,    
\end{equation}
with $G^4$, $G^8$, $\ldots$ defined by
\begin{align}
	&G^{4,j}=\sum_{m=0}^4\sum_{n=-12}^{-4-m}(a^{4,j}_{n,m}+b^{4,j}_{n,m}\log w)w^n\rho^m\\
	&G^{8,j}=\sum_{m=0}^8\sum_{n=-16}^{-m}(a^{8,j}_{n,m}+b^{8,j}_{n,m}\log w)w^n\rho^m\\
	&\phantom{wwwwww}\vdots\nonumber
\end{align}
The upper and lower bounds of the sums were determined in the same way as it was done at the beginning of Section \ref{susucica}.
Ultimately, using the original ansatz (\ref{thetheansatz}) for the off-diagonal sources and the modified one \eqref{thetheansatzmod} for the diagonal ones, allows us  to solve the equation of motion (\ref{es2}). 
The  results are presented in Appendix \ref{AppendixB}.

\subsubsection{$G_{tz,tz}$, $G_{tt,tt}$, $G_{zz,zz}$ and $G_{xx,xx}$}

The action for the sound channel invariant is given by \cite{Kovtun:2005ev}
\begin{equation}
	\begin{split}
		S_2=&-\frac{3\pi^2C_T}{640}\lim_{r\rightarrow\infty}\int\dd t\dd z\frac{r^5\left(1-\frac{\mu}{r^4}\right)}{\left(3\partial_t^2+\partial_z^2\left(3-\frac{\mu}{r^4}\right)\right)^2}\partial_rZ_2(t,z,r)Z_2(t,z,r)\\
		=&-\frac{\pi^2C_T}{1920}\lim_{r\rightarrow\infty}\int\dd t\dd z\left(\frac{r^5}{(\partial_t^2+\partial_z^2)^2}+\mathcal{O}(r^2)\right)\partial_rZ_2(t,z,r)Z_2(t,z,r).
	\end{split}
\end{equation}

Expanding  the bulk-to-boundary propagators for our choices of the sources, eliminating the divergent $\log r$ term and proceeding as above, we eventually obtain

\begin{equation}\label{dufmzeposledna1}
	G^{bulk}_{ab,ab}=\frac{1}{(\partial_t^2+\partial_z^2)^2}D_{ab}\zeta_2^{(ab)},
\end{equation}
where $\zeta_2^{(ab)}$ is the $1/r^4$ term in the near-boundary expansion of the corresponding bulk-to-boundary propagator $\mathcal{Z}_2^{(ab)}$ for the source $\hat{H}_{ab}$ and the operator $D_{ab}$ is given by
\begin{equation}
	{\setlength{\tabcolsep}{11pt}
		\renewcommand{\arraystretch}{1.45}
		\begin{tabular}{c|c|c}\label{table:Dops}
			$a$     &   $b$ &   $D_{ab}$  \\\hline
			$t$     &   $z$ &   $-\frac{\pi^2C_T}{60}\partial_t\partial_z$ \\
			$t$     &   $t$ &   $\frac{\pi^2C_T}{30}\partial_z^2$\\
			$z$     &   $z$ &   $\frac{\pi^2C_T}{30}\partial_t^2$
	\end{tabular}}
\end{equation}

Using the explicit form of the bulk-to-boundary solution we find that the correlation function $G^{(bulk)}_{tz,tz}$ 
is given by
\begin{align}
	G_{tz,tz}^{(bulk)}\Big|_{\mu^0}\!=&-\frac{1}{(\partial_t^2+\partial_z^2)^2}\frac{96 \pi  C_T \left(3 t^4-34 t^2 z^2+3 z^4\right)}{5 \left(t^2+z^2\right)^7}\label{eq:Bulktztz0}\\
	G_{tz,tz}^{(bulk)}\Big|_{\mu^1}\!=&\frac{1}{(\partial_t^2+\partial_z^2)^2}\frac{4 \pi  \mu C_T \left(-t^6+15 t^4 z^2-15 t^2 z^4+z^6\right)}{15 \left(t^2+z^2\right)^6}\label{eq:Bulktztz1}\\
	G_{tz,tz}^{(bulk)}\Big|_{\mu^2}\!=&-\frac{1}{(\partial_t^2+\partial_z^2)^2}\frac{2 \pi  \mu^2 C_T \left(133 t^8-1408 t^6 z^2-110 t^4 z^4+88 t^2 z^6+65 z^8\right)}{1575 \left(t^2+z^2\right)^5},\label{eq:Bulktztz2}
\end{align} 
and analogously for  $G^{(bulk)}_{tt,tt}$ and $G^{(bulk)}_{zz,zz}$ (see Appendix \ref{AppendixB}).

We find that we need to be more careful when analyzing
 the case of $G^{(bulk)}_{xx,xx}$ (and, similarly,  $G^{(bulk)}_{yy,yy}$). 
 If we turn on  the source $\hat{H}_{xx}$ we find a contribution not only from the action $S_2$ but also from $S_3$; the result is
 % In the sound channel this contribution takes the form \eqref{dufmzeposledna1} with $D_{xx}=-\frac{\pi^2C_T}{60}\left(\partial_t^2+\partial_z^2\right)$, while in the scalar channel the sources $\hat{H}_{xx}=-\hat{H}_{yy}\neq0$ lead to the same solution as  for $\hat{H}_{xy}$, which was determined in Section \ref{sekskalal}. Ultimately, we get the boundary correlator as
\begin{equation}\label{dufmzeposledna2}
	G^{bulk}_{xx,xx}=G^{bulk}_{xy,xy}-\frac{\pi^2C_T}{60}\frac{1}{\left(\partial_t^2+\partial_z^2\right)}\zeta_2^{(xx)}.
\end{equation}
The resulting expression for $G^{bulk}_{xx,xx}$ can be found in Appendix \ref{AppendixB}.
In the following section we will compare these results to their CFT counterparts.

\section{Stress tensor thermal two-point functions}\label{CFTside4d}
In this section we study the stress tensor two-point function on $S^1_\beta\times \mathbb{R}^{d-1}$, where $\beta=T^{-1}$ is the inverse temperature, in holographic CFTs, that is, CFTs with large central charge $C_T\gg1$ and a large gap in the spectrum of higher-spin single-trace operators $\Delta_{\rm gap}\gg1$. The case of the purely scalar correlator is reviewed and extended to the integrated correlator in Appendix \ref{AppendixA}, it serves as a useful toy model to study before considering the technically more complicated spinning correlator. Using the stress tensor OPE, we isolate the contribution from multi-stress tensor operators $[T^k]_J$ and read off the CFT data (OPE coefficients, thermal one-point functions and anomalous dimensions) via a comparison to the bulk calculations of metric perturbations around a black hole background in the previous section. In particular, we read off the anomalous dimensions of multi-stress tensor operators of the schematic form $:T_{\mu\nu}T_{\rho\sigma}:$, $:{T_{\mu}}^\rho T_{\rho\nu}:$ and $:T^{\rho\sigma}T_{\rho\sigma}:$ with spin $J=4,2,0$, respectively. We further compute the subleading $\OO(C_T^{-1})$ corrections which determine the near-lightcone behavior of the correlators.

\subsection{OPE expansion and multi-stress tensor contributions}
The contributions of the multi-stress tensor operators to the thermal two-point function of the stress tensor in \eqref{eq:defHTTH} can be computed using the OPE, which can be schematically written as
\begin{equation}\label{eq:TTOPE}\begin{aligned}
		T_{\mu\nu}(x)\times T_{\rho\sigma}(0)\sim \frac{1}{x^{2d}}\Big[&1+x^d\sum_{i=1}^3\lambda^{(i)}_{TTT}A_{\mu\nu\rho\sigma}^{(i),\,\alpha\beta} T_{\alpha\beta}(0)\\
		&+x^{2d}\sum_{J=0,2,4}\sum_{i\in i_J}\lambda^{(i)}_{TT[T^2]_J}B^{(i),\,\mu_1\ldots\mu_J}_{\mu\nu\rho\sigma}[T^2]_{\mu_1\ldots\mu_J}(0)+\ldots\Big],
\end{aligned}\end{equation}
where $[T^k]_{\mu_1\ldots \mu_J}$ are spin-$J$ multi-stress tensor operators, the ellipses denote higher multi-trace operators and their descendants and $i_0=\{1\}$, $i_2=\{1,2\}$ and $i_4=\{1,2,3\}$. On $S^1_\beta \times \mathbb{R}^{d-1}$ only multi-stress tensors $[T^k]_{\mu_1\ldots \mu_J}$ with dimension $\Delta_{k,J}=dk+\OO(C_T^{-1})$ contribute\footnote{In other words, only operators $[T^k]_J$ with no derivatives but various contractions of the indices survive. We therefore denote these operators by the total spin $J$ and the number of stress tensors $k$. Note also that descendants do not contribute to the two-point function on $S^1_\beta\times \mathbb{R}^{d-1}$.}. Here the label $(i)$ denotes the different structures appearing in the OPE of spinning operators. The structures $A_{\mu\nu\rho\sigma}^{(i),\,\alpha\beta}$ and $B^{(i),\,\mu_1\ldots\mu_J}_{\mu\nu\rho\sigma}$ are further fixed by conformal symmetry and depend on $x^{\mu}/|x|$. Upon inserting the OPE \eqref{eq:TTOPE} in the thermal two-point function \eqref{eq:defHTTH}, we find that each term consists of a product of a kinematical piece and the thermal one-point functions $\langle [T^k]_J\rangle_\beta$, weighted by the OPE coefficients $\lambda^{(i)}_{TT[T^k]_J}$. The thermal one-point functions are fixed by symmetry up to an overall coefficient (see $e.g.$ \cite{El-Showk:2011yvt,Iliesiu:2018fao})
\begin{equation}\label{eq:thermalOnePt}
	\langle [T^k]_{\mu_1\ldots\mu_J}\rangle_{\beta} = {b_{[T^k]_{J}}\over \beta^{\Delta_{k,J}}}(e_{\mu_1}\cdots e_{\mu_J}-{\rm traces}),
\end{equation}
where $e_\mu$ is a unit vector on $S^1_\beta$. Rather than using the explicit OPE \eqref{eq:TTOPE} together with the thermal expectation value \eqref{eq:thermalOnePt}, we will use the conformal block expansion in a scalar state and take the OPE limit, see Appendix \ref{app:SpinningBlocks} and \cite{Karlsson:2021duj}. To read off the CFT data we compare this to the bulk computations in a planar black hole background given in Section \ref{Section:Bulk}. The bulk results are shown to be consistent with the OPE expansion and we determine the $\OO(C_T^{-1})$ anomalous dimensions $\gamma^{(1)}_J$ of the double-stress tensor operators of the schematic form $:T_{\mu\nu}T_{\rho\sigma}:$, $:{T_{\mu}}^\rho T_{\rho\nu}:$ and $:T^{\rho\sigma}T_{\rho\sigma}:$. We further determine the product of coefficients $\langle [T^2]_{J=0,2,4}\rangle_\beta\lambda^{(i)}_{TT[T^2]_J}$ to leading order in $C_T^{-1}$ and partially at subleading order. In particular, the leading lightcone behavior of the correlators is determined. 

Let us now review the expected scaling with $C_T$ due to multi-stress tensors appearing in the OPE. Let us remind the central charge $C_T$ is defined by the stress tensor two-point function in the vacuum
\begin{equation}\label{eq:normT}
	\langle T_{\mu\nu}(x)T_{\rho\sigma}(0)\rangle = {C_T\over x^{2d}}\Big[{1\over 2}(I_{\mu\rho}I_{\nu\sigma}+{1\over 2}I_{\mu\sigma}I_{\nu\rho})-{1\over d}\delta_{\mu\nu}\delta_{\rho\sigma}\Big],
\end{equation}
where $I_{\mu\nu}=I_{\mu\nu}(x)=\delta_{\mu\nu}-\frac{2x_\mu x_\nu}{x^2}$.  The CFT data is encoded in a perturbative expansion in $C_T^{-1}$ and a generic $k$-trace operator $[\OO^k]$ with dimension $\Delta_k$ gives the following contribution in the OPE limit\footnote{In general, the OPE expansion is a complicated function of $x^\mu$, below we just keep the scaling with $|x|$. We further suppress the indices of the operators appearing in the OPE.} $|x|/\beta\to0$:
\begin{equation}\label{eq:TkCorr}
	\langle T_{\mu\nu}(x)T_{\rho\sigma}(0)\rangle_\beta|_{[\OO^k]} \propto |x|^{\Delta_{k,J}-2d}\frac{\langle T_{\mu\nu}T_{\rho\sigma}[\OO^k] \rangle\langle[\OO^k]|\rangle_\beta}{\langle [\OO^k][\OO^k]\rangle}.
\end{equation}
Here we are interested in the case of multi-trace stress tensor operators $[\OO^k]=[T^k]_J$ which have a natural normalization
\begin{equation}\label{eq:MultiTNorm}
	\langle [T^k]_J[T^k]_J\rangle \sim C_T^k,
\end{equation}
which follows from the completely factorized contribution. In holographic CFTs dual to semi-classical Einstein gravity, the connected part of correlation functions of stress tensors is proportional to $C_T$:
\begin{equation}\label{eq:CtTTTk}
	\langle T_{\mu\nu}T_{\rho\sigma} [T^{k\neq 2}]_J\rangle \sim C_T.
\end{equation}
An important exception to \eqref{eq:CtTTTk} occurs for $k=2$ where there is a disconnected contribution such that 
\begin{equation}
	\langle T_{\mu\nu}T_{\rho\sigma} [T^2]_J\rangle \sim C_T^2+\ldots,
\end{equation}
where the ellipses refer to subleading corrections in $C_T^{-1}$ which will play an important role later. Lastly, the expectation value of a multi-stress tensor operator in the thermal state has the following scaling with $C_T$
\begin{equation}\label{eq:MultiTOHScaling}
	\langle[T^k]_J\rangle_\beta \sim {C_T^k\over \beta^{dk}},
\end{equation} 
where we also included the dependence on $\beta$ which is fixed on dimensional grounds.

Using \eqref{eq:MultiTNorm}-\eqref{eq:MultiTOHScaling}, we see that the contribution of multi-stress tensor operators $[T^k]_J$ with dimensions $\Delta_{k,J}=dk+{\cal O}(C_T^{-1})$ to the stress tensor two-point function in the thermal state has the following scaling with $C_T$ for $k\neq 2$
\begin{equation}\label{eq:TkContribution}
	\langle T_{\mu\nu}(x)T_{\rho\sigma}(0)\rangle_\beta|_{[T^{k\neq 2}]_J} \propto {1\over x^{2d}}C_T \left({x\over\beta}\right)^{dk},
\end{equation}

Meanwhile, for $k=2$, the double stress tensor contributions $[T^2]_{J=0,2,4}$ to the thermal two-point function give rise to the disconnected part of the correlator due to the fact that the three-point function $\langle T_{\mu\nu}T_{\rho\sigma} [T^2]_J\rangle \sim C_T^2$, compared to the $\OO(C_T)$ contribution from the connected part. The contribution at $\OO(C_T)$ will therefore contain the first subleading correction to the OPE coefficients $\lambda^{(i)}_{TT[T^2]_J}$, the corrections to the thermal one-point functions, as well as the anomalous dimensions of the double-stress tensor operators.

We define coefficients $\rho_{i,J}$ for the double-stress tensor $[T^2]_J$ with dimensions $\Delta_J:=\Delta_{2,J}$ by: 
\begin{equation}\label{eq:defRho}\begin{aligned}
		\hat{G}_{\mu\nu,\rho\sigma}(x)|_{\mu^2}=|x|^{-8}\Big[\rho_{1,0}g_{\Delta_{0},0,\mu\nu,\rho\sigma}(x)+\sum_{i=1,2}\rho_{i,2} g^{(i)}_{\Delta_{2},2,\mu\nu,\rho\sigma}(x)\\
		+\sum_{i=1,2,3}\rho_{i,4} g^{(i)}_{\Delta_{4},4,\mu\nu,\rho\sigma}(x)\Big],
\end{aligned}\end{equation}
where $\hat{G}_{\mu\nu,\rho\sigma}(x):=\langle T_{\mu\nu}(x)T_{\rho\sigma}(0)\rangle_{\beta }$ is the thermal correlator and $g^{(i)}_{\Delta,J,\mu\nu,\rho\sigma}$ can be obtained by taking the OPE limit of the conformal blocks in the differential basis \cite{Costa:2011mg,Costa:2011dw}, see Appendix \ref{app:SpinningBlocks}. The coefficients $\rho_{i,J}$ are therefore products of OPE coefficients and thermal one-point functions, see \eqref{eq:TkCorr}.  The coefficients $\rho_{i,J}$ and the anomalous dimensions $\gamma_J$ have a perturbative expansion in $C_T^{-1}$
\begin{equation}\begin{aligned}\label{eq:OPEDataTT}
		\rho_{i,J} &= \rho_{i,J}^{(0)}\Big[1+{\rho^{(1)}_{i,J}\over C_T}+\OO(C_T^{-2})\Big],\\
		\Delta_{J} &= 2d+{\gamma^{(1)}_J\over C_T}+\OO(C_T^{-2}),
\end{aligned}\end{equation} 
and lead to the following schematic contribution to the stress tensor two-point function from $[T^2]_J$:\footnote{We stress that \eqref{eq:doubleStressCont} only contains the scaling with $|x|\to0$ while the explicit expression has a more complicated dependence on $x^\mu$ captured in \eqref{eq:defRho}.}
\begin{equation}\label{eq:doubleStressCont}\begin{aligned}
		\hat{G}_{\mu\nu,\rho\sigma}(x)|_{[T^2]_J} &\propto \sum_{i\in i_J}\rho_{i,J}|x|^{\gamma_J^{(1)}}\\
		&\propto \sum_{i\in i_J}\rho_{i,J}^{(0)}\Big[1+{1\over C_T}\Big(\rho^{(1)}_{i,J}+\gamma^{(1)}_J\log|x|\Big)+\OO(C_T^{-2})\Big].
\end{aligned}\end{equation}
Note that the number of structures for the three point functions $\langle T_{\mu\nu} T_{\rho\sigma} [T^2]_{J=0,2,4}\rangle$ is (in $d\geq 4$) $1,2,3$ for $J=0,2,4$, respectively, giving a total of $6$ different structures at this order. From now on we will mainly consider $d=4$.
\subsection{Thermalization of heavy states}
The thermal one-point function of an operator $\OO$ with dimension $\Delta$ and spin $J$ on $S_\beta^1\times \mathbb{R}^{d-1}$ is fixed up to an overall coefficient $b_\OO$ \cite{El-Showk:2011yvt,Iliesiu:2018fao}
\begin{equation}\label{e.nudjvakswsfv}
	\langle \OO_{\mu_1\ldots \mu_J}\rangle_{\beta} = {b_\OO\over \beta^\Delta}(e_{\mu_1}\cdots e_{\mu_J}-{\rm traces}),
\end{equation}
where $e^\mu$  is a unit vector along the thermal circle. To leading order in the $C_T^{-1}$ expansion, we expect multi-stress tensor operators to thermalize in heavy states $|\psi\rangle =|\OH\rangle$ with scaling dimension $\Delta_H\sim C_T$: (see \cite{Karlsson:2021duj} for a discussion on the thermalization of multi-stress tensors and \cite{Lashkari:2016vgj,Lashkari:2017hwq} for a discussion on ETH in CFTs.)
\begin{equation}\label{eq:therm}
	\langle [T^k]_J\rangle_H \approx \langle [T^k]_J\rangle_\beta,
\end{equation}
where we have suppressed the indices. This statement holds to leading order in $C_T^{-1}$. In particular, thermalization of the stress tensor $\langle T_{\mu\nu}\rangle_H=\langle T_{\mu\nu}\rangle_{\beta}$\footnote{We will take the large volume limit ${\beta\over R}\to0$ of this equation and further set $R=1$.} leads to the following relation\footnote{See $e.g.$ Eq.\ (6.9) of \cite{Karlsson:2021duj}.} between $\beta$ and the scaling dimension $\Delta_H$ in $d=4$
\begin{equation}\label{eq:Temp}
	{b_{T_{\mu\nu}}\over \beta^4} = -{\mu C_T S_4\over 40},
\end{equation}
where $\mu$ is given by\footnote{Note that the definition of $C_T$ differs by a factor of $S_d^2$ compared to \cite{Kulaxizi:2018dxo}.}
\begin{equation}\label{eq:defMu}
	\mu = {4\Gamma(d+2)\over (d-1)^2\Gamma({d\over 2})^2 S_d^2}{\Delta_H\over C_T}
\end{equation}
and $S_{d}={2\pi^{d\over 2}\over \Gamma({d\over 2})}$. 

To leading order in $C_T^{-1}$, the multi-stress tensor operators are expected to thermalize while the expectation value in the heavy state and the thermal state might differ at subleading order. 
%This was already seen in $d=2$ in Section \ref{Section:2d}.
 As evident from \eqref{eq:OPEDataTT}, the $\OO(C_T\mu^2)$ part of the correlator contains corrections to the dynamical data that are subleading in $C_T^{-1}$ . 
%When  comparing these results to the corresponding bulk results computed in the black hole background these are therefore understood as corrections to the thermal one-point functions of these operators. 
More specifically, $\rho_{i,J}^{(1)}$ contain the following terms
\begin{equation}\label{eq:correction}
	\rho_{i,J}^{(1)}= \lambda^{(i,1)}_{TT[T^2]_J}+b_{[T^2]_J}^{(1)},
\end{equation}
where $\lambda^{(i,1)}_{TT[T^2]_J}$ and $b_{[T^2]_J}^{(1)}$ are the subleading $C_T^{-1}$ corrections to the OPE coefficients and the thermal one-point functions, respectively. 
\subsection{Identity contribution}
In this section we compare the contribution of the identity operator in the $T_{\mu\nu}\times T_{\rho\sigma}$ OPE on the CFT side using \eqref{eq:normT} to the bulk results in Section \ref{Section:Bulk}. To make a comparison to the bulk calculation, we integrate \eqref{eq:normT} over the $xy$-plane 
\begin{equation}\label{eq:Identity}\begin{aligned}
		G_{xy,xy}|_{\mu^0} &=\frac{\pi  C_T}{10 \left(t^2+z^2\right)^3},\\
		G_{tx,tx}|_{\mu^0} &=-\frac{\pi  C_T\left(t^2-5 z^2\right)}{40 \left(t^2+z^2\right)^4},\\
		G_{tz,tz}|_{\mu^0} &=-\frac{\pi  C_T \left(5 t^4-38 t^2 z^2+5 z^4\right)}{60 \left(t^2+z^2\right)^5},
\end{aligned}\end{equation}
where $G_{\mu\nu,\rho\sigma}$ is the integrated correlator defined in \eqref{eq:defHTTH}. The result for $G_{xy,xy}$ in \eqref{eq:Identity} agrees with \eqref{eq:Bulkxyxy0} obtained in the bulk. In order to compare the remaining two polarizations $G_{tx,tx}$ and $G_{tz,tz}$, we further apply the differential operator $(\partial_t^2+\partial_z^2)^p$ with $p=1,2$, respectively, to match these CFT results with their bulk counterparts. Doing so, we find that 
\begin{equation}\label{eq:IdentityWDeriv}\begin{aligned}
		(\partial_t^2+\partial_z^2)G_{tx,tx}|_{\mu^0} &=-\frac{3 \pi  C_T\left(t^2-7 z^2\right)}{5 \left(t^2+z^2\right)^5},\\
		(\partial_t^2+\partial_z^2)^2G_{tz,tz}|_{\mu^0} &=-\frac{96 \pi C_T\left(3 t^4-34 t^2 z^2+3 z^4\right)}{5 \left(t^2+z^2\right)^7},
\end{aligned}\end{equation}
which agree with \eqref{eq:Bulktxtx0} and \eqref{eq:Bulktztz0}, respectively.

\subsection{Stress tensor contribution}
In this section we consider the stress tensor contribution. The stress tensor three-point function is fixed up to three coefficients in $d\geq 4$ \cite{Osborn:1993cr}
\begin{equation}
	\langle T_{\mu\nu}(x_1)T_{\rho\sigma}(x_2)T_{\alpha\beta}(x_3)\rangle =\sum_{i=1,2,3}\lambda_{TTT}^{(i)}{\cal I}^{(i)}_{\mu\nu,\rho\sigma,\alpha\beta},
\end{equation}
for three tensor structures ${\cal I}^{(i)}_{\mu\nu,\rho\sigma,\alpha\beta}(x_j)$ determined by conservation and conformal symmetry. One way to parametrize these coefficients is in terms of $(C_T,t_2,t_4)$, for further details and conventions see Appendix \ref{sec:T}. In particular, in holographic CFTs dual to semi-classical Einstein gravity it is known that $t_2=t_4=0$ \cite{Hofman:2008ar}. This fixes two of the coefficients, with the remaining one being fixed by Ward identities in terms of $C_T$ according to \eqref{eq:WardTTT} \cite{Osborn:1993cr}. 

Using the explicit form of the stress tensor conformal block in the OPE limit together with $t_2=t_4=0$, we can find the explicit contribution of the stress tensor to $G_{\mu\nu,\rho\sigma}$, see Appendix \ref{sec:T} for details. To compare to the corresponding bulk results in Section \ref{Section:Bulk} we further need to integrate the correlator over the $xy$-plane. This is done in Appendix \ref{sec:T} and one finds:
\begin{equation}\begin{aligned}\label{eq:TResult}
		&G_{xy,xy}|_{\mu} = {\pi C_T\mu\over 100}{t^2-z^2\over (t^2+z^2)^2},\\
		&G_{tx,tx}|_{\mu} = {\pi C_T\mu\over 800}{-9t^4+6t^2z^2+7z^4\over (t^2+z^2)^3},\\
		&G_{tz,tz}|_{\mu}= {\pi C_T\mu\over 3600}{-105t^6+3t^4z^2+137t^2z^4+77z^6\over (t^2+z^2)^4}. 
\end{aligned}\end{equation}
The result for $G_{xy,xy}$ in \eqref{eq:TResult} agrees with \eqref{eq:Bulkxyxy1}. For the remaining polarizations we apply the relevant differential operators to find
\begin{equation}\label{eq:TtxtxRes}
	(\partial_t^2+\partial_z^2)G_{tx,tx}|_\mu = {3\pi C_T\mu\over 200}{t^4-6t^2z^2+z^4\over (t^2+z^2)^4}
\end{equation}
and 
\begin{equation}\label{eq:Ttztz}
	(\partial_t^2+\partial_z^2)^2G_{tz,tz}|_{\mu} =-{4\pi C_T\mu\over 15}{t^6-15t^4z^2+15t^2z^4-z^6\over(t^2+z^2)^6}.
\end{equation}
Upon comparing \eqref{eq:TtxtxRes} with \eqref{eq:Bulktxtx1} and \eqref{eq:Ttztz} with \eqref{eq:Bulktztz1} we find perfect agreement between the bulk and the CFT calculation.

\subsection{Double stress tensor contributions}
In this section we consider the contribution due to the double-stress tensor operators of the schematic form $:T_{\mu\nu}T_{\rho\sigma}:$, $:{T_{\mu}}^\rho T_{\rho\nu}:$ and $:T^{\rho\sigma}T_{\rho\sigma}:$. These are captured by \eqref{eq:defRho} with $\Delta_J$ and $\rho_{i,j}$ given by \eqref{eq:OPEDataTT}. Details on the conformal blocks are given in Appendix \ref{app:SpinningBlocks}. At $\OO(C_T^2\mu^2)$ we see from \eqref{eq:doubleStressCont} that there are $6$ undetermined coefficients $\rho_{i,J}^{(0)}$ and at $\OO(C_T\mu^2)$ there is a total of $9$ coefficients, in particular, the $6$ coefficients $\rho_{i,J}^{(1)}$ and the $3$ anomalous dimensions $\gamma_{J}^{(1)}$:
\begin{equation}\label{eq:opeDataToBeDet}
	X= \{\rho_{1,0}^{(1)},\rho_{1,2}^{(1)},\rho_{2,2}^{(1)},\rho_{1,4}^{(1)},\rho_{2,4}^{(1)},\rho_{3,4}^{(1)},\gamma^{(1)}_{0},\gamma^{(1)}_{2},\gamma^{(1)}_{4}\}.
\end{equation}

Note that unlike the conformal data discussed so far, which are largely determined by Ward identities, the
results of this Section follow from the dynamics of the five-dimensional Einstein-Hilbert gravity with a negative
cosmological constant.

\subsubsection{Disconnected part}
As expected from thermalization, the $\OO(C_T^2\mu^2)$ disconnected contribution to the stress tensor two-point function  in the thermal states factorizes and is independent of the position $x$:
\begin{equation}
	\hat{G}_{\mu\nu,\rho\sigma} =\langle T_{\mu\nu}\rangle_{\beta} \langle T_{\rho\sigma} \rangle_{\beta}(1+\OO(C_T^{-1})),
\end{equation}
where $\beta$ is the inverse temperature related to $\mu$ by \eqref{eq:Temp}. In particular, only the diagonal terms of $\langle T_{\mu\nu}\rangle_\beta$ are non-zero:
\begin{equation}\begin{aligned}\label{eq:nonDiagMFT}
		\hat{G}_{xy,xy}=0+\OO({C_T\mu^2}),\\
		\hat{G}_{tx,tx}=0+\OO({C_T\mu^2}),\\
		\hat{G}_{tz,tz}=0+\OO({C_T\mu^2}),
	\end{aligned}
\end{equation}
while 
\begin{equation}\begin{aligned}\label{eq:ttttMFT}
		\hat{G}_{tt,tt}= \left({3\over 4}\right)^2{b_{T_{\mu\nu}}^2\over\beta^{8}}\Big[1+\OO(C_T^{-1})\Big].
	\end{aligned}
\end{equation}
Comparing the conformal block expansion in \eqref{eq:defRho} to \eqref{eq:nonDiagMFT}, we find that $5$ out of $6$ of the leading order coefficients $\rho_{i,J}^{(0)}$ are determined in terms of the remaining undetermined coefficient $\rho_{1,0}^{(0)}$:
\begin{equation}\label{eq:MFTsoln}\begin{aligned}
		\rho^{(0)}_{1,2} =&  {324\over 7}\rho^{(0)}_{1,0},\\
		\rho^{(0)}_{2,2} =& {-1728\over 7}\rho^{(0)}_{1,0},\\
		\rho^{(0)}_{1,4} =& {160\over 7}\rho^{(0)}_{1,0},\\
		\rho^{(0)}_{2,4} =& {-1760\over 7}\rho^{(0)}_{1,0},\\
		\rho^{(0)}_{3,4} =& {-480\over 7}\rho^{(0)}_{1,0}.
	\end{aligned}
\end{equation}

The remaining coefficient is fixed by imposing \eqref{eq:ttttMFT} which gives 
\begin{equation}\label{eq:rhothreefour}
	\rho^{(0)}_{1,0} = {\pi^4 \mu^2 C_T^2\over 480000}.
\end{equation}

\subsubsection{Corrections to double stress tensor CFT data}
At $\OO(C_T\mu^2)$ there is a total of $9$ coefficients that fix $G_{\mu\nu,\rho\sigma}$. The goal of this section is to (partially) determine the CFT data (\ref{eq:opeDataToBeDet}) by comparing the conformal block decomposition at $\OO(C_T\mu^2)$ to the bulk calculations in Section \ref{Section:Bulk}. In particular, our analysis will allow us to extract the anomalous dimensions $\gamma_J^{(1)}$ of double-stress tensors $[T^2]_J$, $J=0,2,4$ as well as the near-lightcone behavior of the correlators.

In order to do so we again need to integrate the correlator over the $xy$-plane. This is divergent, as is manifest from dimensional analysis (see also \eqref{eq:TkContribution}). We will tame this divergence by including a factor of $ |x|^{-\epsilon}$ in the integrals which produces simple poles as $\epsilon\to 0$\footnote{Alternatively, one can introduce an IR cutoff in the integrals and the results for the anomalous dimensions and the coefficients $\rho^{(1)}_{i,J}$ will remain the same.}. These will then be absorbed in the undetermined bulk coefficients, see \eqref{eq:solACoeff}.

We will fix the CFT data by comparing the polarizations, $G_{xy,xy}$, $G_{tx,tx}$ and $G_{tz,tz}$, with the corresponding conformal block decomposition given in \eqref{eq:xyxyCFT}, \eqref{eq:txtxCFT} and \eqref{eq:tztzCFT}, with the bulk results given in \eqref{eq:Bulkxyxy2}, \eqref{eq:Bulktxtx2} and \eqref{eq:Bulktztz2}, respectively. For the latter two polarizations, we apply the differential operators $(\partial_t^2+\partial_z^2)^p$, with $p=1,2$, on the OPE expansion in order to match against the bulk calculations, just as for the identity and stress tensor operator, which give
\begin{equation}\label{eq:matching}
	\begin{aligned}
		G_{xy,xy}^{(CFT)}-G_{xy,xy}^{(bulk)}\Big|_{\mu^2 C_T} = 0,\\
		(\partial_t^2+\partial_z^2)\left[G_{tx,tx}^{(CFT)}-G_{tx,tx}^{(bulk)}\right]\Big|_{\mu^2 C_T}= 0,\\
		(\partial_t^2+\partial_z^2)^2\left[G_{tz,tz}^{(CFT)}-G_{tz,tz}^{(bulk)}\right]\Big|_{\mu^2 C_T}= 0.
	\end{aligned}
\end{equation}

There is a common solution which unambiguously fixes the anomalous dimensions to the values:
\begin{equation}\label{eq:solAnomDim}\begin{aligned}
		\gamma_{0}^{(1)} &= -{2480\over 63\pi^4},\\
		\gamma_{2}^{(1)} &=-{4210\over 189\pi^4},\\
		\gamma_{4}^{(1)} &= -{1982\over 35\pi^4},\\
	\end{aligned}
\end{equation}
where we note that the anomalous dimensions in \eqref{eq:solAnomDim} are all negative. Further, we find the following relations among three out of the six coefficients $\rho^{(1)}_{i,J}$
\begin{equation}\label{eq:solCoeff}
	\begin{aligned}
		\rho^{(1)}_{2,2} &=  -{14465\over 1296\pi^4}+\rho_{1,2}^{(1)},\\
		\rho^{(1)}_{2,4} &= {379\over 210\pi^4}+\rho_{1,4}^{(1)},\\
		\rho^{(1)}_{3,4} &= {3083\over 1260\pi^4}+\rho_{1,4}^{(1)},
	\end{aligned}
\end{equation}
while the remaining CFT data $\{\rho_{1,0}^{(1)},\rho_{1,2}^{(1)},\rho_{1,4}^{(1)}\}$ is undetermined and the bulk coefficients are given in \eqref{eq:solACoeff}. We have further checked that this solution is consistent with several other polarizations such as $G_{zx,zx}, G_{tx,zx},G_{zz,zz}$ and $G_{tt,tt}$ by inserting \eqref{eq:solAnomDim}, \eqref{eq:solCoeff} and \eqref{eq:solACoeff} in the OPE expansion and comparing to the explicit bulk calculations. Comparing $G_{xx,xx}$ from the CFT to the bulk calculation, one finds one more linearly independent equation\footnote{The reason for this can be seen from \eqref{dufmzeposledna1} and the table in  (\ref{table:Dops}), when comparing to the CFT result we only apply a differential operator of degree $2$ for the $G_{xx,xx}$ polarization compared to a degree $4$ operator for other polarizations in the sound channel.} \eqref{eq:Axxxx}. The undetermined coefficients $\{\rho_{1,0}^{(1)},\rho_{1,2}^{(1)},\rho_{1,4}^{(1)}\}$ can then be expressed in terms of the undetermined bulk coefficients, see Eqs.\ \eqref{eq:solACoeff} and \eqref{eq:Axxxx}.

\subsection{Lightcone limit}
In this section we consider the lightcone limit which is obtained by Wick-rotating $t\to i t$ and taking $v\to 0$, with $u=t-z$ and $v=t+z$. Imposing unitarity on the stress tensor contribution leads to the conformal collider bounds, see $e.g.$ 
\cite{Hofman:2008ar,Hofman:2009ug,Kulaxizi:2010jt,Hartman:2015lfa,Li:2015itl,Komargodski:2016gci,Hartman:2016dxc,Hofman:2016awc,Faulkner:2016mzt,Hartman:2016lgu}. Consider now the lightcone limit of the double-stress tensor contribution. One finds the following result for the integrated correlators in the lightcone limit $v\to 0$:
\begin{equation}\label{e.prijdeakpovolan}
	\begin{aligned}
		G_{xy,xy}^{(CFT)}(u,v)|_{\mu^2C_T} &\underset{v\to 0}{=}\pi ^5 \mu ^2 C_T\frac{2 \gamma^{(1)}_4-41\rho^{(1)}_{1,4}+11\rho^{(1)}_{2,4}+30\rho^{(1)}_{3,4}}{48000}{u^3\over v},\cr 
		G_{tx,tx}^{(CFT)}(u,v)|_{\mu^2C_T} &\underset{v\to 0}{=} \pi ^5 \mu ^2 C_T \frac{-113\gamma^{(1)}_4+16(188\rho^{(1)}_{1,4}-77\rho^{(1)}_{2,4}-111\rho^{(1)}_{3,4})}{10752000}{u^4\over v^2},\cr
		G_{tz,tz}^{(CFT)}(u,v)|_{\mu^2C_T} &\underset{v\to 0}{=}\pi ^5 \mu ^2 C_T \frac{ 29\gamma^{(1)}_4-740\rho^{(1)}_{1,4}+308\rho^{(1)}_{2,4}+432\rho^{(1)}_{3,4}}{16128000}{u^5\over v^3},
	\end{aligned}
\end{equation}
where, as expected, only the spin-4 operator of the schematic form $:T_{\mu\nu}T_{\rho\sigma}:$ contributes\footnote{We have dropped the divergent terms from the integration since these do not contain negative powers of $v$ when $v\to 0$.}. Inserting the solution \eqref{eq:solAnomDim}-\eqref{eq:solCoeff} we find 
\begin{equation}
	\begin{aligned}
		G_{xy,xy}^{(CFT)}(u,v)|_{\mu^2C_T} &\underset{v\to 0}{=}-\frac{\pi  \mu ^2 C_T}{2400}{u^3\over v},\cr 
		G_{tx,tx}^{(CFT)}(u,v)|_{\mu^2C_T} &\underset{v\to 0}{=} -\frac{17 \pi  \mu ^2C_T}{1075200}{u^4\over v^2},\cr
		G_{tz,tz}^{(CFT)}(u,v)|_{\mu^2C_T} &\underset{v\to 0}{=}-\frac{11 \pi  \mu ^2 C_T}{6048000}{u^5\over v^3},
	\end{aligned}
\end{equation}
where we note that the undetermined coefficient $\rho_{1,4}^{(1)}$ drops out in the lightcone limit. The solution in \eqref{eq:solAnomDim}-\eqref{eq:solCoeff} obtained from the bulk computations in Section \ref{Section:Bulk} therefore determines completely the lightcone limit of the correlator to this order.

\section{Discussion and Implications}\label{sec:disc}
In this chapter we have examined the thermal two-point function of stress tensors in holographic CFTs. 
In the dual picture, this corresponds to studying metric perturbations around a black hole background. 
The thermal two-point function can be decomposed into  contributions of individual operators using the OPE. 
Important contributions to the OPE of two stress tensors include the identity operator, 
 the stress tensor itself, and composite operators made out of the stress tensor  (multi-stress tensors). 
 
The holographic contribution of the identity reproduces the vacuum result.
We also verify that the  stress tensor contribution to the holographic $TT$ correlator agrees with the CFT result,
which is fixed by the three-point functions of the stress tensor in CFTs dual to Einstein gravity
 (our CFT result  agrees with \cite{Kulaxizi:2010jt}).
The leading contribution from the double-stress tensors  corresponds to the disconnected 
part of the correlator.

The  anomalous dimensions and the corrections to the OPE coefficients and thermal one-point functions contribute at next-to-leading order in the $C_T^{-1}$ expansion. 
Comparing the CFT and  holographic calculations, we are able to read off the anomalous dimensions of the double-stress tensors with spin $J=0,2,4$ and obtain partial relations for the subleading corrections to the 
products of OPE coefficients and thermal one-point functions.
It would be interesting to compare our results with the one-loop results of
\cite{Rastelli:2016nze,Alday:2017xua,Aprile:2017bgs,Rastelli:2017udc}.

We are unable to fully determine the double-stress tensor contribution from the near-boundary analysis in the bulk;
indeed some OPE coefficients remain unfixed, although the leading lightcone behaviour of the $TT$ correlators at this order
is completely determined.
The situation is reminiscent of the scalar case \cite{Fitzpatrick:2019zqz}, where the contributions of double-trace operators of external scalars were not determined by the near-boundary analysis.
It would be interesting to go beyond the near-boundary expansion to further determine this remaining data. 
In contrast to the scalar case considered in Chapters \ref{ch2} and \ref{ch3}, 
in our analysis we further integrated the correlator over a plane to account for different polarizations of the stress tensor. 
This feature introduces some technical complications and it would be interesting to study the correlator without integration. 

Holography provides a powerful tool to study hydrodynamics of strongly coupled quantum field theories and transport coefficients can be read off from the stress tensor two-point function at finite temperature\footnote{The expansion in small momenta compared to the temperature is opposite of the OPE limit and interpolating between the two is challenging. See $e.g.$ \cite{Withers:2018srf,Grozdanov:2019kge,Grozdanov:2019uhi,Abbasi:2020ykq,Jansen:2020hfd,Grozdanov:2020koi,Choi:2020tdj,Baggioli:2020loj,Grozdanov:2021gzh} for recent work on the convergence of the hydrodynamic expansion.}. 
The conformal bootstrap provides another window into strongly coupled phenomena when perturbation theory is not applicable. 
While the bootstrap program for vacuum correlators has led to significant developments in the past decade, the corresponding tools for thermal correlators are still developing, see $e.g.$ \cite{Caron-Huot:2009ypo,El-Showk:2011yvt,Gobeil:2018fzy,Iliesiu:2018fao,Delacretaz:2018cfk,Delacretaz:2020nit,Alday:2020eua,Karlsson:2021duj,Delacretaz:2021ufg,Dodelson:2022eiz,Marchetto:2023xap,Barrat:2024fwq} for related work. In particular, due to an important role played by the stress tensor thermal two-point function, it would be interesting to  better understand the constraints imposed by the conformal bootstrap on this correlator as well as  the implications for a gravitational dual description. 

By the nature of a duality, there are two sides to the same story. In this chapter we have used the structure of the stress tensor two-point functions at finite temperature, imposed by conformal symmetry, in order to read off the CFT data by making a comparison
to  the corresponding calculations in the bulk. At the same time, it would be very interesting to study properties of black holes in AdS by bootstrapping thermal correlators on the boundary. We expect a major role to be played by the stress tensor operator and its composites which are related to the metric degrees of freedom in the bulk.

%\acknowledgments
%This work was supported in part by an Irish Research Council consolidator award.
%We thank K-W. Huang, D. Jafferis, M. Kulaxizi, Y-Z. Li, P. O'Donovan, C. Pantelidou for useful discussions and correspondence. Valentina Prilepina gratefully acknowledges support from the Simons Center for Geometry and Physics, Stony Brook University at which some or all of the research for this chapter was performed.

%% file: 5_chapter/chapter_5.tex
\chapter{Gauss-Bonnet Gravity and ANEC}\label{ch5}

In this chapter, we extend the analysis of the $TT$ correlators to Gauss-Bonnet gravity. We follow the same procedure as in the previous chapter and investigate how higher-derivative corrections modify the extracted CFT data. We then examine a specific causality constraint known as the Averaged Null Energy Condition (ANEC). We show that when ANEC is saturated, the near-lightcone $TT$ correlators take the vacuum form, completely independent of temperature. 

\section{Motivation}

Exploring universal constraints and their consequences in quantum field theories is of great importance. 
This chapter considers questions related to Averaged Null Energy Conditions (ANECs) which  generally hold in unitary QFTs \cite{Faulkner:2016mzt,Hartman:2016lgu}. ANEC requires that the total energy measured along a complete lightlike path is non-negative. More concretely, it states that the integral of $T_{\mu\nu}k^\mu k^\nu$, over an entire null geodesic is a non-negative operator, where $T_{\mu\nu}$ is the stress tensor and $k^\mu$ is the tangent vector to the geodesic. We focus on conformal field theories where important examples of ANECs are conformal collider bounds \cite{Hofman:2008ar}.\footnote{Strictly speaking, the conformal collider bounds (CCB) are equivalent to evaluating the ANEC operator in a stress-tensor state. Throughout this chapter, however, we will often use the terms ANEC and CCB interchangeably, with the precise meaning understood from context.} In this chapter, we shall pay special attention to the situation where ANECs are saturated, and discuss the connection to stress-tensor correlators at finite temperature.

In the setup of \cite{Hofman:2008ar},  localized states are created by the stress tensor
with three independent polarizations. The energy flux is determined by the three numbers specifying the stress-tensor three-point functions and the positivity of  the energy flux 
results in three constraints on the combinations of these couplings.
Recent advances in CFT techniques (see, $e.g.$, \cite{Rychkov:2016iqz,Simmons-Duffin:2016gjk,Poland:2018epd} for reviews) 
 allowed proving conformal  collider bounds in unitarity CFTs \cite{Hofman:2016awc} (see also \cite{Li:2015itl,Komargodski:2016gci}). 
The bootstrap proof focuses on the lightcone limit of a four-point function with two scalars and two stress-tensor insertions, 
 which is dominated by the stress-tensor exchange. 
The same techniques allow making statements about interference effects in conformal collider bounds and higher-spin ANECs 
\cite{Cordova:2017zej,Meltzer:2017rtf,Meltzer:2018tnm}.
(See \cite{Balakrishnan:2017bjg,Cordova:2017dhq,Kravchuk:2018htv,Delacretaz:2018cfk,Cordova:2018ygx,Ceyhan:2018zfg,Belin:2019mnx,Kologlu:2019mfz,Manenti:2019kbl,Belin:2020lsr,Besken:2020snx,Korchemsky:2021okt,Korchemsky:2021htm,Caron-Huot:2022eqs}
for some examples of recent work devoted to the study of ANECs.)

 In  \cite{Kulaxizi:2010jt}, it was pointed out that one can observe conformal collider bounds by studying two-point functions of the stress tensor 
 (the $TT$ correlators) at finite temperature, using the operator product expansion (OPE) and focusing on the contribution of the stress tensor. 
 Symmetries  imply that the stress-tensor two-point functions at finite temperature have three independent polarizations.
 As explained in \cite{Kulaxizi:2010jt}, the coefficients of the 
  stress-tensor contributions in the lightcone limit for these polarizations are
 precisely proportional to the corresponding ANECs.
 When one of these coefficients vanish, the corresponding ANEC gets saturated.
Here, we ask the following question: can this result be generalized to include the contributions from multi-stress tensor exchanges?

In this chapter, via holography  \cite{Maldacena:1997re, Gubser:1998bc, Witten:1998qj}, we adopt Gauss-Bonnet gravity to study ANEC saturations using thermal $TT$ correlators. Gauss-Bonnet gravity and more generally Lovelock theories are useful theoretical laboratories for studying higher-derivative corrections because their equations of motion are of second order. Gauss-Bonnet gravity, despite being a special theory, might allow us to identify some universal features of holographic CFTs regardless of what higher-derivative terms are included. Indeed, ANECs manifest themselves via the superluminal propagation of signals in Gauss-Bonnet gravity \cite{Brigante:2007nu,Brigante:2008gz,deBoer:2009pn,Camanho:2009vw,Buchel:2009sk}. (For more recent developments in the holographic aspects of Gauss-Bonnet gravity, see, $e.g.$, \cite{Buchel:2009tt, 
Buchel:2010wf,Cai:2010cv,Bu:2015bwa,Grozdanov:2016vgg, Andrade:2016yzc, Grozdanov:2016zjj,Andrade:2016rln,Grozdanov:2016fkt, Chen:2018nbh,An:2018dbz,Grozdanov:2021gzh}.)

It is important to mention that the holographic Gauss-Bonnet theory, as a pure-gravity theory with non-zero higher derivative corrections, is not unitary, see \cite{Camanho:2014apa}. However, the breakdown of unitarity for small values of the Gauss-Bonnet coupling happens in the small impact parameter regime, as opposed to the large impact parameter (lightcone) limit relevant for ANECs. This
can be seen by analyzing corresponding CFT four-point functions in the impact parameter space. See, $e.g.$, \cite{Camanho:2014apa,Kulaxizi:2017ixa,Li:2017lmh,Costa:2017twz,Bonifacio:2017nnt,Afkhami-Jeddi:2018own,Kologlu:2019bco,Caron-Huot:2021enk}.\footnote{At finite values of the Gauss-Bonnet coupling,  light higher-spin operators  are needed to restore unitarity. Since we do not have control over the full tower of such higher-spin operators, we do not include them in our analysis.} This is why holographic Gauss-Bonnet gravity allows one to observe conformal collider bounds which have a much larger degree of universality and apply to all unitary CFTs.\footnote{Note that to study the 
 regime of ANEC saturation we need to consider large higher derivative terms in the gravitational Lagrangian.
For generic such terms this would lead to equations of motions which will be higher than second order and will result in a variety of complications. Gauss-Bonnet gravity is special in this regard.}

The results of \cite{Brigante:2007nu,Brigante:2008gz} on superluminal propagation in
 Gauss-Bonnet gravity can be directly connected to the OPE analysis of   \cite{Kulaxizi:2010jt}.
 Consider the integrated $TT$ correlators on $S^1_\beta \times \mathbb{R}^3$: 
\begin{equation}
\label{intcorr}
	G_{\mu\nu,\rho\sigma}(t, z; \beta)=\int_{\mathbb{R}^2}  \dd x\dd y ~ \langle T_{\mu\nu}(t,x,y,z)T_{\rho\sigma}(0)\rangle_\beta,
\end{equation} 
where $\beta=T^{-1}$ is the inverse temperature.
Choosing a particular polarization and expanding the holographic correlator in powers of temperature
one should be able to see that when the corresponding ANEC is saturated, the leading near-lightcone 
$\OO(\beta^{-4})$ term in the expansion vanishes.
We perform the finite-temperature expansion of  (\ref{intcorr}) using the techniques developed in Chapter \ref{ch4} and
confirm this expectation.
We then consider the subsequent $\OO(\beta^{-8})$ terms in the expansion and extract 
 the contribution of the spin-4 double-stress tensor operator.
We observe that when  a spin-2 ANEC is saturated, for the same choice of polarization the spin-4 ANEC is also saturated
 and the leading near-lightcone $\OO(\beta^{-8})$ term in the expansion vanishes as well.

 Does this pattern persist to all orders in the temperature expansion?
 Since multi-stress tensor operators of highest spin (for a given conformal dimension) govern the near-lightcone behavior, to answer this question we need to study the near-lightcone regime.  
We analyze the near-lightcone thermal $TT$ correlators to all orders\footnote{We do this by generalizing the approach of \cite{Fitzpatrick:2019zqz}, where near-lightcone scalar correlators were studied,  to the stress-tensor case.}
and observe that once
a spin-2 ANEC for a certain polarization is saturated,  the leading-lightcone limit of the correlator for this polarization 
takes the vacuum form and is completely independent of the temperature.
Hence, all spin-$2 k$ ANECs for multi-stress tensor operators $[T_{\mu\nu}]^k$ of maximal spin are saturated.

It has been observed that free theories saturate conformal collider bounds \cite{Hofman:2008ar}.
However it is less obvious whether theories which saturate conformal collider bounds are necessarily free, although
some evidence in this direction was presented in \cite{Zhiboedov:2013opa,Meltzer:2018tnm}.
In this chapter we propose a scenario where the theory is ``free'' in a limited sense: correlators of the stress-tensor
take a vacuum form for one particular polarization.
We call this behavior ``freedom near lightcone" and observe it in holographic Gauss-Bonnet gravity.

To make contact with the literature, we  read off the double-stress tensor CFT data to subleading order in the $C_T^{-1}$ expansion by comparing the bulk computations to the OPE  in the dual CFT. The leading order mean field theory (MFT) result needs to satisfy consistency conditions. 
These are due to interference effects of the ANEC in states that are superpositions of the stress tensor and double-stress tensors of spin $0,2,4$. For the spin-$0$ double-stress tensor this was shown to impose no constraint on the OPE coefficient \cite{Cordova:2017zej}, while for spin-$2$ and spin-$4$ double-stress tensors interference effects  impose non-trivial constraints on the OPE coefficients \cite{Meltzer:2017rtf,Meltzer:2018tnm}. 
We verify that the MFT coefficients in holographic CFTs are consistent with such interference effects. 
In addition, following \cite{Meltzer:2018tnm},  from the CFT point of view we verify that, using the data obtained from holographic Gauss-Bonnet gravity, the spin-$4$ ANEC is also saturated when the corresponding spin-$2$ ANEC is saturated.

\subsubsection*{Outline}

In the next section, we %discuss holographic Gauss-Bonnet (GB) gravity. 
write down the  equations of motion in Gauss-Bonnet (GB) gravity and analyze them using a near-boundary expansion.
%up to next-to-next-to leading order in temperature.  
Our calculations are done for the four-dimensional CFT case, but we expect to find similar results in other dimensions. 
In Section \ref{sectionLCCCB}, we show that, when an ANEC is saturated all higher-spin ANECs for the same polarization are saturated as well and the corresponding $TT$ correlator near the lightcone is reduced to the vacuum form. 
We read off CFT data for the double-stress tensors in the context of GB gravity in Section \ref{CFTsection} by performing the conformal block decomposition.  
 Section \ref{ANECsection} is devoted to a discussion of conformal collider bounds for states which are linear combinations of stress tensors and double-stress tensors, as well as the study of the spin-$4$ ANEC. We conclude  in Section \ref{s.sekcie56} with a list of future directions.  

\section{Thermal \texorpdfstring{$TT$}{TT} and Gauss-Bonnet Gravity}\label{Sec:Bulk}

In this section, 
after a brief review of Gauss-Bonnet gravity  we study perturbations of the planar black hole, set up our notations and discuss the near-boundary expansion. We then discuss the  thermal stress tensor two-point functions for different polarizations and analyze the contributions of the identity, the stress tensor and the double-stress tensors. 
The near-lightcone behavior of the stress-tensor correlators, including an all order analysis, will be discussed in the next section.  

\subsection{A Brief Review on Gauss-Bonnet Gravity}

In the Euclidean signature, we write the five-dimensional Gauss-Bonnet action with a negative cosmological constant as
\begin{align}\label{actionGBshort}
S_{GB}& =\frac{1}{16\pi G_5}\int\dd^5x\sqrt{g}\left[\frac{12}{L^2}+\mathcal{R}+\lambda_{GB}\frac{L^2}{2}\big( \mathcal{R}^2_{\mu\nu\lambda\rho}-4\mathcal{R}^2_{\mu\nu}+\mathcal{R}^2  \big)\right]
\end{align}
where $G_5$ is the gravitational constant and $\lambda_{GB}$ is the (dimensionless) Gauss-Bonnet coupling. 
Despite having higher curvature terms, the equations of motion resulting from \eqref{actionGBshort} remain second-order PDEs. 
Similarly as in the previous chapters, we focus on the planar (large radius) AdS black hole solution:
    \begin{equation}\label{gbbh}
        \dd s^2=\frac{r^2}{L^2}\left(\frac{f(r)}{f_\infty}\dd t^2+\dd \vec{x}\,^2\right)+\frac{L^2}{r^2}\frac{\dd r^2}{f(r)} \ ,
    \end{equation}
where $f(r)$ and $f_\infty$ are \cite{Boulware:1985wk, Cai:2001dz}
    \begin{align}
        f(r)&=\frac{1}{2\lambda_{GB}}\left[1-\sqrt{1-4\lambda_{GB}\left(1-\frac{\tilde{\mu}}{r^4}\right)}\right]\label{defefss}\ , \\
        f_\infty&=\lim_{r\rightarrow\infty}f(r)=\frac{1-\sqrt{1-4\lambda_{GB}}}{2\lambda_{GB}}\label{finfdef}  \ . 
    \end{align} 
This solution corresponds to a nonsingular black hole in a ghost-free vacuum.  No AdS vacuum exists if $\lambda_{GB}>1/4$. 
 The normalization of the metric is chosen such that the speed of light is one in the dual CFT. 
The parameter $\tilde{\mu}$ and the Hawking temperature $T$ are related in the following way \cite{Cai:2001dz}:
\begin{align}
T={r_+ \over \pi L^2 \sqrt{f_\infty}} \ , ~~~~~ r_{+}^4= \tilde{\mu}
   \end{align}  
where $r_+$ denotes the location of the black-hole horizon.

Taking $\tilde\mu\rightarrow0$ in \eqref{gbbh}, one recovers the AdS vacuum in the Poincar\'{e} coordinates:
    \begin{equation}
        \dd s^2=\frac{r^2}{L^2}\delta_{ab}\dd x^a\dd x^b+\frac{\tilde{L}^2}{r^2}\dd r^2 \ , ~~~~~ \tilde L= {L\over \sqrt{f_\infty} } 
\end{equation}
where $a,b\in\{t,\,x,\,y,\,z\}$ and $\tilde L$ is the AdS curvature scale.  The metric acquires a simpler form
    \begin{equation}\label{vacGBconven}
        \dd s^2=\tilde L\left(\tilde{r}^2\delta_{ab}\dd x^a\dd x^b+\frac{1}{\tilde{r}^2}\dd \tilde{r}^2\right) \ , ~~~ \tilde r={r\over L\tilde L}  
    \end{equation}    using the rescaled coordinate $\tilde r$. 

The central charge $C_T$ of the CFT dual to Gauss-Bonnet gravity is \cite{Buchel:2009sk}
    \begin{equation}\label{CTdefGB}
    C_T=\frac{5L^3}{\pi^3G_5f_\infty^{3/2}}(1-2f_\infty\lambda_{GB}) \ .
    \end{equation}
One can relate the parameter $\tilde{\mu}$ to the conformal dimension $\Delta_H$
  of the heavy operator that creates a heavy state  \cite{Karlsson:2020ghx}:
    \begin{equation}\label{eq:defMuTilde}
        \tilde{\mu}=\frac{20}{3 \pi ^4}\left(1-4 \lambda_{GB} +\sqrt{1-4 \lambda_{GB} }\right) \frac{\Delta_H}{C_T} f_\infty^4 \tilde L^4\  .
    \end{equation} 
In the following we will often set $\tilde L=1$, in which case $L= \sqrt{f_\infty}$.

    \subsection{Black Hole Perturbations and Ansatz}

We shall consider a small perturbation $h_{\mu\nu}$ of the black-hole metric \eqref{gbbh} and restrict ourselves to the case where $h_{\mu\nu}$ does not depend on the coordinates $x$ and $y$.
Let us remind that, according to the representations under the rotations in the $xy$-plane, the fluctuations can be classified into three channels:
\begin{align}
\text{Scalar channel  (spin 2)}&:\qquad h_{\alpha\beta}-\delta_{\alpha\beta}(h_{xx}+h_{yy})/2 \\
\text{Shear channel (spin 1)}&:\qquad h_{tx},\,h_{ty},\,h_{zx},\,h_{zy},\,h_{rx},\,h_{ry} \\
\text{Sound channel (spin 0)}&:\qquad h_{tt},\,h_{tz},\,h_{zz},\,h_{rr},\,h_{tr},\,h_{zr},\,h_{xx}+h_{yy} \ . 
\end{align}
The linearized equations of motion then can be studied separately for each spin, as different representations do not mix.
For each channel, we adopt a quantity $Z$ invariant under diffeomorphisms \cite{Buchel:2009sk}:
    \begin{align}
    &Z_{\rm{scalar}}=H_{xy},\label{Z3Def}\\
    &Z_{\rm{shear}}=\partial_zH_{tx}-\partial_tH_{xz}\label{Z1Def}\ , \\
    &Z_{\rm{sound}}=\frac{2f}{f_\infty}\partial_z^2H_{tt}-4\partial_t\partial_zH_{tz}+2\partial_t^2H_{zz}\nn\\
&~~~~~~~~~~~ -\left(\left(\frac{f}{f_\infty}+\frac{r\partial_rf}{2f_\infty}\right)\partial_z^2+\partial_t^2\right)\left(H_{xx}+H_{yy}\right),\label{Z2Def} 
    \end{align}
where  
    \begin{equation}
    H_{tt}=\frac{L^2}{r^2}\frac{f_\infty}{f(r)}h_{tt}\ , \quad H_{ti}=\frac{L^2}{r^2}h_{ti}\ , \quad H_{ij}=\frac{L^2}{r^2}h_{ij}\ ,  \quad i,\,j\in\{x,y,z\}     \ . 
    \end{equation}
The equations of motion for all three channels have the following form \cite{Buchel:2009sk}: 
    \begin{equation}\label{eomsschm}
        \partial_{\tilde{r}}^2Z+C^{(1)}\partial_{\tilde{r}}Z+C^{(0)}Z =0 \ ,
    \end{equation}
where $C^{(1)}$ and $C^{(0)}$ are differential operators. 
In the scalar channel, they are given by 
    \begin{align}
        C_{\rm{scalar}}^{(1)}&=\frac{4}{f^2 (\kappa +1)^2 \tilde{r}^4}\partial_t^2+\frac{6 f \left(\kappa ^2-1\right) \left(f \left(\kappa ^2-1\right)+4\right)-16 \kappa ^2+24}{f (\kappa +1) \tilde{r}^4 \left(f \left(\kappa ^2-1\right)+2\right)^2}\partial_z^2\ , \\
        C_{\rm{scalar}}^{(0)}&=\frac{f \left(f \left(\kappa ^2-1\right) \left(5 f \left(\kappa ^2-1\right)+16\right)+4\right)+16}{f \tilde{r} \left(f \left(\kappa ^2-1\right)+2\right)^2} \ .
    \end{align}where we introduce 
    \begin{equation}\label{defLambda}
     \kappa= \sqrt {1- 4\lambda_{GB}} 
    \end{equation} 
which will help simplify various expressions.
The shear channel has
{
\allowdisplaybreaks
    \begin{align}
    C_{\rm shear}^{(1)}=&\frac{\left(f
   \left(\kappa ^2-1\right)+2\right)^2 \left(f \left(f \left(\kappa ^2-1\right) \left(5 f \left(\kappa ^2-1\right)+16\right)+4\right)+16\right)}{f \tilde{r} \left(f \left(\kappa
   ^2-1\right)+2\right)^2 \left({\partial_t}^2 \left(f \left(\kappa ^2-1\right)+2\right)^2+2 f (\kappa +1) \kappa ^2 {\partial_z}^2\right)}{\partial_t}^2\nonumber\\
   &+\frac{2 f^2 (\kappa +1) \kappa ^2 \left(3 f \left(\kappa ^2-1\right) \left(f \left(\kappa ^2-1\right)+4\right)+8 \kappa ^2+12\right)}{f \tilde{r} \left(f \left(\kappa^2-1\right)+2\right)^2 \left({\partial_t}^2 \left(f \left(\kappa ^2-1\right)+2\right)^2+2 f (\kappa +1) \kappa ^2 {\partial_z}^2\right)}\partial_z^2\ , \\
   C_{\rm shear}^{(0)}=&\frac{4}{f^2 (\kappa +1)^2 \tilde{r}^4}\partial_t^2+\frac{8 \kappa ^2}{f (\kappa +1) \tilde{r}^4 \left(f \left(\kappa ^2-1\right)+2\right)^2}\partial_z^2 \ .
    \end{align}
}The corresponding $C^{(1)}$ and $C^{(0)}$ for the sound channel can be obtained by Fourier transforming and Wick rotating the corresponding expressions in Appendix D of \cite{Buchel:2009sk}. Due to their length, we will not present them here. 

The above equations of motion are difficult to analyze in general. However, 
using the  techniques developed in \cite{Fitzpatrick:2019zqz, Karlsson:2022osn}, we can solve these equations focusing on the regime 
    \begin{equation}\label{limitossp2}
	\tilde{r}\rightarrow\infty\qquad\text{with}\qquad \tilde{r}t,\, \tilde{r}z\,\,\,\text{fixed} \ ,
    \end{equation}
which corresponds to the OPE limit on the boundary. Introducing new variables 
    \begin{align}
        \rho=\tilde r z \ , ~~~~~~~~ w^2=1+\tilde{r}^2t^2+\tilde{r}^2z^2 \ ,
    \end{align} 
 the limit \eqref{limitossp2} can be rephrased as $\tilde{r}\rightarrow\infty$ with $\rho$ and $w$ held fixed.
We write the bulk-to-boundary propagators $\mathcal{Z}$ as
\begin{equation}
	Z(t,z,r)=\int\dd t'\dd z'\mathcal{Z}(t-t',z-z',r)\hat{Z}(t',z') 
\end{equation}
where the invariant $\hat{Z}$ is (up to derivatives, as will be explained on separated channels below) the boundary value of $Z$. 
In the near-boundary, OPE expansion, we can solve the equations of motion by taking
\begin{equation}\label{thep2theansatz}
	\mathcal{Z}=\mathcal{Z}^{AdS}\left(1+\frac{1}{\tilde{r}^4}\left(G^{4,1}+G^{4,2}\log \tilde{r}\right)+\frac{1}{\tilde{r}^8}\left(G^{8,1}+G^{8,2}\log \tilde{r}\right)+\ldots\right),    
\end{equation}
\vspace{-0.8cm}
\begin{align}
	&G^{4,j}=\sum_{m=0}^2\sum_{n=-2}^{4-m}(a^{4,j}_{n,m}+b^{4,j}_{n,m}\log w)w^n\rho^m \ , \\
	&G^{8,j}=\sum_{m=0}^6\sum_{n=-6}^{8-m}(a^{8,j}_{n,m}+b^{8,j}_{n,m}\log w)w^n\rho^m \ .
\end{align}

One can check\footnote{See also \cite{Liu:1998bu,Buchel:2009sk}.} that the bulk-to-boundary propagators in pure AdS vacuum  $\mathcal{Z}^{AdS}$ for various choices of sources $\hat{H}_{\mu\nu}$ (which are the boundary values of $H_{\mu\nu}$)
are given by
{
\allowdisplaybreaks
\begin{align}
&\hat{H}_{xy}:~~~~~ \mathcal{Z}_{\rm{scalar}}^{AdS} =\frac{2\tilde{r}^2}{\pi w^6} \ , \\
&\hat{H}_{tx}:~~~~~ \mathcal{Z}_{\rm{shear}}^{AdS}=-\frac{12\tilde{r}^3\rho}{\pi w^8} \ , \\
&\hat{H}_{xz}:~~~~~ \mathcal{Z}_{\rm{shear}}^{AdS} =\frac{12\tilde{r}^3}{\pi w^8} \sqrt{w^2-\rho^2-1} \ ,  \\
&\hat{H}_{tz}:~~~~~ \mathcal{Z}_{\rm{sound}}^{AdS}=-\frac{384\tilde{r}^4\rho}{\pi w^{10}} \sqrt{w^2-\rho^2-1} \ , \\
&\hat{H}_{tt}:~~~~~ \mathcal{Z}_{\rm{sound}}^{AdS}=-\frac{24\tilde{r}^4}{\pi w^{10}} (w^2-8\rho^2) \ ,\\
&\hat{H}_{xx}:~~~~~ \mathcal{Z}_{\rm{sound}}^{AdS}=- \frac{24\tilde{r}^4}{\pi w^{10}} (3w^2-4) \ ,\\
&\hat{H}_{zz}:~~~~~ \mathcal{Z}_{\rm{sound}}^{AdS} =\frac{24\tilde{r}^4}{\pi w^{10}} \big(7w^2-8(1+\rho^2)\big) \ ,
\end{align}}where we have expressed these results in terms of  variables $\rho$ and $w$.  
Inserting \eqref{thep2theansatz} into the equations of motion, we obtain $a_{n,m}^{k,j}$ and $b_{n,m}^{k,j}$ for different channels. 

    \subsection{Holographic Thermal \texorpdfstring{$TT$}{TT} Correlators}

Let us first recall the holographic dictionary before proceeding to the computation of the stress-tensor correlators.   
The quadratic part of the on-shell action for a general perturbation $H_{\mu\nu}$ in the AdS vacuum was calculated in \cite{Buchel:2009sk}.  
By restricting $H_{\mu\nu}$ to be independent of $x$ and $y$, one has\footnote{Note the sign difference compared to Eq. (3.11) in \cite{Buchel:2009sk}, which is related to the presence of a minus sign in the stress-tensor two-point function defined later in \eqref{varCon}.}  
    \begin{equation}\label{finalactiongbda}
       I=\frac{\pi^2C_T}{320}\int_{\partial M}\dd^4x\,\tilde{r}^5H_{\mu\nu}(t,z,\tilde{r})\partial_{\tilde{r}}H_{\mu\nu}(t,z,\tilde{r}) \ .
    \end{equation}
The action for the perturbations $H_{\mu\nu}$ of the black-hole metric \eqref{gbbh} has the form \eqref{finalactiongbda} plus terms higher-order  in $1/\tilde{r}$ that vanish in the $\tilde{r}\rightarrow\infty$ limit.     
Thus, using \eqref{finalactiongbda} and the definitions \eqref{Z3Def}-\eqref{Z1Def}, one finds the corresponding on-shell actions for invariants to be
    \begin{align}
 I_{\rm scalar}&=\frac{\pi^2C_T}{160}\lim_{\tilde{r}\rightarrow\infty}\int\dd ^4x\tilde{r}^5Z_{\rm scalar}(t,z,\tilde{r})\partial_{\tilde{r}}Z_{\rm scalar}(t,z,\tilde{r}) \ ,  \label{relaction3} \\
        I_{\rm shear}&=\frac{\pi^2C_T}{160}\lim_{\tilde{r}\rightarrow\infty}\int\dd^4x\frac{\tilde{r}^5}{\partial_t^2+\partial^2_z}Z_{\rm shear}(t,z,\tilde{r})\partial_{\tilde{r}}Z_{\rm shear}(t,z,\tilde{r})\ ,\label{relaction1}\\
        I_{\rm sound}&=-\frac{\pi^2C_T}{1920}\lim_{\tilde{r}\rightarrow\infty}\int\dd^4x\frac{\tilde{r}^5}{(\partial_t^2+\partial_z^2)^2}Z_{\rm sound}(t,z,\tilde{r})\partial_{\tilde{r}}Z_{\rm sound}(t,z,\tilde{r})\label{relaction2} \ . 
    \end{align}

\subsubsection{Scalar Channel}

In the simplest case with only the source $\hat{H}_{xy}$ turned on, we have 
    \begin{equation}\label{legoZ3}
        Z_{\rm{scalar}}(t,z,\tilde{r})=\int\dd t'\dd z'\mathcal{Z}_{\rm{scalar}}^{(xy)}(t-t',z-z',\tilde{r})\hat{H}_{xy}(t',z') \ ,
    \end{equation}
where the superscript index of the bulk-to-boundary propagator $\mathcal{Z}_{\rm{scalar}}^{(xy)}$ indicates the non-zero sources.

After inserting \eqref{thep2theansatz} into \eqref{eomsschm} for this channel, we determine $a_{n,m}^{k,j}$ and $b_{n,m}^{k,j}$.
 We expand the solution near the boundary:
    \begin{equation}\label{Z3expans}
        \mathcal{Z}_{\rm{scalar}}^{(xy)}(t,z,\tilde{r})=\delta^{(2)}(t,z)+\frac{1}{\tilde{r}^4}\zeta_{\rm{scalar}}^{(xy)}(t,z)+\ldots 
    \end{equation}
where the dots represent contributions analytic in $t$ and $z$ of order $\mathcal{O}(\tilde{r}^{-6})$ and 
subleading contact terms $\sim \mathcal{O}(\tilde{r}^{-2})$ of the schematic form $\partial^n\delta^{(2)}/\tilde{r}^n$. 
Plugging \eqref{legoZ3} and \eqref{Z3expans} into \eqref{relaction3} and taking the limit $\tilde{r}\rightarrow\infty$ gives 
    \begin{equation}
        I_{\rm{scalar}}=-\frac{\pi^2C_T}{40}\int\dd^2x\dd^2x'\zeta_{\rm{scalar}}^{(xy)}(x-x')\hat{H}_{xy}(x)\hat{H}_{xy}(x') \ ,
    \end{equation}
where $x=\{t,z\}$ and $x'=\{t',z'\}$. The CFT correlator can be obtained via 
    \begin{equation}\label{varCon}
        G^{(bulk)}_{xy,xy}=\expval{T_{xy}(t,z)T_{xy}(0,0)}_\beta=-\frac{\delta^2I_{\rm{scalar}}}{\delta\hat{H}_{xy}(t,z)\delta\hat{H}_{xy}(0,0)}=\frac{\pi^2C_T}{20}\zeta^{(xy)}_{\rm{scalar}}(t,z) \ ,
    \end{equation}
where the superscript ``bulk" indicates that these correlators are computed via holography.  Order-by-order in $\tilde{\mu}$, we obtain 
{
\allowdisplaybreaks
\begin{align}
    G_{xy,xy}^{(bulk)}\Big|_{\tilde{\mu}^0}\!=&\frac{\pi  C_T}{10 \left(t^2+z^2\right)^3} \ , \label{Gxy0}\\
	G_{xy,xy}^{(bulk)}\Big|_{\tilde{\mu}^1}\!=&(5 \kappa -4) \frac{\pi  C_T\tilde{\mu }  (t^2-z^2)}{50 \kappa ^2 (\kappa +1) L^8 \left(t^2+z^2\right)^2}\ , \label{mu1res}\\
    G_{xy,xy}^{(bulk)}\Big|_{\tilde{\mu}^2
    }\!=&\frac{\pi  C_T \tilde{\mu }^2}{1050 \kappa ^4 (\kappa +1)^2 L^{16} \left(t^2+z^2\right)} \Big[3 \left(t^2+z^2\right) \big((\kappa  (89 \kappa -206)\nonumber\\
    &+122) t^2+(\kappa  (809 \kappa -1698)+890) z^2\big) \log \left(t^2+z^2\right)\nonumber\\
    &-2 z^2
   \left(15 (\kappa  (89 \kappa -206)+122) t^2+(5 \kappa  (197 \kappa -506)+1606) z^2\right)\Big]\nonumber\\
   &+\frac{1}{10} \pi  C_T \left(a^{8,1(xy)}_{8,0} \left(t^2-7 z^2\right)-6 z^2 a^{8,1(xy)}_{6,0}\right),\label{Gxy2}
    \end{align}
}where, similar to the Einstein gravity case in the previous chapter, the coefficients $a^{8,1(xy)}_{8,0},\,a^{8,1(xy)}_{6,0}$ remain undetermined.  In the limit $\kappa\rightarrow1$, $i.e.$, $\lambda_{GB}\rightarrow0$, these correlator results reduce to those in the Einstein gravity case, as they must. 

    \subsubsection{Shear Channel}\label{shearHoloss}
    
When the source $\hat{H}_{tx}$ is turned on, we have 
    \begin{equation}\label{legoZ1}
        Z_{\rm{shear}}(t,z,\tilde{r})=\int\dd t'\dd z'\mathcal{Z}_{\rm{shear}}^{(tx)}(t-t',z-z',\tilde{r})\hat{H}_{tx}(t',z') \ , 
    \end{equation}
    \begin{equation}\label{Z1expans}
        \mathcal{Z}_{\rm{shear}}^{(tx)}(t,z,\tilde{r})=\partial_z\delta^{(2)}(t,z)+\frac{1}{\tilde{r}^4}\zeta_{\rm{shear}}^{(tx)}+\ldots \ . 
    \end{equation}
After solving for the corresponding equation of motion, we insert the solution in  \eqref{relaction1} and take the second variational derivative with respect to the source $\hat{H}_{tx}$. We have 
    \begin{equation}
        G_{tx,tx}^{(bulk)}=\frac{\pi^2C_T}{20}\frac{\partial_z}{\partial_t^2+\partial_z^2}\zeta_{\rm{shear}}^{(tx)} \ .
    \end{equation}
The explicit results, order-by-order in $\tilde{\mu}$, are given by 
{
\allowdisplaybreaks
\begin{align}
	G_{tx,tx}^{(bulk)}\Big|_{\tilde{\mu}^0}=&-\frac{1}{\partial_t^2+\partial_z^2}\frac{3 \pi  C_T \left(t^2-7 z^2\right)}{5 \left(t^2+z^2\right)^5}\label{eq:Bulktxtx0p2}\  , \\
	G_{tx,tx}^{(bulk)}\Big|_{\tilde{\mu}^1}=&-(\kappa -2)  \frac{1}{\partial_t^2+\partial_z^2}\frac{3 \pi  C_T \tilde{\mu } \left(t^4-6 t^2 z^2+z^4\right)}{100 \kappa ^2 (\kappa +1) L^8 \left(t^2+z^2\right)^4}\label{eq:Bulktxtx1p2}\ , \\
	G_{tx,tx}^{(bulk)}\Big|_{\tilde{\mu}^2}=&-\frac{1}{\partial_t^2+\partial_z^2}\Bigg[\frac{\pi  C_T \tilde{\mu }^2}{2100 \kappa ^4 (\kappa +1)^2 L^{16} \left(t^2+z^2\right)^3}\Big(-6 (\kappa  (105 \kappa \nonumber\\
	&-388)+60) t^4 z^2-24 (\kappa  (33 \kappa -160)+60) t^2 z^4\nonumber\\
	&+3 (\kappa  (97 \kappa -156)+100) \left(t^2+z^2\right)^3 \log \left(t^2+z^2\right)\nonumber\\
	&+2 (\kappa  (55 \kappa
   +212)+4) z^6\Big)+\frac{3}{5} \pi  C_T a^{8,1(tx)}_{8,0}\Bigg]\label{eq:Bulktxtx2p2} \ . 
\end{align}
}The coefficient $a^{8,1(tx)}_{8,0}$ is not determined by the near-boundary analysis. These results  in the limit $\kappa\rightarrow1$ agree with the Einstein gravity case.

    \subsubsection{Sound Channel}

The sound-channel computation becomes rather cumbersome. 
We focus on the case with the source $\hat{H}_{tz}$ turned on.  
An analogous analysis gives 
    \begin{equation}
        G_{tz,tz}^{(bulk)}=-\frac{\pi^2C_T}{60}\frac{\partial_t\partial_z}{(\partial_t^2+\partial_z^2)^2}\zeta_{\rm{sound}}^{(tz)} \ ,
    \end{equation}
where $\zeta_{\rm{sound}}^{(tz)}$ is the $1/\tilde{r}^4$ term in the near-boundary expansion of the bulk-to-boundary propagator.
Explicit results up to double-stress tensors exchanges are 
{
\allowdisplaybreaks
\begin{align}
	G_{tz,tz}^{(bulk)}\Big|_{\tilde{\mu}^0}=&-\frac{1}{(\partial_t^2+\partial_z^2)^2}\frac{96 \pi  C_T \left(3 t^4-34 t^2 z^2+3 z^4\right)}{5 \left(t^2+z^2\right)^7}\label{tzmu0}\ , \\
	G_{tz,tz}^{(bulk)}\Big|_{\tilde{\mu}^1}=& (3 \kappa -4)  \frac{1}{(\partial_t^2+\partial_z^2)^2}\frac{8 \pi  C_T\tilde{\mu } \left(t^6-15 t^4 z^2+15 t^2 z^4-z^6\right)}{15 \kappa ^2 (\kappa +1) L^8 \left(t^2+z^2\right)^6}\label{tzmu1}\ , \\
	G_{tz,tz}^{(bulk)}\Big|_{\tilde{\mu}^2}=&-\frac{1}{(\partial_t^2+\partial_z^2)^2}\frac{8 \pi  C_T \tilde{\mu }^2 }{1575 \kappa ^4 (\kappa +1)^2 L^{16} \left(t^2+z^2\right)^5}\Big[(9 \kappa  (61 \kappa -134)\nonumber\\
	&+790) t^8-4 (9 \kappa  (283 \kappa -685)+3970) t^6 z^2+10 (3 \kappa  (185 \kappa \nonumber\\
	&-552)+1090) t^4 z^4+4 (15 \kappa  (59 \kappa -139)+1222) t^2
   z^6\nonumber\\
	&+(3 (98-25 \kappa ) \kappa -154) z^8\Big] \ .  \label{tzmu2}
\end{align}
}Again, these results in the $\kappa\rightarrow1$ limit are consistent with the Einstein gravity case, see Chapter \ref{ch4}.

\section{Near-Lightcone Dynamics}\label{sectionLCCCB}

In this section we take the near-lightcone limit of the expressions discussed in the previous section.
We observe that when the conformal collider bounds are saturated, the near-lightcone behavior of
$\OO(\beta^{-4})$ terms (coming from the  stress-tensor contribution to the $TT$ OPE) and 
$\OO(\beta^{-8})$ terms (coming from the spin-4 double-stress tensor contribution to the $TT$ OPE) vanishes.
We subsequently provide an all-order analysis by taking the lightcone limit in the bulk equations of motion. 

\subsection{Thermal \texorpdfstring{$TT$}{TT} Correlators near the Lightcone}  

We define the lightcone limit by going to the Lorenzian signature and defining $(x^+, x^-)=(i t+z, i t-z)$, and 
then we take $x^- \to 0$.

First consider  $\OO(\tilde{\mu})$ contribution. When the conformal collider bounds are saturated, the corresponding critical values of the GB coupling are
    \begin{align}
\label{kappaCCB}
        \kappa^*_{\rm scalar}=\frac45 \ , ~~~~~~~ \kappa^*_{\rm shear}=2\ , ~~~~~~~  \kappa^*_{\rm sound}=\frac43 \ .
    \end{align}
We immediately observe that the expression \eqref{mu1res}
vanishes, while  \eqref{eq:Bulktxtx1p2} and \eqref{tzmu1} vanish in the  lightcone limit.

Next, we turn to  $\OO(\tilde{\mu}^2)$ term. 
In a small $x^-$ expansion, we find the thermal correlators \eqref{Gxy2}, \eqref{eq:Bulktxtx2p2}, and \eqref{tzmu2} have the following behaviour:
{\small{
	\allowdisplaybreaks
    \begin{align}
        G_{xy,xy}^{(bulk)}(x^+,x^-)\Big|_{\tilde{\mu}^2} \underset{x^-\to 0}{=}&
-\frac{(5 \kappa -4)^2 \pi  C_T  (x^+)^3 \tilde{\mu }^2}{600 \kappa ^4 (\kappa +1)^2 L^{16} x^-}
-\frac{\pi  C_T (x^+)^2}{2100 \kappa ^4 (\kappa +1)^2 L^{16}} 
\Big( \tilde{\mu }^2  \big(5 \kappa  (197 \kappa -506)\nonumber\\  &-6 (\kappa  (180 \kappa -373)+192) \log (- x^+ x^-)+1606\big)\nonumber\\
        &+105 \kappa ^4 (\kappa +1)^2 L^{16}
\left(3 a^{8,1 (xy)}_{6,0}+4 a^{8,1 (xy)}_{8,0}\right)\Big)
+\order{x^-}\label{bulkLCxy1}\ , \\
        G_{tx,tx}^{(bulk)}(x^+,x^-)\Big|_{\tilde{\mu}^2} \underset{x^-\to 0}{=}&
\frac{1}{\partial_+\partial_-}\bigg(\frac{ (\kappa -2)^2 17 \pi  C_T (x^+)^3 \tilde{\mu }^2}{33600 \kappa ^4 (\kappa +1)^2 L^{16} (x^-)^3}\nonumber\\
        &+\frac{\pi  C_T \big((204-73 \kappa ) \kappa -76\big) (x^+)^2 \tilde{\mu }^2}{11200 \kappa ^4 (\kappa +1)^2
   L^{16} (x^-)^2}+{\cal O}\big({1\over x^-}\big)\bigg)\label{bulkLCtx1}\ , \\
        G_{tz,tz}^{(bulk)}(x^+,x^-)\Big|_{\tilde{\mu}^2} \underset{x^-\to 0}{=}
&\,\,-\frac{1}{\partial_+^2\partial_-^2}\bigg(\frac{ (4-3 \kappa )^2 11 \pi  C_T (x^+)^3 \tilde{\mu }^2}{6300 \kappa ^4 (\kappa +1)^2 L^{16} (x^-)^5}\nonumber\\
        &+\frac{4 \pi  C_T\big(3 \kappa  (52 \kappa -125)+236\big) (x^+)^2 \tilde{\mu }^2}{8400 \kappa ^4 (\kappa
   +1)^2 L^{16} (x^-)^4}+{\cal O}\big({1\over (x^-)^{3}}\big)\bigg) \ .\label{bulkLCtz1}
    \end{align}
}}We see that the leading lightcone contributions all vanish at the corresponding critical values of the GB coupling.  In the expressions above, we keep the subleading lightcone limit terms which remain non-zero.  

\subsection{Reduced Equations of Motion}\label{s.sajhbsfvhbsdd}

To give an all-order proof, we derive {\it the reduced equations of motion}.  This method was developed in the study of the scalar correlator in $d>2$ holographic CFTs \cite{Fitzpatrick:2019zqz}, but the method works also for the stress-tensor correlators. The basic idea is to identify a bulk limit which isolates the largest spin (or lowest-twist) contributions, corresponding to the largest power of $\rho$ in the ansatz \eqref{thep2theansatz}, with $w$ fixed.  
More precisely, starting with the equations of motions written in variables $(\tilde r, w, \rho)$, we perform a change of variables
\begin{align}
(\tilde r, w, \rho)~~ \to  ~~(\tilde r, w, v= {\rho\over \tilde r^2})
\end{align} 
and write
\begin{align}
{\cal Z}(\tilde r,w,\rho) ~~ \to  ~~ {\cal Z}^{\rm AdS} \Big(Q(w,v)+ \bar Q(w,v) \log(\tilde r) \Big) \equiv   {\cal Z}^{\rm AdS}  Q_{\text{tot}}
\end{align} 
where, as before, ${\cal Z}^{\rm AdS}$ is the pure AdS solution. 
In the new variables, the lightcone limit corresponds to taking the large $\tilde r$ limit with $v$ fixed. Subleading terms are suppressed at large $\tilde r$.   
Functions $Q$ and $\bar Q$ determine all the information about the near-lightcone stress-tensor correlators.

\vspace{2mm}

{\noindent{\bf{Scalar Channel:}}}  In this simplest case, we obtain the reduced equation of motion in the form
\begin{align}
\big(\kappa- {4\over 5} \big)  \tilde\mu  \Theta_1+ \Theta_0 = 0 
\end{align} 
where 
{\small{\begin{align}
&{\Theta_1\over 10  v^2 } =  \big(w^2 \del^2_w -13 w \del_w   +48 \big) Q_{\text{tot}}\ , \\
&{\Theta_0\over L^8w^2\kappa^2\left(\kappa+1\right)}=\Big((1-w^2 )w^2\del^2_w - w^2 v^2 \del^2_v +2 (w^2  -2 ) v w  \del_v \del_w+ (24-5 w^2) v \del_v \nn\\
&~~~~~~~~~~~~~~~~~~~~~ +(3 w^2-5)w\del_w \Big)  Q_{\text{tot}}
+ \Big( 2(1-w^2) w \del_w +2  w^2 v \del_v +4(w^2-3)\Big) \bar Q  \ .
\end{align}}}%
Here, without solving the equation of motion, we observe that the $\tilde\mu$ dependence disappears if $\kappa=\frac45$
and the solution takes the vacuum form.   
This phenomenon does not persist in the subleading lightcone limit. 
Near the lightcone, one may define a parameter which vanishes when the corresponding ANEC is saturated: 
\begin{align}
\mu_{\rm{eff}({\rm scalar})}=  \big(\kappa- {4\over 5} \big) \tilde\mu \ .
\end{align}  

\vspace{2mm}

{\noindent{\bf{Shear Channel:}}} There are two sources in the  shear channel. 
In both cases, we find that the reduced  equations of motion can be written as
\begin{align}
\mu^2_{\rm{eff}({\rm shear})}\Theta_{2({\rm shear})} + \mu_{\rm{eff}({\rm shear})}\Theta_{1({\rm shear})}+ \Theta_{0({\rm shear})}= 0  \ , ~~~
\mu_{\rm{eff}({\rm shear})}=  \big(\kappa- 2 \big) \tilde\mu \ .
\end{align}  
For instance, with the source $\hat H_{tx}$ turned on, we obtain 
{\footnotesize{\begin{align}
{ \Theta_{2({\rm shear})}\over 4  v^4  } =& 
\Big[w^2 \left(w^2 \del^4_w-38 w \del^3_w+591 \del^2_w\right)-4431 w \del_w +13440  \Big]Q_{\text{tot}}\ , 
\end{align}}
\vspace{-7mm}
{\footnotesize{\begin{align}
{\Theta_{1({\rm shear})}\over 2  \kappa ^2 (\kappa +1) L^8 v^2}
=&\Big[2 w^3 \Big( \left(w^2-1\right) w^2  \del^3_w  -v w^3  \del_v \del^2_w     +
(27-16 w^2) w \del^2_w +17 v w^2 \del_v \del_w \nn\\
&~~~~~ +  \left(80 w^2-267\right) \del_w-80 v w  \del_v \Big)-160w^2 \left(w^2-12\right)\Big]\bar Q\nn
\end{align}}
\vspace{-7mm}
{\footnotesize{\begin{align}
 &+\Big[\left(w^4-1\right) w^4 \del^4_w    -2  \left(w^2-3\right)v w^5 \del_v \del^3_w +v^2 w^6  \del^2_v \del^2_w  
-2\left(7 w^4+6 w^2-19\right)  w^3  \del^3_w
 \nn\\
& ~~~~~ -17 v^2 w^5  \del^2_v \del_w  
+ 3  \left(11 w^2-54\right) v w^4 \del_v \del^2_w
-\Big(591 - 324 w^2 - 48 w^4 \Big)w^2 \del^2_w \nn\\
&~~~~~+80 w^4   v^2 \del^2_v 
+3  \left(534-59 w^2\right)v w^3 \del_v \del_w  
+3  \left(64 w^4-1068 w^2+1477\right)w \del_w \nn\\
& ~~~~~+240  \left(w^2-24\right)  w^2 v\del_v  -240 \left(5 w^4-48 w^2+56\right)  \Big] Q_{\text{tot}}\ , \\
{\Theta_{0({\rm shear})}\over \kappa ^4 (\kappa +1)^2 L^{16} w^2}
=& \Big[2 (w^2-1)^2 w^3 \del^3_w-4 v^2 w^5 \del_w \del^2_v +2(w^2-1) vw^4 \del_v \del^2_w\nn\\
& -(32 w^6-86 w^4+54 w^2) \del^2_w+32 w^4 v^2  \del^2_v  + 2(17-7 w^2) vw^3 \del_v \del_w \nn\\
& +2 (94 w^4-347 w^2+267) w \del_w-160   w^2 v \del_v -32 (12 w^4-65 w^2+60)\Big]\bar Q\nn
\end{align}
\vspace{-8mm}
\begin{align}
&+ \Big[(w^2-1)^2 w^4 \del^4_w + 2 v^3 w^5 \del_w \del^3_v +(7-3 w^2) v^2 w^4  \del^2_w\del^2_v+4 (w^2-1) v w^3  \del^3_w \del_v \nn\\
&
~~~ -8 \left(2 w^4-5 w^2+3\right) w^3   \del^3_w  -16 v^3 w^4  \del^3_v  + \left(7 w^4-99 w^2+108\right) v w^2 \del^2_w \del_v  + \left(31 w^2-119\right) v^2 w^3\del_w \del^2_v \nn\\
&
~~~ + \left(106 w^4-311 w^2+213\right) w^2 \del^2_w  -80 v^2 w^2 \left(w^2-7\right) \del^2_v -3  \left(27 w^4-305 w^2+356\right)vw \del_w \del_v \nn\\
&
~~~ -(320 w^4-967 w^2+693 ) w \del_w +16 \left(16 w^4-195 w^2+240\right) v \del_v +80 \left(4 w^2-5\right) w^2  \Big] Q_{\text{tot}} \ . 
\end{align}}}}}The $\tilde\mu$ corrections are suppressed in the lightcone limit when $\kappa=2$.  The expressions for another shear-channel source are similar -- see Appendix \ref{Appendixxzxz}.

\vspace{2mm}

{\noindent{\bf{Sound Channel:}}}  The sound-channel reduced  equations of motion are rather complicated and we do not   include them here.  
After a tedious computation, we are able to verify that, when the corresponding ANEC is saturated, $i.e.$, $\kappa={4\over 3}$, the pure AdS solutions for all sources solve the sound-channel reduced  equations of motion.

\section{Conformal Block Decomposition}\label{CFTsection}  

In this section, we decompose the stress-tensor two-point function using the stress-tensor  OPE. By matching against the bulk results in Section \ref{Sec:Bulk}, we extract the corresponding CFT data of multi-stress tensors, including their OPE coefficients. 
This section follows closely Section \ref{CFTside4d} of the precious chapter as well as Appendix \ref{app:SpinningBlocks}.  
In order to compare against the bulk results, we study the $TT$ correlators on $S^1_\beta \times \mathbb{R}^3$ integrated over the $xy$-plane, $i.e.$, \eqref{intcorr}.
We can use the OPE to decompose the stress-tensor two-point function 
on $S^1_\beta\times \mathbb{R}^3$:
\begin{equation}\label{eq:ConfBlockExp}
\hat{G}_{\mu\nu,\rho\sigma}=	\langle T_{\mu\nu}(x)T_{\sigma\rho}(0)\rangle_\beta  = {1\over |x|^8}\sum_{\Delta,J,i_{n_J}}  \rho_{{\cal O},i} ~ g^{(i)}_{\Delta,J,\mu\nu,\rho\sigma}(x^\mu),  ~~~ \rho_{{\cal O},i} = \lambda_{TT{\cal O}}^{(i)}\langle {\cal O}\rangle_\beta,
\end{equation}
where we sum over operators in the $T\times T$ OPE and $i$ labels the different structures in the  OPE. 
For further details on the conformal blocks, 
see Appendix \ref{App:CFT} and also Appendix \ref{app:SpinningBlocks}. 
Integrating over the $xy$-plane, we will compare the OPE \eqref{eq:ConfBlockExp} against the bulk results in Section \ref{Sec:Bulk}.

We consider the OPE up to ${\cal O}(({x\over \beta})^8)$. The operators that contribute are the identity operator, the stress-tensor operator, and the double-stress tensors $[T^2]_J$ of the schematic form $:T_{\alpha\beta}T^{\alpha\beta}:$~, $:T_{\mu\alpha}{T^{\alpha}}_{\nu}:$~, and  $:T_{(\mu\nu}T_{\rho\sigma)}:$~, with spin $J=0,2,4$, respectively.  
For the double-stress tensors, we denote 
\begin{equation}
\rho_{i,J} = \rho_{[T^2]_{J,i}}
\end{equation}
where $i=\{1\}$ for $J=0$, $i=\{1,2\}$ for $J=2$, and $i=\{1,2,3\}$ for $J=4$. Perturbatively in $C_T^{-1}$, the coefficients $\rho_{i,J}$ and the anomalous dimensions $\Delta_J=8+\gamma_J$ are given by
\begin{equation}\label{eq:dataExp}
	\begin{aligned}
		\rho_{i,J} = \rho^{(0)}_{i,J}\Big[1+{\rho^{(1)}_{i,J}\over C_T}+\ldots\Big] \ , ~~~~~ \Delta_{J} = 8+{\gamma^{(1)}_J\over C_T}+\ldots \ .
	\end{aligned}
\end{equation}
The leading terms $\rho^{(0)}_{i,j}\sim C_T^2$ are due to the disconnected contribution to $\langle TT[T^2]_J\rangle$. This, in turn, produces the factorized part of the stress-tensor two-point function. Namely, to leading order in $C_T$, the correlator reads 
\begin{equation}\label{eq:Factorization}
	\langle T_{\mu\nu}(x)T_{\rho\sigma}(0)\rangle_\beta = \langle T_{\mu\nu}\rangle_\beta\langle T_{\rho\sigma}\rangle_\beta+{\cal O}(C_T) \ .
\end{equation}
Imposing factorization \eqref{eq:Factorization} fixes $5$ out of $6$ coefficients $\rho^{(0)}_{i,J}$:
\begin{equation}\label{eq:MFT}
	\begin{aligned}
		\rho_{1,2}^{(0)} = {324\over 7}\rho_{1,0}^{(0)} \ , ~~~ & \rho_{2,2}^{(0)} = {-1728\over 7}\rho_{1,0}^{(0)} \ ,\cr
		\rho_{1,4}^{(0)} = {160\over 7}\rho_{1,0}^{(0)}\ ,~~~ &\rho_{2,4}^{(0)} = {-1760\over 7}\rho_{1,0}^{(0)} \ , ~~~ \rho_{3,4}^{(0)} = {-480\over 7}\rho_{1,0}^{(0)} \ . \cr
	\end{aligned}
\end{equation}
The remaining coefficient $\rho_{1,0}^{(0)}$ is fixed by the non-zero diagonal terms in \eqref{eq:Factorization}. 

The thermal one-point function of a symmetric traceless operator $\OO$ on $S_\beta^1\times \mathbb{R}^3$ is fixed by symmetry up to a coefficient  $b_{\OO}$ defined in equation \eqref{e.nudjvakswsfv}.

In particular, by the thermalization of the stress tensor in a heavy state with $\Delta_H\sim C_T$ we have  
\begin{equation}
	\langle T_{\mu\nu}\rangle_\beta \approx \langle T_{\mu\nu}\rangle_H \ ,
\end{equation}
from which we find 
\begin{equation}
	\frac{b_{T_{\mu\nu}} }{\beta^4}= -\frac{d\Delta_H}{(d-1)S_4} \ ,
\end{equation}
where on the RHS we have inserted the OPE coefficient and $S_d=\frac{2 \pi^{d\over 2}}{\Gamma(d/2)}$. The relation between $\Delta_H$ and the parameter $\tilde{\mu}$ is given in \eqref{eq:defMuTilde} which leads to the relation:
\begin{equation}\label{eq:oneptT}
	\frac{b_{T_{\mu\nu}}}{\beta^4} = -\frac{C_T S_4(1+\kappa)^3\tilde{\mu}}{320\kappa} \ .
\end{equation}
Furthermore, plugging the MFT solution \eqref{eq:MFT} into the conformal block decomposition \eqref{eq:ConfBlockExp} together with factorization, one finds (to leading order in $C_T^{-1}$) 
\begin{equation}
	\langle T_{tt}\rangle_\beta^2 = 675 \rho^{(0)}_{1,0}  \ .
\end{equation} 
Inserting the stress-tensor one-point function in terms of $\tilde{\mu}$ from \eqref{eq:oneptT} gives 
\begin{equation}\label{eq:resOnePt}
	\rho^{(0)}_{1,0} = \frac{\pi^4 C_T^2(1+\kappa)^6\tilde{\mu}^2}{30720000\kappa^2} \ .
\end{equation}

\subsection{Stress-Tensor Contribution}

We first consider the stress-tensor contribution in the $T\times T$ OPE to the thermal two-point function. The stress-tensor three-point function is fixed by conformal symmetry up to three OPE coefficients $(\hat{a},\hat{b},\hat{c})$ in $d=4$ \cite{Osborn:1993cr} and the contribution to the stress-tensor two-point function at finite temperature was studied in, $e.g.$,  \cite{Kulaxizi:2010jt,Karlsson:2022osn}.\footnote{In particular, the contribution to the stress-tensor two-point functions $\hat{G}_{xy,xy}$, $\hat{G}_{tx,tx}$ and $\hat{G}_{tz,tz}$ can be found in Eqs.\ \eqref{e.p1c24} and \eqref{eq:Ttxtx} which, after integrating over the $xy$-plane, is given by Eqs.\ \eqref{eq:TxyxyInt}, \eqref{e.p1c28} and \eqref{e.p1c30}.  We shall not repeat them here due to their lengthy and unilluminating form.} 

Here, we are interested in the values for $(\hat{a},\hat{b},\hat{c})$ computed holographically in Gauss-Bonnet gravity. 
It was found in \cite{Buchel:2009sk} that  
\begin{equation}\label{eq:t2GB}
	t_{2,GB} = \frac{4f_\infty \lambda_{GB}}{1-2 f_\infty \lambda_{GB}}\frac{d(d-1)}{(d-2)(d-3)} \ , ~~~~ t_{4,GB}=0
\end{equation}
with the remaining coefficient fixed by Ward identities. 
The relation to the $(\hat{a},\hat{b},\hat{c})$ and $(t_2, t_4, C_T)$ bases can be found in \eqref{eq:abcGB}. 
We will be interested in the conformal collider bounds \cite{Hofman:2008ar}:
\begin{equation}
	\begin{aligned}
		(1-\frac{t_2}{3}-\frac{2t_4}{15})\geq 0 \ , ~~~~ 2(1-\frac{t_2}{3}-\frac{2t_4}{15})+t_2\geq 0 \ , ~~~~ \frac{3}{2}(1-\frac{t_2}{3}-\frac{2t_4}{15})+t_2+t_4\geq 0 \ , 
	\end{aligned}
\end{equation}
which for $t_2=t_{2,GB}$ and $t_4=0$ reduce to 
\begin{equation}\label{eq:ccbGB}
	\begin{aligned}
              (\kappa-\frac{4}{5})\geq 0 \ , ~~~~~ (2-\kappa)\geq 0 \ , ~~~~~ (\frac{4}{3}-\kappa)\geq 0 \ , 
	\end{aligned}
\end{equation}
where $\kappa=\sqrt{1-4\lambda_{GB}}$. 
The bounds are saturated for $\kappa=\{\frac{4}{5},2,\frac{4}{3}\}$. 

In \cite{Kulaxizi:2010jt}, the stress-tensor two-point function at finite temperature in the OPE expansion was considered in momentum space. In particular, the leading term in the lightcone limit due the stress-tensor contribution in the OPE was proportional to the conformal collider bounds in the respective channel. We now study this in position space after integrating over the $xy$-plane in the context of Gauss-Bonnet gravity. 

Using \eqref{eq:defMuTilde} together with $(\hat{a},\hat{b},\hat{c})$ \eqref{eq:abcGB} relevant for Gauss-Bonnet gravity, we find\footnote{The corresponding results  in terms of $(\Delta_H,\hat{a},\hat{b},\hat{c})$
can be found  in  Eqs.\ \eqref{eq:TxyxyInt}, \eqref{e.p1c28} and \eqref{e.p1c30}.}
{\small{
\begin{equation}
	\begin{aligned}\label{eq:ResultStressTensor}
		G_{xy,xy}|_{\tilde{\mu}} &= \frac{(5\kappa -4) \pi  C_T (1+\kappa)^3\tilde{\mu} (t^2-z^2)
			}{800 \kappa ^2
			\left(t^2+z^2\right)^2} \ , \\
              G_{tx,tx}|_{\tilde{\mu}} &= -\frac{\pi  C_T(1+\kappa)^3\tilde{\mu}  \left((13 \kappa -4)
			t^4+6 (\kappa -2) t^2 z^2+(8-15 \kappa )
			z^4\right)}{6400 \kappa ^2 
			\left(t^2+z^2\right)^3} \ , \\
		G_{tz,tz}|_{\tilde{\mu}} &= \frac{\pi  C_T(1\!+\!\kappa)^3\tilde{\mu}  \left(-21 (3 \kappa \!+\!2)
			t^6\!+\!3 (94-93 \kappa ) t^4 z^2\!+\!(39 \kappa \!+\!98) t^2
			z^4\!+\!(111 \kappa -34) z^6\right)}{28800 \kappa ^2
		\left(t^2+z^2\right)^4} \ . \\
	\end{aligned}
\end{equation}}}%
The result for $G_{xy,xy}$ in \eqref{eq:ResultStressTensor} is in agreement with the bulk computation in \eqref{mu1res}. To compare the remaining two polarizations with the bulk results, we apply the differential operators ${\cal D}^{2p}=(\partial_t^2+\partial_z^2)^p$ with $p=1$ for $G_{tx,tx}$ and $p=2$ for $G_{tz,tz}$. The results are
\begin{equation}
	\begin{aligned}
		{\cal D}^2G_{tx,tx}|_{\tilde{\mu}} &= -\frac{ (\kappa -2) 3 \pi  C_T(1+\kappa)^3 \tilde{\mu}
			\left(t^4-6 t^2
			z^2+z^4\right)}{1600 \kappa ^2
			 \left(t^2+z^2\right)^4} \ ,  \cr 
		{\cal D}^4G_{tz,tz}|_{\tilde{\mu}} &= \frac{(3 \kappa -4)  \pi  C_T(1+\kappa)^3 \tilde{\mu} 
			\left(t^6-15 t^4 z^2+15 t^2
			z^4-z^6\right)}{30 \kappa ^2
			\left(t^2+z^2\right)^6} \ ,
	\end{aligned}
\end{equation}
which agree with the bulk results in \eqref{eq:Bulktxtx1p2} and \eqref{tzmu1}. It follows that when $\kappa=\{\frac{4}{5},2,\frac{4}{3}\}$, the stress-tensor contribution to $G_{xy,xy}$, ${\cal D}^2 G_{tx,tx}$ and ${\cal D}^4 G_{tz,tz}$ vanishes.

\subsection{Double-Stress Tensor Contributions}

In the previous section, we saw that when a conformal collider bound is saturated, the contribution due to the stress-tensor operator to the $TT$ correlators at finite temperature vanishes for the corresponding polarization. 
In the lightcone limit at $\OO(({x\over \beta})^{4 k})$, the only operator that contributes is the multi-stress tensor operators on the leading Regge trajectory $[T^k]_{\mu_1\mu_2\ldots\mu_{2k}}$ (with spin $J=2k$). The bulk computation shows that not only does the stress-tensor contribution vanish when the conformal collider bounds are saturated, but the full contribution from the leading Regge trajectory also vanishes for the same choice of polarization. 

Below, we will read off the conformal data of double-stress tensors by comparison to the bulk results in Section \ref{Sec:Bulk}, following closely Chapter \ref{ch4}. 
With this, we will see how the leading terms in the lightcone limit vanish when ANECs are saturated, which we further relate to the saturation of higher-spin ANECs in the next section.

We now consider the contribution due to double-stress tensors $[T^2]_J$ with $J=0,2,4$ to $G_{\mu\nu,\rho\sigma}$. 
We again use the OPE \eqref{eq:ConfBlockExp} and expand the dynamical data \eqref{eq:dataExp} to subleading order in $C_T$. 
The disconnected contribution was discussed above which gave the MFT coefficients $\rho_{i,J}^{(0)}$ in \eqref{eq:MFT} and \eqref{eq:resOnePt}.\footnote{The expression for the integrated conformal blocks expanded to subleading order in $C_T^{-1}$ can be found in \ref{as.labelik}. Only the overall normalization differs due to different values for $\rho^{(0)}_{1,0}$.}
Note that we need to regulate the integrals over the $xy$-plane which we do by inserting a factor of $(t^2+x^2+y^2+z^2)^{-{\epsilon\over 2}}$. As in Chapter \ref{ch4}, we determine the double-stress tensor CFT data by imposing that the conformal block decomposition in terms of the CFT data agrees with the bulk results obtained in Section \ref{Sec:Bulk}:
\begin{equation}
	\begin{aligned}
		G_{xy,xy}^{(CFT)}-G_{xy,xy}^{(bulk)}\Big|_{\tilde{\mu}^2 C_T} = 0 \ , \\
		{\cal D}^2\Big[G_{tx,tx}^{(CFT)}-G_{tx,tx}^{(bulk)}\Big]\Big|_{\tilde{\mu}^2 C_T} = 0 \ , \\
		{\cal D}^4\Big[G_{tz,tz}^{(CFT)}-G_{tz,tz}^{(bulk)}\Big]\Big|_{\tilde{\mu}^2 C_T} = 0 \ . 
	\end{aligned}
\end{equation}

Using the bulk results \eqref{Gxy2}, \eqref{eq:Bulktxtx2p2} and \eqref{tzmu2} together with the conformal block expansion \eqref{eq:ConfBlockExp}, we find
\begin{equation}\label{eq:anomDimGB}
	\begin{aligned}
		\gamma^{(1)}_0 &= -\frac{80 \left(2103 \kappa ^2-4464 \kappa +2392\right)}{63 \pi ^4 \kappa ^2} \ ,\cr
		\gamma^{(1)}_2&= \frac{10 \left(19563 \kappa ^2-39996 \kappa +20012\right)}{189 \pi ^4 \kappa ^2}\ , \cr
		\gamma^{(1)}_4 &= -\frac{2 \left(24157 \kappa ^2-51412 \kappa +30228\right)}{105 \pi ^4 \kappa ^2},
	\end{aligned}
\end{equation}
and 
\begin{equation}\label{eq:OPEGB}
	\begin{aligned}
		&\rho_{2,2}^{(1)} = \frac{5 \left(157699 \kappa ^2-323228 \kappa +162636\right)}{1296 \pi ^4 \kappa ^2}+\rho _{1,2}^{(1)}  \ ,\cr
		&\rho_{2,4}^{(1)} =\frac{108521 \kappa ^2-170036 \kappa +65684}{2310 \pi ^4 \kappa ^2}+\rho _{1,4}^{(1)} \ ,\cr
		&\rho_{3,4}^{(1)} = \frac{-4053 \kappa ^2-14652 \kappa +21788}{1260 \pi ^4 \kappa ^2}+\rho _{1,4}^{(1)}\ ,
	\end{aligned}
\end{equation}
which reduce to the pure Einstein gravity results when $\kappa=1$. The remaining coefficients $(\rho_{1,0}^{(1)},\rho_{1,2}^{(1)},\rho_{1,4}^{(1)})$ are undetermined in the near-boundary analysis in the bulk, as mentioned in Section \ref{Sec:Bulk}.

Consider now the lightcone limit $(x^+,x^-)=(it+z,it-z)$ with $x^-\to 0$. Doing so, we find 
\begin{equation}\label{eq:TsqLC}
	\begin{aligned}
		G_{xy,xy}^{(CFT)}(x^+,x^-)\Big|_{\tilde{\mu}^2C_T} &\underset{x^-\to 0}{=} -\frac{ (4-5 \kappa )^2  \pi (1+\kappa)^6 C_T\tilde{\mu }^2}{153600 \kappa ^4 }\frac{(x^+)^3}{x^-} \ , \cr
		G_{tx,tx}^{(CFT)}(x^+,x^-)\Big|_{\tilde{\mu}^2C_T} &\underset{x^-\to 0}{=} -\frac{(\kappa -2)^2 17 \pi(1+\kappa)^6   C_T  \tilde{\mu }^2}{68812800 \kappa ^4  }\frac{(x^+)^4}{(x^-)^2} \ , \cr
		G_{tz,tz}^{(CFT)}(x^+,x^-)\Big|_{\tilde{\mu}^2C_T} &\underset{x^-\to 0}{=} -\frac{(4-3 \kappa )^2  11 \pi  (1+\kappa)^6 C_T \tilde{\mu }^2}{387072000 \kappa ^4 }\frac{(x^+)^5}{(x^-)^3} \ , 
	\end{aligned}
\end{equation}
where we note that this contribution comes solely from the spin-$4$ operator.\footnote{This property can be seen in Eq.\ \eqref{e.prijdeakpovolan}.}
Moreover, the near-lightcone behaviour is completely determined by the data in \eqref{eq:anomDimGB} and \eqref{eq:OPEGB}.  

In the lightcone limit,  when the conformal collider bounds are saturated, $i.e.$, $\kappa=\{\frac{4}{5},2,\frac{4}{3}\}$, both the stress-tensor and the spin-$4$ double-stress tensor contributions vanish.  As we will see in the following section, this is related to the saturation of the spin-$4$ ANEC, where the spin-$4$ operator is the double-stress tensor of the schematic form $:T_{(\mu\nu}T_{\rho\sigma)}:$.

\section{ANEC Interference Effects and Spin-4 ANEC}\label{ANECsection}

In this section, we study interference effects of the ANEC as well as the spin-4 ANEC. Interference effects in large-$C_T$ CFTs impose strong constraints on the MFT OPE coefficients.  We will see explicitly that the MFT OPE coefficients for the double-stress tensors, \eqref{eq:MFT}, are consistent with interference effects. 
In particular, we  verify that when the spin-$2$ ANEC is saturated
the spin–$4$ ANEC, the null-integrated $[T^2]_{J=4}$ double-stress tensor in holographic Gauss-Bonnet gravity, is also saturated in a stress-tensor state. 

Assuming a holographic CFT with a large $C_T$ and no light scalars, 
 the leading Regge trajectory of the $d=4$ stress-tensor OPE takes the following schematic form:
\begin{equation}
	T(x)T(0) = x^{-8}\Big[1+x^4 T(0)+ x^{\Delta_{[T^2]_4}}[T^2]_{J=4}(0)+\ldots\Big],
\end{equation}
where the ellipses denote higher-spin operators on the leading Regge trajectory, $i.e.$, multi-stress tensors $[T^k]_{J=2k}$ as well as all other operators. 
When integrated over a light-ray, the operators on the leading Regge trajectories $\OO^{(J)}$ are positive operators, see, $e.g.$,\ \cite{Komargodski:2016gci,Hartman:2016lgu,Meltzer:2017rtf,Meltzer:2018tnm}:
\begin{equation}
	\EE^{(J)} = \int_{-\infty}^\infty dx^- \OO^{(J)}_{-,-,\ldots,-}(x^-,0) \ , ~~~~ J=2,4,6, \ldots \ .
\end{equation}
In putative holographic CFTs dual to pure gravity 
in the bulk, the operators on the leading Regge trajectory are the multi-stress tensors $\OO^{(J)} = [T^k]_{J=2k}$. 
These are the ones that we will study. 
In particular, by studying matrix elements of $\EE^{(J)}$ in states that are superpositions of the stress tensor and multi-stress tensors, the positivity of the ANEC and higher-spin ANECs impose constraints on the stress-tensor OPE. 

To begin with, we consider the ANEC $\EE^{(2)}>0$ following \cite{Cordova:2017zej,Meltzer:2017rtf,Meltzer:2018tnm} and verify that it is satisfied in states of the schematic form $|\psi_J\rangle = v_1|T\rangle + v_2|[T^2]_{J}\rangle$ with $J=0,2,4$. 
This leads to a positive definite matrix schematically given by 
\begin{equation}\label{eq:ANEC}
	\langle\psi_J| \EE^{(2)}|\psi_J\rangle^{(i)} = v^\dagger \begin{pmatrix}
		\langle T|\EE^{(2)}|T\rangle & \langle T| \EE^{(2)}|[T^2]_J\rangle \\
		\langle [T^2]_J| \EE^{(2)}|T\rangle &\langle [T^2]_J| \EE^{(2)}|[T^2]_J\rangle
	\end{pmatrix}^{(i)}v\geq  0\,
\end{equation}
where the superscript $(i)$ labels different structures. Note that the entries are in general matrices. One then obtains bounds of the schematic form:  
\begin{equation}\label{eq:boundsSpin2ANEC}
	f^{(i)}(\{\Delta\},\{J\}) (\langle TT[T^2]_J\rangle^{(i)})^2 \leq \langle [T^2]_JT[T^2]_J\rangle^{(i)} \langle TTT\rangle^{(i)} 
\end{equation}
where $f^{(i)}(\{\Delta\},\{J\})$ is some function which depends on the scaling dimensions, spins,  
and the kinematical structure independent of the details of a theory. 

We expect that \eqref{eq:ANEC} in holographic CFTs 
has a $C_T$ scaling like follows
\begin{equation}
	v^\dagger\begin{pmatrix}
		\langle \tilde{T}| \EE^{(2)}|\tilde{T}\rangle & \langle \tilde{T}| \EE^{(2)}|[\tilde{T}^2]_J\rangle \\
		\langle [\tilde{T}^2]_J| \EE^{(2)}|\tilde{T}\rangle &\langle [\tilde{T}^2]_J| \EE^{(2)}|[\tilde{T}^2]_J\rangle
	\end{pmatrix}^{(i)}v = v^\dagger\begin{pmatrix}
		m_1 & C_T^{1/2} m_2\\
		C_T^{1/2} m_3 & m_4
	\end{pmatrix}^{(i)}v\geq 0 \ ,
\end{equation}  
for some $\OO(1)$ matrices $m_i$.  Here $\tilde{T}$ and $\tilde{T}^2$ denote unit-normalized operators/states.
By an appropriate choice of $v$, the above matrix requires positivity of any $2\times2$ submatrix. By a suitable choice of $v$, one can obtain terms of $\OO(C_T^{1/2})$ from the off-diagonal part and $\OO(1)$ terms from the diagonal part; this leads to potential positivity violations. Below, we will explicitly examine the spin-$2$ ANEC in the states $|\psi_J\rangle$ and show that the solution \eqref{eq:MFT} is consistent with positivity.

In what follows, we use the following three-point function basis \cite{Costa:2011mg,Costa:2011dw}:
{\small
\begin{equation}\label{eq:spinningBasis}
	\langle \OO_{\Delta_1,J_1} \OO_{\Delta_2,J_2} \OO_{\Delta_3,J_3}\rangle = \!\!\!\!\!\sum_{n_{12},n_{13}n_{23}}\!\!\!\!\! c^{(123)}_{n_{23},n_{13},n_{12}}{V_1^{J_1-n_{12}-n_{13}} V_2^{J_2-n_{23}-n_{12}} V_3^{J_3-n_{13}-n_{23}} H_{12}^{n_{12}}H_{13}^{n_{13}}H_{23}^{n_{23}} \over x_{12}^{\beta_{123}} x_{13}^{\beta_{132}} x_{23}^{\beta_{231}}},
\end{equation}}with $\beta_{ijk}=\beta_i+\beta_j-\beta_k$ and $\beta_i=\Delta_i+J_i$.   
This notation will be convenient to compare the data in the differential basis used in this chapter with the results of 
\cite{Meltzer:2017rtf,Meltzer:2018tnm}.

\subsection{Spin-0 Double-Stress Tensor Interference}

Interference effects between the stress-tensor state and a scalar was considered in \cite{Cordova:2017zej} which found that the function $f(\Delta)$ appearing in \eqref{eq:boundsSpin2ANEC} has (double) zeroes at $\Delta=2d+n$, 
where $\Delta$ refers to the dimension of the $T^2$ operator in $|\psi_0\rangle$.   
Due to the double-zero, there's no violation of the ANEC when considering interference effects in the state $|\psi_0\rangle$ to leading order in $C_{T}^{-1}$.

\subsection{Spin-2 Double-Stress Tensor Interference}

In \cite{Meltzer:2017rtf}, the positivity of the ANEC operator in a mixed state of a stress tensor and a spin-$2$ operator was studied. To this end, consider the state 
\begin{equation}
	|\psi_2\rangle = v_1|T\rangle +v_2 |[T^2]_{2}\rangle \ . 
\end{equation}
Due to the large-$C_T$ expansion, there is again a potential issue with the ANEC for the mixed stress tensor and spin-2 double-stress tensor state.  
It was explained in \cite{Meltzer:2017rtf} that
 if one parameterizes the three-point function $\langle TT[T^2]_2\rangle $ by $c^{(T[T^2]_2T)}_{0,0,0}$ and $c^{(T[T^2]_2T)}_{1,0,1}$ in the basis \eqref{eq:spinningBasis}, 
and imposes conservation, the ANEC positivity implies that 
$c^{(T[T^2]_2T)}_{0,0,0}=0$ while $c^{(T[T^2]_2T)}_{1,0,1}$ is unconstrained. 

Translating between the basis $(c^{(T[T^2]_2T)}_{0,0,0},c^{(T[T^2]_2T)}_{1,0,1})$ and the differential basis$(\rho_{1,2}^{(0)},\rho_{2,2}^{(0)})$, we find\footnote{Note that coefficients $\rho$ is a product of OPE coefficients and the thermal one-point function.} 
\begin{equation}\label{eq:spin2Coeffs}
	\begin{aligned}
		c^{(T[T^2]_2T)}_{0,0,0} = -\frac{96}{7} \left(16 \rho_{1,2}^{(0)}+3 \rho_{2,2}^{(0)}\right) \ , ~~~ c^{(T[T^2]_2T)}_{1,0,1} = \frac{1}{63} \left(2108 \rho_{1,2}^{(0)}+89\rho_{2,2}^{(0)}\right). 
	\end{aligned}
\end{equation} The superscript denotes the leading $C_T$ expressions, corresponding to $\Delta_{T^2}=8$ in $d=4$. 
Inserting the MFT solution \eqref{eq:MFT} in \eqref{eq:spin2Coeffs} gives
\begin{equation}
	\begin{aligned}
		c^{(T[T^2]_2T)}_{0,0,0} = 0 \ , ~~~ c^{(T[T^2]_2T)}_{1,0,1} =1200\rho_{1,0}^{(0)} \ .  
	\end{aligned}
\end{equation}
We see that $c^{(T[T^2]_2T)}_{0,0,0}=0$ while $c^{(T[T^2]_2T)}_{1,0,1}$ is unconstrained, showing consistency with the ANEC to leading order in $C_T^{-1}$ as discussed in \cite{Meltzer:2017rtf}.

\subsection{Spin-4 Double-Stress Tensor Interference}

Interference effects of both the ANEC and the spin-4 ANEC was studied in \cite{Meltzer:2018tnm}. 
There is again a potential issue with off-diagonal term that gives the leading large $C_T$ contribution when the minimal-twist spin-$4$ operator has dimension $\Delta=8+\OO(C_T^{-1})$ in holographic CFTs. This potentially leads to violations of the ANEC, but we will show that this is not the case based on the solution \eqref{eq:MFT}. 

Below, we define $\Theta=[T^2]_4$ and also denote the matrix elements of $\langle \OO_1|\EE^{(2)}|\OO_2\rangle^{(j)}$ by $\EE^{(2,j)}_{\OO_1\OO_2}$. 
Based on the results obtained in  \cite{Meltzer:2018tnm}, we obtain\footnote{More precisely, we take Eq.\ (C.9) in \cite{Meltzer:2018tnm} to obtain $\mathcal{E}^{(2,i)}_{T\Theta}$ in terms of $\mathcal{E}^{(4,j)}_{TT}$  and then use Eq.\ (C.2)-(C.4) in \cite{Meltzer:2018tnm} to express $\mathcal{E}^{(2,i)}_{T\Theta}$ in terms of the OPE coefficients $(c_{0,0,2}^{(T\Theta T)},c_{0,1,1}^{(T\Theta T)},c_{1,0,1}^{(T\Theta T)})$ for the basis \eqref{eq:spinningBasis}.  We refer the reader to \cite{Meltzer:2018tnm} for more details.} 
\begin{equation}
	\begin{aligned}
		\mathcal{E}^{(2,0)}_{T\Theta} &= \frac{1053\mathcal{E}^{(4,0)}_{TT} +748 \mathcal{E}^{(4,1)}_{TT}+128 \mathcal{E}^{(4,2)}_{TT}}{2419200} \ ,\cr
		\mathcal{E}^{(2,1)}_{T\Theta} &= \frac{319\mathcal{E}^{(4,0)}_{TT} +1284 \mathcal{E}^{(4,1)}_{TT}+204 \mathcal{E}^{(4,2)}_{TT}}{3225600} \ ,\cr
		\mathcal{E}^{(2,2)}_{T\Theta} &= \frac{217\mathcal{E}^{(4,0)}_{TT} +852 \mathcal{E}^{(4,1)}_{TT}+1752 \mathcal{E}^{(4,2)}_{TT}}{9676800} \ .
	\end{aligned}
\end{equation}
Due to the large-$C_T$ scaling, we need to impose $\mathcal{E}^{(2,i)}_{T\Theta }=0$ to leading order in $C_T^{-1}$; otherwise we would find violations of the ANEC. However, each $\mathcal{E}^{(4,i)}_{TT}$ is non-negative which implies that $\mathcal{E}^{(4,i)}_{TT}=0$  to leading order in $C_T$. In terms of $(c_{0,0,2}^{(T\Theta T)},c_{0,1,1}^{(T\Theta T)},c_{1,0,1}^{(T\Theta T)})$, we find the only solution is 
\begin{equation}
	\begin{aligned}
		c_{0,0,2}^{(T \Theta  T)}=0 \ , ~~~ c_{0,1,1}^{(T \Theta  T)}=0 \ , ~~~ c_{1,0,1}^{(T \Theta  T)}=0 \ , ~~~
	\end{aligned}
\end{equation}
which seems to imply that $\Theta=[T^2]_{J=4}$ cannot appear in the stress-tensor OPE.  
But this is not the case due to the behavior of the OPE coefficients as we now explain. 
Solving conservation and the permutation symmetry  in terms of the three coefficients $(c_{0,0,2}^{(T \Theta  T)},c_{0,1,1}^{(T \Theta  T)},c_{1,0,1}^{(T \Theta  T)})$, we find that all the coefficients are regular as $\Delta\to 8$ except for 
\begin{equation}
	c_{2,0,2}^{(T \Theta  T)} \sim {1\over\Delta-8}p(c_{0,0,2}^{(T \Theta  T)},c_{0,1,1}^{(T \Theta  T)},c_{1,0,1}^{(T \Theta  T)}) \ ,
\end{equation}
where $p(c_{0,0,2}^{(T \Theta  T)},c_{0,1,1}^{(T \Theta  T)},c_{1,0,1}^{(T \Theta  T)})$ is a linear function of the OPE coefficients $(c_{0,0,2}^{(T \Theta  T)},$ $c_{0,1,1}^{(T \Theta  T)},c_{1,0,1}^{(T \Theta  T)})$. 
Requiring that the three-point function is regular as $\Delta=\Delta_\Theta\to 8$, we write\footnote{Including the anomalous dimensions would lead to the coefficients having different scaling with $C_T$.} 
\begin{equation}
	\begin{aligned}
		\lim_{\Delta\to 8}c_{0,0,2}^{(T \Theta  T)} &= (\Delta-8)\tilde{c}_{0,0,2}^{(T \Theta  T)},\cr
		\lim_{\Delta\to 8}c_{0,1,1}^{(T \Theta  T)} &= (\Delta-8)\tilde{c}_{0,1,1}^{(T \Theta  T)},\cr
		\lim_{\Delta\to 8}c_{1,0,1}^{(T \Theta  T)} &= (\Delta-8)\tilde{c}_{1,0,1}^{(T \Theta  T)},
	\end{aligned}
\end{equation}
with constants $\tilde{c}$'s that are finite as $\Delta \to 8$. 
This does not imply that the three-point function is trivial due to the simple pole in $c_{2,0,2}^{(T \Theta  T)}$.  In particular, the three-point function is 
\begin{equation}\label{eq:attt}
	\langle T(P_1)[T^2]_4(P_2)T(P_3)\rangle = {\alpha H_{12}^2H_{23}^2\over (P_1\cdot P_2)^6(P_3\cdot P_2)^6} \ ,
\end{equation}
for some coefficient $\alpha$. As the three coefficients $(c_{0,0,2}^{(T \Theta  T)},c_{0,1,1}^{(T \Theta  T)},c_{1,0,1}^{(T \Theta  T)})$ all vanish as $\Delta\to 8$, the solution is consistent with $\EE_{T\Theta}^{(2,i)}=0$ to leading order in $C_T$.

Note that the leading Regge trajectory obey the inequalities 
\begin{equation}
	d-2\leq \tau_{J,min}<2(d-2) \ ,
\end{equation}
in interacting CFTs. Therefore including anomalous dimensions of $\OO(C_T^{-1})$ such that $\tau_4<4$, the coefficients  $(c_{0,0,2}^{(T \Theta  T)},c_{0,1,1}^{(T \Theta  T)},c_{1,0,1}^{(T \Theta  T)})$ can become non-zero and not violate the spin-$4$ interference effects.

We find that the solution \eqref{eq:attt} agrees with that of MFT \eqref{eq:MFT}. This can be seen by inserting \eqref{eq:MFT} into the explicit expressions for the three-point function in the differential basis, giving $\alpha={75\rho_{1,0}^{(0)}\over 26}$ in the three-point function \eqref{eq:attt}.\footnote{It can also be seen by solving for $c_{n_{23},n_{13},n_{12}}^{(T\Theta T)}=c_{n_{23},n_{13},n_{12}}^{(T \Theta  T)}(\rho_{1,4}^{(0)},\rho_{2,4}^{(0)},\rho_{3,4}^{(0)})$, from which one finds that all coefficients vanish except for $c_{2,0,2}^{(T \Theta  T)}$ (to leading order in $C_T^{-1}$).} 
Therefore, we conclude that the MFT coefficients \eqref{eq:MFT} are consistent with positivity of the ANEC in the state which is a superposition of the stress tensor and the spin-$4$ double-stress tensor $[T^2]_{J=4}$.

\subsection{Spin-4 ANEC in Stress-Tensor State}

We have seen how the MFT solution is consistent with the ANEC in states $|\psi_J\rangle$ that are superpositions of a stress-tensor and double-stress tensor state. We now move on to consider the spin-$4$ ANEC and study it when the  spin-$4$ operator is the double-stress tensor $[T^2]_{J=4}$ with the OPE data obtained in holographic Gauss-Bonnet theory. We will show that the saturation of the spin-$4$ ANEC happens precisely when the corresponding contribution to the near-lightcone $TT$ correlators at finite temperature vanishes, generalizing the results for the stress tensor in \cite{Kulaxizi:2010jt}. Note this analysis is sensitive to the subleading terms in the $C_T^{-1}$ expansion of the double-stress tensor data.

One can obtain the spin-4 ANEC in a stress-tensor state $\EE^{(4,j)}_{TT}$ using the results from \cite{Meltzer:2018tnm}.\footnote{See Eq.\ (4.4)-(4.6) in \cite{Meltzer:2018tnm}.} We change basis from $(c_{0,0,2}^{(T \Theta  T)},c_{0,1,1}^{(T \Theta  T)},c_{1,0,1}^{(T \Theta  T)})$ to the differential basis $(\rho_{1,4},\rho_{2,4},\rho_{3,4})$ used in this chapter and perform the $C_T^{-1}$ expansion.\footnote{The results are proportional to the leading lightcone expressions in Eq.\ \eqref{e.prijdeakpovolan}.} 
Using the values in Gauss-Bonnet gravity given in \eqref{eq:anomDimGB} and \eqref{eq:OPEGB}, we obtain 
\begin{equation}\label{eq:spin4ANECGB}
	\begin{aligned}
		0\leq\EE^{(4,0)}_{TT} &=  \frac{11 \pi ^4 C_T(4-3 \kappa )^2 (\kappa +1)^6 \tilde{\mu }^2}{2211840000 \kappa ^4},\cr
		0\leq\EE^{(4,1)}_{TT} &=  \frac{17 \pi ^4 C_T(\kappa -2)^2 (\kappa +1)^6 \tilde{\mu }^2}{98304000 \kappa ^4},\cr
		0\leq\EE^{(4,2)}_{TT} &=  \frac{7 \pi ^4 C_T(4-5 \kappa )^2 (\kappa +1)^6 \tilde{\mu }^2}{3072000 \kappa ^4},
	\end{aligned}
\end{equation}
which saturates when $\kappa=\{{4\over 3},2,{4\over 5}\}$.

\section{Discussion and Implications}\label{s.sekcie56}

In this chapter, we studied  thermal $TT$ correlators and explore their connections to  ANECs.
One can use the OPE between two stress tensors and expand the correlator in powers of the temperature.
The contributions from a single-stress tensor in the lightcone limit are proportional to
the corresponding spin-2 ANECs.
To go beyond it, we consider holographic Gauss-Bonnet gravity, where the breakdown of spin-$2$ ANECs is related to superluminal signal propagation.
We analyze the multi-stress tensor contributions to the $TT$ correlators in the dual $d=4$ CFT with a large central charge.
Our chief finding in this chapter is that, when an  ANEC is saturated in a state created by the stress tensor, all higher-spin ANECs are saturated in this state as well -- the corresponding near-lightcone thermal $TT$ correlator takes the vacuum form.
 
 Note that the statement about ANEC saturation is really a statement about the OPE of
 the stress tensors, so instead of a thermal state one may consider any other suitable state in the theory.
 One may ask how general our observation is -- does it apply beyond holographic models
 and beyond the large $C_T$ limit?  Below, we discuss related questions and possible future directions.

\begin{itemize}

\item  {\it Scope of the result and possible proof:}

It was argued in \cite{Zhiboedov:2013opa,Meltzer:2018tnm}, that ANEC saturation implies that the theory is, in some sense, free. In particular, by studying ANECs in the states created by linear combinations of spin-2 and spin-4 operators, \cite{Meltzer:2018tnm} argued that the spin-4 operator must be a conserved current and hence the theory is free. However we found that things can be more subtle when the spin-4 operator has dimension eight, which is the case for the minimal-twist double-stress tensors in CFTs with a large $C_T$. In this case the theory is not free, and only thermal correlators with certain polarization simplify in the near-lightcone regime.

\qquad Are there  examples of unitary interacting CFTs which are ``free'' near the lightcone, like holographic GB gravity we studied here? That would be an interesting question to investigate.  Once the scope of this phenomenon becomes more clear, it would be natural to search for a proof as well.

\item {\it Free theories and their large $N$ limit:} 

Free theories (bosons, fermions and gauge fields in four spacetime dimensions) saturate conformal collider bounds, so it is natural to ask what happens with the higher-spin ANECs in this case.
Of course, the near-lightcone behavior in free theories is governed by the conserved, higher-spin currents.
Nevertheless, it would be interesting to see if there are any patterns of the type we observed in this chapter.
It seems that studying the large $C_T$ (or large $N$) limit of free theories might be particularly interesting.

\item {\it Relation to experiment and to lattice computations:} 

One may wonder if there are CFTs which are interacting and at the same time saturate ANECs, like the holographic model we considered in this chapter. Presumably a spin-four operator with conformal dimension close to eight might be necessary for this to happen. It would be interesting to check, how far, $e.g.$, QCD at finite temperature is from this regime and to compare our results with the lattice computations of the $TT$ correlators (see, $e.g.$, \cite{Meyer:2011gj} for a review).

\item  {\it Anomalous dimensions of the spin-2 $[T_{\mu\nu}]^2$ operator:}

We note that the anomalous dimension for the spin-$2$ double-stress tensor, given by the second equation in (\ref{eq:anomDimGB}), is negative for $\lambda_{GB}=0$ (Einstein gravity) but changes sign and becomes positive for values of $\lambda_{GB}$ inside the conformal collider bounds.  
It would be interesting to understand the meaning of these values of $\lambda_{GB}$ where this happens.

\item {\it Minimal-twist multi-stress tensors with derivatives and spherical black holes:} 

For technical reasons, we restrict our discussion to a black hole with a planar horizon. This corresponds to considering multi-stress tensor operators without additional derivatives appearing in the OPE.  
It would be interesting to study the role of operators with derivatives, although this would be technically more involved than the analysis we did in this chapter.

\item  {\it Near-lightcone $TT$ correlators %; 
and 
higher-derivative gravities:}

On a related note, one may ask if one can make progress in computing the near-lightcone behavior of holographic correlators for generic holographic models.

\qquad Much recent progress has been made in understanding the multi-stress tensor sector of the $d=4$ thermal {\it scalar} two-point functions and related heavy-heavy-light-light (HHLL) correlators \cite{Fitzpatrick:2019zqz, Karlsson:2019qfi,Li:2019tpf,Huang:2019fog,Kulaxizi:2019tkd,Fitzpatrick:2019efk,Karlsson:2019dbd,Haehl:2019eae,Li:2019zba,Karlsson:2019txu,Huang:2020ycs,Karlsson:2020ghx,Li:2020dqm,Parnachev:2020fna,Fitzpatrick:2020yjb,Berenstein:2020vlp,Parnachev:2020zbr,Karlsson:2021duj,Rodriguez-Gomez:2021pfh,Huang:2021hye,Rodriguez-Gomez:2021mkk,Krishna:2021fus,Bianchi:2021yqs,Karlsson:2021mgg,Huang:2022vcs,Dodelson:2022eiz,Dodelson:2022yvn, Parisini:2022wkb}. 
As was observed in \cite{Huang:2021hye, Karlsson:2021mgg}, the structure of the $d=4$ thermal scalar two-point correlator in the lightcone limit has certain similarity with the ${\cal W}_3$ vacuum blocks in $d=2$ CFT \cite{Z1985}.  While the reasons for this remain unclear,  one may wonder whether a similar story exists for the $TT$ correlators.

\qquad For example, is there a universality of the near-lightcone $TT$ correlators similar to 
the one exhibited by the near-lightcone HHLL holographic correlators?
The addition of higher-derivative terms to the bulk gravitational Lagrangian leads to the variation of the $TTT$ couplings, but is the near-lightcone behaviour of the holographic $TT$ correlators fixed (and universal) in terms of these couplings? Can the bootstrap techniques of \cite{Karlsson:2019dbd} be applied to compute the full $TT$ correlator? We leave these questions for future investigation.

\qquad Note that the model we consider, Gauss-Bonnet gravity,  can be regarded as the simplest type of the Lovelock theories \cite{lovelock1971}. We expect that the techniques used here can be used to deal with other higher-derivative corrections to the bulk Lagrangian.  Additional parameters present in such theories can also be useful for studying possible universality of the holographic thermal $TT$ correlators. 

\qquad An immediate example to consider could be quasi-topological gravity (QTG) \cite{Myers:2010ru,Myers:2010jv},\footnote{Note that the quasi-topological gravity is not a Lovelock theory. Despite this, the linearized field equation turns out to be second order in AdS$_5$ background and the theory admits an exact black hole solution of type \eqref{bhmetrics}.}
{
\allowdisplaybreaks{\small{
\begin{align}
S &= -\frac{1}{16\pi G} \int d^5x \sqrt{g} \left( \mathcal{R} + \frac{12}{L^2} + \lambda \,\mathcal{L}^{(\lambda)} + \gamma \,\mathcal{L}^{(\gamma)} \right), \\
\mathcal{L}^{(\lambda)} &= \frac{L^2}{2} \left( \mathcal{R}_{\mu\nu\rho\sigma}^2 - 4 \mathcal{R}_{\mu\nu}^2 + \mathcal{R}^2 \right), \\
\mathcal{L}^{(\gamma)} &= \frac{7 L^4}{8} \bigg[ \mathcal{R}_{\mu\nu}{}^{\rho\sigma} \mathcal{R}_{\rho\sigma}{}^{\alpha\beta} \mathcal{R}_{\alpha\beta}{}^{\mu\nu}
+ \frac{1}{14} \Big(21 \mathcal{R}_{\mu\nu\rho\sigma}^2 \mathcal{R} - 120 \mathcal{R}_{\mu\nu\rho\sigma} \mathcal{R}^{\mu\nu\rho}{}_\alpha \mathcal{R}^{\sigma\alpha} \nonumber\\
&\quad + 144 \mathcal{R}_{\mu\nu\rho\sigma} \mathcal{R}^{\mu\rho} \mathcal{R}^{\nu\sigma} + 128 \mathcal{R}_\mu{}^\nu \mathcal{R}_\nu{}^\rho \mathcal{R}_\rho{}^\mu
- 108 \mathcal{R}_\mu{}^\nu \mathcal{R}_\nu{}^\mu \mathcal{R} + 11 \mathcal{R}^3 \Big) \bigg].
\end{align}}}}%
where $\lambda$ and $\gamma$ are the couplings.

\item  {\it Finite-gap corrections:}

In the case of the stress-tensor sector of holographic HHLL correlators, the finite-gap corrections have been 
investigated in \cite{Fitzpatrick:2020yjb} and were shown to lead to the loss of universality.
It would be interesting to repeat this analysis for the $TT$ correlators.

\item {\it  Higher-point correlators:}  

Another natural extension of this chapter is to go further and investigate the thermal properties of $n$-point ($n>2$) stress-tensor correlators near the lightcone. 

\item {\it  Going beyond double-stress tensors:}  

In this chapter, as well as in Chapter \ref{ch4}, the conformal block decomposition of the holographic thermal $TT$ correlators has been performed up to the double-stress tensors.  It would be interesting to go beyond this and study the $k$-stress tensor contributions for generic values of $k$.

\end{itemize}

%% file: 6_chapter/chapter_6.tex
\chapter{Universality Near Lightcone}\label{ch6}

Motivated by the results of the previous chapter, we consider thermal stress-tensor correlators in the near-lightcone regime and analyse them in momentum space using the OPE. In the limit we study, only the leading-twist multi-stress tensor operators contribute, and the correlators depend on a specific combination of lightcone momenta. We argue that such correlators are in any holographic theory described by three universal functions, which can be holographically computed in Einstein gravity. Higher-derivative terms in the gravitational Lagrangian enter the arguments of these functions via the cubic stress-tensor couplings and the thermal stress-tensor expectation value in the dual CFT. We analyse both the real and imaginary parts of these universal functions using perturbative and non-perturbative techniques.

\section{Motivation}
\label{sec.introductionp3}

%Understanding the non-perturbative structure of quantum field theories (QFTs) at finite temperature is an important challenge of theoretical physics. 
%In particular, thermal fluctuations cannot be ignored when studying real-world, out-of-equilibrium phenomena in, $e.g.$, the quark-gluon plasma or strongly coupled condensed matter systems.  
%Conformal field theories (CFTs) provide a natural starting point for investigating QFTs more generally.  
%The additional symmetries in conformal theories impose powerful constraints on physical observables, even at strong coupling. 
%Achieving a better understanding of general thermal field theories should therefore begin with an investigation of CFTs at finite temperature. 

A basic probe in any local field theory is the stress tensor, $T_{\mu\nu}$. 
The importance of the stress tensor, and correlators thereof, to CFT is hard to overstate.  The stress-tensor sector is completely universal in two-dimensional CFT, where the infinite-dimensional Virasoro algebra allows determining physical observables based on symmetries.  In higher dimensions, although the structure of conformal correlators is less constrained and in general they are determined via case-by-case computations, the conformal invariance still fixes two- and three-point stress-tensor correlators uniquely up to a few constants \cite{Osborn:1993cr}.   In particular, the two-point stress-tensor correlator at zero temperature is fixed up to an overall coefficient, the central charge $C_T$. 

When considering the theory at finite temperature, the thermal $TT$ correlators are generally theory dependent:  they depend on, for instance, the coefficients appearing in the zero-temperature three-point correlators of stress tensors. 
It is desirable to identify physical limits that make some universal aspects of the thermal correlators  manifest, and then devote future efforts to computing non-universal corrections to the correlators.

In this chapter, using the AdS/CFT correspondence \cite{Maldacena:1997re, Gubser:1998bc, Witten:1998qj}, we  analyse thermal $TT$ correlators in a class of holographic CFTs  in four dimensions, focusing on a certain universal, near-lightcone regime.   
Our analysis of the near-lightcone $TT$ correlators  
is in part motivated by the large body of recent work on  thermal scalar correlators and their near-lightcone behaviour in spacetime dimensions greater than two 
\cite{Kulaxizi:2018dxo,Fitzpatrick:2019zqz, Karlsson:2019qfi,Li:2019tpf,Huang:2019fog,Kulaxizi:2019tkd,Fitzpatrick:2019efk,Haehl:2019eae,Karlsson:2019dbd,Li:2019zba,Karlsson:2019txu,Huang:2020ycs,Karlsson:2020ghx,Li:2020dqm,Parnachev:2020fna,Fitzpatrick:2020yjb,Berenstein:2020vlp,Parnachev:2020zbr,Belin:2020lsr,Besken:2020snx,Karlsson:2021duj,Rodriguez-Gomez:2021pfh,Huang:2021hye,Rodriguez-Gomez:2021mkk,Krishna:2021fus,Korchemsky:2021htm,Bianchi:2021yqs,Karlsson:2021mgg,Huang:2022vcs,Dodelson:2022eiz,Dodelson:2022yvn, Bhatta:2022wga, Bajc:2022wws}.
In the context of AdS$_3$/CFT$_2$ such correlators have been well-studied in the literature, $e.g.$, \cite{Fitzpatrick:2014vua, Asplund:2014coa, Fitzpatrick:2015zha, Hijano:2015qja, Fitzpatrick:2015foa, Galliani:2016cai, Balasubramanian:2017fan, Galliani:2017jlg,  Faulkner:2017hll, Giusto:2018ovt, Giusto:2019pxc, Giusto:2020mup, Ceplak:2021wak, Bufalini:2022wyp, Berenstein:2022ico}.  
In $d=4$ holographic CFTs, heavy-heavy-light-light (HHLL) correlators were compared to thermal two-point functions in \cite{Kulaxizi:2018dxo} and 
several operator product expansion (OPE) coefficients of multi-stress tensor exchanges were computed in \cite{Fitzpatrick:2019zqz} which also observed that the scalar correlator in the near-lightcone limit is unaffected by higher-derivative interactions if one assumes a minimally coupled scalar in the bulk.  
Corrections to such universality due to non-minimally coupled interactions were discussed in \cite{Fitzpatrick:2020yjb}. 
In  \cite{Kulaxizi:2019tkd,Karlsson:2019dbd,Karlsson:2020ghx} the bootstrap procedure for computing HHLL correlators was developed.
Subsequently, it was pointed out  \cite{Huang:2021hye, Karlsson:2021mgg} that   higher-dimensional scalar correlators near the lightcone 
share certain similarities with the two-dimensional Virasoro vacuum blocks.  
Although the underlying mechanism responsible for this remains to be better understood, the time is ripe for investigation of a parallel story for the thermal correlators of stress tensors.  

An initial step towards this direction was made in Chapter \ref{ch4}, where we computed the thermal two-point correlators of stress tensors in $d=4$ holographic CFTs dual to Einstein gravity and read off conformal data beyond the leading order in the large $C_T$ expansion. 
 It was also observed that some OPE coefficients cannot be determined in the near-boundary analysis of the bulk equations of motion but these coefficients do not affect the near-lightcone $TT$ correlators.  
Subsequently, in Chapter \ref{ch5} we included
the Gauss-Bonnet (GB) higher-derivative term in the gravitational action in $AdS_5$
to study what happens when the conformal collider bounds  \cite{Hofman:2008ar} in the dual CFT are saturated.
We shown that the thermal stress-tensor correlators near the lightcone take the vacuum form when this happens.

In this chapter we point out that the stress-tensor correlators computed in Gauss-Bonnet gravity  suggest a certain near-lightcone universality, which does {\it not} require ANEC saturation.  
We shall elaborate on this observation in more detail, but the main message is the following: thermal $TT$ correlators near the lightcone are completely determined by three universal functions (depending on polarization).
The Gauss-Bonnet term in the bulk Lagrangian only affects the arguments of these functions via corrections to the cubic stress-tensor couplings and the thermal stress-tensor one-point function.
We hypothesize that this remains the case in more general higher-derivative gravitational theories.\footnote{We expect similar results for the thermal correlators of conserved currents of spin one, with
two (instead of three) universal functions. }

The analysis in this chapter involves  correlators in momentum space, and we mostly focus on retarded correlators in the near-lightcone regime.
To show that they have a universal structure, we identity a suitable  limit in momentum space
and show that the equations of motion of gravitational fluctuations in this limit take the same form as those in Einstein gravity.   
(These reduced bulk  equations of motion isolate the contributions of the leading-twist operators to the thermal $TT$ correlators in the dual CFTs.)
The near-lightcone $TT$ correlators depend on a single parameter, $\alpha \sim  q^+ (q^-)^3/T^4$, where $q^{\pm}$ are the lightcone momenta and $T$ is the temperature.
The expansion in the inverse powers of $\alpha$ is essentially an OPE
 (where only the leading-twist multi-stress tensor operators contribute because of the near-lightcone limit).
The reduced equations can be solved perturbatively in $1/\alpha$  and one can read off the corresponding OPE data.
Using  the WKB approximation, we  observe a non-perturbative imaginary term $\sim i e^{-\alpha^{1\over4}}$ in the retarded correlator.   
Such non-perturbative terms take the same  form irrespective of  polarization.

\subsubsection*{Outline}

The rest of this chapter is organized as follows.  
In Section \ref{sec.unistat}, we   argue for the universality of thermal $TT$-correlators in the near-lightcone regime.
We then adopt a momentum-space approach in Section \ref{sec:Momentum}, where 
we show that the momentum-space equations of motion in Gauss-Bonnet gravity in a suitable limit take the same form as the ones in Einstein gravity.  
By studying the on-shell action, we find that the near-lightcone $TT$ correlators (with three independent polarizations) in the holographic Gauss-Bonnet theory are completely determined by three universal functions. 

In Section \ref{sec.lceq}, we compute  the near-lightcone correlators in momentum space perturbatively in $1/\alpha$.
In Section  \ref{sec.nonperts}, we  compute the near-lightcone correlators numerically and analyse the large $\alpha$ behavior using the WKB approach.
We extract the non-perturbative  term in the retarded correlators.
We discuss our results and pose some questions for future work  in Section \ref{sec.discussion}. 

In Appendix \ref{ap.pertpos}, we discuss the position-space correlators and perform the Fourier transform to check  the momentum-space results. 
We also estimate the radius of convergence in momentum space (for the scalar channel) of the near-lightcone correlators and find that the radius of convergence approaches zero, $i.e.$, the series is asymptotic.
Thermal conformal blocks in momentum space, discussed in Appendix  \ref{apB}, provide additional checks.  We list the equations of motion in Gauss-Bonnet gravity in Appendix \ref{ap.eomsGB}. 
%In Appendix \ref{wkbap} we provide some further details on the WKB approximation used to estimate the non-perturbative imaginary part. 

%%%%%%%%%%%%%%%%%%%%%%
%%%%%%%%%%%%%%%%%%%%%%

\section{Position-Space Correlators}\label{sec.unistat}

In this section, we point out a universality of the near-lightcone $TT$ correlators based on the position-space computation performed in the previous chapter. 

Let us first recall that the vacuum stress-tensor three-point function in $d=4$ CFT in general depends on three coefficients $(\hat{a},\hat{b},\hat{c}$)  \cite{Osborn:1993cr}.  
 Depending on three different channels, $i.e.$, scalar, shear, and sound channels (polarisations),  the ``conformal collider bounds'' place constraints 
 on some linear combinations of these coefficients \cite{Hofman:2008ar}: 
\begin{align}  \label{eq:CCB1}
		{\cal{C}}_{\rm scalar}&\equiv\frac{5\pi^2}{3C_T} \big(-7\aa-2\bb+\cc \big)\geq 0 \ , \\
  \label{eq:CCB2}
                  {\cal{C}}_{\rm shear}&\equiv \frac{10\pi^2}{3C_T}\big(16\aa+5\bb-4\cc \big) \geq 0\ , \\
  \label{eq:CCB3}
		{\cal{C}}_{\rm sound}&\equiv\frac{15\pi^2}{C_T} \big(-4\aa-2\bb+\cc\big)\geq0 \ ,
 \end{align}
where the central charge can be expressed as
\begin{align}
\label{CTabc}
C_T= \frac{\pi^2}{3}\big(14\aa-2\bb-5\cc\big) \ . 
\end{align} 

In this chapter we are interested in thermal stress-tensor two-point functions on spatial $\mathbb{R}^3$ which involve three independent polarizations, mapping separately to the three different conformal collider bounds. 
The single-stress-tensor exchange contribution to thermal $TT$ correlators near the lightcone,  as shown by \cite{Kulaxizi:2010jt} using the stress-tensor OPE, are directly proportional to the same combinations of $(\hat{a},\hat{b},\hat{c})$ appearing in the conformal collider bounds.\footnote{One can use this fact and some minor assumptions to prove conformal collider bounds \cite{Komargodski:2016gci}.} At higher orders in the OPE, in large-$C_T$ CFTs, the thermal $TT$ correlators receive contributions from multi-stress-tensor exchanges.  
 In Chapter \ref{ch4}, the OPE limit of the thermal $TT$ correlators in holographic CFTs dual to Einstein gravity was studied using the thermal conformal blocks, including double-stress tensors $\sim [T^2]_J$ with dimension $8+ \OO(1/C_T)$  and the corresponding CFT data was read off by comparison with a bulk computation.  
 
 It is interesting to ask what happens when one includes higher-derivative corrections to Einstein-gravity which modifies, in particular, the stress-tensor OPE coefficients $(\aa,\bb,\cc)$. In the context of holographic CFTs dual to Gauss-Bonnet gravity,  the thermal $TT$ correlators were examined in Chapter \ref{ch5} in position space.  
We observed that, in the near-lightcone regime, the saturation of a CCB, $i.e.$, ANEC saturation, implies that the corresponding correlator takes the vacuum form, independent of temperature. 
To gain a broader understanding of the near-lightcone dynamics of the thermal correlators and their possible universal behavior, in this chapter we study these thermal correlators away from  ANEC saturation.\footnote{In particular, the results of this chapter are applicable for small values of the Gauss-Bonnet coupling, which one expects for a unitary theory with a large gap   \cite{Camanho:2014apa}.}  

Before proceeding, let us remind some notations. 
In any CFT, the thermal one-point function of an operator ${\cal{O}}_{\Delta,J}$ on $S^1_\beta\times\mathbb{R}^3$ with dimension $\Delta$ and spin $J$ is fixed, see $e.g.$ \cite{El-Showk:2011yvt,Iliesiu:2018fao}  
\begin{equation}
	\langle {\cal{O}}_{\Delta,J}\rangle_\beta = {b_{{\cal{O}}_{\Delta,J}}\over \beta^{\Delta}}(e_{\mu_1}\ldots e_{\mu_J}-\text{traces}) 
\end{equation}
where $\beta$ is the inverse temperature and $e^\mu$ is a unit vector on the thermal circle $S^1_\beta$. For our discussion of the near-lightcone correlators, it will be further useful to define the following quantities: 
\begin{align}
	\label{hatC}
	{\hat{\cal{C}} }_{\rm scalar}\equiv \frac{b_{T
	}}{C_T} \cal{C}_{\rm scalar}  \ , ~~~ 
	{\hat{\cal{C}} }_{\rm shear}\equiv \frac{b_{T
	}}{C_T}  \cal{C}_{\rm shear}\ , ~~~
	{\hat{\cal{C}} }_{\rm sound}\equiv \frac{b_{T
	}}{C_T} \cal{C}_{\rm sound} \ ,
\end{align} where  $\cal{C}_{\rm scalar, shear, sound}$ are defined in \eqref{eq:CCB1} -- \eqref{eq:CCB3}. 

We  will study the stress-tensor correlators integrated over the $xy$-plane:%\footnote{See \cite{Karlsson:2022osn, Huang:2022vet} for related discussions on the integrated correlators.} 
\begin{equation}
	G_{\mu\nu, \rho\sigma}(t,z)\equiv \int_{\mathbb{R}^2}dxdy~ \langle T_{\mu\nu}(t,z, x,y)T_{\rho\sigma}(0)\rangle_\beta  \ . 
\end{equation}   The stress-tensor correlators 
can be classified into three independent channels, see, $e.g.$, \cite{Kovtun:2005ev}.  
 In the lightcone coordinates $x^{\pm}=t\pm z$ in Lorentzian signature, we consider the limit  $x^-\to 0$, with fixed $x^-(x^+)^3 \beta^{-4}$.
  In any four-dimensional CFT with no operators with twist less than or equal to two other than the stress-tensor, the thermal $TT$ correlators in the near-lightcone limit are given by 
{\small
	\begin{align} \label{eq:univ1}
	G_{xy,xy}(x^+,x^-)&=\frac{-\pi C_T}{10(x^-)^3(x^+)^3}\Big(1-\frac{{\hat{\cal{C}} }_{\rm scalar}}{\pi^2}\frac{x^-(x^+)^3}{\beta^4}+\ldots\Big) \ ,\\ 
\label{eq:univ2}
		G_{tx,tx}(x^+,x^-) &= \frac{-3\pi C_T}{80(x^-)^4(x^+)^2}\Big(1-\frac{{\hat{\cal{C}} }_{\rm shear}}{3\pi^2}\frac{x^-(x^+)^3}{\beta^4}+\ldots\Big)\ ,\\
\label{eq:univ3}
		G_{tz,tz}(x^+,x^-) &= \frac{-\pi C_T}{20(x^-)^5x^+}\Big(1-\frac{{\hat{\cal{C}} }_{\rm sound}}{12\pi^2}\frac{x^-(x^+)^3}{\beta^4}+\ldots\Big) \ .
\end{align}}The leading and subleading terms  in \eqref{eq:univ1} -- \eqref{eq:univ3} are universal, but the higher-order terms  {\it a priori} are model-dependent.  These higher-order terms contain contributions from the operators with larger dimensions -- in holographic CFTs, they include multi-stress tensor operators denoted as $[T^k]_J$.  

\subsubsection*{Position-Space Correlators in Holographic Gauss-Bonnet Theory}

Here we make an observation based on the  thermal $TT$ correlators obtained holographically using Gauss-Bonnet gravity. 
Denote the dimensionless Gauss-Bonnet coupling as $\lambda_{\rm GB}$.\footnote{Gauss-Bonnet gravity is discussed in more detail in Section \ref{sec:Momentum}.} We introduce a parameter 
\begin{equation}
\label{definek}
\kappa= \sqrt{1-4 \lambda_{\rm GB}}\ , 
\end{equation} which will help simplify expressions. The limit $\kappa \to 1$ recovers Einstein gravity. 
The coefficients $(\hat{a}, \hat{b}, \hat{c})$ can be related to the Gauss-Bonnet coupling: 
\begin{align} 
\label{abckappa}
\hat{a}=\frac{8C_T}{45\pi^2}(-6+\frac{5}{\kappa})\ ,~~~ \hat{b}=\frac{C_T}{90\pi^2}(33-\frac{50}{\kappa})\ , ~~~ \hat{c}=\frac{2C_T}{45\pi^2}(-84+\frac{61}{\kappa}) \ .
\end{align} 
The conformal collider bounds, \eqref{eq:CCB1} -- \eqref{eq:CCB3} translate to, respectively,  
\begin{align} 
\label{CCBkappa}
(5\kappa-4)\geq 0 \ ,~~~ (2-\kappa)\geq0 \ ,~~~  (4-3\kappa)\geq0 \ . 
\end{align}   

Now we make the following observation: the near-lightcone $TT$ correlators in holographic CFTs dual to Gauss-Bonnet gravity computed in Chapter \ref{ch5} can be recast in the following way:
{\small\begin{align}
	&G_{xy,xy}=\frac{-\pi C_T}{10(x^-)^3(x^+)^3}\left(1+\frac{1}{\pi^2}\bar{\alpha}_{\rm scalar}+\frac{5}{3\pi^4}\bar{\alpha}_{\rm scalar}^2+\ldots\right)\label{ob1} \ , \\        
	&G_{tx,tx}=\frac{-3\pi C_T}{80(x^-)^4(x^+)^2}\left(1+\frac{1}{3\pi^2}\bar{\alpha}_{\rm shear}+\frac{85}{504 \pi ^4}\bar{\alpha}_{\rm shear}^2+\ldots\right)\label{ob2} \ , \\
	&G_{tz,tz}=\frac{-\pi C_T}{20(x^-)^5x^+}\left(1+\frac{1}{12\pi^2}\bar{\alpha}_{\rm sound}+\frac{11}{756 \pi ^4}\bar{\alpha}_{\rm sound}^2+\ldots\right)\label{ob3} \ ,
\end{align}}where
\begin{align}\label{def:alphabar}
	\Big(  \bar{\alpha}_{{\rm scalar}},  \bar{\alpha}_{{\rm shear}}, \bar{\alpha}_{{\rm sound}} \Big)\equiv& \Big({\hat{\cal{C}} }_{\rm scalar},{\hat{\cal{C}} }_{\rm shear},{\hat{\cal{C}} }_{\rm sound} \Big )~ \frac{(-x^-(x^+)^3)}{\beta^4} \ .
\end{align}
While the first and second term are fixed by conformal symmetry as was shown in \eqref{eq:univ1} -- \eqref{eq:univ3}, one can see that, even at  $\cal O(\bar{\alpha}^2)$, the  dependence on the Gauss-Bonnet coupling can be  absorbed into the parameters $\bar{\alpha}$, which depend on the thermal one-point function of the stress tensor ($\sim b_T \beta^{-4}$) and $(\hat{a},\hat{b},\hat{c})$ through the particular combinations appearing in the conformal collider bounds.\footnote{The coefficients appearing in \eqref{ob1}-\eqref{ob3} in the $\bar{\alpha}$ expansion can also be verified to be the same in Einstein gravity (see Chapter \ref{ch4}) and Gauss-Bonnet gravity (see Chapter \ref{ch5}).} 
This observation hints at the following intriguing possibility: {\it multi-stress tensor contributions to the near-lightcone $TT$ correlators in holographic CFTs might be fixed by (the $k$-th power of) the single-stress-tensor contribution.}  
Exploring such a possibility and its consequences is the underlying motivation of the present chapter. 

To see this universality in position space we note that near the lightcone the corresponding reduced EoMs in Gauss-Bonnet gravity obtained in Sec.\ \ref{s.sajhbsfvhbsdd} are identical to the ones in Einstein gravity, after performing suitable rescalings of the coordinates. 
This is easy to see in the scalar and shear channels, while the sound channel is technically more complicated when analysed in position space.  
In the next section we shall analyse the equations of motion in  momentum space and show the universality in all channels.

Before moving to the momentum-space analysis, let us conclude this section with a technical remark: 
the structure of the higher-order terms, denoted by dots in \eqref{ob1} -  \eqref{ob3},  
in fact slightly differs from the first three terms we listed -- besides the corresponding dependence of $\bar{\alpha}_{\rm scalar}$, $\bar{\alpha}_{\rm shear}$ and $\bar{\alpha}_{\rm sound}$, the higher-order terms are multiplied by a $\log(-x^+x^-)$ piece. These arise for two different reasons, one is a contribution due to the anomalous dimensions of the multi-stress tensor operators and the other is because we consider the integrated correlator. We discuss this more in Appendix \ref{ap.pertpos} where we perform the Fourier transformation of the position-space correlators.
% In momentum space, the only logarithmic term is due to the identity operator, while for all other operators we see explicitly that there are none. 

%%%%%%%%%%%%%%%%%%%%%%%%%%
%%%%%%%%%%%%%%%%%%%%%%%%%%

\section{Momentum-Space Correlators}\label{sec:Momentum}

In this section, we will study momentum-space  thermal $TT$ correlators and 
show that the near-lightcone correlators computed in the holographic Gauss-Bonnet 
theory are universal: they are determined by three universal functions, depending on three polarizations.
We will hypothesize that this might be the case for all holographic theories. In subsequent sections, we will compute these functions.

Let us give quickly remind the basic aspects of Gauss-Bonnet gravity. The action in five dimensions is given by
{\small\begin{align}
S_{GB} = \frac{1}{16 \pi G} \int d^5 x \sqrt{-g} \left[{12\over L^2} + \mathcal{R}  + \lambda_{\rm GB}   \frac{L^2}{2} 
\left( \mathcal{R}^2 - 4 \mathcal{R}_{\mu\nu} \mathcal{R}^{\mu\nu} + \mathcal{R}_{\mu\nu\rho\sigma} \mathcal{R}^{\mu\nu\rho\sigma} \right) \right] \ .
\end{align}}The theory admits a black-hole solution \cite{Boulware:1985wk, Cai:2001dz}:
    \begin{equation}\label{gbbhp3}
        \dd s^2=\frac{r^2}{L^2}\left(-\frac{f(r)}{f_\infty}\dd t^2+\dd x^2+\dd y^2+\dd z^2\right)+\frac{L^2}{r^2}\frac{\dd r^2}{f(r)} 
    \end{equation}
where $f(r)$ and $f_\infty$ are 
    \begin{equation}\label{eq.fdefs}
        f(r)=\frac{1}{2\lambda_{GB}}\left[1-\sqrt{1-4\lambda_{GB}\left(1-\frac{r_+^4}{r^4}\right)}\right]\quad\text{and}\quad
        f_\infty=\frac{1-\sqrt{1-4\lambda_{GB}}}{2\lambda_{GB}} \ .
    \end{equation} 
The parameter $r_+$ is the location of the black-hole horizon. We focus on a planar horizon. In the following we set the AdS radius $L/\sqrt{f_\infty}$ to 1, $i.e.$ $L=\sqrt{f_\infty}$.

To study the equations of motion of gravitational fluctuations, we consider the metric perturbation
$h_{\mu\nu} = h_{\mu\nu} (r) e^{-i\omega t + i q z}$, with the momentum along the $z$-direction, 
and adopt gauge-invariant quantities following the recipe in \cite{Kovtun:2005ev}. The radial gauge $h_{r\mu} = 0$ is used. 
The fluctuations are classified into three independent channels; in Gauss-Bonnet gravity, the corresponding gauge invariants in momentum space are given by \cite{Buchel:2009sk}:
\begin{align}
  &Z^{\rm (GB)}_{\rm{scalar}} = \frac{1}{r^2}h_{xy}\ ,   \\
  &Z^{\rm (GB)}_{\rm{shear}} = \frac{\qfr}{r^2}h_{tx} + \frac{\wfr}{r^2}h_{zx} \ , \\
 &Z^{\rm (GB)}_{\rm {sound}} = \frac{2\qfr^2}{r^2}h_{tt}+\frac{4\wfr\qfr}{r^2}h_{tz}+\frac{2\wfr^2}{r^2}h_{zz}+\qfr^2\left(\frac{2f+rf'}{2f_{\infty}}-\frac{\wfr^2}{\qfr^2}\right)\left(h_{xx}+h_{yy}\right) \ , 
\end{align}
where $(\wfr, \qfr) = {1\over {2\pi T}} (\omega, q)$ are the dimensionless frequency and momentum and $T$ is the Hawking temperature. 
The linearized equations of motion of gravitational fluctuations in Gauss-Bonnet gravity were worked out in \cite{Buchel:2009sk}. 
(See \cite{Brigante:2007nu,Brigante:2008gz,Buchel:2009tt,deBoer:2009pn,Camanho:2009vw,Buchel:2010wf,Cai:2010cv,Bu:2015bwa,Grozdanov:2016vgg, Andrade:2016yzc, Grozdanov:2016zjj,Andrade:2016rln,Grozdanov:2016fkt,Casalderrey-Solana:2017zyh,Chen:2018nbh,An:2018dbz,Grozdanov:2021gzh}
for more recent applications of Gauss-Bonnet holographic gravity.) 
In this chapter, we are  interested in the near-lightcone regime of the correlators.

Defining a coordinate $u = r_+^2/r^2$,  the equation of motion in each of the three channels can be written as a second-order differential equation:
\begin{align}
\label{ZEoM}
Z''(u) + A ~Z'(u) + B~ Z(u) = 0 \ .
\end{align}
The channel-dependent coefficients $A$ and $B$ are given in Appendix \ref{ap.eomsGB}.

\subsection{Equations of motion in the near-lightcone limit}

In this chapter, we are interested in the near-lightcone limit.  Denote $\mathfrak{q}^{\pm}= \wfr \pm \qfr$. We consider the following  limit:\footnote{We use the minus sign in the expression for $\alpha$ to make calculations in the space-like regime (which we will be interested in) more convenient.}
\be
\label{LCbulklimit}
   \mathfrak{q}^+ \ra \infty,  \qquad \mathfrak{q}^- \ra 0, \qquad \alpha = - \mathfrak{q}^+(\mathfrak{q}^-)^3 ~{\rm fixed},  \qquad \tilde u = u/ (\mathfrak{q}^-)^2  ~{\rm fixed}.
\ee
This limit isolates contributions from the leading twist operators, which corresponds to zooming in on the
near-boundary region in the bulk. For the position-space stress-tensor correlators, the corresponding near-lightcone limit in the bulk was discussed in the previous chapter.

We next show that, in the limit \eqref{LCbulklimit}, the equations of motion in Gauss-Bonnet gravity reduce to those in Einstein gravity. 
\\

\noindent {\bf Scalar Channel:} First we derive the reduced equation of motion in the scalar channel. 
In the limit \eqref{LCbulklimit}, the equation of motion at  leading order can be written as 
\begin{align}
\label{scalarint}
\alpha  (\kappa +1)^2 \Big(\alpha  \left(5 \kappa ^2+\kappa -4\right) \tilde u^2-8 \kappa ^2\Big)Z_{\rm scalar}(\tilde u)  -32 \kappa ^2 \left(Z'_{\rm scalar}(\tilde u)-\tilde u Z''_{\rm scalar}(\tilde u)\right)= 0 \ .
\end{align}
The observation is that, after rescaling the variables
\begin{align}
\label{rescalescalar}
\tilde u  = \frac{\kappa ^2 (\kappa +1)}{2 (5 \kappa -4)}  \tilde u_r   \ , ~~~
\alpha =  \frac{8 (5 \kappa -4)}{\kappa ^2 (\kappa +1)^3}  \alpha_r \ ,
\end{align} 
the equation of motion \eqref{scalarint} becomes 
\begin{align}
Z''_{\rm scalar}(\tilde u_r)  -\frac{1}{\tilde u_r} Z'_{\rm scalar}(\tilde u_r)+  \big(   \frac{ \alpha_r^2  \tilde u_r}{4 }  - { \alpha_r \over  \tilde u_r}  \big)Z_{\rm scalar}(\tilde u_r) = 0   \ .
\end{align}
Since this equation is completely independent of $\kappa$, we conclude it is identical to the equation of motion in Einstein gravity in the same limit.   
\\\\
\noindent {\bf Shear Channel:}  The equation of motion in the limit \eqref{LCbulklimit} is 
{\small
\begin{equation}\label{uteq1}
\begin{split}
&\alpha  (\kappa +1)^2  \left(8 \kappa ^2+\alpha  (\kappa -2) (\kappa +1) \tilde u^2\right)^2 Z_{\rm shear}(\tilde u)+ 32 \kappa ^2 \Big( (8 \kappa ^2+3 \alpha  (\kappa\\
&-2) (\kappa +1) \tilde u^2) Z'_{\rm shear}(\tilde u)+ \left(\alpha  \left(\!-\kappa ^2+\kappa +2\right) \tilde u^2\!-8 \kappa ^2\right) \tilde u  Z''_{\rm shear}(\tilde u)\Big) \!= 0\,.
\end{split}
\end{equation}}After performing the rescalings
\begin{align}
\label{rescaleshear}
\tilde u  = -\frac{ \kappa ^2 (\kappa +1)}{2 (\kappa -2)} \tilde u_r   \ , ~~~ \alpha =  -\frac{8 (\kappa -2)}{\kappa ^2 (\kappa +1)^3} \alpha_r \ , 
\end{align} 
we find 
{\small
\begin{align}
Z''_{\rm shear}(\tilde u_r)
- {1\over \tilde u_r }  \Big(\frac{3 \alpha_r \tilde u_r^2-4}{\alpha_r \tilde u_r^2-4 }  \Big) Z'_{\rm shear}(\tilde u_r)
+ \big(   \frac{ \alpha^2  \tilde u_r}{4 }  - { \alpha_r \over  \tilde u_r}  \big)  Z_{\rm shear}(\tilde u_r) = 0   
\end{align}}which is identical to the equation of motion in Einstein gravity in the same limit.
\\\\
\noindent{\bf Sound Channel:}  In the limit \eqref{LCbulklimit}, the sound-channel equation becomes  
{\small
\begin{equation}\label{uteq2}
\begin{split}
&\alpha  (\kappa +1)^2 \left(8 \kappa ^2+\alpha  (\kappa +1) (3 \kappa -4)  \tilde u^2\right) \left(24 \kappa ^2+\alpha  (\kappa +1) (3 \kappa -4) \tilde u^2\right) Z_{\rm sound}(\tilde u)\\
&+32 \kappa ^2 \Big(\left(24 \kappa ^2+5 \alpha  (\kappa +1) (3 \kappa -4)  \tilde u^2\right) Z'_{\rm sound}(\tilde u)-\big(\alpha  (\kappa+1) (3 \kappa -4) \tilde u^2\\
&\hspace{8.6cm}+24 \kappa ^2\big) \,\tilde u  \,Z''_{\rm sound}(\tilde u)\Big) = 0  \ .
\end{split}
\end{equation}}Performing the rescalings
\begin{align}
\label{rescalesound}
\tilde u   = \frac{\kappa ^2 (\kappa +1)}{8-6 \kappa }  \tilde u_r   \,\,\qq{and}\,\,
\alpha =  -\frac{8 (3 \kappa -4)}{\kappa ^2 (\kappa +1)^3} \alpha_r  \ , 
\end{align} 
we obtain the same equation as in Einstein gravity:
{\small
\begin{align}
Z''_{\rm sound}(\tilde u_r)
- {1\over \tilde u_r }  \Big(   \frac{5 \alpha_r  \tilde u_r^2-12}{\alpha_r  \tilde u_r^2-12} \Big) Z'_{\rm sound}(\tilde u_r)
+\big(   \frac{ \alpha_r^2  \tilde u_r}{4 }  - { \alpha_r \over  \tilde u_r}  \big)Z_{\rm sound}(\tilde u_r) = 0   \ .
\end{align}}In the Einstein gravity case, $\kappa=1$, one can verify that $\tilde u= \tilde u_r$, $\alpha= \alpha_r $ in all channels.

In summary, the momentum-space reduced equations of motion in the three different channels can be written as
\begin{align}
\label{rEoMmi}
&Z''(\tilde u_r) - {K(\alpha_r, \tilde u_r) \over \tilde u_r }  Z'(\tilde u_r) +\big(   \frac{ \alpha_r^2  \tilde u_r}{4 }  - { \alpha_r \over \tilde u_r}  \big) Z (\tilde u_r) = 0 
\end{align}
where  $K(\alpha_r, \tilde u_r)$ is channel-dependent:
\begin{align}
&~~~~~~~ \Big( K_{\rm scalar}, K_{\rm shear}, K_{\rm sound} \Big) = \Big(1,   \frac{3 \alpha_r \tilde u_r^2-4}{\alpha_r \tilde u_r^2-4 },    \frac{5 \alpha_r  \tilde u_r^2-12}{\alpha_r  \tilde u_r^2-12} \Big)  \ .
\end{align}
Hence, we have observed that the reduced equations of motion in the holographic Gauss-Bonnet theory take the same form as the ones obtained in Einstein gravity.

We will next analyze the action and show that the near-lightcone $TT$ correlators are determined by three universal functions which correspond to the three independent polarizations.

\subsection{Thermal correlators from holography}

Here we compute holographic thermal $TT$ correlators $G_{\mu\nu,\rho\lambda}$  in four spacetime dimensions.  
The symmetries of the theory imply that the momentum-space retarded correlator 
has the following form \cite{Kovtun:2005ev}:
    \begin{equation}
    G_{\mu\nu,\rho\lambda}(k)=L_{\mu\nu,\rho\lambda}G_{\rm scalar}(k)+ S_{\mu\nu,\rho\lambda}G_{\rm shear}(k)+Q_{\mu\nu,\rho\lambda}G_{\rm sound}(k) \ ,
    \end{equation}
where $G_{\rm scalar}$, $G_{\rm shear}$ and $G_{\rm sound}$ are three independent scalar functions of momenta and the tensor structures $L_{\mu\nu,\rho\lambda}$, $S_{\mu\nu,\rho\lambda}$ and $Q_{\mu\nu,\rho\lambda}$ are fixed by the symmetries, see, $e.g.$, \cite{Kovtun:2005ev}.
If not stated otherwise, we use  Minkowski signature.

Following the Lorentzian AdS/CFT dictionary \cite{Policastro:2001yc,Son:2002sd,Policastro:2002se,Policastro:2002tn,Kovtun:2004de}
we impose incoming boundary conditions  at the horizon, $\tilde u \ra \infty$.
To compute the correlators we need the $\OO(Z^2)$ on-shell action in all three channels, which  is given by \cite{Buchel:2009sk} 
{\small  
    \begin{align}
        I_{\rm scalar}&=-\frac{\pi^2C_T r_+^4}{80 f_\infty^2}\lim_{u\rightarrow0}\int\frac{\dd\omega\dd q}{(2\pi)^2}\frac{1}{u}\pdv{u}\Zt(u,k)\Zt(u,-k) \ , \\
        I_{\rm shear}&=-\frac{\pi^2C_T r_+^4}{80 f_\infty^2}\lim_{u\rightarrow0}\int\frac{\dd\omega\dd q}{(2\pi)^2}\frac{1}{u(\wfr^2-\qfr^2)}\pdv{u}\Zv(u,k)\Zv(u,-k) \ , \\
        I_{\rm sound}&=\frac{3\pi^2C_T r_+^4}{320 f_\infty^2}\lim_{u\rightarrow0}\int\frac{\dd\omega\dd q}{(2\pi)^2}\frac{1}{u(3\wfr^2-\qfr^2(3-u^2))^2}\pdv{u}\Zs(u,k)\Zs(u,-k) \ ,
    \end{align}}
where $C_T$ is the central charge of the holographic Gauss-Bonnet theory.

The correlators are given by  
\begin{equation}\label{eq.origG}
    G_{\rm scalar}=\frac{\pi^2C_T r_+^4}{10 f_\infty^2}\frac{\mathcal{B}_{\rm scalar}}{\mathcal{A}_{\rm scalar}} \ ,\,
    G_{\rm shear}=-\frac{\pi^2C_T r_+^4}{10 f_\infty^2}\frac{\mathcal{B}_{\rm shear}}{\mathcal{A}_{\rm shear}} \ ,\,
    G_{\rm sound}=-\frac{\pi^2C_T  r_+^4}{10 f_\infty^2}\frac{\mathcal{B}_{\rm sound}}{\mathcal{A}_{\rm sound}} 
    \end{equation}
where $\mathcal{A}$ and $\mathcal{B}$ are the coefficients in the near-boundary expansion:\footnote{This is the standard Frobenius expansion around $u=0$, $i.e.$, $Z=\mathcal{A}u^{\Delta_1}(1+\ldots)+\mathcal{B}u^{\Delta_2}(1+\ldots)$, where $\Delta_1$ and $\Delta_2$ are the leading exponents. For equations \eqref{scalarint}, \eqref{uteq1} and \eqref{uteq2} one finds $\Delta_1=0$ and $\Delta_2=2$. Since these differ by an integer, one has to include log-terms in the $\Delta_2$-part of the solution. Inserting the resulting ansatz into the corresponding equation, one can determine the subleading coefficients in terms of $\mathcal{A}$ and $\mathcal{B}$. } 
    \begin{equation}\label{eq.ten.schemufrom}
        Z(u)=\mathcal{A}-\frac{\alpha\mathcal{A}}{f^2_\infty(\mathfrak{q}^-)^2}u+\mathcal{B}u^2-\frac{\alpha^2\mathcal{A}}{2f^4_\infty(\mathfrak{q}^-)^4}u^2\log u+\ldots
    \end{equation}
Let us adopt a new variable  
    \begin{equation}
        x \equiv \tilde{u}_r \alpha_r   = f_\infty^{-2} \tilde u \alpha=-f_\infty^{-2}\mathfrak{q}^+\mathfrak{q}^- u\ ,
    \end{equation}
 where we used (\ref{rescalescalar}), (\ref{rescaleshear}), (\ref{rescalesound}) and
 the relation 
 \be
 \label{kappafinfty}
 \kappa = \frac{2}{f_\infty} -1 \ .
 \ee
We can now rewrite the equations of motion \eqref{rEoMmi} as 
\begin{align}
\label{eq.reommnew}
&Z''(x)- B (x, \alpha_r) Z'(x)+\frac{x^2-4\alpha_r}{4\alpha_r x}Z(x)=0 \ ,\\
\nonumber &~~~~~~~~~~\big(B_{\rm scalar}, B_{\rm shear} , B_{\rm sound} \big)= \Big( \frac{1}{x},   \frac{3x^2-4\alpha_r}{x^3-4\alpha_r x},\frac{5x^2-12\alpha_r}{x^3-12\alpha_r x}  \Big) \ .
\end{align}
Note that $x \to 0$ is the boundary limit while $x \to \infty$ corresponds to  the black hole horizon.

Before solving the equation \eqref{eq.reommnew}, let us consider a formal analysis of the near-boundary structure of the correlators. 
The near-boundary expansion up to quadratic order is the same in all channels and reads
    \begin{equation}\label{eq.ten.schemxfrom}
    Z(x)=a-ax+bx^2-\frac{a}{2}x^2\log x+\ldots 
    \end{equation} 
where $a$ and $b$ are functions of $\alpha_r$.  
The coefficients $\mathcal{A}$ and $\mathcal{B}$ can be related to $a, b$ in the following way:
    \begin{equation}\label{abexpansion}
        \mathcal{A}=a \ , ~~~  \mathcal{B}=(\mathfrak{q}^+\mathfrak{q}^-)^2 f_\infty^{-4} \left(b-\frac{a}{2}\log(-f_\infty^{-2}\mathfrak{q}^+\mathfrak{q}^-)\right).
    \end{equation}
We can now write the correlator in $e.g.$ the scalar channel as
    \begin{equation}\label{eq.G3defab}
        G_{\rm scalar}=\frac{\pi^2C_T}{160} (q^+ q^-)^2  \left( \mathfrak{f}_{\rm scalar}(\alpha_r)-\frac{1}{2}\log(-q^+q^-)\right) \ ,
~~~     \mathfrak{f}_{\rm scalar}(\alpha_r)=\frac{b}{a} \ ,
    \end{equation}
where we have ignored terms analytic in momenta.\footnote{These correspond to contact terms in position space.} Note that the ratio $\mathfrak{f}=\frac{b}{a}$ only depends on 
$q^{\pm}$ through $\alpha_r$.  
The function $\ff_{\rm scalar}$ can be obtained via
    \begin{equation}\label{eq.fscalar}
        \mathfrak{f}_{\rm scalar}\left(\alpha_r\right) =\lim_{x\rightarrow0}\pdv[2]{x}\left(\frac{\Zt(x)}{2\lim_{x'\rightarrow0}\Zt(x')}+\frac{x^2}{4}\log x\right) 
    \end{equation} after we solve for $Z(x)$ from the corresponding equation of motion.

Analogously, in the other channels we obtain
    \begin{align}
        G_{\rm shear}&=-\frac{\pi^2C_T}{160} (q^+q^-)^2  \left( \mathfrak{f}_{\rm shear}(\alpha_r)-\frac{1}{2}\log(-q^+q^-)\right)\label{eq.G1defab}\ , \\
        G_{\rm sound}&=-\frac{\pi^2C_T}{160} (q^+q^-)^2  \left( \mathfrak{f}_{\rm sound}(\alpha_r)-\frac{1}{2}\log(-q^+q^-)\right)\label{eq.G2defab} \ .
    \end{align}
The functions $\ff_{\rm shear}(\alpha_r)$ and $\ff_{\rm sound}(\alpha_r)$ are again defined as the ratios of the corresponding coefficients in the near-boundary expansion of  $Z(x)$. 

In the next two sections, we will compute the functions $\ff_{{\rm scalar}, {\rm shear}, {\rm sound}}(\alpha_r)$  first perturbatively in $1/\alpha_r$ and then numerically. A few comments are in order:  

\begin{itemize}

\item We again emphasize that  the near-lightcone $TT$ correlators in the holographic Gauss-Bonnet theory are expressed 
in terms of the same functions (there are three of them, which corresponds to the three independent polarizations) as the ones that 
appear in pure Einstein gravity.  It is possible that this universality holds true more generally, going beyond the holographic Gauss-Bonnet theory. We rephrase it in the language of the 
stress-tensor three-point couplings below.

\item The function $\ff$ has a perturbative expansion in $1/\alpha$  
which is basically an OPE, and also non-perturbative terms of the type $e^{-\alpha^{1\over4}}$. 
The non-perturbative terms correspond to  tunneling under the potential barrier in the Schr\"odinger equation  
which can be obtained from (\ref{eq.reommnew})
and are sensitive to the boundary conditions at the horizon ($x{\ra} \infty$), as we explain in Section \ref{sec.nonperts}.

\item The perturbative expansion is not sensitive to the horizon boundary conditions -- it is equivalent to the OPE which can also be performed
in  position space.
In Appendix \ref{ap.pertpos}, we explicitly match several terms between the position- and momentum-space expansions.

\end{itemize}

Let us now point out that the first term in the perturbative expansion of  $\ff$ is a non-physical (cutoff-dependent) number, while
the second term (proportional to $\beta^{-4}$) is fixed by the $TTT$ three-point couplings and the coefficient $b_T$  \cite{Kulaxizi:2010jt}
\be
\label{reminderalpha}
\ff_{\rm scalar} (\alpha_r) |_{\beta^{-4} } \sim \frac{7 \aa +2 \bb -\cc}{14 \aa- 2\bb -5 \cc} \ \frac {{\tilde b}_T }{\alpha}
\ee  
where ${\tilde b}_T $ is defined by 
\be
\label{defbt}
\tilde b_T = \frac{b_T}{C_T} \ .
\ee
The ratio of  (\ref{reminderalpha}) to the corresponding term in Einstein gravity is given by
\be
\label{fratio}
\frac{   \ff_{\rm scalar} (\alpha_r) |_{\beta^{-4}}  }{   \ff_{\rm scalar} (\alpha) |_{\beta^{-4} } }=-5  \frac{7 \aa +2 \bb -\cc}{14 \aa- 2\bb -5 \cc}  \ \frac{\tilde b_T}{\tilde b_{T,0}}= \frac{8 (5 \kappa-4)}{ \kappa^2 (\kappa+1)^3}
\ee
where the zero in the subscript indicates the corresponding value for Einstein gravity.
The first equality in  (\ref{fratio}) follows from (\ref{reminderalpha}) and to get the second equality
we have used expressions for $\aa,\bb,\cc$ from (\ref{abckappa}). We also used\footnote{This relation can be obtained from, $e.g.$, Eq.\ \eqref{eq:defMuTilde}.  
	We do not need to know the value of $\tilde b_{T,0}$, but it is available -- for example it can be read off from Eqs.\ \eqref{eq:Temp}-\eqref{eq:defMu}.}  
\be
\label{btgb}
\frac{\tilde b_T }{\tilde b_{T,0}}= \frac{8}{\kappa (\kappa+1)^3 } \ .
\ee 
Note that the first equality in (\ref{fratio}) was derived from the Ward identity.
Eq.\ (\ref{fratio})  is consistent with the relation between $\alpha$ and $\alpha_r$ in (\ref{rescalescalar}), as it should.
It is tempting to propose that in any holographic theory the function that enters the correlator in the
scalar channel is
\be
\label{genscalar}
\ff_{\rm scalar}  (\alpha_r) =  \ff_{\rm scalar}\left(\left[ -5  \ \frac{7 \aa +2 \bb -\cc}{14 \aa- 2\bb -5 \cc}  \ \frac{\tilde b_T}{\tilde b_{T,0}}  \right]^{-1} \alpha \right) \ .
\ee
This statement means that the near-lightcone correlator for all holographic CFTs  
is completely fixed in terms of basic conformal data (such as $\aa,\bb,\cc, \tilde b_T$) and the function $\ff$, which we compute in the next two sections. 
For the other channels, similar logic implies 
\be
\label{genshear}
\ff_{\rm {shear}}  (\alpha_r) =  
\ff_{\rm {shear}} \left( \left[   10\ \frac{ 16 \aa +5 \bb -4 \cc  }{ 14 \aa- 2\bb -5 \cc }   \ \frac{\tilde b_T}{\tilde b_{T,0}} \right]^{-1}  \alpha \right) \ ,
\ee
\be
\label{gensound}
\ff_{{\rm sound}}(\alpha_r) = 
\ff_{\rm {sound}}\left( \left[  -45\ \frac{  4 \aa +2 \bb -\cc  }{ 14 \aa- 2\bb -5 \cc  }  \ \frac{\tilde b_T}{\tilde b_{T,0}} \right]^{-1}   \alpha \right) \ .
\ee
Note that the combinations of  $\aa,\bb,\cc$ in the numerators are proportional to the corresponding ANECs.
Hence, one obtains the vacuum result once an ANEC gets saturated, reproducing the results of Chapter \ref{ch5}.

\section{Perturbative Analysis}\label{sec.lceq}

Here  we shall focus on computing the 
near-lightcone thermal $TT$ correlators assuming Einstein gravity in the bulk, setting $ \alpha_r = \alpha$.
We focus on the
scalar channel where the perturbative expansion reads
    \begin{equation}
        \ff_{\rm scalar}(\alpha) =\sum_{n=0}^\infty ~{\ff^{(n)}_{\rm scalar} \over \alpha^n} .
    \end{equation}
The computation in the other two channels is analogous, so we will simply list the corresponding results.  

    \subsection{Leading order (vacuum correlators) }
    
We start with the $\OO(1/\alpha^0)$ term, $\ff_{\rm scalar}^{(0)}$, in the large $\alpha$ expansion, which corresponds to the vacuum solution.  
As $\alpha\to\infty$ the reduced equation of motion becomes
    \begin{equation}\label{eq.0reomm3}
        \left(\pdv[2]{x}-\frac1x\pdv{x}-\frac1x \right)\Zt^{(0)}=0 \ ,
    \end{equation}
and admits an analytic solution in terms of the Bessel functions:
    \begin{equation}\label{homog}
\Zt^{(0)}(x)=2c_1xI_2\left(2\sqrt{x}\right)+2c_2xK_2\left(2\sqrt{x}\right) 
    \end{equation}  with  two coefficients $c_1$ and $c_2$.  
Regularity in the bulk requires $c_1=0$, while $c_2$ remains arbitrary. 
This remaining coefficient corresponds to the  norm of $\Zt^{(0)}$ which does not affect the value of the correlator; without loss of generality we require $a=1$ in \eqref{abexpansion}, thus $c_2=1$. 
The near-boundary expansion is then given by 
    \begin{equation}\label{kekeska}
       \Zt^{(0)}(x)=1-x+\frac{1}{4}x^2(3-4\gamma-2\log(x))+\frac{1}{36}x^3(17-12\gamma-6 \log (x))+\mathcal{O}\left(x^4\right) \ , 
    \end{equation}
where $\gamma$ is Euler's constant. From \eqref{kekeska} we find the ratio $\ff^{(0)}_{\rm scalar}= {b\over a} = \frac{1}{4}\left(3-4\gamma\right)$.
Thus, the $\OO(1/\alpha^0)$ contribution to the correlator $G_{\rm scalar}$ is
    \begin{equation}\label{eq.0corrG3}
       G^{(0)}_{\rm scalar}=\frac{\pi^2C_T}{160} (q^+q^-)^2\left(\frac{1}{4}\left(3-4\gamma\right)-\frac{1}{2}\log(-q^+q^-)\right) \ .
    \end{equation}
A similar calculation in the shear and sound channels yields the same result: 
    \begin{equation}\label{eq.analytic123}
\ff^{(0)}_{\rm scalar}=\ff^{(0)}_{\rm shear}=\ff^{(0)}_{\rm sound}=\frac{1}{4}\left(3-4\gamma\right).
    \end{equation}
Note that in all channels $\ff^{(0)}$ corresponds to the contact term.

    \subsection{Subleading order}\label{ssec.firstthorderm}

To proceed with perturbative expansion for the scalar channel, it will be useful to convert the corresponding reduced equation of motion \eqref{eq.reommnew} into  the following Schrödinger form:
    \begin{equation}\label{shok}
        \pdv[2]{x}Y(x) + \left(-\frac{3}{4x^2} - \frac{1}{x} + \frac{x}{4 \alpha}\right) Y(x) = 0 \ , ~~ Y(x)=\sqrt{\frac{\alpha}{x}}Z(x) \ .
    \end{equation}
The expansion in $1/\alpha$ reads 
    \begin{equation}\label{yeq}
        Y(x) = Y^{(0)}(x) + \frac{1}{\alpha} Y^{(1)}(x) + \dots
    \end{equation}
where  $Y^{(0)}(x)=\sqrt{\frac{\alpha}{x}}\Zt^{(0)}(x)$ was computed before. 
Expanding the equation \eqref{shok} to  $\OO(1/\alpha)$ gives 
\begin{equation}
 \pdv[2]{x}Y^{(1)}(x)-\left(\frac{1}{x} + \frac{3}{4 x^2} \right) Y^{(1)}(x) = - \frac{x}{4} Y^{(0)}(x) \ .
\end{equation}
The solution can be written in terms of the MeijerG functions:  
\begin{equation}
\begin{split}
Y^{(1)}(x)=
&\frac{x^3}{4 \sqrt{\pi }} \Bigg[\!K_2\left(2 \sqrt{x}\right) G_{1,3}^{2,1}\Bigg(4x\Bigg|
\begin{array}{c}
 1 \\
 \frac{1}{2},\frac{5}{2},-\frac{5}{2} \\
\end{array}
\Bigg)\!-\!\pi  I_2\left(2 \sqrt{x}\right) G_{2,4}^{3,1}\Bigg(4x\Bigg|
\begin{array}{c}
 -\frac{3}{2},1 \\
 -\frac{3}{2},\frac{1}{2},\frac{5}{2},-\frac{5}{2} \\
\end{array}
\Bigg)\Bigg]\\
&-2 i c_3 \sqrt{x} I_2\left(2 \sqrt{x}\right)+2 c_4 \sqrt{x} K_2\left(2 \sqrt{x}\right) \ . 
\end{split}
\end{equation}
Expanding the solution near the horizon, regularity restricts the coefficient $c_3 = \frac{i}{10}$. Setting $a=1$ in the expansion \eqref{abexpansion} leads  to $c_4 = 0$, which completely fixes the $\OO(1/\alpha)$ solution.
This gives the following contribution to the correlator
\begin{equation}
    G_{\rm scalar}=  G^{(0)}_{\rm scalar}+\frac{\pi^2 C_T}{1600} \ {(q^+q^-)^2 \over \alpha} + \OO(\alpha^{-2}).
\end{equation}
Following the same path, we also obtain the results in the shear and sound channels. 

Extracting the functions $\ff_{\rm scalar}$, $\ff_{\rm shear}$ and $\ff_{\rm sound}$, 
we find 
    \begin{align}\label{eq:TExch}
        \ff^{(1)}_{\rm scalar}=\frac{1}{10} \ , ~~~~
        \ff^{(1)}_{\rm shear}=-\frac{1}{40} \ , ~~~~
        \ff^{(1)}_{\rm sound}=\frac{1}{60}\ .
    \end{align} 

\subsection{Higher orders}

The general equation satisfied by the higher-order terms is given by
\begin{equation}
 \pdv[2]{x}Y^{(n)}(x)-\left(\frac{1}{x} + \frac{3}{4 x^2} \right) Y^{(n)}(x) = - \frac{x}{4} Y^{(n-1)}(x)\ .
\end{equation}
The homogeneous solution is given by
    \begin{equation}
Y_H^{(n)}(x) = -2 i c_{2n+1} \sqrt{x} I_2\left(2 \sqrt{x}\right)+2 c_{2n+2} \sqrt{x} K_2\left(2 \sqrt{x}\right),
    \end{equation}
while a particular solution can be expressed via the Green's function method
described in, $e.g.$, \cite{Romatschke_2009}, as
\begin{equation}
\begin{split}\label{ypn}
Y_P^{(n)}(x) = & (-2 i \sqrt{x} I_2(2\sqrt{x})) \int_0^x \dd y \frac{i}{2} (2 \sqrt{y} K_2(2\sqrt{y})) \frac{-y}{4} Y^{(n-1)}(y)\\
& + (2 \sqrt{x} K_2(2\sqrt{x})) \int_x^\infty \dd y \frac{i}{2} (-2 i \sqrt{y} I_2(2\sqrt{y})) \frac{-y}{4} Y^{(n-1)}(y) \ .
\end{split}
\end{equation}
Although it is not easy to perform the integrals in \eqref{ypn} explicitly, one can examine the near-horizon behaviour of the particular solution
which will be used to impose  regularity at the horizon
    \begin{equation}\label{obecnaformula}
Y_P^{(n)}(x \to \infty) = \left(\lim_{x\to \infty}-2 i \sqrt{x} I_2(2\sqrt{x})\right) \int_0^\infty \dd y \frac{i}{2} (2 \sqrt{y} K_2(2\sqrt{y})) \frac{-y}{4} Y^{(n-1)}(y)      \ .    
    \end{equation}

Focus now on the  $\OO(1/\alpha^2)$ term. Regularity of $Y^{(2)}(x)$ near the horizon leads to $c_5=\frac{i}{20}$, while requiring $a=1$ in the expansion \eqref{abexpansion} fixes the remaining coefficient $c_6 = 0$. 
This yields the $\OO(1/\alpha^2)$ contribution
\begin{equation}
 G_{\rm scalar}=  G^{(0)}_{\rm scalar}+\frac{\pi^2 C_T}{1600} \ {(q^+q^-)^2 \over \alpha} + \frac{\pi^2 C_T}{3200} \ {(q^+q^-)^2\over \alpha^2}+\OO(\alpha^{-3}).    
\end{equation}

In the same way, we can obtain the results in the remaining two channels. After extracting the function $\ff$, we find the following results:   
    \begin{align}
        \ff^{(2)}_{\rm scalar}=\frac{1}{20} \ , ~~~~
        \ff^{(2)}_{\rm shear}=-\frac{1}{40}\ , ~~~
        \ff^{(2)}_{\rm sound}=\frac{11}{420} \ . 
    \end{align} 
    
The same method in principle allows one to work out higher-order terms. 
In  Appendix \ref{ap.pertpos} we also verify the above results using the position-space approach.

{\it Radius of convergence:}  
\label{ssec.radiuscon}
Let us now estimate the radius of convergence of the perturbative expansion. We focus on the scalar channel. Define
\begin{equation}\label{rok}
	r_n=\abs{\frac{\ff^{(n)}_{\rm scalar}}{\ff^{(n+1)}_{\rm scalar}}} \ .
\end{equation}
We plot $r_n(n)$ in Fig.\ \ref{fig.rc} in Appendix \ref{ap.pertpos}.\footnote{The higher-order perturbative terms in this plot are computed in position space and then Fourier transformed to momentum space.}
The radius of convergence is defined as $\lim_{n\to\infty}  r_n$ and we find that it seems to be zero.

%%%%%%%%%%%%%%%%%%%%%%
%%%%%%%%%%%%%%%%%%%%%%

\section{Non-Perturbative Behavior} \label{sec.nonperts} 

In this section we analyze the stress-tensor correlators in all channels by solving the reduced equations of motion \eqref{eq.reommnew} numerically.
Note that the retarded correlators in general have an  imaginary part, which represents a purely non-perturbative contribution.
Using a WKB approximation we analyse this contribution explicitly and show that all three channels decay exponentially at the same rate. 

    \subsection{Numerical solution}

\begin{figure}[t!]
\begin{center}
\hspace*{-1.5cm}
\includegraphics[width=0.85\textwidth]{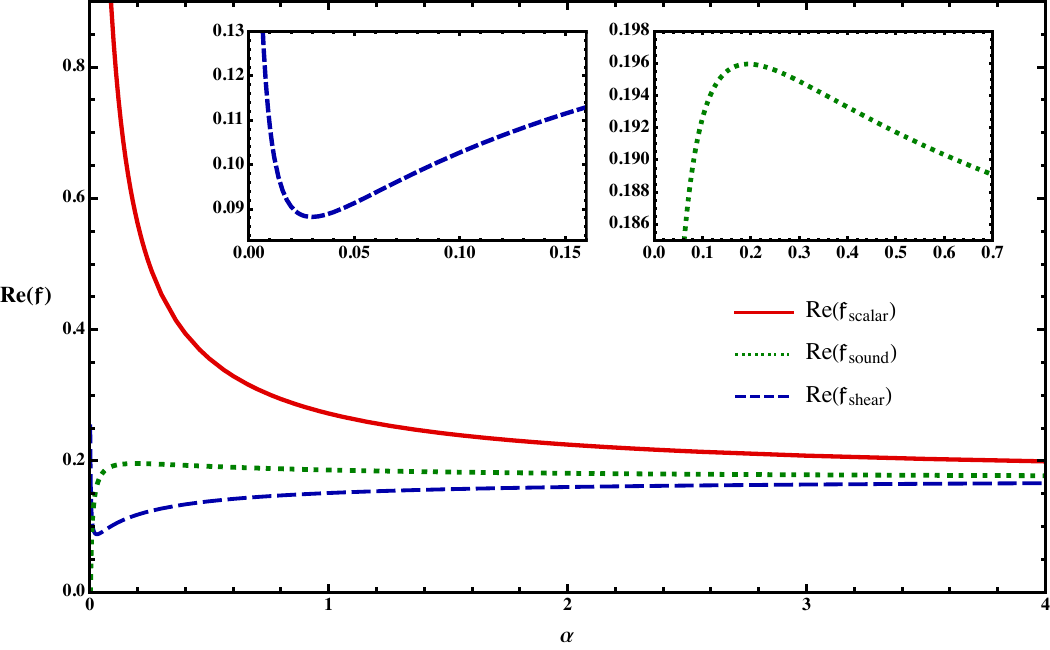}
\caption{The real part of $\mathfrak{f}(\alpha)$. 
The upper (solid, red) line corresponds to the scalar channel. 
The middle (dotted, green) line corresponds to the sound channel. 
The bottom (dashed, blue) line corresponds to the shear channel. 
 At large $\alpha$, all three lines approach the expected value, $\frac{3}{4}-\gamma$ $\approx 0.173$, where $\gamma$ is Euler's constant. 
The additional  smaller figures show the local minimum and maximum that appear in the shear and sound channels, respectively, in a small $\alpha$ region.}
\label{fig.re}
\end{center}
\end{figure}

\begin{figure}[ht]
\begin{center}
\hspace*{-1.5cm}
	\includegraphics[scale=0.59]{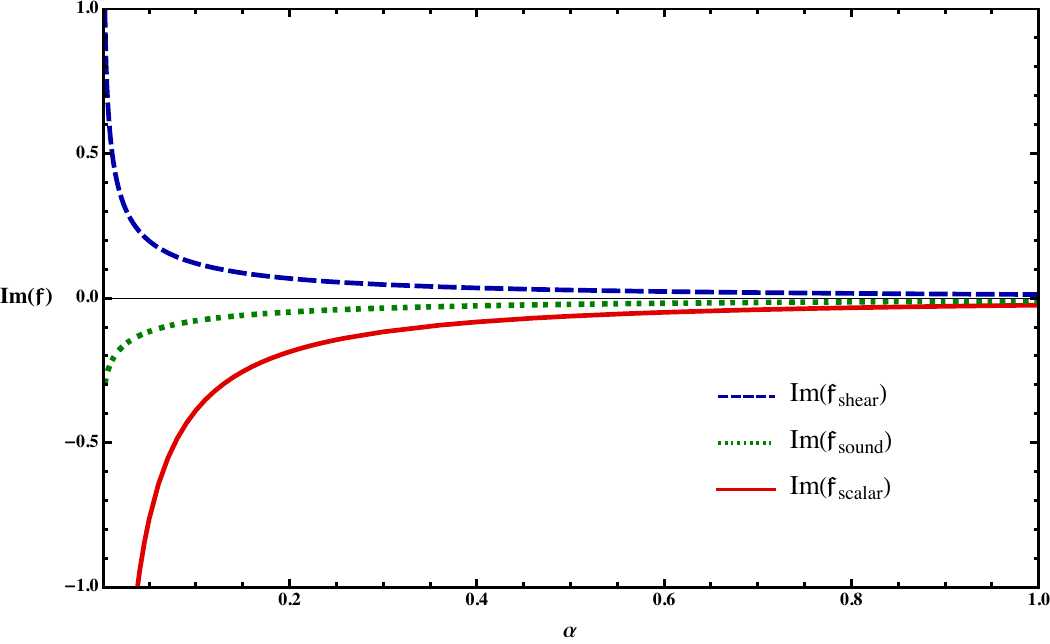}
    \caption{The imaginary part of $\mathfrak{f}(\alpha)$. 
The upper (dashed, blue) line corresponds to the shear channel. 
The middle (dotted, green) line corresponds to the sound channel. 
The bottom (solid, red) line corresponds to the scalar channel. }
	\label{fig.imAll}
\end{center}
\end{figure} 

In what follows, we again focus on space-like momenta where $\alpha$ (and thus $x$) is positive. In general, the solution in the limit $x\rightarrow\infty$ is a regular function multiplied by a superposition of incoming and outgoing waves. 
The natural choice is to pick the incoming-wave condition as discussed in \cite{Policastro:2001yc,Policastro:2002se,Policastro:2002tn,Son:2002sd}. 
With this choice one obtains the retarded correlators in the dual CFT. 

Consider the scalar channel.  Near the horizon the corresponding reduced equation \eqref{eq.reommnew} reduces to 
    \begin{equation}\label{eq.nearh.reomm3}  
       \left( \pdv[2]{x}-\frac1x\pdv{x}+\frac{x}{4\alpha} \right)\Zt(x\!\rightarrow\!\infty)=0 
    \end{equation}
where $\Zt(x\!\rightarrow\!\infty)$ denotes the solution deep in the bulk.
This equation can be solved analytically in terms of the (differentiated) Airy functions. Expanding the solution for large $x$ and picking the incoming wave, one finds
    \begin{equation}\label{eq.rawsol3hor}
        \Zt(x\!\rightarrow\!\infty)=x^{\frac14}e^{-\frac{i x^{3\over 2}}{3 \sqrt{\alpha}}} \left(\alpha^{-{1\over12}}-\frac{7 i \alpha^{5\over 12}}{24x^{3\over 2}}+\ldots\right)\ .
    \end{equation}
We use this expression to  numerically solve equation \eqref{eq.reommnew} starting from  large values of $x$ all the way to the boundary at $x=0$.
Then, using  \eqref{eq.fscalar} we compute the function $\ff_{\rm scalar}$ from this numerical solution.
We compute $\ff_{\rm shear}$ and $\ff_{\rm sound}$ in a similar way.

We next present the numerical results of both the real and imaginary parts of the correlators, for all three channels.

{\bf{Real part:}} We plot the real part of the function $\ff(\alpha)$ in Fig.\ref{fig.re}. 
For large $\alpha$, in all channels the numerical solutions quickly converge to the value $\frac{3}{4}-\gamma$, $i.e.$, the leading order in the perturbative expansion. We have also verified that, for every $n\in\mathbb{N}$, there exists a value $\alpha_n$, such that for all $\alpha>\alpha_n$ the $n$-th order perturbative expansion approximates the numerical solution better than any expansion with $n-1$ (or less) terms.  
%On the other hand, we remark that local extrema appear only in the shear and sound channels: we find that the function $\ff_{\rm shear}$ has a minimum near $\alpha^*_{\rm shear}\approx0.03$ with the value $\ff^*_{\rm shear}\approx0.88$. In the sound channel this extremum is a maximum, located at $\alpha^*_{\rm sound}\approx0.2$ with $\ff^*_{\rm sound}\approx0.196$.

{\bf{Imaginary part:}} We present the numerical solutions of the imaginary part of the function $\ff(\alpha)$ in Fig.\ \ref{fig.imAll}.
The imaginary part of the correlators is {\it purely} non-perturbative in the $1/\alpha$ expansion. Its concrete form is sensitive to the boundary condition at the horizon.  
If one instead imposes the outgoing-wave condition in the bulk and computes an advanced correlator, one finds that the correlator has the same real part but the imaginary part differs by a sign
(this also  follows from the general properties of Green's functions).
\subsection{Imaginary part of  correlators from WKB }  \label{sec.wkbim} 

An interesting question is how to estimate the non-perturbative behavior of the thermal correlators calculated numerically above. Let us answer this question by calculating the decay rate of $\Im G_R$ using the WKB analysis. To do so, 
we shall transform the reduced equations of motion \eqref{eq.reommnew} to the following form:
    \begin{equation}
    \label{forQM}
    \hbar^2\dv[2]{Z(\xi)}{\xi}=V(\xi)Z(\xi)\ .
    \end{equation} 
%which may be interpreted as a Schr{\"o}dinger equation.
Compared to the previous Schr{\"o}dinger-like equation \eqref{shok}, here the $V(\xi)$ term is independent of the expansion parameter, 
allowing us to perform a standard WKB analysis. 
Starting from \eqref{eq.reommnew},
we first rescale $x\rightarrow\sqrt{\alpha}y$ to have $B(x)=B(y)\alpha^{-1/2}$ where $B(y)$ is independent of $\alpha$. 
We next introduce $\xi(y)$ which satisfies the relation
    \begin{equation}\label{eqrk}
        \partial_y\log\big(\partial_y\xi\big)=B(y)\ .
    \end{equation}
The equations of motion can be written as
    \begin{equation}\label{fulrk}
        \alpha^{-\frac{1}{2}}Z''(\xi)=-\frac{y^2-4}{4y(\xi'(y))^2}Z(\xi)\ .
    \end{equation}
The parameter $\alpha$ plays the role of $\hbar$; the precise identification is $\alpha^{-{1\over 4}}=\hbar$.  
For simplicity we omit the channel index in $\xi$ and $Z$. 
Note that the potential  in \eqref{fulrk} is positive in the  region $y\in(0,2)$
which corresponds to a classically forbidden region. 

It may be useful to list  explicit expressions relating $y$ and $\xi$ for different channels:\footnote{Note \eqref{eqrk} is a second-order differential equation. We have chosen two integration constants such that these  expressions look simple.}\vspace{-0.2cm}
{\allowdisplaybreaks{
    \small\begin{align}
    \text{Scalar}&:\quad y=\sqrt{\xi} \ , \label{pozor0}\\
    \text{Shear}&:\quad y=\sqrt{4\pm\abs{\xi}^\frac12}\label{pozor} \ , \\
    \text{Sound}&:\quad y=\sqrt{12+\xi^{1\over 3}}\ . \label{pozor2}
    \end{align}}}In \eqref{pozor} we use plus for $\xi>0$ and minus for $\xi<0$. We omit the channel index in $\xi$. Transformations \eqref{pozor0}-\eqref{pozor2} map the conformal boundary to the points $0$, $-16$ and $-1728$ in the scalar, shear, and sound channel, respectively.
In all three channels the horizon corresponds to $\xi=\infty$.
The transformed equations of motion have the form \eqref{forQM}, with the identifications $\hbar=\alpha^{-{1\over 4}}$ and 
  {\small{\begin{equation}\label{pot}
        \left\{V_{\rm scalar},V_{\rm shear},V_{\rm sound}\right\}=\left\{ \frac{4-\xi}{16\,\xi^{3\over2}},-\frac{\pm1}{64 \left(4\pm\abs{\xi}^\frac12\right)^{3\over2} \abs{\xi}^\frac12},-\frac{8+\xi^\frac13}{144 \left(12+\xi^\frac13\right)^{3\over2} \xi ^{4\over3}} \right\}\ ,
    \end{equation}}}where in $V_{\rm shear}$ we use the plus signs for $\xi>0$ and the minus ones for $\xi<0$. 
In all channels the potential forms a barrier ($V>0$) in the near-boundary region:
     \begin{align}
    \text{Scalar}&:\quad \xi\in(0,4) \ , \label{neg1}\\
    \text{Shear}&:\quad \xi\in(-16,0) \ , \label{neg2}\\
    \text{Sound}&:\quad \xi\in(-1728,-512) \ .\label{neg3}
    \end{align}
The potential becomes negative for large $\xi$. In addition, in the sound channel we find a singularity in the classically allowed region at $\xi=0$.

One may now follow the standard WKB analysis. Considering the ansatz of the form
    \begin{equation}
        Z(\xi)=e^{i\alpha^{-\frac14}(W_0(\xi)+\alpha^{-\frac14} W_1(\xi)+\ldots)} 
    \end{equation}
and plugging this into the transformed equations, one determines the functions $W_i(\xi)$. The leading term is
    \begin{equation}\label{lead}
        W_0(\xi)=\pm\int^\xi\sqrt{-V(\xi')}\dd \xi'\ .
    \end{equation}
In the classically forbidden region we pick the sign corresponding to the decaying exponential, while deep in the bulk we select the oscillating solution that satisfies the incoming-wave condition near the horizon $\xi=\infty$.
Validity of WKB is restricted to the region where
    \begin{equation}
\left|\dv{\xi}\sqrt{V(\xi)}\right|\ll\alpha^{1\over 4}\abs{V(\xi)}\ .
    \end{equation}
One cannot use the WKB ansatz close to the boundary and around the turning point -- in these regions one has to solve the  equations of motion and connect solutions inside and outside the barrier.
 
The imaginary part of the correlator, using \eqref{eq.origG},  can be expressed as the following tunnelling probability\footnote{See also Appendix B in \cite{Son:2002sd} for a related discussion.} 
    \begin{equation}\label{imgr}
        \Im\ff\sim\Im G_R\sim \exp\left[-2\alpha^{\frac14}\int_{\text{barrier}}\!\!\!\!\sqrt{V(\xi')}\,\dd \xi'\right]\ .
    \end{equation}
%Taking the potential in \eqref{fulrk}, we find that 
In all three channels, the imaginary part of the correlator decays exponentially at the same rate\footnote{Note the overall prefactors depend on the channels, as is apparent already from Fig.\ \ref{fig.imAll}. We verified that the WKB results agree with the numerics.} 
    \begin{equation}\label{wkbres}
        \Im\mathfrak{f}\sim\exp\left[-2\alpha^{\frac14}\int_0^2\sqrt{\frac{4-y^2}{4y}}\dd y\right]=\exp\left[-\frac{\sqrt{\frac{\pi }{2}} \Gamma \left(\frac{1}{4}\right)}{\Gamma \left(\frac{7}{4}\right)}\alpha^{1\over 4}\right]\approx e^{-4.94~ \alpha^{1\over 4}}\ .
    \end{equation} 
It would be interesting to see if this  behavior holds more generally.
Note that the exponent in the  retarded thermal correlator of a scalar field in the large momentum  limit was computed in \cite{Son:2002sd,Dodelson:2022yvn}.   
Although the near-lightcone limit we consider here is different, the power of momenta in the exponent is the same, \textit{i.e}.\ $\alpha^{1/4}\sim q$, while  the multiplicative constants in the exponent differ. 

%%%%%%%%%%%%%%%%%%%%%%
%%%%%%%%%%%%%%%%%%%%%%

\section{Discussion and Implications}\label{sec.discussion} 

In this chapter we point out that the near-lightcone thermal correlators of stress tensors in holographic  Gauss-Bonnet gravity take the same form as those in Einstein gravity.\footnote{This chapter focuses on $d=4$ and we expect similar results in other $d>2$ dimensions.}   More precisely, we observe that the thermal two-point correlators of stress tensors are rather constrained in the near-lightcone limit: they are given by three 
universal functions ($\ff_{\rm scalar},\ \ff_{\rm shear},\ \ff_{\rm sound}$) whose arguments involve the combination of $\alpha \sim q^+ (q^-)^3/T^4$  and 
the three coefficients  $\aa, \bb, \cc$ which determine the stress-tensor three-point functions.
The correlator in a given channel takes the vacuum form when the corresponding ANEC is saturated,
as we already noticed in Chapter \ref{ch5}.

The correlators  admit a perturbative expansion in powers of $1/\alpha$.
This is essentially the OPE  combined with the near-lightcone limit, where only the leading-twist multi-stress tensors contribute.
One can read off the OPE coefficients of the two stress tensors and multi-stress tensors.
%The momentum space approach might be  more convenient than the one which employs the near-boundary ansatz in position space and substitutes it into the equations of motion, $e.g.$,  \cite{Karlsson:2022osn, Huang:2022vet}.
Note that the power series in momentum space seems to have zero radius of convergence, $i.e.$, it's an asymptotic series.\footnote{This was recently discussed in, $e.g.$, \ \cite{Caron-Huot:2009ypo,Iliesiu:2018fao,Manenti:2019wxs,Dodelson:2023vrw}.}

Depending on whether we want to compute retarded or advanced correlators, we need to impose appropriate boundary conditions at the horizon (which in our variables corresponds to the behavior at large $x$). 
Perturbatively, the correlator is completely determined by the OPE (as we explain in Section \ref{sec:Momentum} and Appendix \ref{ap.pertpos}).
However the boundary conditions at the horizon affect the solution non-perturbatively in $\alpha$.
This is because a general solution decays exponentially under the barrier, as discussed in Section \ref{sec.wkbim}.
It would be interesting to understand the significance of such non-perturbative terms.\footnote{In two spacetime dimensions, similar terms 
appear after the Fourier transform of the HHLL Virasoro vacuum block, see, $e.g.$, \cite{Son:2002sd,Manenti:2019wxs}.
(See also \cite{Haehl:2018izb,Datta:2019jeo,Jensen:2019cmr,Ramirez:2020qer} for explicit expressions of  retarded $TT$ correlators.) In four-dimensional holographic CFTs on the sphere, the spectrum of quasinormal modes contains contributions that are non-perturbative in spin \cite{Dodelson:2022eiz} (see also \cite{Festuccia:2008zx}). It would be interesting to understand to what extent the non-perturbative terms are determined by the OPE in $d=4$.}

In the near-lightcone limit, because of the universality of $TT$-correlators discussed above,
we can simply focus on the analysis based on pure Einstein gravity. 
Note that for the transverse polarization of the stress tensor, the equation of motion for the metric fluctuation is the same as that for 
a minimally coupled scalar.
Hence the two point functions must be identical.
Naively one may find this surprising, given that the OPE coefficients for a scalar contain poles at integer values of the scalar's conformal dimension \cite{Fitzpatrick:2019zqz}. This corresponds to the mixing with the double-trace operators and fixes the residue of the OPE coefficient of the latter, which ensures
the divergence is cancelled.
How does this work in the stress-tensor correlator case and how is this reflected in momentum space?
The answer to this question is that the logarithmic terms in position space which are produced by the cancellation of
the poles at $\Delta=4$ get Fourier transformed to the rational functions of momenta (or, in our limit, 
$\alpha$) in momentum space.
Indeed, in Appendix  \ref{apB}, we verify explicitly that the OPE coefficients of the two scalars and multi-stress tensors,
together with the thermal expectation values of the latter, multiplied by the corresponding conformal blocks in momentum space reproduce
the perturbative expansion of the transverse $TT$ correlator.

It is useful to examine the large-$N$ counting 
of  thermal  $TT$ correlators (where $N\sim \sqrt{C_T}$). Consider a finite temperature connected $TT$-correlator on a sphere 
above the confinement-deconfinement phase transition.
The disconnected component scales like $N^4$ and this behavior is entirely due to the double-stress tensor operators $[T_{\mu\nu}]^2$, as explained
in Chapter \ref{ch4}. 
Indeed, the MFT OPE coefficients $\lambda_{T_{\mu\nu} T_{\alpha\beta} [T_{\mu\nu}]^2} \sim 1$ while $\langle [T_{\mu\nu}]^2 \rangle \sim N^4$.
The subleading corrections to the OPE coefficients and to the anomalous dimensions of $[T_{\mu\nu}]^2$ contribute to the connected correlator.
It is easy to extract the large-$N$ behavior of the OPE coefficients with the $k$-stress tensors and convince oneself that they all contribute to 
the connected correlator with the expected $N^2$ scaling. 

On the other hand, in the low-temperature phase $\langle [T_{\mu\nu}]^k \rangle $ scales like $N^0$, while the leading large-$N$ behavior of the $TT$ correlator scales like $ N^2$.
In holographic theories such correlators are simply given by the sum of the vacuum correlators over the thermal images.
One can immediately see how this is reproduced by multi-stress tensors.
The only contributions that survive in addition to the identity are the double-stress tensors.

There are various extensions to consider. Let us mention a few of them: (i) Extend the analysis of near-lightcone correlators to the charged black hole case. (ii) Understand more precisely how non-universal coefficients affect the thermal $TT$ correlators when moving slightly away from the lightcone limit. (iii) It may be useful to further study the thermal $TT$ correlators using CFT techniques developed in, $e.g.$, 
\cite{Caron-Huot:2017vep, Simmons-Duffin:2017nub, Iliesiu:2018fao, Gobeil:2018fzy, Petkou:2018ynm,Karlsson:2019dbd, Manenti:2019wxs,Delacretaz:2020nit,Alday:2020eua,Rodriguez-Gomez:2021pfh,Engelsoy:2021fbk,Parisini:2022wkb,Dodelson:2022yvn,Bhatta:2022wga,Avdoshkin:2022xuw,Fortin:2023czq,Dodelson:2023vrw}. 
(iv) It would also be interesting to see if anything useful can be said about the regime $|k|\gg|\omega|$ \cite{Banerjee:2019kjh,Caron-Huot:2022akb}. 

In this chapter, we speculate that the universality observed in the holographic Gauss-Bonnet theory remains valid  in more general holographic theories.  It would be interesting to study the near-lightcone $TT$ correlators using different gravity models  to see if this universality persists. On the other hand, understanding this directly from the CFT point of view would be a more ambitious but very interesting goal. In this spirit a possible step is to study the lightcone limit of heavy-heavy-light-light correlators, with the light operators being stress tensors, from the bootstrap point of view. This would extend recent progress for scalar correlators in $e.g.$ \cite{Karlsson:2019dbd,Li:2019zba} and related works. In the latter, the Lorentzian inversion formula was used to get the OPE data for multi-stress tensors in the scalar case and it would be interesting to generalize this to stress tensor correlators, or spinning correlators more generally. A CFT bootstrap approach would likely shed light on the regime of universality beyond the cases explored in this chapter and is therefore of great interest.

%% file: 7_conclusion/conclusion.tex
\chapter{Conclusion}\label{ch7}

The aim of this thesis was to study thermal correlation functions in conformal field theories with gravitational duals. This investigation, carried out within the framework of AdS/CFT correspondence, focused on two interrelated themes: (1) identifying the imprint of black hole singularities in thermal boundary correlators and (2) understanding the structure and universality of thermal two-point functions of the stress tensor. While both of these topics are of fundamental theoretical interest, they also shed light on the deep and intricate connections between gravity, thermodynamics, quantum field theory, and causality.

The guiding philosophy throughout this thesis has been that CFT two-point functions at finite temperature -- especially their stress-tensor sector -- can serve as a powerful diagnostic of both: boundary and bulk physics. From a methodological perspective, the central tool employed was the \emph{near-boundary expansion} of bulk equations of motion. This technique constructs solutions as systematic expansions in inverse powers of the radial AdS coordinate near the conformal boundary. Importantly, this expansion mirrors the structure of the Operator Product Expansion (OPE) in the boundary CFT, and hence provides a natural bridge between gravitational physics in the bulk and conformal data on the boundary.

\section*{Black Hole Singularity from Boundary CFT}

One of two mayor themes of this thesis was the question of whether, and how, boundary correlators can encode information about the spacetime region behind the black hole horizon -- most notably, the black hole singularity. This question strikes at the heart of the holographic principle, since the singularity represents a breakdown of the classical spacetime description, whereas the boundary CFT is manifestly unitary and well-defined at all times.

To address this question, we considered scalar two-point function in a thermal state dual to AdS-Schwarzschild black holes. Using geodesic approximation and a detailed OPE analysis at finite conformal dimension, we demonstrated that the stress-tensor sector of the CFT two-point function develops \emph{bouncing singularities} in the complex time plane, see Fig.\ \ref{fig:Polesd4}. These singularities arise from bulk geodesics that traverse the black hole interior, reflect off the singularity, and return to the boundary of the maximally extended spacetime \cite{Fidkowski:2003nf,Festuccia:2005pi}. We provided robust evidence that these features are present at finite conformal dimension, indicating that the boundary CFT is, in a precise sense, sensitive to the presence of the black hole singularity.

Importantly, we also showed that the \emph{double-trace sector} of the OPE plays a critical role in restoring analyticity and enforcing the Kubo-Martin-Schwinger (KMS) condition. While the singularities arise from the stress-tensor sector, they are cancelled in the full correlator due to compensating singularities in the double-trace contributions. This intricate interplay emphasizes the non-trivial nature of the CFT decomposition.

We generalized these results to higher dimensions (specifically $d = 6,\,8$) and showed that similar bouncing singularities persist in the analytic structure of the stress-tensor sector, with their positions precisely matching those predicted by the geodesic analysis \cite{Fidkowski:2003nf,Festuccia:2005pi}. For $d = 2$, in contrast, we found that the absence of curvature singularities in the bulk BTZ black hole geometry is reflected in the analyticity of the boundary correlators up to the KMS singularity.

Furthermore, we checked that, due to a subtle non-commutativity of limits, the singularities in the correlator disappear when $\Delta\rightarrow\infty$, as expected. In this limit, one instead recovers the contribution from a complex geodesic. This allowed us to establish a concrete map between CFT contributions to the thermal correlator and individual bulk geodesics, see Eq.\ \eqref{eq:RealCorr}.

These findings provide an example of bulk singularity data being encoded in CFT and also point to the possibility of using boundary diagnostics to characterize and potentially classify different types of spacetime singularities based on their analytic fingerprints in the dual theory.

\section*{\texorpdfstring{$TT$}{TT} Correlators and Near-Lightcone Universality}

The second major focus of this thesis was the structure of thermal two-point functions of the stress tensor in holographic CFTs. These correlators are essential for understanding thermalization, transport phenomena, and causal constraints in strongly coupled quantum systems.

We began by analysing these correlators in theories dual to Einstein gravity, computing them in all polarization channels (scalar, shear, sound) using a suitably modified near-boundary ansatz. We then extended our analysis to Gauss-Bonnet gravity to assess the impact of higher-curvature corrections. The choice of Gauss-Bonnet gravity was motivated both by its tractability and by its ability to capture key features of higher-derivative bulk dynamics while remaining free of ghosts in specific regimes.

A key outcome of this study was the discovery of \emph{near-lightcone universality} in the behaviour of thermal stress-tensor correlators. In particular, we found that the correlators, when examined near the lightcone in position or momentum space, exhibit a structure that is insensitive to the details of the bulk theory, including the presence of higher-curvature terms. More precisely, thermal $TT$ correlators near the lightcone are completely determined by three universal functions $\mathfrak{f}_{\rm scalar}$, $\mathfrak{f}_{\rm shear}$ and $\mathfrak{f}_{\rm sound}$. The bulk Lagrangian only affects the arguments of these functions via corrections to the cubic stress-tensor couplings and the thermal stress-tensor one-point function.

To make this structure explicit, we decomposed the correlators into conformal blocks corresponding to the exchange of multi-stress tensor operators. We found that, near the lightcone, the coefficient of the $n$-stress tensor exchange is proportional to the $n$-th power of the corresponding conformal collider bound. An immediate consequence of this is that saturation of the averaged null energy condition (ANEC) automatically implies saturation of an infinite tower of its higher-spin generalizations, leading to the correlator being temperature independent.

This part of the thesis culminates in a detailed perturbative and non-perturbative analysis of the near-lightcone correlators in momentum space, including WKB approximations and numerical analysis for the imaginary part of the universal functions $\mathfrak{f}_{\rm scalar}$, $\mathfrak{f}_{\rm shear}$ and $\mathfrak{f}_{\rm sound}$.

\section*{Broader Implications and Future Directions}

The results presented in this thesis open several promising avenues for future research. A few of the most significant ones are listed below:

\begin{itemize}
    \item \textbf{Singularities in More General Spacetimes:} The current analysis focused primarily on planar AdS-Schwarzschild black holes. A natural extension is to study spherical or charged black holes, time-dependent geometries with curvature singularities, or spacetimes with back-reacting matter content. These may modify or regularize the bouncing singularities and could be relevant for understanding the fate of singularities in more realistic holographic settings.
    
    \item \textbf{Connection with Thermalization and Chaos:} The structure of thermal correlators is closely tied to the process of thermalization and the onset of quantum chaos. Future work could explore connections between the stress-tensor correlators in holographic CFTs and chaotic properties such as out-of-time-ordered correlators (OTOCs), Lyapunov exponents, and butterfly velocities.
    
    \item \textbf{Bootstrapping the Stress-Tensor Sector:} While the near-boundary meth\-od provides access to multi-stress tensor exchanges, recent developments in the conformal bootstrap suggest that these structures can also be obtained via non-perturbative numerical or analytic bootstrap techniques. An integrated approach combining holographic methods with bootstrap constraints may yield even deeper insights.
    
    \item \textbf{Subleading Corrections and Finite Coupling Effects:} Understanding how the observed singularities and universal features get corrected at finite coupling ($i.e.$, away from the strict large-$\lambda$ or large-$N$ limit) remains a challenging but important goal. Such corrections may reveal the limitations of the semi-classical gravity approximation and provide a glimpse into the stringy or quantum gravitational resolution of singularities.
    
    \item \textbf{Implications for Real-World Systems:} While the theories studied in this thesis are not directly realized in nature, they share many structural similarities with systems in high-energy and condensed matter physics. The techniques and results developed here may help interpret phenomena in quark-gluon plasma, condensed matter holography, and early-universe cosmology.
    
    \item \textbf{Geometric Interpretation of Conformal Data:} One of the most powerful aspects of holography is its ability to translate geometric concepts into field-theoretic language. The results presented here suggest that information about the region beyond the black hole horizon can be read off from the OPE data and analytic structure of thermal correlators. Further developing this dictionary could be instrumental for both holography and CFT. In particular, it would be interesting to understand the boundary imprints of the inner horizon in the cases of charged and rotating black holes.
\end{itemize}

\section*{Final Reflections}

This thesis has sought to advance our understanding of how geometry and quantum field theory are connected in the framework of holography. By studying thermal correlators through the lens of the near-boundary expansion and OPE, we were able to recover deep geometric features of the bulk -- such as black hole singularities -- and examine causality constraints and universal features of the boundary theory.

Much remains to be discovered, but this work affirms that the holographic correspondence provides both a powerful conceptual framework and an effective calculational tool for exploring the frontiers of high-energy physics.

%% file: A_appendices/appendix.tex
\appendix

%%%%%%%%%%%
%%%%%%%%%%%
%%%%%%%%%%%
%%%%%%%%%%%
%%%%%%%%%%% p4 appendices:
%%%%%%%%%%%
%%%%%%%%%%%
%%%%%%%%%%%
%%%%%%%%%%%

\chapter{Structure of Scalar Correlator}\label{a.cbd}

In this appendix we discuss the stress-tensor sector of the thermal two-point function of scalar operators \eqref{eq:DefCorrFunagain}. The contribution of the $n$-stress tensor $[T^n]_J$ to the correlator is, up to a constant, the $[T^n]_J$  conformal partial wave\footnote{Here, we are referring to conformal partial waves as defined in \cite{Dolan:2000ut,Dolan:2003hv}. Note that in modern CFT literature this term is sometimes used in a different context.} (CPW) of the heavy-heavy-light-light (HHLL) correlator on $\mathbb{R}\times S^{d-1}_R$ \cite{Karlsson:2021duj}. Light and heavy refer to how the conformal dimensions of the inserted operators scale with the central charge, $C_T$, of the CFT. The conformal dimension of the light operator, $\Delta$, does not scale with the central charge, while for the heavy operator it does,  $\Delta_H \sim \coo{C_T}$.

Starting in flat space, the HHLL four-point function can be expanded in the corresponding CPWs $\widehat{W}_{\Delta',J'}$ as
    \begin{equation}\label{e.cordef}
G(z,\overline z) \equiv \expval{\mathcal{O}_H(0)\phi(z,\overline{z})\phi(1)\mathcal{O}_H(\infty)}=\sum_{\Delta',J'}c_{\Delta',J'}\widehat{W}_{\Delta',J'}\ ,
    \end{equation}
where the sum runs over all primaries and $c_{\Delta',J'}$ is a combination of OPE coefficients.
The flat space CPW are given by \cite{Dolan:2000ut,Dolan:2003hv}%
\footnote{Compared to \cite{Dolan:2000ut} we omit the factor $(-2)^{-J}$ in the conformal waves.}
{\allowdisplaybreaks{\small{
    \begin{align}
        d=2\quad\,&\widehat{W}_{\Delta',J'}=\frac{1}{(Z\Zb)^\Delta}\left(k_{\Delta'+J'}(Z)k_{\Delta'-J'}(\Zb)+k_{\Delta'-J'}(Z)k_{\Delta'+J'}(\Zb)\right)\label{e.2db}\\
        d=4\quad\,&\widehat{W}_{\Delta',J'}=\frac{1}{(Z\Zb)^\Delta}\frac{Z\Zb}{Z-\Zb}\left(k_{\Delta'+J'}(Z)k_{\Delta'-J'-2}(\Zb)-k_{\Delta'-J'-2}(Z)k_{\Delta'+J'}(\Zb)\right)\label{e.4db}\\   
        d=6\quad\,&\widehat{W}_{\Delta',J'}=\frac{1}{(Z\overline{Z})^\Delta}\Big(\mathcal{F}_{0,0}(Z,\overline{Z})-\frac{J'+3}{J'+1}\mathcal{F}_{-1,1}(Z,\overline{Z})\nonumber\\
        &\hspace{1.5cm}-\frac{(\Delta'-4)(\Delta'+J')^2}{16(\Delta'-2)(\Delta'+J'+1)(\Delta'+J'-1)}\mathcal{F}_{1,1}(Z,\overline{Z})\label{e.6db}\\
        &\hspace{1.5cm}+\frac{(\Delta'-4)(J'+3)}{(\Delta'-2)(J'+1)}\frac{(\Delta'-J'-4)^2}{16(\Delta'-J'-5)(\Delta'-J'-3)}\mathcal{F}_{0,2}(Z,\overline{Z})\Big)\ ,\nonumber
    \end{align}}}}where
{\allowdisplaybreaks{\small{
    \begin{align}
    &k_\eta(\xi)=\xi^{\frac{\eta}{2}}\!\!\!\!\phantom{F}_2F_1(\tfrac{\eta}{2},\tfrac{\eta}{2},\eta,\xi)\,,\\
    &\mathcal{F}_{n,m}(Z,\overline{Z})=\frac{(Z\overline{Z})^{\frac12 (\Delta'-J')}}{(\overline{Z}-Z)^3}(\mathbb{F}_{n,m}(Z,\overline{Z})-\mathbb{F}_{n,m}(\overline{Z},Z))\,,\\
    &\mathbb{F}_{n,m}(Z,\overline{Z})=\overline{Z}^{J'+n+3}Z^m\!\!\!\!\phantom{F}_2F_1\Big(\frac{\Delta'+J'}{2}+n,\frac{\Delta'+J'}{2}+n;\Delta'+J'+2n;\overline{Z}\Big)\nonumber\\
    &\hspace{2.2cm}\times\!\!\!\!\phantom{F}_2F_1\Big(\frac{\Delta'-J'}{2}-3+m,\frac{\Delta'-J'}{2}-3+m;\Delta'-J'-6+2m;Z\Big)\ .
    \end{align}
}}}%
We will use that in the T-channel the relation between $(Z,\Zb)$ and the cross-ratios $(z,\overline{z})$ is \mbox{$(Z,\Zb)=(1-z,1-\overline{z})$}.

We now map the flat space to $\mathbb{R}\times S^{d-1}_R$ using 
    \begin{equation}\label{e.cylmap}
            z=1-Z=e^{-\frac{\tau}{R}-i\frac{\abs{\vec{x}}}{R}}\,, \qquad \overline{z} = 1- \overline{Z} = e^{-\frac{\tau}{R}+i\frac{\abs{\vec{x}}}{R}}\,, 
    \end{equation}
where $R$ is the sphere radius and $\vec{x}$ schematically denotes all coordinates besides $\tau$.
This transformation introduces an overall prefactor in conformal partial waves
    \begin{equation}\label{e.cylcpw}
        W_{\Delta',J'}=R^{-2\Delta}(z\overline{z})^\frac{\Delta}{2}\widehat{W}_{\Delta',J'}=R^{-2\Delta}\left((1-Z)(1-\overline{Z})\right)^\frac{\Delta}{2}\widehat{W}_{\Delta',J'}\,.
    \end{equation}
The $W_{\Delta',J'}$ are the CPW on $\mathbb{R}\times S^{d-1}_R$, where the two light operators are inserted at 0 and $(\tau,\vec{x})$ while the heavy operators sit at $-\infty$ and $+\infty$.
Via the operator-state correspondence, the correlator on $\mathbb{R}\times S^{d-1}_R$ can be seen as a two-point function in a heavy state $\ev{\phi(\tau,x)\phi(0,0)}{\mathcal{O}_H}$. 

Let us now focus on the exchanges of $n$-stress tensors, which have conformal dimension $\Delta'=d\,n$ and spin $J'=0,2,4,\ldots,2n$.
Since these thermalize in heavy states \cite{Karlsson:2021duj}, their expansion in CPW \eqref{e.cylcpw} precisely matches the corresponding expansion in terms of thermal conformal blocks \cite{Iliesiu:2018fao}.
We are interested in the OPE limit\footnote{In this limit we effectively work on $\mathbb{R}^d$. It is however important to distinguish this from the original flat space-time on which correlator \eqref{e.cordef} was formulated.} $z,\overline{z}\rightarrow1$ (or equivalently $Z,\Zb\rightarrow0$), which in $\mathbb{R}\times S^{d-1}_R$ corresponds to $\tau,\abs{\vec{x}}\ll R$.

First consider the case with $x=\abs{\vec{x}}=0$.  In the regime $\tau\ll R$,
the corresponding CPWs simplify to 
    \begin{align}
    d=2\qquad&W_{2n,J'}\approx2\tau^{-2\Delta}\left(\frac{\tau}{R}\right)^{2n}\label{e.assu}\\
    d=4\qquad&W_{4n,J'}\approx(1+J')\tau^{-2\Delta}\left(\frac{\tau}{R}\right)^{4n}\label{e.for4d}\\
    d=6\qquad&W_{6n,J'}\approx\frac16(2+J')(3+J')\tau^{-2\Delta}\left(\frac{\tau}{R}\right)^{6n}\ .   
    \end{align}
Each $n$-stress tensor exchange is multiplied by a factor of $\mu^n$, where%
\footnote{In holographic theories, $\mu$ is also the mass parameter appearing in the metric of the AdS-Schwarzschild black hole, see for example \eqref{eq:d=4fFunction}.
For a black brane in $d+1$ dimensions, the relation between $\mu$ and the (inverse) Hawking temperature is 
    \begin{equation*}
        \mu=\left(\frac{4\pi R}{d\,\beta}\right)^d\ ,
    \end{equation*}
    which scales with $R/\beta$ exactly as in \eqref{eq:Mutemp}.}
\begin{align}
\label{eq:Mutemp}
    \mu \propto \frac{\Delta_H}{C_T}
    \propto \varepsilon\,\frac{R^d}{C_T}
    \propto \left(\frac{R}{\beta}\right)^d\,.
\end{align}
In the above, we used that the energy density $\varepsilon$ is proportional to $C_T\,T^{d}$ \cite{El-Showk:2011yvt, Iliesiu:2018fao}.
It is convenient to isolate these dimensionful factors in the OPE coefficients and define
\begin{equation}
        c_{d\,n,J'} \equiv  \lambda_{n,J'}\left(\frac{R}{\beta}\right)^{d\,n}\,,
\end{equation}
as in this case all dependence on the sphere radius $R$ completely cancels out in the stress-tensor sector of the thermal correlator%
\footnote{The subscript in $G_T(\tau)$ denotes that this is the stress-tensor sector and \emph{not} the full correlator.}
{\allowdisplaybreaks{\small{
    \begin{align}
        d=2\quad\,&G_T(\tau)=\frac{1}{\tau^{2\Delta}}\left(1+2\lambda_{1,2}\left(\frac{\tau}{\beta}\right)^2+\sum_{n=2}^\infty\left[\sum_{J'}\,2\lambda_{n,J'}\right]\left(\frac{\tau}{\beta}\right)^{2n}\right)\,,\\
        d=4\quad\,&G_T(\tau)=\frac{1}{\tau^{2\Delta}}\left(1+3\lambda_{1,2}\left(\frac{\tau}{\beta}\right)^4+\sum_{n=2}^\infty\left[ \sum_{J'}\,\lambda_{n,J'}(1+J')\right]\left(\frac{\tau}{\beta}\right)^{4n}\right)\,,\label{e.lebara}\\
        d=6\quad\,&G_T(\tau)=\frac{1}{\tau^{2\Delta}}\left(1+\frac{10}{3}\lambda_{1,2}\left(\frac{\tau}{\beta}\right)^6+\sum_{n=2}^\infty\left[ \sum_{J'}\,\lambda_{n,J'}\frac{(2+J')(3+J')}{6}\right]\left(\frac{\tau}{\beta}\right)^{6n}\right)\,,\label{e.6dcb}
    \end{align}}}}%
where the sums over $J'$ run over $J'=0,2,\ldots 2\,n$.
Finally, it is useful to introduce summed coefficients $\Lambda_n$ through
    \begin{equation}\label{e.topeL}
        G_T(\tau)=\frac{1}{\tau^{2\Delta}}\sum_{n=0}^\infty\Lambda_n\left(\frac{\tau}{\beta}\right)^{d\,n}\,,
    \end{equation}
which we use in Chapter \ref{ch3}.

Now let $x$ be  small, but non-vanishing, such that \mbox{$x\ll\tau\ll R$}.
For concreteness, we focus only on $d=4$.
Mapping the flat space CPW \eqref{e.4db} to $\mathbb{R}\times S^{d-1}_R$, taking the OPE limit, and then expanding in $x\ll\tau$, we get the stress-tensor sector of the correlator as
    \begin{equation}\label{e.extendedg}
    \begin{split}
        G_T(\tau,x)&=\frac{1}{\tau^{2\Delta}}\left(1-\frac{\Delta\,x^2}{\tau^2}\right) \sum_{n=0}^\infty \left[\Lambda_n+\left(2\,n\,\Lambda_n +{\widetilde{\Lambda}^{(1)}_n}\right)\frac{x^2}{\tau^2}\right]\left(\frac{\tau}{\beta}\right)^{4\,n}\!\!+\coo{x^4},
    \end{split}
    \end{equation}
with $\Lambda_n$ defined in \eqref{e.topeL} and
    \begin{equation}\label{e.dufamzeposledna}
        {\widetilde{\Lambda}^{(1)}_n}=-\frac{1}{6}\sum_{J'=0,2,\ldots,2n}\!\!\!J'\,(1+J')\,(2+J')\,\lambda_{n,J'}\ .
    \end{equation}
Note that $n=0$ corresponds to the identity contribution and thus $\widetilde \Lambda^{(1)}_0=0$, while at  $n=1$ the only non-vanishing contribution comes from the stress tensor with $J'=2$.

To make the notation for $x\neq0$ more transparent, we generalise equation\ \eqref{e.topeL} to
    \begin{equation}\label{e.exnenula}
        G_T(\tau,x)=\frac{1}{\tau^{2\Delta}}\sum_{n=0}^\infty\left[\Lambda^{(0)}_n+\frac{x^2}{\tau^2}\,\Lambda^{(1)}_n+\coo{\frac{x^4}{\tau^4}}\right]\left(\frac{\tau}{\beta}\right)^{dn}\ ,
    \end{equation}
where $\Lambda^{(0)}_n\equiv\Lambda_n$ and $\Lambda^{(1)}_n\equiv(2n-\Delta)\Lambda_n+{\widetilde{\Lambda}^{(1)}_n}$. In the same way one can define $\Lambda_n^{(m)}$ for any $m$. 
Note that the holographic method we used to determine the stress-tensor contributions extracts $\lambda_{n,J'}$ \cite{Fitzpatrick:2019zqz}.
This means that in principle one can use this method  to obtain the coefficients $\Lambda^{(m)}_n$ to arbitrary high orders in $m$ and $n$.

Let us conclude this appendix by comparing the above analysis with
the expansion of the correlator using the thermal conformal blocks formulated  on $\mathbb{S}_\beta^1\times\mathbb{R}^{d-1}$  \cite{Iliesiu:2018fao}
    \begin{equation}\label{e.fgege}
        G_T(\tau,x)=\frac{1}{\abs{\tau^2+x^2}^\Delta}\sum_{n=0}^\infty\sum_{J'=0,2,\ldots,2n}\!\!\!\hat{\lambda}_{n,J'}\frac{\abs{\tau^2+x^2}^\frac{dn}{2}}{\beta^{dn}}C_{J'}^{\left(\frac{d+2}{2}\right)}\left(\frac{\tau}{\sqrt{\tau^2+x^2}}\right)\ ,
    \end{equation}
where $C_{J'}^{(\nu)}(\eta)$ are Gegenbauer polynomials and $\hat{\lambda}_{n,J'}$ are dimensionless coefficients. Imposing $x=0$ and restricting to $d>2$, \eqref{e.fgege} simplifies to
    \begin{equation}\label{e.fgeg2}
        G_T(\tau)=\frac{1}{\tau^{2\Delta}}\sum_{n=0}^\infty\sum_{J'=0,2,\ldots,2n}\!\!\!\hat{\lambda}_{n,J'}\smqty(d-3+J \\ J)\left(\frac{\tau}{\beta}\right)^{dn}\ .
    \end{equation}
This can now be compared with the equations \eqref{e.lebara} and \eqref{e.6dcb}. For $d=4$ one finds $\hat{\lambda}_{n,J'}=\lambda_{n,J'}$, while in $d=6$ there is a conventional difference by a factor $(J+1)$ in the coefficients $\lambda_{n,J'}$ and $\hat{\lambda}_{n,J'}$.

%%%%%%%%%%%%%%%%%%%%%%%%%%%%%%%%%%%%%%%%%%%%%%%%%%%%%
%%%%%%%%%%%%%%%%%%%%%%%%%%%%%%%%%%%%%%%%%%%%%%%%%%%%%

\chapter{Scalar Equation of Motion}\label{app:eomcoefz}
In this appendix we give the explicit form of the equation of motion
{{\small
    \begin{equation}\label{e.wew}
        (\Box-m^2)\phi=0\, ,\qquad m^2=\Delta(\Delta-d)
    \end{equation}}}%
in the background of ($d+1$)-dimensional  planar Euclidean Schwarzschild-AdS black hole after the coordinate  transformation 
{\small{    
    \begin{equation}
    \rho^2=r^2\vec{x}^2\qq{and}w^2=1+r^2(\tau^2+\vec{x}^2)\ . 
    \end{equation}}}%
Introducing the following ansatz for the scalar bulk field
{\small{
    \begin{equation}\label{e.anz1again}
        \phi(w,\rho,r)=\left(\frac{r}{w^2}\right)^\Delta \psi(w,\rho,r)\ ,
    \end{equation}}}%
where $(r/w^2)^\Delta$ is the solution in the pure AdS space, ``reduces'' the Klein-Gordon equation \eqref{e.wew} to  \cite{Fitzpatrick:2019zqz}
{\small{
    \begin{equation}\label{e.theequation}
    \left(\partial_r^2\!+\!C_1\partial_w^2\!+\!C_2\partial_\rho^2\!+\!C_3\partial_r\partial_w\!+\!C_4\partial_r\partial_\rho\!+\!C_5\partial_w\partial_\rho\!+\!C_6\partial_r\!+\!C_7\partial_w\!+\!C_8\partial_\rho\!+\!C_9\right)\psi=0\ ,
    \end{equation}}}with
the coefficients $C_i$ given by
{\allowdisplaybreaks{\footnotesize{   
    \begin{align}
        C_1&=\frac{f\left(\rho^2+(w^2-1)^2f\right)+w^2-\rho^2-1}{r^2w^2f^2}\,,\\
        C_2&=\frac{1+f\rho^2}{r^2f}\,,\\
        C_3&=\frac{2}{rw}(w^2-1)\,,\\
        C_4&=\frac{2\rho}{r}\,,\\
        C_5&=\frac{2\rho}{r^2wf}\left(1+(w^2-1)f\right)\,,\\
        C_6&=\frac{1}{f}\dv{f}{r}+\frac{w^2(10-4\Delta)+8\Delta}{2rw^2}+\frac{d-4}{r}\,,\\
        C_7&=\left(\frac{1}{rwf}\dv{f}{r}-\frac{w^2(2\Delta-5)-4\Delta-1}{r^2w^3}\right)\left(w^2-1\right)\nonumber\\
        &\phantom{=}+\frac{3w^2-\rho^2(1+4\Delta)}{r^2w^3f}+\frac{1+\rho^2+4\left(1-w^2+\rho^2\right)\Delta}{r^2w^3f^2}+(d-4)\frac{C_5}{2\rho}\,,\\
        C_8&=\frac{2(w^2-2\rho^2\Delta)+\rho^2\left(w^2(5-2\Delta)+4\Delta\right)f}{r^2w^2\rho f}+\frac{\rho}{rf}\dv{f}{r}+(d-4)\frac{C_2}{\rho}\,,\\
        C_9&=\frac{\Delta}{w^2}\Bigg(\frac{(w^2-2)^2\Delta+4(1+w^2-w^4)}{r^2w^2}+\frac{4\rho^2(\Delta+1)-w^4(\Delta-4)-6w^2}{r^2w^2f}\nonumber\\
        &\phantom{=}+\frac{2w^2(1+2\Delta)-4(1+\rho^2)(1+\Delta)}{r^2w^2f^2}-\frac{w^2-2}{rf}\dv{f}{r}-(d-4)\frac{(w^2-2)(f-1)}{r^2f}\Bigg)\ ,
    \end{align}}}}where
$f=1-\frac{\mu}{r^d}$.

%%%%%%%%%%%%%%%%%%%%%%%%%%%%%%%%%%%%%%%%%%%%%%%%%%%%%
%%%%%%%%%%%%%%%%%%%%%%%%%%%%%%%%%%%%%%%%%%%%%%%%%%%%%

\chapter{Approximating OPE by Integral}\label{a.sumz}

In Chapter \ref{ch3} we have approximated the OPE sums by an integral, such as in \eqref{e.sin1}. In this appendix we now discuss the validity of this approximation.

We want to resum the OPE \eqref{e.resumL} where the coefficients are given by their asymptotic form \eqref{eq:Lambdand=4Ansatz}. Assume that this asymptotic form is a good approximation to the actual OPE coefficients after a certain value $n = n_*$ and that one can neglect terms with $n<n_*$. The first assumption comes from our analysis of OPE data. We can justify the second assumption by noting that 
near the critical points $\tau \approx \tau_c$, which  are at the radius of convergence of the stress-tensor OPE, all terms in the $\tau$ expansion contribute. 
One sees from \eqref{eq:Lambdand=4Ansatz} that for large enough $\Delta$ the OPE coefficients are increasing with $n$ and thus close enough to the critical points, the large-$n$ terms will be the most important. All in all, to calculate the stress-tensor contribution to the correlator, one has to evaluate a sum of the type
    \begin{equation}\label{e.vzorovasume}
        \sum_{n=n_*}^\infty n^{a\Delta+b}y^{d n}=y^{d\,n_*}\,\Phi\left(y^d,-a\Delta-b,n_*\right)\ ,
    \end{equation}
where  $\Phi$ is the Hurwitz-Lerch transcendent, $\abs{y}\leq1$, and $a>0$ and $b$ are constants. We want to compute this sum in the limit $y\rightarrow1$, which corresponds to the correlator near $\tau \approx \tau_c$. For $\Delta>-\frac{a+b}{2}$, one can expand
    \begin{equation}
    \begin{split}
        &y^{d\,n_*}\,\Phi(y^d,-a\Delta-b,n_*)=\\
        &\hspace{1.3cm}\Gamma(1+a\Delta+b)(-\log y^d)^{-(1+a\Delta+b)}+\sum_{k=0}^\infty\zeta(-a\Delta-b-k,n_*)\frac{\left(\log y^d\right)^k}{k!}\ ,
    \end{split}
    \end{equation}
where $\zeta$ is the generalised Riemann zeta function. 
The second term is regular at $y=1$ for all $k$, so it gives a subleading contribution to the correlator near the critical point.
Therefore, near $\tau \approx \tau_c$ we can approximate the sum \eqref{e.vzorovasume} by
    \begin{equation}
    \sum_{n=n_*}^\infty n^{a\Delta+b}y^{dn}\approx\Gamma(1+a\Delta+b)(-\log y^d)^{-(1+a\Delta+b)}=\int_0^\infty n^{a\Delta+b}y^{dn}\,dn\ .        
    \end{equation}
This justifies the exchange of the sum for an integral from 0 to $\infty$.

%%%%%%%%%%%%%%%%%%%%%%%%%%%%%%%%%%%%%%%%%%%%%%%%%%%%%
%%%%%%%%%%%%%%%%%%%%%%%%%%%%%%%%%%%%%%%%%%%%%%%%%%%%%

\chapter{The KMS pole and OPE in \texorpdfstring{$d=2$}{d=2} }\label{s.kms2d}

In this appendix we show how the KMS pole emerges from the OPE in $d=2$. The finite temperature two-point function is known in a closed form and  the KMS pole can be seen explicitly without the use of the OPE.
Nevertheless, the recovery of the KMS pole from the OPE can serve as a guideline for the analysis in higher dimensional cases, where the thermal two-point functions are not known exactly.

\section{KMS pole}

Consider a scalar two-point function at finite temperature $T=\beta^{-1}$ 
    \begin{equation}\label{e.def}
        G(\tau,x)=\expval{\phi(\tau,x)\phi(0,0)}_\beta\ .
    \end{equation}
This can be rewritten, using the periodicity of the trace, as
    \begin{equation}
    \begin{split}
    \label{e.kmsc}
G(\tau,x)=\frac{1}{Z}\Tr e^{-\beta H}\phi(\tau,x)\phi(0,0)
=\frac{1}{Z}\Tr e^{-\beta H}\phi(\beta,0)\phi(\tau,x)=G(\beta-\tau,-x)\ ,
    \end{split}
    \end{equation}
which is the KMS condition \cite{Kubo:1957mj,Martin:1959jp}. Let us consider the case where $x=0$. For a unit-normalised scalar with scaling dimension $\Delta$, the small $\tau$ behaviour is
    \begin{equation}
G(\tau)\xrightarrow{\tau\rightarrow0}\frac{1}{\abs{\tau}^{2\Delta}}\ .
    \end{equation}
From \eqref{e.kmsc}, it then follows that $G$ also has a \textit{KMS pole} at
    \begin{equation}\label{e.pole}
      G(\tau) \xrightarrow{\tau\rightarrow\beta}\frac{1}{\abs{\beta-\tau}^{2\Delta}}\ .
    \end{equation}
This analysis is valid for any quantum field theory at finite temperature. 
When dealing with CFTs one has the additional tool of the OPE with a non-vanishing radius of convergence, which is typically determined by the first singular point encountered in the complex plane.
Thus, if there are no singularities closer to the origin, the KMS pole should be encoded in the asymptotic behaviour of the OPE.

\section{OPE analysis}

Let us consider a CFT in two dimensions. 
Since $S_\beta\times\mathbb{R}$ is conformally equivalent to $\mathbb{R}^2$, the two-point correlation function  is known in a closed form \cite{Iliesiu:2018fao}, and in the case of two identical scalar operators it is given by
    \begin{equation}\label{e.expl}
G(\tau, x)=\left[\frac{\beta}{\pi}\sinh\left(\frac{\pi(x-i\tau)}{\beta}\right)\right]^{-\Delta}\left[\frac{\beta}{\pi} \sinh\left(\frac{\pi(x+i\tau)}{\beta}\right)\right]^{-\Delta}\ ,
    \end{equation}
where we have chosen the normalisation such that in the zero-temperature limit we recover unit norm, $i.e.$ $\lim_{\beta\rightarrow\infty}G(\tau,x)=(\tau^2+x^2)^{-\Delta}$.

\begin{figure}
\includegraphics[width=0.8\textwidth]{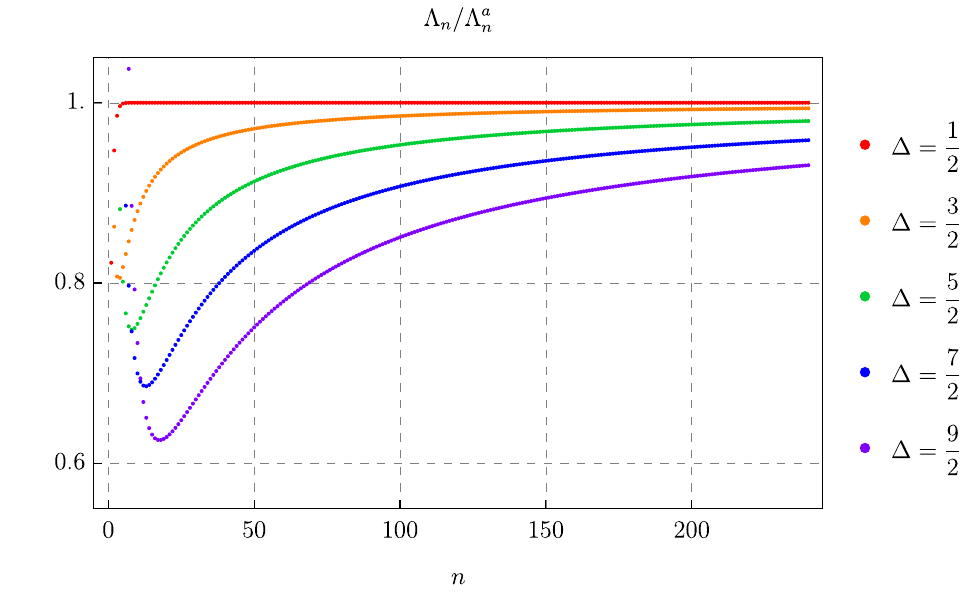}
	\centering
    \caption{Ratio of the explicit results for $\Lambda_n$ to the leading large-$n$ prediction $\Lambda_n^a$, given in \eqref{eq:AsyFormd2}, for different values of $\Delta$ in $d=2$.}
    \label{fig.Lambdand=2}
\end{figure}

Let us first set $x=0$. In a two-dimensional CFT, only the Virasoro vacuum module contributes to the thermal correlator. In other words,  the only non-zero contribution comes from the multi-stress tensors and thus the full correlator can be expanded as
    \begin{equation}\label{e.sum}
G(\tau)=\frac{1}{\tau^{2\Delta}}\sum_n\Lambda_n\left(\frac{\tau}{\beta}\right)^{2n}\ .
    \end{equation}
The coefficients $\Lambda_n$ can simply be read off from the  expansion of the correlator \eqref{e.expl} near $\tau =0$. We are interested in the behaviour of the OPE near its convergence radius, where we expect all terms in the expansion to be of similar magnitude. As such, we are interested in the behaviour of $\Lambda_n$ for large values of $n$. One finds that for large enough $n$ these can be written in a $1/n$ expansion (see figure~\ref{fig.Lambdand=2})
\begin{equation}\label{eq:LambdaNd=2}
    \begin{split}
    \Lambda_n &= \frac{2^{2\Delta}}{\Gamma(2\Delta)}\,n^{2\Delta-1}\sum_{k=0}^\infty\frac{c_k(\Delta)}{n^k}
    = \frac{2^{2\Delta}}{\Gamma(2\Delta)}\,n^{2\Delta-1}\left(1+ \frac{c_1(\Delta)}{n} + \ldots  \right)\,,
    \end{split}
\end{equation}
where we choose the overall prefactor in such a way that $c_0(\Delta) = 1$, while all remaining  $c_k(\Delta)$  can be arbitrary functions of $\Delta$.
We see that as $n \to \infty$ the coefficients tend to 
\begin{align}
    \label{eq:AsyFormd2}
    \Lambda_n^a &= \frac{2^{2\Delta}}{\Gamma(2\Delta)}\,n^{2\Delta-1}\,,
\end{align}
which determines the leading order behaviour of the correlator.
For $\tau$ near the radius of convergence of the OPE \eqref{e.sum}, we can approximate the sum by an integral
\begin{equation}\label{eq.Corr2dLead}
\begin{split}
    G(\tau) &\approx  \frac{1}{\tau^{2\Delta}}\int_0^\infty \Lambda_n^a\,\left(\frac{\tau}{\beta}\right)^{2n} \,d n
    =\left[-\frac{\tau}{2}\log\left(\frac{\tau^2}{\beta^2}\right)\right]^{-2\Delta}\,,
\end{split}
\end{equation}
where we have taken \eqref{eq:AsyFormd2} for the OPE coefficients.
We encounter a singularity when the argument of the logarithm is equal to 1, which happens at $\tau = \pm \beta$ where we find
    \begin{equation}
        G(\tau) \xrightarrow{\tau\rightarrow\pm \beta}\frac{1}{|\beta-\tau|^{2\Delta}}\,,
    \end{equation}
which are exactly the KMS poles \eqref{e.pole}.
This explicitly shows that the asymptotic OPE analysis reproduces the first non-trivial poles of the full correlator \eqref{e.expl} in the complex-$\tau$ plane.
However, as may be expected, the OPE analysis does not contain any information about the higher order poles  at $|\tau| > \beta $.

\paragraph{Subleading analysis:}

\begin{figure}[t]
    \centering
    \includegraphics[width=\textwidth]{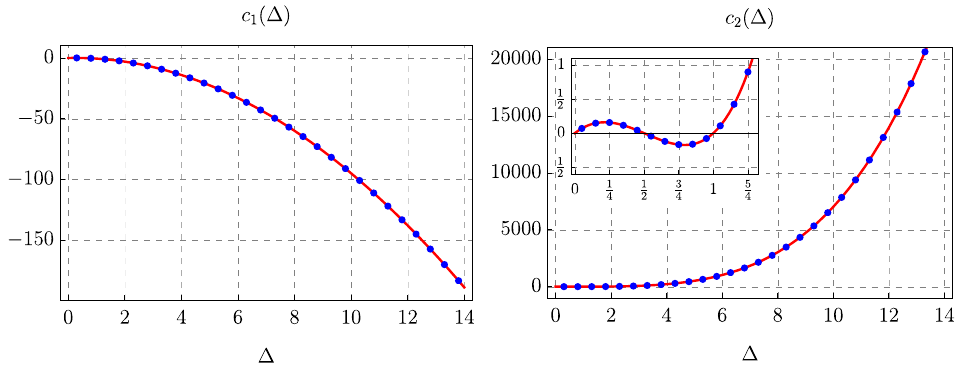}
    \vspace*{-0.6cm}
    \caption{The comparison between $c_1(\Delta)$ (left) and $c_2(\Delta)$ (right) as obtained from expansion of the exact correlator (red curve) and the results from the direct analysis of $\Lambda_n$ for large $n$ (blue markers).}
    \label{fig:c1c22d}
\end{figure}

The above analysis shows that KMS pole is already contained in $\Lambda_n^a$, the leading behaviour of the OPE coefficients at large $n$. 
Let us now analyse the $1/n$ corrections in \eqref{eq:LambdaNd=2}.
The coefficients $c_k(\Delta)$ are determined by carefully analysing $\Lambda_n$ as a function of $\Delta$,\footnote{We discuss how to determine the form of \eqref{eq:LambdaNd=2} and the values $c_k(\Delta)$ from $\Lambda_n$ in more detail in Appendix~\ref{app:DeltaTau}, where we analyse  the correlator in $d=4$.} see blue markers in Figure\ \ref{fig:c1c22d}. 

Alternatively, one can follow a different approach to obtain coefficients $c_k(\Delta)$:
Insert the full expansion \eqref{eq:LambdaNd=2} into the sum \eqref{e.sum} and again approximate it with an integral
\begin{equation}\label{eq.Gd2SubDeltaTauInt}
    \begin{split}
    G(\tau) =& \frac{2^{2\Delta}}{\Gamma(2\Delta)\, \tau^{2\Delta}}\int_0^\infty \!n^{2\Delta-1}\left(\frac{\tau}{\beta}\right)^{2n}\left(1+ \frac{c_1(\Delta)}{n} +\frac{c_2(\Delta)}{n^2}+ \ldots  \right)d n\\
    =& \left[-\tau\log\left(\frac{|\tau|}{\beta}\right)\right]^{-2\Delta}\,\Bigg[1+ \frac{2\,c_1(\Delta)}{2\Delta-1}\,\left(-\log\left(\frac{|\tau|}{\beta}\right)\right) \\
    &\hspace{2.4cm}+ \frac{4\,c_2(\Delta)}{(2\Delta-1)(2\Delta-2)}\,\left(-\log\left(\frac{|\tau|}{\beta}\right)^2\right)+ \ldots\Bigg]\,.
    \end{split}
\end{equation}
Expanding this result around%
\footnote{One can equally expand around $\tau = - \beta$, but we will focus on this pole without the loss of generality.}
$\tau = \beta$ then gives
\begin{align}
    \label{eq:Gtaud2Exp}
    G(\tau) &\xrightarrow{\tau\rightarrow\beta}\frac{1}{(\beta-\tau)^{2\Delta}}\Bigg[1 + \frac{1}{\beta}\left(\Delta + \frac{2\,c_1(\Delta)}{2\Delta-1}\right) (\beta -\tau)\\
    &+ \frac{1}{\beta^2}\left(\frac{\Delta(6 \Delta +7)}{12}+ \frac{(2\Delta+1)c_1(\Delta)}{2\Delta-1}+ \frac{2\,c_2(\Delta)}{(2\Delta-1)(\Delta-1)}\right) (\beta -\tau)^2 + \ldots \Bigg]\nonumber\,.
\end{align}
We thus see that the $1/n$ corrections in \eqref{eq:LambdaNd=2} translate to $(\beta-\tau)$ corrections to the correlator near the KMS pole.

We can now use the exact form of the correlator \eqref{e.expl} to  determine the coefficients $c_k(\Delta)$.
Expand \eqref{e.expl} around $\tau = \beta$ to first subleading order
\begin{align}
    \label{eq.G2dExpSubTau}
    G(\tau) = \frac{1}{(\beta - \tau)^{2\Delta}}\left(1+ \frac{\Delta\,\pi^2}{3\,\beta^2}\,(\beta - \tau)^2+ \coo{(\beta-\tau)^3}\right)\,.
\end{align}
This expression and \eqref{eq:Gtaud2Exp} should match, which leads to
\begin{align}
    \label{eq:2dCkPred}
    c_1(\Delta) =\Delta\,\left(\frac12-\Delta\right)\,,\qquad
    c_2(\Delta) = \frac{1}{24}\Delta\,(\Delta-1)(2\Delta-1)(6\Delta + 4\,\pi^2 -1)\,,
\end{align}
which can be continued to arbitrary $k$.
One can compare these expressions with the direct data obtained from analysing the asymptotic form of $\Lambda_n$ and find perfect agreement, see Figure~\ref{fig:c1c22d}.

\paragraph{Nonzero $x$:}
Let us now set $x \neq 0$, so that $x\ll \tau$. 
Essentially, we  work in the limit where $x$ is the smallest length scale in the expression.
Expanding the exact correlator \eqref{e.expl} in small $x$ gives to leading order
\begin{align}
\label{eq:GtxExp}
    G(\tau, x) = \left[\frac{\beta}{\pi}\sin\left(\frac{\pi\,\tau}{\beta}\right)\right]^{-2\Delta} + \Delta\,x^2\,\left[\frac{\beta}{\pi}\sin\left(\frac{\pi\,\tau}{\beta}\right)\right]^{-2(\Delta+1)}+ \coo{x^4}\,,
\end{align}
where each term contains non-trivial poles at $\tau = m\,\beta$, with $m\in \mathbb{Z}$.
We expand each term in the above series individually around $\tau = \beta$, focusing only on the leading behaviour 
\begin{align}
\label{eq:d2SmallxPoles}
    G(\tau, x) \approx \left(\beta - \tau\right)^{-2\Delta}\left[ 1 + \frac{1}{(\beta-\tau)^2}\,\frac{\Delta\,x^2}{\beta^2}\right]\,.
\end{align}
Naively, one would think  that the $x^2$ term  is more divergent than the leading order term.
However, this expansion is only valid if $x$ is the smallest parameter in the expression, $i.e.$  $x^2\ll (\beta -\tau)^2$, in which case all corrections to the leading divergent behaviour in \eqref{eq:d2SmallxPoles} are small. 

Similar to the analysis of the subleading $(\beta-\tau)$ contributions we would like to reproduce this result using the OPE analysis. 
For that we take \eqref{eq:GtxExp} and expand each term individually in a series in $\tau$
\begin{align}
\label{eq:GtauxExp2}
    G(\tau, x) = \frac{1}{\tau^{2\Delta}}\left[\sum_{n=0}^\infty \Lambda_n^{(0)}\,\left(\frac{\tau}{\beta}\right)^{2n} + \frac{\Delta \,x^2}{\beta^2} \sum_{n=0}^\infty \Lambda_n^{(1)}\,\left(\frac{\tau}{\beta}\right)^{2n} + \coo{x^4} \right]\,.
\end{align}
Again, we can make an ansatz for asymptotic behaviour of the OPE coefficients
\begin{align}
    \Lambda_n^{(\alpha)} = c_0^{(\alpha)}(\Delta)\, n^{a\, \Delta +b}\,,
\end{align}
where we ignore all $1/n$ corrections, since we are only interested in the leading behaviour at each order in $x$. We find 
\begin{align}
    \Lambda_n^{(0)} \approx \frac{2^{2\Delta}}{\Gamma(2\Delta)}\,n^{2\Delta-1}\,,\qquad \Lambda_n^{(1)} \approx \frac{2^{2(\Delta+1)}}{\Gamma(2(\Delta+1))}\,n^{2\Delta+1}\,.
\end{align}
Inserting these values into \eqref{eq:GtauxExp2} and replacing the sum with the integral leads to 
\begin{align}
    G(\tau,x) &\approx \frac{1}{\tau^{2\Delta}}\int_0^\infty\left[\frac{2^{2\Delta}}{\Gamma(2\Delta)}\,n^{2\Delta-1}+\frac{\Delta\,x^2}{\beta^2} \,\frac{2^{2(\Delta+1)}}{\Gamma(2(\Delta+1))}\,n^{2\Delta+1}\right]\left(\frac{\tau}{\beta}\right)^{2n}d n\,\nonumber\\*
    &= \frac{1}{\tau^{2\Delta}}
\left[\left(-\log\frac{|\tau|}{\beta}\right)^{-2\Delta} +\frac{\Delta\,x^2}{\beta^2} \,\left(-\log\frac{|\tau|}{\beta}\right)^{-2(\Delta+1)}\right]\,.
\end{align}
By expanding this near $\tau=\beta$ we exactly reproduce \eqref{eq:d2SmallxPoles} -- the asymptotic OPE analysis reproduces the KMS poles of the exact two-point correlation function for $x\neq0$.

Let us conclude this appendix by noting that in the main text and in the appendices that follow  we repeat this analysis for thermal correlators in higher dimensions. 
In these cases we can still extract $c_k(\Delta)$ (respectively $\Lambda_n^{(1)}$) by careful analysis of the OPE coefficients. However, since a closed form expression of the thermal correlator is not known, to cross-check these results we compare them with the geodesic analysis at subleading $\delta\tau$ (respectively $x\neq0$).

%%%%%%%%%%%%%%%%%%%%%%%%%%%%%%%%%%%%%%%%%%%%%%%%%%%%%
%%%%%%%%%%%%%%%%%%%%%%%%%%%%%%%%%%%%%%%%%%%%%%%%%%%%%

\chapter{Subleading \texorpdfstring{$\delta\tau$}{delta tau} Analysis}
\label{app:DeltaTau}

In this appendix we analyse the contributions to $\mathcal{L}_T(\tau)$ subleading in $\delta \tau$ and how these arise as $1/n$ corrections to the asymptotic form of the OPE coefficients. This analysis is valid for any finite $\Delta$.
When $\Delta\rightarrow\infty$ the situation becomes more subtle. As we argued in Section\ \ref{s.abintldl}, in this limit, the cross-over point also goes to infinity, $n^*= \Delta/2\to \infty$, which  leads to disappearance of the bouncing singularity in $\mathcal{L}_T(\tau)$.  However, as we saw in \eqref{eq:ProperLengthNearLim},  taking $\Delta$ large \emph{after} expanding the logarithm of the correlator near the bouncing singularity leads to the behaviour expected from the bouncing geodesic. Below we examine if this match persists beyond the leading order singularity.
We also detail the procedure that we used in Chapter \ref{ch3} to determine the asymptotic form of the OPE coefficients $\Lambda_n$.

The main object of interest are the OPE coefficients $\Lambda_n$, as defined in \eqref{e.resumL}. In particular, we are interested in their behaviour when $n$ is large. 
Ideally, one would find the exact expressions of these OPE coefficients as functions of $\Delta$, however, in practice, finding such expressions for large $n$ is computationally too expensive. It is  more efficient to first fix $\Delta$ and then calculate $\Lambda_n$ for that specific value. In this way, we are able to calculate $\Lambda_n$ up to $n \approx 50$ in about 5 days on a standard desktop machine.

\begin{figure}[t]
     \centering
    \includegraphics[width=\textwidth]{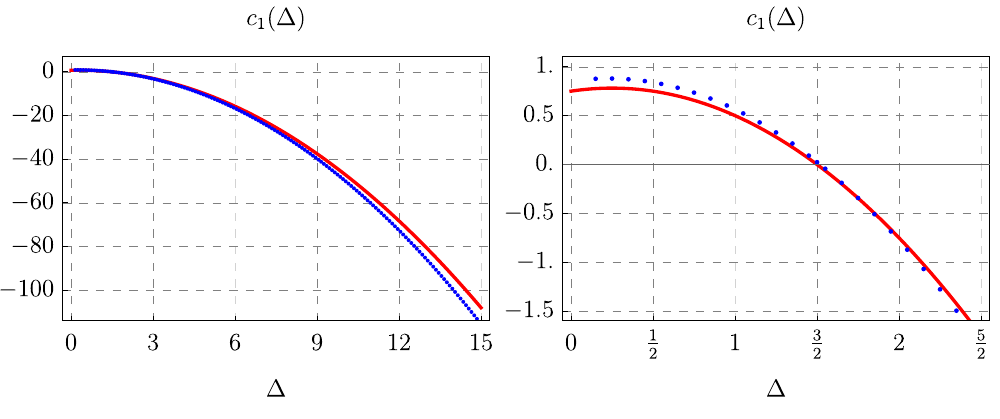}
    \vspace*{-0.7cm}
    \caption{The numerical values for $c_1(\Delta)$ (blue) compared with the data predicted by the geodesic analysis (red). On the right, the close-up shows that even at small $\Delta$ the geodesic results seems to match the data to a relatively high degree. This suggests that even in the full correlator, there is no term linear in $\delta\tau=\tau_c-\tau$.}
    \label{fig:c1s}
\end{figure}

\begin{figure}[t]
    \centering
    \includegraphics[width=\textwidth]{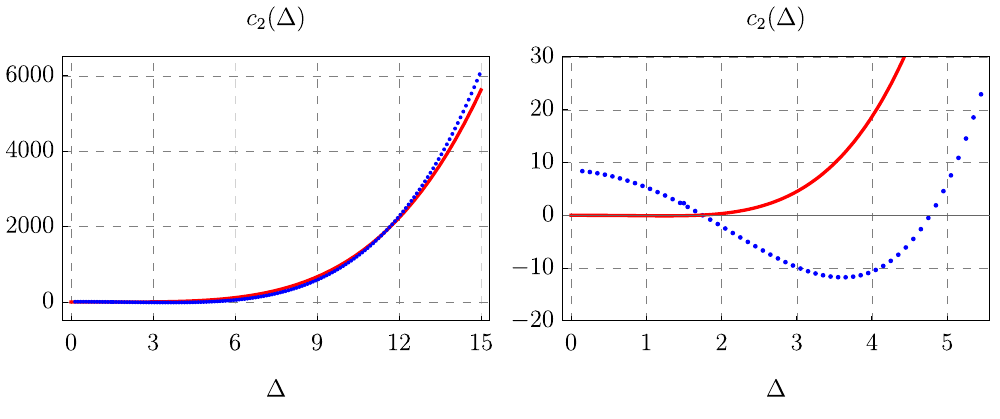}
        \vspace*{-0.7cm}
    \caption{The numerical values for $c_2(\Delta)$ (blue) compared with the data predicted by the geodesic analysis (red). On the right, the close-up shows that at small $\Delta$ the geodesic prediction disagrees with the numerical data. This is expected, since there is no reason for the geodesic result to be applicable outside the regime where $\Delta\gg 1$.}
    \label{fig:c2s}
\end{figure}

We find that for large enough $n$ the OPE coefficients can be described by
\begin{equation}\label{eq:rybana}
    \Lambda_n = c(\Delta)\frac{n^{2\Delta-3}}{\left(\frac{1}{\sqrt{2}}\right)^{4n}e^{i\pi \,n}}\,\sum_{k=0}^\infty\frac{c_k(\Delta)}{n^k}\ ,
\end{equation}
where we choose $c_0(\Delta)=1$, so that this expression is in accord with the dominant contribution \eqref{eq:Lambdand=4Ansatz} used in the main text. 
Let us here briefly explain how we obtained this expression. 
First, one draws inspiration from the two-dimensional analysis \eqref{eq:LambdaNd=2} and considers an ansatz for the dominant contribution 
\begin{equation}
\Lambda_n^a = e^{-i\pi\,n} \widetilde k^n\,c(\Delta)n^{a\,\Delta+b}\ .
\end{equation}
The first factor comes from the observed oscillating sign of $\Lambda_n$, while $\widetilde k$, $a$, and $b$ are constants and $c(\Delta)$ is a function which all need to be determined by our analysis. The values of the three constants can be determined by analysing different ratios of $\Lambda_n$ as $n$ and $\Delta$ are varied. The expressions for $c(\Delta)$ and $c_k(\Delta)$ are then obtained numerically by analysing $\Lambda_n$  as $n$ is varied for fixed values of $\Delta$. One first assumes the $1/n$ expansion \eqref{eq:rybana} and compares it with the values of $\Lambda_n$ as a function of $n$. This allows us to read off the coefficients $c_k(\Delta)$. This procedure gives, for example, the values given in the blue markers in Figure~\ref{fig:c(Delta)}, Figure~\ref{fig:c1s}, and Figure~\ref{fig:c2s}. 

Let us note that the accuracy of our analysis is limited by the number of $\Lambda_n$ we can calculate: The larger the $n$, the more accurate the asymptotic form \eqref{eq:rybana} will be. A higher $n$ also means that we can include, in practice, a higher number of $1/n$ corrections, which in turn allow for a more accurate values of $c_k(\Delta)$. An estimate of the value of $n$ necessary for an accurate determination of $\Lambda_n$ for a given $\Delta$ is given at the end of this appendix.

The $1/n$ terms in \eqref{eq:rybana} are mapped to the ($\tau_c - \tau$) corrections of the correlator. To see this, we insert the full $1/n$ expansion of $\Lambda_n$ into \eqref{e.resumL} and replace the sum with an integral
\begin{align}
    \label{eq:CorrFullSum}
    G_T(\tau)&\approx
    \frac{1}{\tau^{2\Delta}}\int_0^\infty\Lambda_n\left(\frac{\tau}{\beta}\right)^{4n}d n =
    \frac{c(\Delta)}{\tau^{2\Delta}}\int_0^\infty n^{2\Delta-3}\left(\frac{\tau}{\tau_c}\right)^{4n}\sum_{k=0}^\infty\frac{c_k(\Delta)}{n^k}\,d n\nonumber\\
    &= \frac{c(\Delta)}{\tau^{2\Delta}}\sum_{k=0}^\infty c_k(\Delta)\,\Gamma(2\Delta-2-k)\,\left[-\log\left(\frac{\tau^4}{\tau_c^4}\right)\right]^{-(2\Delta-2-k)}\,,
\end{align}
where $\tau_c$ is schematically one of the critical points defined in \eqref{e.poles}.
We see that near $\tau\approx \tau_c$ we get a sum of diverging terms with a pole of order $2\Delta-2-k$: The higher the $k$, the milder the singularity.

In four dimensions, we lack an exact expression for the correlator (or the stress-tensor sector of the correlator), so we cannot determine $c_k(\Delta)$ in a similar manner as in the two-dimensional case. 
However, we can take the large-$\Delta$ limit and compare these results with the expectation from the geodesic analysis. In particular, we know that the correlator obtained from the bouncing geodesic expanded around the lightcone singularity receives the first correction to the leading result at fourth order \eqref{eq:GeoCOrrSmallDeltaTauExp}.
Therefore, we  insert \eqref{eq:CorrFullSum} into \eqref{eq:LogOfCorr} and  expand in  $\delta\tau = \tau_c-\tau$
\begin{align}
    \label{eq:LogCorrNearPoleExp}
     \cL_T \approx -\frac{1}{\Delta}\log\left[\frac{c(\Delta)\,\Gamma(2\Delta-2)}{4^{(2\Delta-2)}}\,\frac{1} {\tau_c^{2}}\right]+ \frac{2\Delta-2}{\Delta}\log\delta\tau+\sum_{k=1}^{\infty}\gamma_k(\Delta)\left(\frac{\delta\tau}{\tau_c}\right)^k\,,
\end{align}
where $\gamma_k(\Delta)$ are non-trivial combinations of $c_k(\Delta)$.
Let us check if the $\Delta\to \infty$ limit reproduces the geodesic result \eqref{eq:PropLenDeltat}, which means that 
\begin{equation}\label{e.geodude}
    \gamma_1(\Delta) = 0\,, \qquad  \gamma_2(\Delta) = 0\,,\qquad  \gamma_3(\Delta) = 0\,,\qquad \gamma_4(\Delta) = \frac{\pi^4}{160}\,,\quad \ldots\,,
\end{equation}
up to $1/\Delta$ corrections. We only focus on $\gamma_1(\Delta)$ and $\gamma_2(\Delta)$. The former is given by
\begin{align}
\label{eq:Gamma1}
    \gamma_1(\Delta) = 1+ \frac{1}{\Delta}+ \frac{4\,c_1(\Delta)}{\Delta(2\Delta-3)}\,,
\end{align}
which can be expanded in large-$\Delta$ to first few orders
\begin{align}
\label{eq:Gamma1Exp}
    \gamma_1(\Delta) = 1 + \frac{1}{\Delta} + \frac{2\,c_1(\Delta)}{\Delta^2} + \frac{3\,c_1(\Delta)}{\Delta^3}+\coo{\Delta^{-4}}\,.
\end{align}
For this expression to be consistent with the bouncing geodesic, the leading behaviour of $c_1(\Delta)$ has to be given by 
\begin{equation}
\label{eq:c1exp}
    c_1(\Delta) \sim -\frac{\Delta^2}{2} + \ldots\,,
\end{equation}
which cancels out the leading order term in \eqref{eq:Gamma1Exp}. However, we can be slightly bolder and assume that $\gamma_1(\Delta) =0$ for all $\Delta$, which ensures absence of a linear term in the proper length and the correlator. In this case, we can solve \eqref{eq:Gamma1} directly and find
\begin{align}
\label{eq:c1}
    c_1(\Delta) &= -\frac{1}{4} (\Delta +1) (2 \Delta -3)\,,
\end{align}
whose large $\Delta$ behaviour agrees with \eqref{eq:c1exp}. In Figure~\ref{fig:c1s}, we plot the numerical data (in blue) against this prediction (in red). We find good agreement not only at large $\Delta$, but for all values which strongly suggests that $c_1(\Delta)$ is given by \eqref{eq:c1}.
The expression for $\gamma_2(\Delta)$ is more complicated
\begin{subequations}
    \begin{align}
        \gamma_2(\Delta) &= -\frac{7}{12} - \frac{5}{12\,\Delta} - \frac{2\,c_1(\Delta)}{\Delta(2\Delta-3)}+ \frac{8\,c_1(\Delta)^2}{\Delta(2\Delta-3)^2}- \frac{8\,c_2(\Delta)}{\Delta(6-7\Delta+2\Delta^2)}\label{eq:Gamma2}\\
        & \approx -\frac{7}{12} - \frac{5}{12\,\Delta} - \frac{c_1(\Delta)}{\Delta^2} -\frac{4\,c_2(\Delta)- 2c_1(\Delta)^2 +\frac32\,c_1(\Delta)}{\Delta^3}+ \coo{\Delta^{-4}}\,.
        \label{eq:Gamma2exp}
    \end{align}
\end{subequations}
Above we have not yet inserted the value of $c_1(\Delta)$ because it will help us illustrate an important point as to why higher $c_k(\Delta)$ are inaccessible. As seen in \eqref{eq:c1exp}, even  in the most agnostic estimate $c_1(\Delta)$ scales at most quadratically at large values of $\Delta$. Therefore, the $c_1(\Delta)^2$ factor in the fourth term will contribute at order $\Delta$ in \eqref{eq:Gamma2exp}.%
\footnote{We see from \eqref{eq:Gamma2} that $c_1(\Delta)$ appears at most quadratically in $\gamma_2$ and thus no higher order term in the $1/\Delta$ expansion will contribute at order $\Delta$ when the value of $c_1(\Delta)$ is inserted.} Such term is incompatible with a smooth large-$\Delta$ limit, independent of whether the geodesic is the bouncing geodesic or a non-singular complex geodesic.
It therefore needs to be cancelled by the leading order term in $c_2(\Delta)$, which in this case gives
\begin{align}
    c_2(\Delta) \sim \frac{\Delta^4}{8}+ \ldots.
\end{align}
It needs to be stressed that this behaviour does not contain any information about whether the correlator is given by the bouncing geodesics or by the combination of two complex geodesics. This is contained in the subleading behaviour of $c_2(\Delta)$, which can be determined uniquely only if one knows the subleading behaviour of $c_1(\Delta)$.
To illustrate this, let us introduce an expansion
    \begin{equation}
        c_1(\Delta) = c_1^{(2)}\,\Delta^2 + c_1^{(1)}\,\Delta + c_1^{(0)} + \ldots\,,\quad 
        c_2(\Delta) = c_2^{(4)}\,\Delta^4 + c_2^{(3)}\,\Delta^3 + c_2^{(2)}\,\Delta^2 + \ldots\,,
    \end{equation}
and insert it into \eqref{eq:Gamma2exp}
\begin{align}
    \gamma_2(\Delta) &= 2\Delta\left[\left(c_1^{(2)}\right)^2-2\,c_2^{(4)}\right] \\*
    &\quad + \frac{1}{12}\left[-7 - 12\,c_1^{(2)}+ 48 \,c_1^{(1)}\,c_1^{(2)}+ 72\,\left(c_1^{(2)}- 48 c_2^{(3)}- 168 c_2^{(4)}\right)^2\right] + \coo{\frac1{\Delta}}\,.\nonumber
\end{align}
The term that scales with $\Delta$ can be determined as explained above. But the term that does not scale with $\Delta$, and gives a finite contribution in the large-$\Delta$ limit, depends on the subleading terms of both $c_1(\Delta)$ and $c_2(\Delta)$. Since we believe that $c_1(\Delta)$ can be determined exactly \eqref{eq:c1}, one can determine $c_2^{(3)}$ and write
\begin{align}
    c_2(\Delta) = \frac{\Delta^4}{8}- \frac{5\,\Delta^3}{24} + \coo{\Delta^2}\,,
\end{align}
but other, lower order terms are out of reach, since they are not constrained by the geodesic result. We can compare this result with the data obtained from solving the bulk equations of motion, which is pictured in Figure~\ref{fig:c2s}. We see that while the asymptotic behaviours match, we cannot say anything about the behaviour of $c_2(\Delta)$ for small values of $\Delta$. In fact, one can directly solve \eqref{eq:Gamma2} and determine $c_2(\Delta)$ using \eqref{eq:c1} in the process. Even in this case one find significant deviation from the data obtained from the holographic calculation.

Of course, the  geodesic analysis is expected to hold only at large values of the conformal dimensions. We observed this already in Section~\ref{ss:NCLim}, where the geodesic analysis failed to reproduce all $1/\Delta$ corrections at higher orders of the $\tau =0$ OPE. However, $c_1(\Delta)$ predicted from the analysis of the bouncing geodesic is in good agreement with the holographic data for all values of $\Delta$. This suggests that the linear correction to the singularity, $\gamma_1(\Delta)$, vanishes for all conformal dimensions and not just in the large-$\Delta$ limit. This is generically not true for all other $\gamma_k(\Delta)$, starting with $\gamma_2(\Delta)$, and suggest that for finite values of $\Delta$, the contribution to the correlator from the stress-tensor sector near the bouncing singularity goes as
    \begin{align}
    \label{eq:CorrNearPoleExp}
    G_T(\tau \approx \tau_c) \sim \frac{c(\Delta)\,\Gamma(2\Delta-2)}{2^{2(2\Delta-2)}}\,\frac{1} {\tau_c^{2}}\frac{1}{\left(\tau_c-\tau\right)^{2\Delta-2}}\left[1+ \gamma_2(\Delta)\left(\frac{\tau_c-\tau}{\tau_c}\right)^2+\ldots\right]\,,
\end{align}
where $\gamma_2(\Delta)$ is some non-trivial function of $\Delta$.

Higher order $c_k(\Delta)$ cannot be uniquely determined because we are unable to determine $c_2(\Delta)$ to all orders -- the reasoning is the same as with the subleading order in $c_2(\Delta)$ only being accessible if we know $c_1(\Delta)$ to subleading order.
One is only  able to determine that the leading behaviour at large $\Delta$ is
\begin{align}
\label{eq:ckLargeDelta}
    c_k(\Delta)\sim \frac{(-1)^k\,\Delta^{2k}}{2^{k}\,k!}\,.
\end{align}
One can show that this behaviour is compatible with the holographic data obtained from solving the bulk equations of motion.
However, as already discussed above, \eqref{eq:ckLargeDelta} does not  contain any information about the specific geodesic that contributes to the correlator -- this information is encoded in the subleading terms which we are unable to access with out current precision.

The large-$\Delta$ behaviour \eqref{eq:ckLargeDelta} serves as a useful rough estimate for the amount of terms in the $1/n$ expansion of $\Lambda_n$ that give a considerable contribution at a fixed value of $\Delta$. Let us assume that we can determine the $\Lambda_n$ coefficients up to some number $n_{\rm max}$. At this point, the $1/n$ terms in \eqref{eq:rybana} scale roughly as $\Delta^2/n_{\rm max}$. For $n_{\rm max} \lesssim \Delta^2$ the apparent $1/n$ expansion will not look convergent, since successive terms in the expansion will increase. In other words, the analysis can only be trusted for $\Delta\lesssim\sqrt{n_{\rm max}}$. In practice, we are able to reach $n_{\rm max} \approx 50$, which would suggest that we can fully trust the results up to $\Delta \approx 7$. It would thus be important if the method of obtaining $\Lambda_n$ could be optimised so that $n_{\rm max}$ would be increased.

The main conceptual goal of this appendix was to check whether taking the large-$\Delta$ limit of the stress-tensor sector \textit{after} we expanded the
correlator near the bouncing singularity  reproduces the results predicted by the bouncing geodesic. In particular, we focused on the subleading behaviour in $\delta \tau$. While we found no obvious disagreement, we have also not found any conclusive evidence that confirms a relation. Therefore, a more thorough analysis is needed to establish a definite connection between these two results.

%%%%%%%%%%%%%%%%%%%%%%%%%%%%%%%%%%%%%%%%%%%%%%%%%%%%%
%%%%%%%%%%%%%%%%%%%%%%%%%%%%%%%%%%%%%%%%%%%%%%%%%%%%%

\chapter{Non-Zero \texorpdfstring{$x$}{x} Analysis}
\label{app:XCorr}

In this appendix we consider thermal correlation functions of scalar operators where the operators are inserted at a finite spatial distance, $x\neq 0$. We begin by discussing spacelike geodesics connecting the two insertion points and the effect of non-zero spatial distance on the bouncing geodesics. We then perform the OPE analysis for the stress-tensor sector at $x\neq 0$, expand the logarithm of the correlator near the bouncing singularity and \textit{then} take the large-$\Delta$ limit. We find that due to a slower convergence, even the leading correction to the bouncing singularity at non-vanishing $x$ cannot be conclusively matched between the geodesic and stress-tensor OPE results. We conclude that a more detailed analysis is needed to make a definite statement. 

\section{Semi-classical analysis}
\label{ss:SCApp}

In contrast to the main part of the Chapter \ref{ch3}, we work here directly in the Lorentzian signature and consider spacelike geodesics  in \eqref{eq:d=4BlackBranemet}. Let us parameterise the geodesics as $(t(s), r(s), x(s))$, where $s\in \mathbb{R}$ is the affine parameter. We can introduce conserved charges 
\begin{align}
\label{eq:ConservedQuantities}
    E~ \equiv~ r^2\,f(r) \, \dot t\,, \qquad P_i ~\equiv~ r^2\,\dot x_i\,,
\end{align}
where the dot denotes the derivative with respect to the affine parameter, which reduce finding the geodesics to a one-dimensional problem. We refer to $P_i$ as the momentum and $E$ as the (imaginary) energy, though we will omit the imaginary adjective in this appendix. The signs in \eqref{eq:ConservedQuantities} are chosen in such a way that in Patch I of the complexified spacetime (Figure~\ref{fig:Penrose}) the time increases for $E>0$. Finally, we use the isometry of $\mathbb{R}^3$ and rotate the spacetime such that only one component of the momentum is non-vanishing, for example $P_1 \equiv P$. 
Generalising to having all momenta non-trivial is straightforward.

\begin{figure}
        \centering
        \includegraphics[width = 0.7\textwidth]{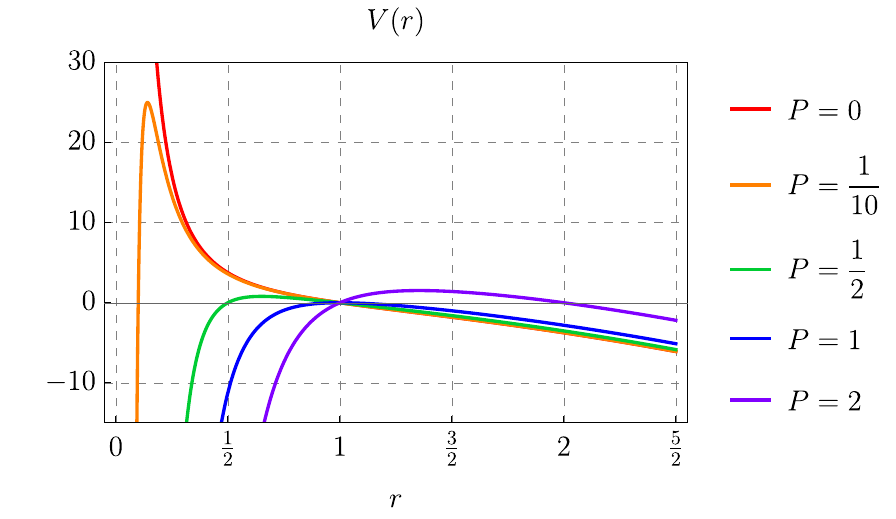}
        \caption{The potential $V(r)$ for different values of momentum $P$. When $P>1$, the turning points are always outside the horizon and such geodesics do not probe the singularity. When $0<P<1$, the potential has a maximum inside the horizon. In our analysis, we consider the case where $P$ (and thus $x$) is the smallest parameter of the problem and analyse perturbative corrections to the leading bouncing singularity due to non-vanishing spatial separation. }
        \label{fig:Potential1}
    \end{figure}

Spacelike geodesics in \eqref{eq:d=4BlackBranemet} have to satisfy the local constraint
\begin{align}
    \label{eq:SpacelikeConstraint}
 -r^2\,f(r) \, \dot t^2 + \frac{\dot r^2}{r^2\,f(r)} + r^2 \, \dot x^2 = 1\,,
\end{align}
which can be rewritten, using the expressions for the conserved charges, as
\begin{align}
    \label{eq:SpacelikeScattering}
    \dot r^2 -r^2\, f(r)\left(1 -\frac{P^2}{r^2}\right)  = E^2\,.
\end{align}
This effectively reduces the problem to classical scattering of a point particle in the potential (see Figure~\ref{fig:Potential1})
\begin{align}
\label{eq:ScatteringPotential}
    V(r) = -r^2\, f(r)\left(1 -\frac{P^2}{r^2}\right)\,.
\end{align}
The nature of the geodesics depends on the values of the conserved charges, which can  be seen from the turning point determined by the solution of $E^2 = V(r)$.
When $P=0$, the potential is monotonically decreasing and goes to positive infinity at $r\to 0$. As $E\to \pm\infty$, we get close to the singularity -- these are the bouncing geodesics discussed in the main text and in \cite{Fidkowski:2003nf, Festuccia:2005pi}. When $P\neq 0$, we have to distinguish between $P>1$ and $P<1$ regimes, as was already pointed out in \cite{Fidkowski:2003nf}. For $P>1$, there exist one real turning point, which is always outside the horizon. Such geodesics never probe the region near the singularity and will not be of interest in our analysis. For $P<1$, we can see that the behaviour near the origin changes -- the potential now reaches a maximal value inside the horizon before going to negative infinity. While this drastically alters the strict $|E|\to\infty$ limit, the behaviour does not change if we consider the regime where $P\,E \ll 1$. Namely, this regime is where we can think of spatial displacement $x$ (which is, as we will show, linearly related to the momentum $P$) as the smallest scale in the problem and treat it as an expansion parameter.%
\footnote{Another justification as to why this is a valid expansion comes from the WKB analysis of the correlator \cite{Festuccia:2005pi}. There it was shown that the Fourier conjugate to the spatial distance $x$ is actually $Q= -i\,P$. For purely imaginary momentum,  the potential is monotonically decreasing as a function of $r$. In that case the behaviour of bouncing geodesics does not change as we increase $x$.} As such, we can compare the small $x$ expansion on the geodesic side with the same expansion in the stress-tensor OPE \eqref{e.exnenula} -- in analogy of what was done in the two-dimensional example discussed in Appendix~\ref{s.kms2d}.

We want to calculate its regularised proper length and express it in terms of time and position displacements. The latter are given by
\begin{subequations}
\label{eq:Displacements}
    \begin{align}
        \label{eq:deltat}
           t &\equiv t_f - t_i = 2\int_{r_t}^{\infty}\frac{E\,dr}{r^2\,f(r)\sqrt{E^2 +r^2\, f(r) \left(1-\frac{P^2}{r^2}\right)}}\,,
        \\
        \label{eq:deltax}
        x &= x_f - x_i = 2\int_{r_t}^{\infty}\frac{P\,dr}{r^2\,\sqrt{E^2 + r^2\,f(r) \left(1-\frac{P^2}{r^2}\right)}}\,.
    \end{align} 
\end{subequations}
whereas the proper length integral is
\begin{align}
\label{eq:LengthInt}
    \cL = 2\int_{r_t}^{r_{\rm max}}\,\frac{dr}{\sqrt{E^2 + r^2\,f(r) \left(1-\frac{P^2}{r^2}\right)}}\,.
\end{align}
In the above, $r_t$ denotes the turning point of the geodesic, which is the largest real root of $E^2 = V(r)$, and we have used a cut-off parameter $r_{\rm max}$ that will help us to regularise the integral. 

For  $P\neq 0$ these integrals can be evaluated in terms of incomplete elliptic functions \cite{festucciathesis}. To express the solutions, we first define $q \equiv r^2$ and note that the turning point equation is a cubic equation in $q$. Let us denote the three, in general complex, solutions as $q_1$, $q_2$, and $q_3$, where $q_1$ corresponds to the largest real root and is identified with the turning point. Then one can show that the integrals \eqref{eq:Displacements} become
\begin{subequations}
\label{eq:EllipticalDisplacements}
    \begin{align}
         t &= \frac{E}{\sqrt{q_1(q_3-q_2)}}\left[\frac{(q_3-q_1)}{(q_1-1)(q_3-1)}\Pi\left(\frac{c}{a}, \phi, s\right)\right.\nonumber\\*
    &\left.\quad - \frac{(q_3-q_1)}{(q_1+1)(q_3+1)}\Pi\left(\frac{\bar c}{a}, \phi, s\right)+ \frac{2q_3^2}{(q_3-1)(q_3+1)}F\left(\phi, s\right)\right]\,,
        \\
         x &= \frac{2\,P}{\sqrt{q_1\left(q_3-q_2\right)}}\, F(\phi, s)\,,
    \end{align}
\end{subequations}
where $F(\phi,s)$ and $\Pi(z,\phi, s)$ are incomplete elliptic integrals of the first and third kind respectively, and we have used
\begin{gather}
\label{eq:Parameters}
    a = \frac{q_3-q_2}{q_1-q_2}\,,
    \qquad c = \frac{q_3-1}{q_1-1}\,,
    \qquad \bar c = \frac{q_3+1}{q_1+1}\,, 
    \quad s = \frac{q_3}{q_1\,a}\,,\qquad 
    %\nonumber\\
    %
    \phi = \arcsin{\sqrt{a}}\,.
\end{gather}
Finally, the proper length integral is given by
\begin{align}
    \label{eq:ProperLengthElliptic}
    \cL = \lim_{r_{\rm max}\to \infty}\Bigg\{2\sqrt{\frac{q_1}{q_3-q_2}}\,\left[s\,a\,F(\tilde \phi, s) + (1-s\,a) \Pi\left(\frac1{a},\tilde \phi, s\right)\right]- 2\,\log r_{\rm max} \Bigg\}\,,
\end{align}
where
\begin{align}
    \tilde \phi \equiv\arcsin\left( \sqrt{a\frac{r_{\rm max}^2- q_1}{r_{\rm max}^2-q_3}}\right)\,.
\end{align}
One can check that when $P=0$, these expressions reduce to \eqref{eq:LorentzianIntegrals}.

We now use these expressions to calculate the leading $x$ correction to the correlator near the bouncing singularity. Our method is to expand the expressions in $P$ and take the $E\to \infty$ limit order by order. For example, at first order in $P$, the turning point \eqref{eq:P=0Turning} is given by
\begin{align}
    q_1 &= \frac12\left(\sqrt{4 + E^4}- E^2\right) - P^2\,\frac{E^2}{\sqrt{4+ E^2}}\,,+ \coo{P^4}\,,
\end{align}
which, when taking $E\to \infty$ at each term separately, gives
\begin{align}
    q_1 &= \frac{1}{E^2}\Big[1- \varepsilon^2 - \coo{\varepsilon^4}\Big]\,.
\end{align}
In the above, we have used
\begin{align}
\label{eq:EpsilonParameter}
    \varepsilon \equiv P\,E\,,
\end{align}
which is a convenient expansion parameter in this double limit, since this calculation only gives sensible results if $ x$ (or $P$ in this instance) is the smallest parameter in the expansion. As such, $P \ll 1/E$ or equivalently $\varepsilon = P\,E \ll 1$, which is why this combination appears naturally in this limit.

One can then expand \eqref{eq:EllipticalDisplacements} in $\varepsilon$  and express the conserved charges as
\begin{align}
    \label{eq:EllipticalDisplacementsExpansion}
     E = \frac{2}{\delta t}\,, \qquad \varepsilon = \frac{2}{\pi}\,\frac{ x}{\delta t}\,,
\end{align}
where $\delta t = -i\delta \tau =  -t- i\tau_c$. The proper length integral expanded in $\varepsilon$ is
\begin{align}
\label{eq:cLExp}
    \cL \approx 2\log2 -2\log E + \frac{\pi}{2E^2}\,\varepsilon^2 \,,
\end{align}
which, when expressed in terms of position space coordinates, gives
\begin{align}
\label{eq:SummaryProperLength}
    \cL = 2 \log(\delta t) +\frac{ x^2}{2\pi}\,.
\end{align}
In a saddle point approximation, the contribution from such a geodesic to the correlator would be, to leading order in $x$
\begin{align}
\label{eq:SummaryCorrelator1}
    e^{-\Delta \,\cL} \sim \frac{1}{(\delta t)^{2\Delta}}\,\left(1-\frac{\Delta}{2\pi}\, x^2\right) = \frac{1}{(\delta t)^{2\Delta}}\,\left(1-\frac{\Delta\,\pi}{2\,\beta^2}\, x^2\right)\,,
\end{align}
where in the last expression we reinstated the inverse temperature. This is the result that we will compare to the OPE analysis.

\section{OPE analysis}
\label{ss:OPENonZeroX}

Let us now discuss the CFT side of the $x\neq0$ story. As discussed in Appendix~\ref{a.cbd}, for $0<x\ll\tau$ the decomposition of the correlator generalises to
        \begin{equation}\label{e.xxcoco}
        G(\tau,x)=\frac{1}{\tau^{2\Delta}}\sum_{n=0}^\infty\left[\Lambda^{(0)}_n+\frac{x^2}{\tau^2}\Lambda^{(1)}_n+\coo{\frac{x^4}{\tau^4}}\right]\left(\frac{\tau}{\beta}\right)^{4n}\ ,
    \end{equation}
where $\Lambda_n^{(0)}\equiv\Lambda_n$ is the contribution at $x=0$ that we examined in Chapter \ref{ch3}. The coefficients $\Lambda^{(1)}$  can be further decomposed as
\begin{align}
\label{eq:Lambda1Decomp}
    \Lambda^{(1)}_n\equiv(2n-\Delta)\,\Lambda_n+\widetilde{\Lambda}^{(1)}_n\,,
\end{align}
where ${\widetilde{\Lambda}^{(1)}_n}$ are given by \eqref{e.dufamzeposledna} and can thus be determined using the same holographic calculation as $\Lambda_n^{(0)}$. Since the latter have already been analysed, we focus here on the asymptotic behaviour of ${\widetilde{\Lambda}^{(1)}_n}$. We find that this data is well described by 
\begin{align}
\label{eq:tLambda1Exp}
    {\widetilde{\Lambda}^{(1)}_n}  = c(\Delta)\, \frac{n^{2\Delta -2}}{\left(\frac{1}{\sqrt{2}}\right)^{4n}e^{i\pi\,n}}\sum_{k=0}^{\infty}\frac{d_k(\Delta)}{n^k}\,.
\end{align}
The most important difference compared to \eqref{eq:Lambdand=4Ansatz} is the power of the factor of $n$. As we will see, this has significant effects on the degree of the singularity at first non-trivial correction in $x$ near the bouncing singularity. The prefactor $c(\Delta)$ is chosen to be the same as in $\Lambda_n^{(0)}$, in which case $d_0(\Delta)$ is not necessarily equal to 1. Let us note that the convergence of the numerical data obtained from solving the bulk equation of motion to \eqref{eq:tLambda1Exp} is slower compared to the convergence in the $x=0$ case. We can trace this back to more prefactors in the expression for ${\widetilde{\Lambda}^{(1)}_n}$ \eqref{e.dufamzeposledna} compared to $\Lambda_n^{(0)}$  \eqref{e.lebara}. As such, the results in this appendix are less reliable. Nonetheless, one is able to extract some information even with $n\approx 50$ data points at each value of $\Delta$.

\begin{figure}
\centering
    \includegraphics[scale=0.8]{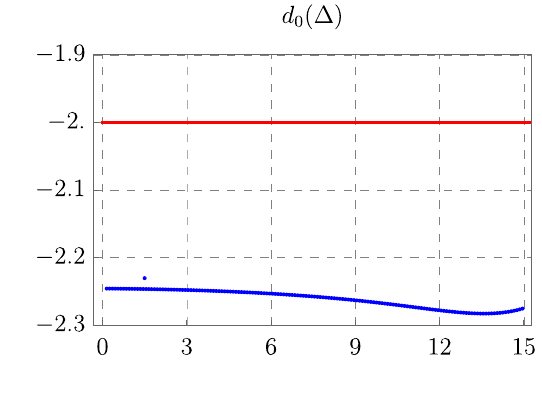}
    \vspace*{-0.8cm}
    \caption{The geodesic prediction for $d_0(\Delta)$ (red) compared to the numerical data obtained using the holographic method (blue). We see that there is significant mismatch between the two results. We relate this to the slower convergence of the OPE data to the asymptotic form.}
    \label{fig:new}
\end{figure}

Combining \eqref{eq:tLambda1Exp} with the expansion for $\Lambda^{(0)}_n$ gives
\begin{align}
    \Lambda_n^{(1)} =  c(\Delta)\, \frac{n^{2\Delta -2}}{\left(\frac{1}{\sqrt{2}}\,e^{\frac{i\pi}{4}}\right)^{4n}}\sum_{k=0}^{\infty}\frac{d_k(\Delta) + 2\,c_k(\Delta) - \Delta\,c_{k-1}(\Delta)}{n^k}\,,
\end{align}
where we define $c_{-1}(\Delta) = 0$. This, combined with the asymptotic form of $\Lambda^{(0)}_n$, can be inserted into the correlator expansion
{\allowdisplaybreaks{\small{
    \begin{align}
 G_T(\tau, x)&\approx \frac{c(\Delta)}{\tau^{2\Delta}}\int_0^{\infty}\!\!\!dn\left[n^{2\Delta-3} + \frac{x^2}{\tau^2}\,n^{2\Delta-2}\left(d_0(\Delta)+2 + \frac{d_1(\Delta)+ 2c_1(\Delta) -\Delta}{n}\right)\right]\left(\frac{\tau^4}{\tau_c^4}\right)^{n}\nonumber\\
 &= \frac{c(\Delta)\,\Gamma(2\Delta-2)}{\tau^{2\Delta}} \left[-\log\left(\frac{\tau^4}{\tau_c^4}\right)\right]^{2-2\Delta}\\
 &\quad\qquad  \times\Bigg\lbrace1 + \frac{x^2}{\tau^2}\bigg[
 \frac{(2\Delta-2)(d_0(\Delta)+2)}{-\log\left(\frac{\tau^4}{\tau_c^4}\right)}
 + d_1(\Delta)+ 2c_1(\Delta) -\Delta\bigg] \nonumber\Bigg\rbrace\label{eq:Gtx1}\,,
\end{align}
}}}%
where we have again replaced the sum with an integral, inserted 
$\tau_c$ using \eqref{e.poles}, and used that $c_0(\Delta)=1$. In this expression we work only at leading order in $n$ in the $x^0$ term while keeping the subleading contribution in the $x^2$ term. We will shortly show why the subleading correction is important. We can now expand the correlator around $\tau \approx \tau_c$, which gives the bouncing singularity together with several corrections
\begin{equation}
    G(\tau, x)\approx 
    \frac{c(\Delta)\,\Gamma(2\Delta-2)}{4^{2\Delta-2}}\,\frac{1} {\tau_c^{2}\,\delta\tau^{2\Delta-2}} \Bigg[1  + \frac{x^2}{\tau_c^2} \tilde \gamma_1(\Delta)+ \frac{x^2}{\tau_c\,\delta\tau}\,\frac{(2\Delta-2)(2+d_0(\Delta))}{4}\Bigg]\,,
\end{equation}
where we defined
\begin{align}
    \tilde \gamma_1(\Delta) \equiv d_1(\Delta) +\frac{-5 + 8\Delta+ 4 \Delta^2}{8}\,d_0(\Delta) + 2\,c_1(\Delta) - \frac54+ \Delta + \Delta^2\,.
\end{align}
To compare this analysis with the proper length of the bouncing geodesic we use \eqref{eq:LogOfCorr} and take the large-$\Delta$ limit
\begin{align}
\label{eq:Geodesicx2}
    \lim_{\Delta \to \infty}\cL_T &= 2 \log \delta\tau + \frac{x^2}{\tau_c^2}\Bigg[\frac{3-2\Delta}{4\Delta} - \frac{d_1(\Delta)}{\Delta}- \frac{2c_1(\Delta)}{\Delta}- \frac{3(2\Delta-1)d_0(\Delta)}{8\Delta}\Bigg]\nonumber\\
    & \quad +\frac{x^2}{\tau_c\,\delta\tau}\,\frac{(2\Delta-2)(2+d_0(\Delta))}{4\Delta} \,,
\end{align}
where it should be understood that one needs to take $\Delta\to \infty$ in all terms in this expression. We immediately notice that in contrast to \eqref{eq:SummaryProperLength}, the leading correction at $x^2$ comes at order $1/\delta\tau$. For the two expressions to be compatible, we would have to find
\begin{align}
\label{eq:d0value}
    d_0(\Delta) = - 2 + \coo{\frac{1}{\Delta}}\,.
\end{align}
We compare this prediction with the numerical data coming from solving the bulk equations of motion in Figure~\ref{fig:new}.
We note that in the range $0\lesssim \Delta \lesssim15$, the numerical data is roughly constant, but misses the value predicted by the geodesic analysis by about 10\%. 
While we believe that this difference is due to the slow convergence of data toward the asymptotic OPE ansatz \eqref{eq:tLambda1Exp} and that when the number of OPE coefficients calculated is increased the numerical data will converge to the geodesic result, we currently cannot show that this is the case. 

Since we cannot conclusively say something about the most divergent term at order $x^2$ in \eqref{eq:Geodesicx2}, we cannot compare the second term (with the square bracket) which has a counterpart in \eqref{eq:SummaryProperLength}.
However, we immediately run into an interesting problem: Note that in the geodesic result, the first correction is real, while in the result from the OPE analysis, the analogous correction is multiplied by $\tau_c^2$, which is imaginary. 
Not only that, this value differs based on which of the four poles was chosen to expand the correlator around. 
We currently do not have a good understanding of this discrepancy and leave this interesting avenue for future work. 
Let us also remark that from the CFT point of view having $x\neq 0$ does not significantly alter the physics, while the behaviour of the (real) bouncing geodesic changes \cite{Fidkowski:2003nf} . It would be interesting to understand whether the difference in the prefactors is related to this phenomenon. 

%%%%%%%%%%%%%%%%%%%%%%%%%%%%%%%%%%%%%%%%%%%%%%%%%%%%%
%%%%%%%%%%%%%%%%%%%%%%%%%%%%%%%%%%%%%%%%%%%%%%%%%%%%%

\chapter{Black Hole Singularity in \texorpdfstring{$d=6$}{d=6} and \texorpdfstring{$d=8$}{d=8}}\label{a.ope6d}

In this appendix we show that the stress-tensor sector in six and eight dimensional holographic CFTs also contains singularities at the critical values predicted by bouncing geodesics. In $d+1$ spacetime dimensions the value of $\tau_c$ is given by \cite{Fidkowski:2003nf}
\begin{equation}\label{tauc}
  \tau_c^{(d+1)}  = {\beta \over 2} \pm i {\beta \over 2}  \ {\cos {\pi\over d} \over \sin {\pi\over d}} =\pm i  {\beta e^{\mp {i \pi\over d}} \over 2 \sin {\pi\over d}}\ . 
\end{equation}
Below we reproduce $\tau_c$ from the asymptotic analysis of the stress tensor OPE in $d=6$ and $d=8$.

\section{Singularity in six dimensions}\label{a.6dcase}

\begin{figure}[t]
\centering
\includegraphics[width = \textwidth]{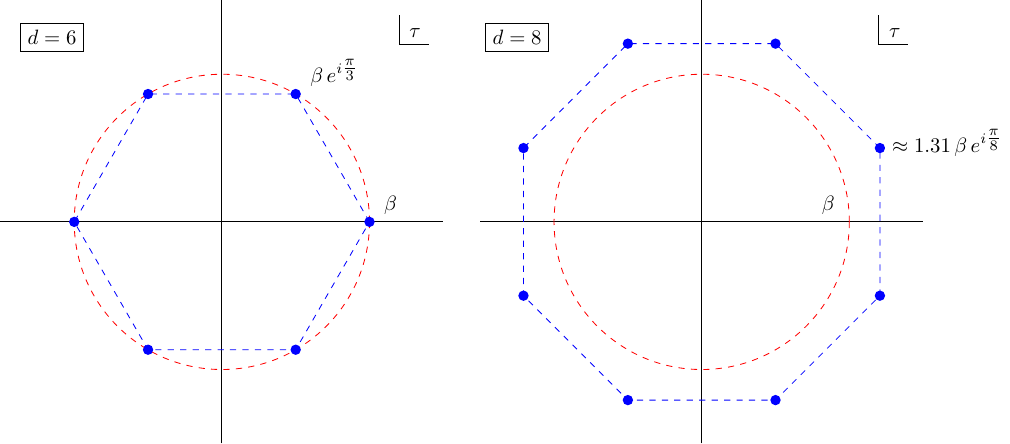}
\caption{Poles in the complex $\tau$ plane in $d=6$ and $d=8$.}
\label{fig:Polesd6d8}
\end{figure}

We start by solving the 7-dimensional scalar equations of motion in the black-hole background using the ansatz \eqref{e.theansatz}. By expanding the results and comparing them with the CPW expansion \eqref{e.resumL}, we extract the CFT data from the dual bulk theory. For example the first two OPE coefficients are
{\allowdisplaybreaks{\small{
    \begin{align}
    \Lambda_1&=\frac{8 \pi ^6 \Delta }{15309}\\
    \Lambda_2&=\frac{32 \pi ^{12} \Delta  \left(715 \Delta ^5-6930 \Delta ^4+17204 \Delta ^3-9323 \Delta ^2+26334 \Delta +9000\right)}{167571318915 (\Delta -6) (\Delta -5) (\Delta -4) (\Delta -3)}\,.
    \end{align}}}}%
Again we are interested in the large-$n$ behaviour of $\Lambda_n$.
We find that as $n$ grows, the OPE coefficients tend to
    \begin{equation}\label{e.pole6}
        \Lambda_n^a=j(\Delta)n^{2\Delta-4}\ ,
    \end{equation}
up to $1/n$ corrections which we do not here. In the above, $j(\Delta)$ is an undetermined function of the conformal dimension. An interesting difference compared to $d=4$ is the absence of an oscillating sign in $\Lambda_n^a$. Inserting these coefficients into the OPE and approximating the sum with an integral gives
    \begin{equation}\label{e.6sumint}
        G_T(\tau)=\frac{1}{\tau^{2\Delta}}\int_0^\infty\Lambda^a_n\left(\frac{\tau}{\beta}\right)^{6n}d n
        =\frac{j(\Delta)\,\Gamma(2\Delta-3)}{\tau^{2\Delta}}\left(-\log\left( \frac{\tau^6}{\beta^6}\right)\right)^{3-2\Delta}\,.
    \end{equation}
This expression has a singularity whenever the argument of the logarithm is equal to 1, which is precisely at (see left panel of Figure~\ref{fig:Polesd6d8})
	\begin{equation}\label{viva}
	\tau_c=\beta e^{i\frac{k\pi}{3}}\qq{for}k\in\mathbb{Z}\,.
	\end{equation}
Interestingly, we note that all singularities are located on a circle of radius $\beta$ and that two of these critical points coincide with the  positions where one expects KMS poles.
When comparing these critical values to the geodesic result \eqref{tauc}
    \begin{equation}
      \tau_c^{(7)}=\frac{\beta}{2}\pm i\frac{\beta}{2}\sqrt{3}=\beta e^{\pm i\frac{\pi}{3}}\ ,
    \end{equation}
we find a precise match.

\section{Singularity in eight dimensions}\label{a.8dcase}

We follow the same approach in $d=8$. The first two coefficients are
{\small{
    \begin{align}
    \Lambda_1&=\frac{\pi ^8 \Delta }{184320}\\
    \Lambda_2&=\frac{\pi ^{16} \Delta}{1156266432921600 (\Delta -8) (\Delta
   -7) (\Delta -6) (\Delta -5) (\Delta -4)}\times \big(17017 \Delta ^6\nonumber\\
   &\phantom{=}-211666 \Delta ^5+681619 \Delta ^4-881554 \Delta ^3+3831472 \Delta ^2+2284352 \Delta +1317120\big)\,.
    \end{align}}}Overall
we find that the leading  asymptotic behaviour of the OPE coefficients is
	\begin{equation}\label{e.osemrozmerovLb}
	\Lambda^a_n=Q(\Delta)\frac{n^{2\Delta-5}}{W^{8n}\,e^{i\pi\,n}}\ ,
	\end{equation}
where $Q(\Delta)$ is a function of the conformal dimension and $W$ is a constant 
	\begin{equation}\label{e.Wval8d}
	W\approx1.3066\ .
	\end{equation}
We note that the $\Lambda_n^a$ have an oscillating sign,  similar to the four-dimensional case. Inserting them in the OPE and performing the integral gives
    \begin{equation}\label{e.8sumint}
        G_T(\tau)=\frac{1}{\tau^{2\Delta}}\int_0^\infty\Lambda^a_n\left(\frac{\tau}{\beta}\right)^{8\,n}d n
        =\frac{Q(\Delta)\,\Gamma(2\Delta-5)}{\tau^{2\Delta}}\left(-\log\left( \frac{\tau^8}{W\,\beta^8\,e^{i\pi}}\right)\right)^{3-2\Delta}\,.
    \end{equation}
$I.e.$ the stress-tensor sector has a singularity at 
	\begin{equation}
	\tau_c=\beta\, W\, e^{\frac{i\pi}{8}+k\frac{i\pi}{4}}\qq{for}k\in\mathbb{Z}\ ,
	\end{equation}
which are pictured on the right in Figure~\ref{fig:Polesd6d8}. This agrees with the geodesic analysis which predicts the singularities at
    \begin{equation}
       \tau_c^{(9)}=\frac{\beta}{2}\pm i\frac{\beta}{2}\cot{\frac{\pi}{8}}=\beta \frac{e^{\pm i\frac{3\pi}{8}}}{2\sin\frac{\pi}{8}}\,,
    \end{equation}
since
	\begin{equation}
	\frac{1}{2\sin\frac{\pi}{8}}\approx1.3066\approx W\ .
	\end{equation}
It is interesting to note that in $d=8$ these singularities are all located further away than where we would expect to find the KMS poles. This further strengthens the argument that the stress-tensor sector does not contain the information about the KMS poles and the double-traces are needed to recover their location.

%%%%%%%%%%%%%%%%%%%%%%%%%%%%%%%%%%%%%%%%%%%%%%%%%%%%%
%%%%%%%%%%%%%%%%%%%%%%%%%%%%%%%%%%%%%%%%%%%%%%%%%%%%%

\chapter{Lowest-Twist Asymptotic Analysis}\label{app:LTanal}
   
In Chapter \ref{ch3} of this thesis we examined the \textit{summed coefficients} $\Lambda_n$ defined by the expansion \eqref{e.topeL}. It is also interesting to study the individual coefficients $\lambda_{n,J'}$ corresponding to the contributions from multi-stress tensors with different spin $J'$ (respectively twist $\Delta'-J'$). We are mainly interested in the behaviour of the coefficients $\lambda_{n,2n}$ that correspond to the multi-stress tensors with the lowest twist. These coefficients are universal in holographic theories \cite{Fitzpatrick:2019zqz} and one can calculate them using bootstrap techniques (see $e.g.$ \cite{Karlsson:2019dbd}).

Alternatively, one can develop an effective method\footnote{This way one get access to the leading-twist OPE coefficients $\lambda_{n,2n}$ up to order $n\sim10000$.} to calculate $\lambda_{n,2n}$ by combining the bulk recursion relation found in \cite{Fitzpatrick:2019zqz}, with the fact that only one coefficient of the bulk ansatz contributes to the near-lightcone correlator at each order in $1/r$ expansion.\footnote{This is an analogous approach as the one we will use in Appendix \ref{ap.pertpos}.} Such coefficient can be then mapped to the lowest-twist coefficient $\lambda_{n,2n}$. In practice we get this coefficient as
    \begin{equation}\label{e.baf}
        \lambda_{n,2n}=\left(-\frac{\pi^4}{4}\right)^{n}b_{n,n}\ .
    \end{equation}
where $b_{n,n}$ can be systematically calculated by solving the recursion relation \cite{Fitzpatrick:2019zqz}:
    \begin{equation}
        4(j-4i)b_{i,j}=\frac{4(1-j+i)(i-j-1+\Delta)}{j-\Delta}b_{i,j-1}-4(1+j-\Delta) b_{i-1,j+1}
    \end{equation}
where $i,j\in\mathbb{Z}$, $b_{i,j}=0$ for $j\notin[-i,i]$ and for all $i<0$, and $b_{0,0}=1$.

Analysing the coefficients $\lambda_{n,2n}$ we find their leading large-$n$ behaviour to be
    \begin{equation}\label{e.ltfo}
        \lambda_{n,2n}=A(\Delta)\frac{e^{i\pi n}n^{2\Delta-5/2}}{B^{4n}}\ ,
    \end{equation}
where $A(\Delta)$ is a function of the scaling dimension and value of the constant $B\approx0.8968...$.

To compare this with the asymptotic form of $\Lambda_n\sim n^{2\Delta-3}$, we consider the coefficient $\lambda_{n,2n}$ multiplied by $(1+J)$, yielding the behaviour $\sim n^{2\Delta-3/2}$ which grows faster than the overall summed coefficient $\Lambda_n$. Thus, terms with the subleading twists are crucial to ensure that $\Lambda_n$ has the correct overall scaling as
observed in Chapter \ref{ch3} of the thesis. It would be interesting to study this in more detail.

%%%%%%%%%%%
%%%%%%%%%%%
%%%%%%%%%%%
%%%%%%%%%%%
%%%%%%%%%%% p1 appendices:
%%%%%%%%%%%
%%%%%%%%%%%
%%%%%%%%%%%
%%%%%%%%%%%

    \chapter{Integrated Scalar}\label{AppendixA}

In Chapters \ref{ch4}-\ref{ch6} we analyse integrated spinning correlators. As several new phenomena emerge due to integration, we will first discuss a toy model -- ($d=4$) scalar field, that will serve as a consistency check. We will show that one is able to extract the same OPE data when working with correlators integrated over the $xy$-plane, as in the original approach. 

This appendix is divided into two parts: the first subsection focuses on the case of a scalar field with non-integer scaling dimension, while the second one studies the $\Delta=4$ case, which is more relevant for the stress tensor calculations.

In both subsections we begin by solving the bulk equations of motion where two spatial dimensions are integrated out. We find the solution using the ansatz introduced in \cite{Fitzpatrick:2019zqz, Li:2019tpf} and discussed in Chapter \ref{ch2}, naturally adapted for the integrated case.

On the CFT side we examine the integrated conformal blocks in the OPE limit. In the integer case we explain the emergence of the log term as a result of mixing of the scalar and stress tensor sectors. We also find that further regularization is needed due to integration.

Finally we extract the OPE coefficients\footnote{To the leading order in the large $C_T$ limit.} from the comparison of the bulk calculations and the CFT analysis. 
We conclude that we can extract the same amount of the OPE data in the integrated and non-integrated cases.

\section{Scalar field with non-integer scaling dimension}\label{sec:IntegratedScalarNoninteger}

\subsubsection{Bulk-side}
Our aim is to calculate the bulk-to-boundary propagator satisfying the scalar field equation
\begin{align}
	(\Box-m^2)\phi&=0\label{eom1}\\
	\Delta(\Delta-4)-m^2&=0,
\end{align}
on the planar Euclidean AdS-Schwarzschild black hole background
\begin{equation}
	\dd s^2 = r^2(1-\frac{\mu}{r^4})\dd t^2+r^2\dd \vec{x}^2+\frac{1}{r^2(1-\frac{\mu}{r^4})}\dd r^2,\label{metrika}
\end{equation}
where $\vec{x}=(x,y,z)$.

According to the AdS/CFT dictionary we obtain the thermal two-point function as
\begin{equation}
	\expval{\mathcal{O}_L(x_1)\mathcal{O}_L(x_2)}_{\beta}=\lim_{r\rightarrow\infty}r^\Delta\phi(r,x_1,x_2).\label{comp}
\end{equation}
In this subsection we consider the conformal dimension of the scalar operator $[\mathcal{O}_L]=\Delta\notin\mathbb{Z}$\,.\footnote{The subscript index $L$ refers to the scalar being light, $i.e.$ $\Delta\ll C_T$.}

We now integrate over the $xy$-plane, hence we work with the integrated bulk-to-boundary propagator
\begin{equation}
	\Phi(t,z,r)=\iint_{\mathbb{R}^2}\dd x\dd y\,\phi(t,\vec{x},r)\,.
\end{equation}
Equation (\ref{eom1}) in the background (\ref{metrika}) then takes the form
\begin{equation}
	\left[\Delta(\Delta-4)-r(4+f)\partial_r-r^2f\partial^2_r-\frac{1}{r^2}\partial^2_z-\frac{1}{r^2f}\partial^2_t\right]\,\Phi=0,    
\end{equation}
where $f=1-\frac{\mu}{r^4}$.

To solve this equation, we first transform coordinates $(t,z,r)$ to $(w,\rho,r)$ defined by 
\begin{align}
	\rho&\coloneqq rz\label{defofFPanalogues1}\\
	w^2&\coloneqq 1+r^2t^2+r^2z^2\,.\label{defofFPanalogues2}
\end{align}
These are the natural integrated analogues of the variables introduced in \cite{Fitzpatrick:2019zqz}. In these coordinates we have the following equation for $\Phi$:
\begin{align}
	\big[C_1+C_2\partial_r&+C_3\partial_\rho+C_4\partial_w+C_5\partial^2_r+C_6\partial^2_\rho\nonumber\\
	&+C_7\partial^2_w+C_8\partial_r\partial_\rho+C_9\partial_\rho\partial_w+C_{10}\partial_w\partial_r\big]\,\Phi=0,\label{eom2}
\end{align}
where
\begin{align}
	C_1&=-r^4w^3(\Delta-4)\Delta(r^4-\mu)\\
	C_2&=rw^3(5r^8-6r^4\mu+\mu^2)\\
	C_3&=\rho w^3(5r^8-6r^4\mu+\mu^2)\\
	C_4&=w^2(w^2-1)(5r^8-6r^4\mu+\mu^2)+r^8(1+\rho^2)\nonumber\\
	&\phantom{wwwwwwwwww}+(r^4-\mu)^2(w^2-1)+r^4(r^4-\mu)(w^2-\rho^2)\\
	C_5&=(r^4-\mu)^2r^2w^3\\
	C_6&=(r^4-\mu)^2w^3\rho^2+r^4(r^4-\mu)w^3\\
	C_7&=r^8w(w^2-\rho^2-1)+(r^4-\mu)^2w(w^2-1)^2+r^4(r^4-\mu)w\rho^2\\
	C_8&=2rw^3\rho(r^4-\mu)^2\\
	C_9&=2(r^4-\mu)^2w^2(w^2-1)\rho+2r^4(r^4-\mu)w^2\rho\\
	C_{10}&=2rw^2(r^4-\mu)^2(w^2-1)\ .
\end{align}
Here, using the same logic as in \cite{Fitzpatrick:2019zqz}, we assume the ansatz (focusing only on the solution that corresponds to 
the stress-tensor sector on the CFT side) as
\begin{equation}
	\Phi=\Phi_{AdS}\left(1+\frac{G_4}{r^4}+\frac{G_8}{r^8}+\ldots\right),
\end{equation}
where 
\begin{align}
	&G_4=\sum_{m=0}^2\sum_{n=-2}^{4-m}a^4_{n,m}w^n\rho^m\\
	&G_8=\sum_{m=0}^6\sum_{n=-6}^{8-m}a^8_{n,m}w^n\rho^m\\
	&\phantom{wwwwww}\vdots
\end{align}
The vacuum propagator $\Phi_{AdS}$ can be obtained by integrating the known vacuum bulk-to-boundary propagator for the scalar field: 
\begin{equation}
	\Phi_{AdS}(t,z,r)=\iint\dd x\dd y\left[\frac{r}{1+r^2(t^2+x^2+y^2+z^2)}\right]^\Delta =\frac{\pi r^{\Delta-2}}{\Delta-1}\left(1+r^2(t^2+z^2)\right)^{1-\Delta}.
\label{phiads}
\end{equation}
Changing the coordinates to $(w,\rho,r)$, we get
\begin{equation}
	\Phi_{AdS}(w,\rho,r)\propto\frac{r^{\Delta-2}}{w^{2-2\Delta}}\ .
\end{equation}

Substituting the ansatz into Eq.\ (\ref{eom2}) we can determine the coefficients $a^j_{n,m}$ as functions of $\Delta$ and $\mu$. In the non-integer case all coefficients $a_{n,m}^4$ and $a_{n,m}^8$ can be found explicitly. 
Here we list the nonzero parameters that appear at order $\mu^1$:
\begin{align}
	a^4_{-2,0}&=\frac{2\mu(1-\Delta)}{5}\\
	a^4_{0,0}&=\frac{\mu(\Delta-1)}{5}\\
	a^4_{2,0}&=\frac{3\mu\Delta(\Delta-1)}{20(\Delta-2)}\\
	a^4_{4,0}&=\frac{\mu\Delta(\Delta-1)(3\Delta-10)}{120(\Delta-3)(\Delta-2)}\\
	a^4_{-2,2}&=-\frac{\mu(\Delta-1)}{5}\\
	a^4_{0,2}&=-\frac{\mu\Delta}{10}\\
	a^4_{2,2}&=-\frac{\mu\Delta(\Delta-1)}{30(\Delta-2)}\ .
\end{align}

Finally, we have to take the boundary limit (\ref{comp}); in general it can be written as
\begin{equation}
	\mathcal{G}_\Delta=\lim_{r\rightarrow\infty}r^\Delta\Phi_{AdS}\left(1+G^T+G^\phi\right)\ ,\label{tcomp}
\end{equation}
where $G^T=\frac{G^4}{r^4}+\frac{G^8}{r^8}+\ldots$ corresponds to the stress-tensor sector
and $G^\phi$ corresponds to the double-trace scalars. As  mentioned earlier, for $\Delta\notin\mathbb{Z}$ the two sectors decouple. Therefore, focusing on the stress-tensor sector, at order $\mu^1$ we get
\begin{equation}\label{jezibaba}
	\mathcal{G}_{\Delta}\Big|_{\mu^1}\!=\lim_{r\rightarrow\infty}\frac{\pi r^{2\Delta-6}}{\Delta-1}\frac{G^4(t,z,r)}{(1+r^2(t^2+z^2))^{\Delta-1}}=\frac{\pi(t^2\!+\!z^2)^{2-\Delta}(t^2(3\Delta-10)\!+\!z^2(2-\Delta))\Delta \mu}{120(\Delta-3)(\Delta-2)}\ .
\end{equation}
Similarly one obtains terms $\OO(\mu^2)$.

\subsubsection{CFT-side}

On the boundary, the integrated scalar correlator can be decomposed in conformal blocks; one obtains 
\begin{equation}
	\mathcal{G}_{\Delta}\equiv\iint\dd x\dd y\expval{\mathcal{O}_L\mathcal{O}_L}_\beta=\iint\dd x\dd y\sum_{\Delta_i,J}C_{\Delta_i,J}\frac{g_{\Delta_i,J}(Z,\overline{Z})}{(Z\overline{Z})^\Delta}\ ,\label{ego}
\end{equation}
where $Z$ and $\overline{Z}$,\footnote{We use this notation, to distinguish the cross ratios and the space coordinate $z$.} are the cross ratios defined in terms of $t$, $x$, $y$ and $z$ as:
\begin{align}
	Z&=-t-i\sqrt{x^2+y^2+z^2}\\
	\overline{Z}&=-t+i\sqrt{x^2+y^2+z^2}\ ,
\end{align}
and $C_{\Delta_i,J}$ is the product of the OPE data corresponding to the exchanged primary with the conformal dimension $\Delta_i$ and spin $J$.

The first nontrivial contribution to the correlator (\ref{ego}) comes from the exchange of the  stress tensor. In the OPE limit\footnote{Here by the OPE limit of the conformal blocks we mean the large-volume limit described in the Appendix \ref{a.cbd}.} the corresponding conformal block is
\begin{equation}
	g_{4,2}(Z,\overline{Z})\approx Z\overline{Z}(Z^2+Z\overline{Z}+\overline{Z}^2)\ .
\end{equation}        

\noindent Hence, at this order Eq.\ (\ref{ego}) becomes
\begin{equation}\label{CCC1}
	\mathcal{G}_{\Delta}\Big|_{\mu^1}=-C_{4,2}\frac{\pi(t^2+z^2)^{2-\Delta}(t^2(10-3\Delta)+z^2(\Delta-2))}{(\Delta-3)(\Delta-2)}\ .
\end{equation}

Following the same approach, it is straightforward to obtain the corresponding integrated conformal blocks for the multi-trace stress tensors.

\subsubsection{Comparison}

In order to determine the OPE coefficients, we compare the bulk results with the expected sum of the integrated conformal blocks on the CFT side. 

Comparing (\ref{CCC1}) and \eqref{jezibaba} we extract the OPE coefficient:
\begin{equation}
	C_{4,2}=\frac{\Delta \mu}{120}\ ,
\end{equation}
which agrees with  Eq.\ \eqref{e.dalekyskokik}.

The OPE coefficients at higher orders in $\mu$ can be determined in an analogous way.

\section{Scalar field with \texorpdfstring{$\Delta=4$}{Delta=4}}
\label{sec:IntegratedScalarInteger}

\subsubsection{Bulk-side}

We now consider $\Delta=4$.
The setup for this case is identical to the one above; that is, 
we again need  to solve the bulk equation of motion (\ref{eom2}). 

Here, however, the situation becomes more subtle as some of the OPE coefficients are expected to be singular for $\Delta=4$, see $e.g.$ \eqref{eaestedakn}. On the other hand, for integer $\Delta$ the multi-stress tensor sector and double-trace scalar sector are no longer decoupled. We expect the contribution from operators $[\mathcal{O}_L\mathcal{O}_L]$ to compensate for these divergent parts in the $[T^n]_J$ OPE coefficients. As a result,  log terms will appear in the solution. We  explain this in more detail in the next subsection. 

In the bulk this leads to a slightly modified ansatz \cite{Li:2019tpf}:
\begin{equation}
	\Phi=\Phi_{AdS}\left(1+\frac{1}{r^4}\left(G^{4,1}+G^{4,2}\log r\right)+\frac{1}{r^8}\left(G^{8,1}+G^{8,2}\log r\right)+\ldots\right),
\end{equation}
where $\Phi_{AdS}$ is the  vacuum propagator (\ref{phiads}) and $G^{4,j}$ and $G^{8,j}$ are given by
\begin{align}
	&G^{4,j}=\sum_{m=0}^2\sum_{n=-2}^{4-m}(a^{4,j}_{n,m}+b^{4,j}_{n,m}\log w)w^n\rho^m\\
	&G^{8,j}=\sum_{m=0}^6\sum_{n=-6}^{8-m}(a^{8,j}_{n,m}+b^{8,j}_{n,m}\log w)w^n\rho^m\\
	&\phantom{wwwwww}\vdots\nonumber
\end{align}
Inserting this ansatz into   (\ref{eom2}), we can determine the coefficients $a_{n,m}^{k,j}$ and $b_{n,m}^{k,j}$.

The result (in the $w$, $\rho$ and $r$ coordinates) is
\begin{equation}\label{sbulkyscalar}
	\begin{split}
		\Phi&=\frac{\pi}{25200r^6w^{10}}\big[8400w^4(r^8+w^6((1-6\rho^2)a^{8,1}_{6,0}+(w^2-8\rho^2)a^{8,1}_{8,0}))\\
		&\phantom{=}+840r^4w^2(-12+6w^2+w^4+w^6-2(3+2w^2+w^4)\rho^2)\mu\\
		&\phantom{=}+(8064-12656w^2+3136w^4+448w^6+655w^8-4(-2016\\
		&\phantom{=}+448w^2+476w^4+345w^6+40w^8+750w^{10})\rho^2\\
		&\phantom{=}+56(36+44w^2+35w^4+20w^6+10w^8)\rho^4\\
		&\phantom{=}+120w^{10}(-6+5w^2-4\rho^2)(\log r+\log w))\mu^2\big]+\mathcal{O}(\mu^3)
	\end{split}
\end{equation}
For  the stress tensor exchange  all log terms vanish and we are also able to determine all of the corresponding coefficients in the ansatz. 
As expected, we obtain the same results as in the non-integer case. However, we find that for the double-stress tensor exchange (order $\mu^2$) the coefficients $a^{8,1}_{6,0}$ and $a^{8,1}_{8,0}$ are not fixed by the near-boundary analysis.

Taking the boundary limit (\ref{tcomp}) and focusing on the double-trace sector
\begin{equation}
	\mathcal{G}_{4}\Big|_{\mu^2}=\lim_{r\rightarrow\infty}r^4\Phi_{AdS}\frac{G_{8,1}+G_{8,2}\log r}{r^8}\ ,\label{comp3}
\end{equation}
we get
\begin{equation}
	\begin{split}
		\mathcal{G}_{4}\Big|_{\mu^2}&=\frac{\pi}{1260}\Bigg[420\Big(\!-6z^2a^{8,1}_{6,0}+(t^2-7z^2)a^{8,1}_{8,0}\Big)\\
		&+\mu^2\bigg(-\frac{2(75t^2z^2+61z^4)}{t^2+z^2}+3(5t^2+z^2)\log\left(t^2+z^2\right)\bigg)\Bigg]\ .\label{RHSf}
	\end{split}
\end{equation}

\subsubsection{CFT-side}\label{sscalcft}

At  $\mathcal{O}(\mu^0)$ and $\mathcal{O}(\mu^1)$ the contribution for $\Delta=4$ will be the same as  for $\Delta\notin\mathbb{Z}$. Let us therefore focus on the $\mu^2$ terms.

Here the double-trace stress tensors mix with the double-trace scalar $[\mathcal{O}_L\mathcal{O}_L]$. We thus have to consider four contributions to the correlator at  $\OO(\mu^2)$ -- three from the double stress tensor (we label them by the conformal dimension and the spin: $(\Delta_i,J)$): 
\begin{align}
	T_{\mu\nu}T^{\mu\nu}\quad&\Longleftrightarrow\quad (8,0)\\
	T_{\mu\nu}T^\nu_{\phantom{\mu}\alpha}\quad&\Longleftrightarrow\quad (8,2)\\
	T_{\mu\nu}T_{\alpha\beta}\quad&\Longleftrightarrow\quad (8,4)
\end{align}        
and one contribution from the double-trace scalar:
\begin{equation}
	[\mathcal{O}_L\mathcal{O}_L]\quad\Longleftrightarrow\quad(8,0),
\end{equation}
which will mix with the $(8,0)$ contribution from $[T^2]$. This agrees with the fact that it is only the coefficient $C_{8,0}^{\text{\tiny\textit{TT}}}$, that is expected to diverge.

Let us have a closer look at the divergent terms that appear in the corresponding OPE coefficients.
 First, as the coefficient $C_{8,0}^{\text{\tiny\textit{TT}}}$ has a pole in $\Delta=4$ \cite{Fitzpatrick:2019zqz} we can write it as
\begin{equation}
	C_{8,0}^{\text{\tiny\textit{TT}}}=\frac{C_{\text{\tiny sing}}}{\Delta-4}+C^{\text{\tiny\textit{TT}}}_{\text{\tiny reg}},\label{res1}
\end{equation}
where the term $C^{\text{\tiny\textit{TT}}}_{\text{\tiny{reg}}}$ is regular in $\Delta=4$ and $C_{\text{\tiny sing}}$ is the residue. In order to cancel the singular part, the OPE coefficient of the double-trace scalar must also have a pole at $\Delta=4$ with the same residue but opposite sign:
\begin{equation}
	C_{8,0}^{\text{\tiny{$\mathcal{OO}$}}}=-\frac{C_{\text{\tiny{sing}}}}{\Delta-4}+C^{\text{\tiny{$\mathcal{OO}$}}}_{\text{\tiny{reg}}}\label{res2}
\end{equation}
Now, as the conformal block for $J=0$ in the large-volume limit is $g_{\Delta',0}\approx(Z\overline{Z})^{\Delta'}$, we can study what happens if the contributions from $[T^2]$ and $[\mathcal{O}_L\mathcal{O}_L]$ mix. 
Setting $\Delta=4+\delta$, summing the contributions from $[T^2]$ and $[\mathcal{OO}]$ and then taking the limit $\delta\rightarrow0$ we find
\begin{equation}\label{int}
    \begin{split}
	\mathcal{G}_{4}\Big|_{\mu^2}=\iint\dd x\dd y\Bigg(&C^{\text{\tiny\textit{TT}}}_{\text{\tiny{reg}}}+C^{\text{\tiny{$\mathcal{OO}$}}}_{\text{\tiny{reg}}}-C_{\text{\tiny{sing}}}\log Z\overline{Z}+\\ &C_{8,2}\frac{Z^2+Z\overline{Z}+\overline{Z}^2}{Z\overline{Z}}+C_{8,4}\frac{Z^4+Z^3\overline{Z}+Z^2\overline{Z}^2+Z\overline{Z}^3+\overline{Z}^4}{Z^2\overline{Z}^2}\Bigg).
    \end{split}
\end{equation}
where we used  Eqs.\ (\ref{res1}) and (\ref{res2}).

As can be seen, the integral \eqref{int} is divergent and thus needs to be regulated. In practise we can do this using a
form of dimensional regularization: we multiply the integrand by a factor $|x|^{-\epsilon}=\left(t^2+x^2+y^2+z^2\right)^{-\frac{\epsilon}{2}}$, integrate and then expand the resulting expression around $\epsilon=0$. We get
\begin{align}
	\mathcal{G}_{4}\Big|_{\mu^2}=&\frac{8\pi t^2(C_{8,2}-3C_{8,4})}{\epsilon}+\pi\bigg[C_{8,4}\frac{\left(15t^4-2t^2z^2-z^4+12t^2 \left(t^2+z^2\right)\log\left(t^2+z^2\right)\right)}{t^2+z^2}\nonumber\\
	&+C_{8,2}\big(t^2+z^2-4t^2\log\left(t^2+z^2\right)\big)+(t^2+z^2)\Big(C_{\text{\tiny{sing}}}(\log(t^2+z^2)-1)\nonumber\\
	&-C_{\text{\tiny{reg}}}^{\text{\tiny\textit{TT}}}-C_{\text{\tiny{reg}}}^{\text{\tiny{$\mathcal{OO}$}}}\Big)\bigg]+\mathcal{O}(\epsilon^1)\label{cermak}
\end{align}
%In the limit $\epsilon\rightarrow0$ the $\mathcal{O}(\epsilon^1)$ drops out and we are only left with the regular part and the divergent piece that is roughly related to the size of the plane we integrate over.

\subsubsection{Comparison}

Comparing (\ref{cermak}) and (\ref{RHSf}) we can extract the coefficients $C_{8,2}$, $C_{8,4}$ and $C_{\text{\tiny{sing}}}$:
\begin{align}
	C_{8,2}&=\frac{\mu^2}{560}\label{val1}\\
	C_{8,4}&=\frac{\mu^2}{720}\label{val2}\\
	C_{\text{\tiny{sing}}}&=\frac{\mu^2}{420},\label{val3}
\end{align}
while for the coefficients $C_{\text{\tiny{reg}}}^{TT}$, $C_{\text{\tiny{reg}}}^{\text{\tiny{$\mathcal{OO}$}}}$ and the parameter $\epsilon$ we get the following relations
\begin{align}
	C_{\text{\tiny{reg}}}^{\text{\tiny{$\mathcal{OO}$}}}+C_{\text{\tiny{reg}}}^{\text{\tiny\textit{TT}}}&=2 a^{8,1}_{6,0}+\frac{7}{3} a^{8,1}_{8,0}+\frac{239 \mu ^2}{2520}\label{sht1}\\
	\frac{1}{\epsilon}&=-\frac{420 (3 a^{8,1}_{6,0}+4 a^{8,1}_{8,0})+47 \mu ^2}{12 \mu ^2}.\label{sht2}
\end{align}

To conclude, our double-trace results for $C_{8,2}$, $C_{8,4}$ and the residual part of $C_{8,0}^{\text{\tiny\textit{TT}}}$ are in perfect agreement with the results for the non-integer case extrapolated to $\Delta=4$, see \cite{Fitzpatrick:2019zqz}, while the remaining CFT data is related to the undetermined coefficients on the bulk side by  eqs (\ref{sht1})-(\ref{sht2}).

Using the same approach in the non-integrated $\Delta=4$ case, one gets the relations \eqref{val1}-\eqref{val3} for $C_{8,2}$, $C_{8,4}$ and $C_{\text{\tiny{sing}}}$, while $C_{\text{\tiny{reg}}}^{\text{\tiny{$\mathcal{OO}$}}}+C_{\text{\tiny{reg}}}^{\text{\tiny\textit{TT}}}$ is related to a single undetermined bulk coefficient by a relation analogous to \eqref{sht1}.

\chapter{List of Bulk Results for %\texorpdfstring{$Z_1$}{Z1} and \texorpdfstring{$Z_2$}{Z2}
Einstein Gravity}\label{AppendixB}

In this appendix we list some expressions for the invariants  in the shear and sound channels relevant for Chapter \ref{ch4}.

\section{Results in the shear channel}

For the source $\hat{H}_{tx}$ we find the following solution of Eq.\ (\ref{es1}) at $\mathcal{O}(\mu^1)$:
\begin{equation}
	\mathcal{Z}_1^{(tx)}\Big|_{\mu^1}=\frac{\mu \rho  \left(96 \left(\rho ^2+2\right)+3 w^6+\left(6-4 \rho ^2\right) w^4-12 \left(\rho ^2+8\right) w^2\right)}{10 \pi  r w^{10}}     
\end{equation}
and at $\mathcal{O}(\mu^2)$:
\begin{equation}
	\begin{split}
		\mathcal{Z}_1^{(tx)}\Big|_{\mu^2}=&\frac{\mu^2 \rho}{8400 \pi  r^5 w^{12}} \Big[-40320 \left(\rho ^2+2\right)^2-4920 w^{12} \log (w)+\left(6920-7280 \rho ^2\right) w^{10}\\
		&+5 \left(272 \rho ^4-2880 \rho ^2+271\right) w^8+40 \left(136 \rho ^4-331 \rho
		^2-154\right) w^6\\
		&+280 \left(33 \rho ^4+26 \rho ^2-268\right) w^4+896 \left(\rho ^4+140 \rho ^2+262\right) w^2\Big]\\
		&-\frac{12 \rho  \left(a^{8,2(tx)}_{8,0} \log (r)+a^{8,1(tx)}_{8,0}\right)}{\pi  r^5},       
	\end{split}
\end{equation}
where $a^{8,1(tx)}_{8,0}$ and $a^{8,2(tx)}_{8,0}$ are undetermined coefficients.

Choosing a source $\hat{H}_{xz}$, we get the bulk result:
\begin{equation}\label{propxz1}
	\mathcal{Z}_1^{(xz)}\Big|_{\mu^1}=-\frac{f_0 \sqrt{-\rho ^2+w^2-1} \left(96 \left(\rho ^2+2\right)+w^6+\left(2-4 \rho ^2\right) w^4-12 \left(\rho ^2+6\right) w^2\right)}{10 \pi  r w^{10}}
\end{equation}
and
\begin{equation}\label{propxz2}
	\begin{split}
		\mathcal{Z}_1^{(xz)}\Big|_{\mu^2}=&\frac{\mu^2 \sqrt{-\rho ^2+w^2-1}}{8400 \pi  r^5 w^{12}} \Big[40320 \left(\rho ^2+2\right)^2-4200 w^{12} \log (w)+120 \left(38 \rho ^2+17\right) w^{10}+\\
		&+5 \left(-272 \rho ^4+1792 \rho ^2+437\right) w^8+8 \left(-680 \rho
		^4+885 \rho ^2+448\right) w^6-\\
		&-168 \left(55 \rho ^4+46 \rho ^2-284\right) w^4-896 \left(\rho ^4+122 \rho ^2+226\right) w^2\Big]+\\
		&+\frac{12 \sqrt{-\rho ^2+w^2-1} \left(a^{8,2(xz)}_{8,0} \log (r)+a^{8,1(xz)}_{8,0}\right)}{\pi  r^5}\,.
	\end{split}
\end{equation}

Using the results for the bulk-to-boundary propagator $\mathcal{Z}_1^{(xz)}$ \eqref{propxz1} and \eqref{propxz2} we obtain the correlator $G_{xz,xz}^{(bulk)}$:

\begin{align}
	G_{xz,xz}^{(bulk)}\Big|_{\mu^0}\!=&-\frac{1}{\partial_t^2+\partial_z^2}\frac{3 \pi  C_T \left(z^2-7 t^2\right)}{5 \left(t^2+z^2\right)^5}\label{GxzxzBulk0}\\
	G_{xz,xz}^{(bulk)}\Big|_{\mu^1}\!=&-\frac{1}{\partial_t^2+\partial_z^2}\frac{3 \pi  \mu C_T \left(t^4-6 t^2 z^2+z^4\right)}{200 \left(t^2+z^2\right)^4}\label{GxzxzBulk1}\\
	G_{xz,xz}^{(bulk)}\Big|_{\mu^2}\!=&\frac{1}{\partial_t^2+\partial_z^2}\Bigg[\frac{\pi  \mu^2 C_T}{8400 \left(t^2+z^2\right)^3} \Big(210 t^6+648 t^4 z^2+6 t^2 z^4-160 z^6\nonumber\\
	&+105 \left(t^2+z^2\right)^3 \log \left(t^2+z^2\right)\Big)-\frac{3}{5} \pi  a^{8,1(xz)}_{8,0} C_T\Bigg]\label{GxzxzBulk2}.
\end{align}

\section{Results in the sound channel}

First we list the solutions of the sound channel equations of motion (\ref{es2}) for various polarizations.
For the source $\hat{H}_{tz}$ we get
\begin{equation}
	\mathcal{Z}_2^{(tz)}\Big|_{\mu^1}=\frac{16 \mu \rho  \sqrt{-\rho ^2+w^2-1}}{5 \pi  w^{12}} \Big[-w^6-3 w^4-96 w^2+2 \rho ^2 \left(w^4+4 w^2+60\right)+240\Big]
\end{equation}
and
\begin{equation}
	\begin{split}
		\mathcal{Z}_2^{(tz)}\Big|_{\mu^2}=&-\frac{4 \mu^2 \rho  \sqrt{-\rho ^2+w^2-1}}{315 \pi  r^4 w^{14}} \Big[18144 \left(\rho ^2+2\right)^2+798 w^{12}+\left(1356-584 \rho ^2\right) w^{10}\\
		&+\left(176 \rho ^4-2240 \rho ^2+1779\right) w^8+12
		\left(88 \rho ^4-384 \rho ^2+147\right) w^6\\
		&+336 \left(9 \rho ^4-17 \rho ^2+58\right) w^4+672 \left(7 \rho ^4-55 \rho ^2-131\right) w^2\Big].
	\end{split}
\end{equation}
For the source $\hat{H}_{tt}$ we get
\begin{equation}
	\begin{split}
		\mathcal{Z}_2^{(tt)}\Big|_{\mu^1}=&-\frac{2 \mu}{5 \pi  w^{12}} \Big[8 \rho ^4 \left(w^4+4 w^2+60\right)-8 \rho ^2 \left(w^6+3 w^4+66 w^2-120\right)\\
		&+w^2 \left(w^6+2 w^4+48 w^2-96\right)\Big]
	\end{split}
\end{equation}
and
\begin{equation}
	\begin{split}
		\mathcal{Z}_2^{(tt)}\Big|_{\mu^2}=&\frac{\mu^2}{3150 \pi  r^4 w^{14}} \Big[ 362880 \rho ^2 \left(\rho ^2+2\right)^2+15960 w^{14} \log (w)+120 \left(279 \rho ^2-113\right) w^{12}\\
		&-15 \left(1072 \rho ^4-4048 \rho ^2+593\right) w^{10}+20
		\left(176 \rho ^6-3120 \rho ^4+4083 \rho ^2-294\right) w^8\\
		&+120 \left(176 \rho ^6-1083 \rho ^4+651 \rho ^2-406\right) w^6+1344 (5 \left(9 \rho ^4-24 \rho ^2+91\right) \rho
		^2\\
		&+131) w^4+3360 \left(28 \rho ^6-265 \rho ^4-632 \rho ^2-36\right) w^2\Big]+\frac{a_{0,0}^{8,1(tt)}+a_{0,0}^{8,2(tt)} \log (r)}{r^4}
		.
	\end{split}
\end{equation}
For the source $\hat{H}_{zz}$ we get
\begin{equation}
	\begin{split}
		\mathcal{Z}_2^{(zz)}\Big|_{\mu^1}=&\frac{2 \mu}{5 \pi  w^{12}} \Big[480 \left(\rho ^4+3 \rho ^2+2\right)+w^8+\left(2-8 \rho ^2\right) w^6\\
		&+8 \left(\rho ^4-3 \rho ^2+31\right) w^4+16 \left(2 \rho ^4-43 \rho ^2-72\right)
		w^2\Big]
	\end{split}
\end{equation}
and
\begin{equation}
	\begin{split}
		\mathcal{Z}_2^{(zz)}\Big|_{\mu^2}=&\frac{\mu^2}{630 \pi  r^4 w^{14}} \Big[-72576 \left(\rho ^2+1\right) \left(\rho ^2+2\right)^2+1560 w^{14} \log (w)-24 \left(99 \rho ^2+16\right) w^{12}\\
		&+3 \left(720 \rho ^4-1776 \rho ^2-37\right)
		w^{10}-4 \left(176 \rho ^6-2240 \rho ^4+1791 \rho ^2+75\right) w^8\\
		&-24 \left(176 \rho ^6-781 \rho ^4+177 \rho ^2-1806\right) w^6-4032 \left(3 \rho ^6-5 \rho ^4+37 \rho ^2+81\right)
		w^4\\
		&-672 \left(28 \rho ^6-255 \rho ^4-1032 \rho ^2-848\right) w^2\Big]+\frac{a_{0,0}^{8,1(zz)}+a_{0,0}^{8,2(zz)}(0,0) \log (r)}{r^4}.
	\end{split}
\end{equation}
For the source $\hat{H}_{xx}$ we get
\begin{equation}
	\mathcal{Z}_2^{(xx)}\Big|_{\mu^1}=-\frac{8 \mu}{5 \pi  w^{12}} \Big[60 \left(\rho ^2+2\right)+25 w^4-4 \left(5 \rho ^2+33\right) w^2\Big]
\end{equation}
and
\begin{equation}
	\begin{split}
		\mathcal{Z}^{(xx)}_2\Big|_{\mu^2}=&\frac{\mu^2}{3150 \pi  r^4 w^{14}} \Big[181440 \left(\rho ^2+2\right)^2-11880 w^{14} \log (w)+180 \left(43-60 \rho ^2\right) w^{12}\\
		&+15 \left(176 \rho ^4-1136 \rho ^2+315\right) w^{10}+10 \left(880 \rho
        ^4-2292 \rho ^2+369\right) w^8\\
        &+120 \left(151 \rho ^4-237 \rho ^2-700\right) w^6+672 \left(45 \rho ^4+100 \rho ^2+1084\right) w^4\\
        &+3360 \left(5 \rho ^2 \left(\rho
        ^2-40\right)-406\right) w^2\Big]+\frac{a_{0,0}^{8,1(xx)}+a_{0,0}^{8,2(xx)} \log (r)}{r^4}.
	\end{split}
\end{equation}

Using the prescription (\ref{dufmzeposledna1}) for the sound channel and the solutions above, 
we find  that the correlator order-by-order in $\mu$ for the source $\hat{H}_{tt}$ is given by        
\begin{align}
	G_{tt,tt}^{(bulk)}\Big|_{\mu^0}\!=&\frac{1}{(\partial_t^2+\partial_z^2)^2}\frac{96 \pi  C_T \left(t^4-18 t^2 z^2+21 z^4\right)}{5 \left(t^2+z^2\right)^7}\label{GttttBulk0}\\
	G_{tt,tt}^{(bulk)}\Big|_{\mu^1}\!=&\frac{1}{(\partial_t^2+\partial_z^2)^2}\frac{4 \pi  \mu C_T \left(t^6-15 t^4 z^2+15 t^2 z^4-z^6\right)}{15 \left(t^2+z^2\right)^6}\label{GttttBulk1}\\
	G_{tt,tt}^{(bulk)}\Big|_{\mu^2}\!=&-\frac{1}{(\partial_t^2+\partial_z^2)^2}\frac{2 \pi  \mu^2 C_T \left(-691 t^8+1900 t^6 z^2+1910 t^4 z^4+860 t^2 z^6+133 z^8\right)}{1575 \left(t^2+z^2\right)^5},\label{GttttBulk2}
\end{align}
and for the source $\hat{H}_{zz}$ as
\begin{align}
	G_{zz,zz}^{(bulk)}\Big|_{\mu^0}\!=&\frac{1}{(\partial_t^2+\partial_z^2)^2}\frac{96 \pi  C_T \left(21 t^4-18 t^2 z^2+z^4\right)}{5 \left(t^2+z^2\right)^7}\label{GzzzzBulk0}\\
	G_{zz,zz}^{(bulk)}\Big|_{\mu^1}\!=&\frac{1}{(\partial_t^2+\partial_z^2)^2}\frac{4 \pi  \mu C_T \left(t^6-15 t^4 z^2+15 t^2 z^4-z^6\right)}{15 \left(t^2+z^2\right)^6}\label{GzzzzBulk1}\\
	G_{zz,zz}^{(bulk)}\Big|_{\mu^2}\!=&\frac{1}{(\partial_t^2+\partial_z^2)^2}\frac{2 \pi  \mu^2 C_T \left(-65 t^8-724 t^6 z^2+810 t^4 z^4+140 t^2 z^6+79 z^8\right)}{1575 \left(t^2+z^2\right)^5}.\label{GzzzzBulk2}
\end{align}

Finally, using the relation (\ref{dufmzeposledna2}) we get the $G^{bulk}_{xx,xx}$ in the form
\begin{align}
	G_{xx,xx}^{(bulk)}\Big|_{\mu^0}\!=&\frac{1}{\partial_t^2+\partial_z^2}\frac{24 \pi  C_T}{5 \left(t^2+z^2\right)^4}\label{GxxxxBulk0}\\
	G_{xx,xx}^{(bulk)}\Big|_{\mu^1}\!=&\,0\label{GxxxxBulk1}\\
	G_{xx,xx}^{(bulk)}\Big|_{\mu^2}\!=&\frac{1}{\partial_t^2+\partial_z^2}\Bigg[\frac{\pi  \mu^2 C_T}{3150} \left(126 \log \left(t^2+z^2\right)+\frac{-135 t^4+90 t^2 z^2-71 z^4}{\left(t^2+z^2\right)^2}\right)\nonumber\\
	&-\frac{1}{60} \pi  C_T \left(72 \left({a}_{6,0}^{8,1(xy)}+{a}_{8,0}^{8,1(xy)}\right)+\pi  a_{0,0}^{8,1(xx)}\right)\Bigg],\label{GxxxxBulk2}
\end{align}
where the undetermined coefficients ${a}_{6,0}^{8,1(xy)}$ and ${a}_{8,0}^{8,1(xy)}$ come from the scalar channel contribution and $a_{0,0}^{8,1(xx)}$ from  the sound channel.

\chapter{Details on Spinning Correlators}\label{app:SpinningBlocks}
In this appendix we summarize our conventions and provide some details on spinning conformal correlators in the embedding space that are used in this thesis following \cite{Costa:2011dw,Costa:2011mg}. The basic building blocks are 
\begin{equation}\label{eq:EmbeddSpaceHV}\begin{aligned}
		V_{i,jk}&= {(Z_i\cdot P_j)(P_i\cdot P_k)-(Z_i\cdot P_k)(P_i\cdot P_j)\over P_j\cdot P_k},\\
		H_{ij}&= -2[(Z_i\cdot Z_j)(P_i\cdot P_j)-(Z_i\cdot P_j)(Z_j\cdot P_i)],
\end{aligned}\end{equation}
where $V_1\equiv V_{1,23}$, $V_2\equiv V_{2,31}$ and $V_3\equiv V_{3,12}$. Here $P_i$ and $Z_i$ are null vectors in $\mathbb{R}^{1,d+1}$.

One possible basis for the three-point function of two stress tensors and a spin-$J$ operator with dimension $\Delta$ is given by $(P_{ij}=-2P_i\cdot P_j)$
\begin{equation}\label{eq:GenthreePt}
	\langle T(P_1,Z_1)T(P_2,Z_2)\OO(P_3,Z_3)\rangle = {\sum_{p=1}^{10} x^{(TT\OO)}_{p} Q_p\over  (P_{12})^{d+2-{\Delta+J\over 2}}(P_{23})^{\Delta+J\over 2}(P_{31})^{\Delta+J\over 2}},
\end{equation}
where 
\begin{equation}\label{eq:defQ}\begin{aligned}
		&Q_1 = V_1^2V_2^2V_3^J,\\
		&Q_2 = (H_{23}V_1^2V_2+H_{13}V_2^2V_1)V_3^{J-1},\\
		&Q_3 = H_{12}V_1V_2V_3^J,\\
		&Q_4 = (H_{13}V_2+H_{23}V_1)H_{12}V_3^{J-1},\\
		&Q_5 = H_{13}H_{23}V_1V_2V_3^{J-2},\\
		&Q_6 = H_{12}^2V_3^J,\\
		&Q_7 = (H_{13}^2V_2^2+H_{23}^2V_1^2)V_3^{J-2},\\
		&Q_8 = H_{12}H_{13}H_{23}V_3^{J-2},\\
		&Q_9 = (H_{13}H_{23}^2V_1+H_{23}H_{13}^2V_2)V_3^{J-3},\\
		&Q_{10} = H_{13}^2H_{23}^2V_3^{J-4}.
\end{aligned}\end{equation}
Conservation of the stress tensor further reduces the number of independent structures. In particular, when $\OO=T$ there are $3$ independent structures while for non-conserved operators of dimension $\Delta$ and spin $J=0,2,4$, there are $1,2$ and $3$ independent structures, respectively. However, we will mainly consider the differential basis introduced in \cite{Costa:2011dw} since this is useful when considering the four-point conformal blocks. It is based on multiplication by $H_{12}$ as well as the differential operators
\begin{equation}
	\begin{aligned}
		D_{11} =& (P_1\cdot P_2)(Z_1\cdot {\partial\over \partial P_2})-(Z_1\cdot P_2)(P_1\cdot {\partial\over \partial P_2})\cr
		&-(Z_1\cdot Z_2)(P_1\cdot {\partial\over \partial Z_2})+(P_1\cdot Z_2)(Z_1\cdot {\partial\over \partial Z_2}),\cr
		D_{12} =& (P_1\cdot P_2)(Z_1\cdot {\partial\over \partial P_1})-(Z_1\cdot P_2)(P_1\cdot {\partial\over \partial P_1})+(Z_1\cdot P_2)(Z_1\cdot {\partial\over \partial Z_1}),
	\end{aligned}
\end{equation}
and $D_{22}$ and $D_{21}$ obtained from $D_{11}$ and $D_{12}$ by $1\leftrightarrow 2$. We further define the following differential operators:
\begin{equation}\label{eq:DiffOp}
	\begin{aligned}
		D_1 &= D_{11}^2D_{22}^2\Sigma_L^{2,2},\\
		D_2 &= H_{12}D_{11}D_{22}\Sigma_L^{2,2},\\
		D_3 &= D_{21}D_{11}^2D_{22}\Sigma_L^{3,1}+D_{12}D_{22}^2D_{11}\Sigma_L^{1,3},\\
		D_4 &= H_{12}(D_{21}D_{11}\Sigma_L^{3,1}+D_{12}D_{22}\Sigma_L^{1,3}),\\
		D_5 &= D_{12}D_{21}D_{11}D_{22}\Sigma_L^{2,2},\\
		D_6 &= H_{12}^2\Sigma_L^{2,2},\\
		D_7 &= D_{21}^2D_{11}^2\Sigma_L^{4,0}+D_{12}^2D_{22}^2\Sigma_L^{0,4},\\
		D_8 &= H_{12}D_{12}D_{21}\Sigma_L^{2,2},\\
		D_9 &= D_{12}^2D_{21}^2\Sigma_L^{2,2},\\
		D_{10} &= D_{12}D_{21}^2D_{11}\Sigma_L^{3,1}+D_{21}D_{12}^2D_{22}\Sigma_L^{1,3},
	\end{aligned}
\end{equation}
where $\Sigma_{L}^{m,n}$ denotes the shifts $\Delta_{1}\to\Delta_1+m$ and $\Delta_{2}\to\Delta_2+n$. The three-point functions in the differential basis are then given by 
\begin{equation}\begin{aligned}\label{eq:SpinningThreePt}
		\langle T(P_1,Z_1)&T(P_2,Z_2)\OO_{\Delta,J}(P_3,Z_3)\rangle \cr
		&= \sum_{i=1}^{10}\lambda_{TT\OO_{\Delta,J}}^{(i)}D_i {V_3^J\over P_{12}^{\Delta_1+\Delta_2-\Delta-J}P_{23}^{\Delta+\Delta_2-\Delta_1+J}P_{13}^{\Delta+\Delta_1-\Delta_2+J}},
\end{aligned}\end{equation}
where we kept $\Delta_{1,2}$ to keep track of the action of $\Sigma_L^{m,n}$ in \eqref{eq:DiffOp}.

The spinning conformal partial waves can be obtained from the scalar partial waves $W_{\OO}$: 
\begin{equation}\label{eq:PartialWave}
	W_{\OO} = \left({P_{24}\over P_{14}}\right)^{\Delta_{12}\over 2} \left({P_{14}\over P_{13}}\right)^{\Delta_{34}\over 2}{g_{\Delta,J}^{(\Delta_{12},\Delta_{34})}(u,v)\over P_{12}^{\Delta_1+\Delta_2\over 2}P_{34}^{\Delta_3+\Delta_4\over 2}}
\end{equation}
with $\Delta_{ij}=\Delta_i-\Delta_j$ and the cross-ratios $(u,v)$ are given by 
\begin{equation}\begin{aligned}
		u&= {P_{12}P_{34}\over P_{13}P_{24}},\\
		v&={P_{14}P_{23}\over P_{13}P_{24}}.
\end{aligned}\end{equation}
The scalar conformal blocks are normalized as follows in the limit $u\to 0, v\to 1$:
\begin{equation}
	g_{\Delta,J}^{(\Delta_{12},\Delta_{34})}(u,v) \underset{u\to 0,v\to 1}{\sim} {J!\over (-2)^J({d\over 2}-1)_J}u^{\Delta\over 2}C_J^{({d\over 2}-1)}\Big({v-1\over 2\sqrt{u}}\Big),
\end{equation}
where $C_{J}^{({d\over 2}-1)}$ are Gegenbauer polynomials and $(a)_J$ denotes the Pochammer symbol. The spinning conformal partial waves are then obtained via
\begin{equation}\label{eq:diffRep}
	W_{\OO}^{\{i\}} = {\cal D}_{L}{\cal D}_{R} W_{\OO},
\end{equation}
where 
\begin{equation}\label{eq:DLeft}
	{\cal D}_{L} = H_{12}^{n_{12}}D_{12}^{n_{10}}D_{21}^{n_{20}}D_{11}^{m_1}D_{22}^{m_2}\Sigma_L^{m_1+n_{20}+n_{12},m_2+n_{10}+n_{12}},
\end{equation}
where $i$ labels the structure in the scalar partial wave and ${\cal D}_{R}$ is similarly defined with $1\to3$ and $2\to 4$. The integers $n_{ij}\geq 0$ and $m_{i}$ that labels the structure are determined by the solutions to the following equations ensuring the correct homogeneity under $P\to\alpha P$ and $Z\to\beta Z$:
\begin{equation}
	\begin{aligned}
		m_1 &= J_1-n_{12}-n_{10}\geq 0,\cr
		m_2 &= J_2-n_{12}-n_{20}\geq 0,\cr
		m_0 &= J_0-n_{10}-n_{20}\geq 0,
	\end{aligned}
\end{equation}
where $J=J_0$ is the spin of the exchanged operator. In the case of two spin-$2$ operators at $P_1$ and $P_2$ and scalar operators at $P_3$ and $P_4$, the possible combinations appearing in \eqref{eq:DLeft} can be taken to be the ones given in \eqref{eq:DiffOp}.

We are interested in the OPE limit of the contribution of individual blocks to
\begin{equation}
	\hat{G}(P_i,Z_i):=P_{34}^{\Delta_H}\langle  T(P_1,Z_1)T(P_2,Z_2)\OH(P_3)\OH(P_4)\rangle,
\end{equation}
where $\OH$ is a scalar operator with dimension $\Delta_H$. Using \eqref{eq:diffRep}, we find in this case
\begin{equation}\label{eq:SpinningBlocks}
	\begin{aligned}
		\hat{G}(P_i,Z_i)|&_{\OO_{\Delta,J}} = \sum_{i=1}^{10}\lambda_{TT\OO_{\Delta,J}}^{(i)}\lambda_{\OH\OH \OO_{\Delta,J}}D_i \left({P_{24}\over P_{14}}\right)^{\Delta_{12}\over 2} {g^{(\Delta_{12},0)}_{\Delta,J}(u,v)\over P_{12}^{\Delta_1+\Delta_2\over 2}},
	\end{aligned}
\end{equation}
where the differential operators $D_i$ are given by \eqref{eq:DiffOp} and $\Delta_1=\Delta_2=d$.

The spinning correlator in embedding space with indices is then obtained using 
\begin{equation}\begin{aligned}
		\hat{G}_{MN,PS}(P_i)= {1\over 2^2({d\over 2}-1)^2}&\hat{D}^{(1)}_M \hat{D}^{(1)}_N\hat{D}^{(2)}_P \hat {D}^{(2)}_S \hat{G}(P_i,Z_i)
	\end{aligned}
\end{equation}
where $\hat{D}^{(i)}_M$ is given by 
\begin{equation}\label{eq:deffDhat}
	\hat{D}^{(i)}_M = \left({d-2\over 2}+Z_i\cdot {\partial\over \partial Z_i}\right){\partial \over \partial Z_i^M}-{1\over 2}Z_{i M} {\partial^2\over \partial Z^2}. 
\end{equation}
In order to project down to physical space we impose $P^M_{i} = (1,x_i^2,x_i^\mu)$ and contract  indices in embedding space with ${\partial P_i^M\over \partial x^\nu}=(0,2x^{(i)}_\nu,\delta^\mu_\nu)$ \cite{Costa:2011dw,Costa:2011mg}. We then set $x_1^\mu = (1,\vec{0})$, $x_2^{\mu}=(1+t,\vec{x})$, $x_3^\mu =(0,\vec{0})$ and $x_4\to \infty$ with $|x_{21}|\ll 1$ in the OPE limit, such that $u\to 0$ and $v\to 1$. 

\section{Stress tensor block}\label{sec:T}
The relations between different bases for the stress tensor three-point function can be found in $e.g.$ Appendix C.1 in \cite{Hofman:2016awc}, some of which we summarize here for convenience. In the embedding space formalism \cite{Costa:2011mg,Costa:2011dw} the stress tensor three-point function can be built from \eqref{eq:GenthreePt}
\begin{equation}\label{eq:HVbasis}
	\langle T(P_1,Z_1)T(P_2,Z_2)T(P_3,Z_3)\rangle = {\sum_{p=1}^8 x_{p} Q_p\over  P_{12}^{d+2\over 2}P_{23}^{d+2\over 2}P_{31}^{d+2\over 2}},
\end{equation}
where the coefficients $x_p\equiv x_p^{(TTT)}$ are constrained by permutation symmetry and conservation to satisfy
\begin{equation}\label{eq:xp}\begin{aligned}
		x_1 &= 2x_2+{1\over 4}(d^2+2d-8)x_4-{1\over 2}d(d+2)x_7,\\
		x_8 &={1\over {d^2\over 2}-2}\Big[x_2-\Big({d\over 2}+1\Big)x_4+2dx_7\Big],\\
		x_2 &= x_3,\\
		x_4 &= x_5,\\
		x_6 &= x_7.
\end{aligned}\end{equation}

The stress tensor three-point function can be parameterized in terms of $(\hat{a},\hat{b},\hat{c})$ \cite{Osborn:1993cr} where one of these can further be traded for $C_T$ using the Ward identity 
\begin{equation}\label{eq:WardTTT}
	C_T = 4S_d {(d-2)(d+3)\hat{a}-2\hat{b}-(d+1)\hat{c}\over d(d+2)}.
\end{equation}
For the relation between the $x_p$ basis, $(\hat{a},\hat{b},\hat{c})$ and the $(t_2,t_4)$ coefficients that are natural when considering a conformal collider setup, we refer the reader to App.\ C in \cite{Hofman:2016awc}. However, we recall Eq.\ (C.10) in \cite{Hofman:2008ar} that relates these to $t_2$ and $t_4$ in $d=4$:
\begin{equation}
	\begin{aligned}
		t_2=& \frac{30(13\hat{a}+4\hat{b}-3\hat{c})}{14\hat{a}-2\hat{b}-5\hat{c}}\\
		t_4=& -\frac{15(81\hat{a}+32\hat{b}-20\hat{c})}{2(14\hat{a}-2\hat{b}-5\hat{c})}
	\end{aligned}
\end{equation}
and for $t_2=t_4=0$ one finds $\hat{a}={4\hat{c}\over 23}$ and $\hat{b}={17\hat{c}\over 92}$. On the other hand, the ratio of the anomaly coefficients $a,c$ is given by Eq.\ (C.12) in \cite{Hofman:2008ar}
\begin{equation}
	{a\over c} = {9\hat{a}-2\hat{b}-10\hat{c}\over 3(14\hat{a}-2\hat{b}-5\hat{c})}, 
\end{equation}
with $a=c$ when $t_2=t_4=0$. We further need the stress tensor three-point function with two heavy scalar operators
\begin{equation}
	\langle \OH(x_1)\OH(x_2)T_{\mu\nu}(x_3)\rangle = \lambda_{\OH\OH T_{\mu\nu}}\frac{W_\mu W_\nu-{1\over d}W^2\delta_{\mu\nu}}{ x_{12}^{2\Delta_H-2}x_{23}^{2}x_{31}^{2}},
\end{equation}
where $W^\mu = {x_{13}^\mu\over x_{13}^2}-{x_{23}^\mu\over x_{23}^2}$. Conformal Ward identities fix $\lambda_{\OH\OH T}$ to be
\begin{equation}
	\lambda_{\OH\OH T_{\mu\nu}} = -\frac{d}{d-1}{\Delta_H\over S_d},
\end{equation}
where $S_{d}={2\pi^{d\over 2}\over \Gamma({d\over 2})}$, which is related to $\mu$ and $\beta$ according to \eqref{eq:Temp} and \eqref{eq:defMu}.

From now on we consider $d=4$. For the stress tensor block we work with parametrization in terms of $(\hat{a},\hat{b},\hat{c})$.  In the channel $\hat{G}_{xy,xy}$, following the procedure described above, we obtain
\begin{equation}\label{e.p1c24}
	\begin{aligned}
		&\hat{G}_{xy,xy}|_T=\frac{\Delta_H }{2\pi^4(14 \hat{a}-2 \hat{b}-5 \hat{c})(t^2+\vec{x}^2)^5} \times\\
		&\times\Big[4 \hat{b} (-5 t^4 (x^2+y^2)+\vec{x}^2(x^4+6 x^2 y^2+y^4+(x^2+y^2) z^2)-4t^2 (x^4+9 x^2 y^2+y^4\\
		&+(x^2+y^2) z^2))+\hat{c} (-3 t^6+t^4 (13
		(x^2+y^2)-5 z^2)-\vec{x}^2(5 x^4-6 x^2 y^2+5 y^4\\
		&+4 (x^2+y^2) z^2-z^4)+t^2 (11 x^4+102
		x^2 y^2+11 y^4+10 (x^2+y^2) z^2-z^4))\\
		&+4 \hat{a} (t^6+t^4 (-17 (x^2+y^2)+3 z^2)-t^2 (13 x^4+106 x^2
		y^2+13 y^4+10 (x^2+y^2) z^2\\
		&-3 z^4)+\vec{x}^2(5 x^4-6 x^2 y^2+5 y^4+6 (x^2+y^2) z^2+z^4))\Big],
	\end{aligned}
\end{equation}
where $\vec{x}=(x,y,z)$ which after integrating over $x$ and $y$ gives 
\begin{equation}\label{eq:TxyxyInt}
	\begin{aligned}
		G_{xy,xy}|_T=\int dx dy \hat{G}_{xy,xy}|_T=-\frac{2 (7 \hat{a}+2 \hat{b}-\hat{c})\Delta_H (t^2-z^2)}{3\pi^3(14 \hat{a}-2 \hat{b}-5 \hat{c})(t^2+z^2)^2}.
	\end{aligned}
\end{equation}
and for $t_2=t_4=0$:
\begin{equation}\label{eq:TxyxyIntEinstein}
	G_{xy,xy}|_{T} = {2\Delta_H (t^2-z^2)\over 15\pi^3 (t^2+z^2)^2}.
\end{equation}
The results \eqref{eq:TxyxyInt} and \eqref{eq:TxyxyIntEinstein} are in agreement with \cite{Kulaxizi:2010jt}, where the explicit OPE was used to evaluate the stress tensor two-point function in a thermal state. 

Next, we find that $\hat{G}_{tx,tx}|_T$ is given by 
\begin{equation}\label{eq:Ttxtx}
	\begin{aligned}
		\hat{G}_{tx,tx}|_T=&\frac{\Delta_H }{2\pi^4(14 \hat{a}-2 \hat{b}-5 \hat{c})(t^2+\vec{x}^2)^5}\times\\
		&\times\Big[4\hat{b}(-t^6-x^2\vec{x}^4+t^4(-23x^2+4(y^2+z^2))\\
		&+t^2\vec{x}^2(9x^2+5(y^2+z^2)))+\hat{c}(15t^6-(5x^2-y^2-z^2)\vec{x}^4+t^4(41x^2\\
		&+7(y^2+z^2))-t^2\vec{x}^2(43x^2+7(y^2+z^2)))\\
		&+4\hat{a}(-13t^6+(3x^2-y^2-z^2)\vec{x}^4-3t^4(13x^2+y^2+z^2)+t^2\vec{x}^2(41x^2\\
		&+9(y^2+z^2)))\Big],
	\end{aligned}
\end{equation}
which becomes 
\begin{equation}\label{e.p1c28}
	\begin{aligned}
		G_{tx,tx}|_T=-\Delta_H{(64\hat{a}+14\hat{b}-19\hat{c})t^4-12(16\hat{a}+5\hat{b}-4\hat{c})t^2z^2+3(2\hat{b}+\hat{c})z^4\over 12(14\hat{a}-2\hat{b}-5\hat{c})\pi^3(t^2+z^2)^3},
	\end{aligned}
\end{equation}
after integrating over $x$ and $y$.
For $t_2=t_4=0$ this reduces to
\begin{equation}\label{eq:TtxtxIntEinstein}
	G_{tx,tx}|_T= \Delta_H{-9t^4+6t^2z^2+7z^4\over 60\pi^3 (t^2+z^2)^3}.
\end{equation}

Lastly, we consider $\hat{G}_{tz,tz}|_{T}$. Prior to integration, there is a $SO(3)$ rotational symmetry which allows us to obtain $\hat{G}_{tz,tz}|_{T}$ from \eqref{eq:Ttxtx} by $x\leftrightarrow z$. Integrating over the $xy$-plane we find
\begin{equation}\label{e.p1c30}
	\begin{aligned}
		G_{tz,tz}|_T={\Delta_H\over(14\hat{a}-2\hat{b}-5\hat{c})\pi^3(t^2+z^2)^4 }\Big[(-6\hat{a}+\hat{b}+2\hat{c})t^6+(-10\hat{a}-7\hat{b}+3\hat{c})t^4z^2\\
		+(30\hat{a}+7\hat{b}-8\hat{c})t^2z^4+(2\hat{a}-\hat{b}-\hat{c})z^6\Big].
	\end{aligned}
\end{equation}
For $t_2=t_4=0$ this reduces to
\begin{equation}\label{eq:TtztzIntEinstein}
	G_{tz,tz}|_T= \Delta_H{-105t^6+3t^4z^2+137t^2z^4+77z^6\over 270\pi^3 (t^2+z^2)^4}.
\end{equation}

\section{Spin-\texorpdfstring{$0$}{0} double-stress tensor block}
The simplest double-stress tensor operator is the scalar $[T^2]_{J=0}$ with dimension $\Delta_0$. The differential operators from the differential basis relevant here are $D_1$, $D_2$ and $D_6$ from \eqref{eq:DiffOp}, while the three-point function is given by \eqref{eq:SpinningThreePt} with $\lambda_{i,0}\equiv\lambda^{(i)}_{TT[T^2]_{J=0}}$. In order to impose conservation we demand that ${\partial\over \partial P_M}\hat{D}_M$ acting on \eqref{eq:SpinningThreePt} is $0$ \cite{Costa:2011mg}, where $\hat{D}_M$ is given by \eqref{eq:deffDhat}. This implies that the number of structures is reduced to just one
\begin{equation}\label{eq:consSpin0}
	\begin{aligned}
		\lambda_{2,0} &= -\frac{3}{4} (\Delta_0 -6) (\Delta_0 +2) \lambda_{1,0},\\
		\lambda_{6,0} &= \frac{3}{32} (\Delta_0 -6) (\Delta_0 -4) \Delta_0  (\Delta_0 +2)\lambda_{1,0},
	\end{aligned}
\end{equation}
and we are left with a single coefficient $\lambda_{1,0}$. The corresponding contribution to the correlator $\hat{G}(P_i,Z_i)$ (in embedding space) is given by 
\begin{equation}\label{eq:SpinningBlocksSpin0}
	\begin{aligned}
		\hat{G}(P_i,Z_i)|&_{[T^2]_0} =\sum_{i=1,3,6}\rho_{i,0}D_i W_{[T^2]_0},
	\end{aligned}
\end{equation}
where the conformal partial wave $W_{[T^2]_0}$ is given by \eqref{eq:PartialWave}. Note that the coefficients $\rho_{i,0}$ are related to $\lambda_{i,0}$ by an overall factor of the one-point function in the scalar state. They therefore, satisfy the same conservation condition as the $\lambda$'s in \eqref{eq:consSpin0}. The projection to the physical space and the relevant kinematics are described in the first part of this appendix.

\section{Spin-\texorpdfstring{$2$}{2} double-stress tensor block}
Because the spin-$2$ double-stress tensor $[T^2]_{J=2}$ is not conserved  there will be only two structures in the three-point function as compared to three for the stress tensor, even though they both have $J=2$. In the differential basis these can be labeled $\lambda_{i,2}\equiv \lambda^{(i)}_{TT[T^2]_{J=2}}$ with $i=1,2,\ldots 8$ in  \eqref{eq:SpinningThreePt}, which are reduced to two coefficients by imposing conservation:

\begin{align}
	\lambda_{3,2} &=\frac{\left(\Delta _2+2\right) \left(192 \lambda _{2,2}-\left(\Delta _2-4\right) \Delta _2 \left(\left(3 \Delta _2-16\right) \left(3 \Delta _2+4\right) \lambda _{1,2}+20 \lambda
		_{2,2}\right)\right)}{6 \Delta _2 \left(\Delta _2 \left(\Delta _2 \left(\left(\Delta _2-8\right) \Delta _2+2\right)+56\right)+96\right)},\nonumber\\
	\lambda_{4,2} &={\left(\Delta _2-4\right) \left(\Delta _2+2\right) \over 16 \left(\Delta _2 \left(\Delta _2 \left(\left(\Delta _2-8\right) \Delta
		_2+2\right)+56\right)+96\right)}\Big[(((\Delta _2-4) \Delta _2 (3 (\Delta _2-4) \Delta _2\nonumber\\
	&-52)-64) \lambda
	_{1,2}+4 (\Delta _2-8) (\Delta _2+4) \lambda _{2,2})\Big],\nonumber\\
	\lambda_{5,2} &=\frac{\left(\Delta _2-4\right) \Delta _2 \left(\left(15 \left(\Delta _2-4\right) \Delta _2+52\right) \lambda _{1,2}+52 \lambda _{2,2}\right)-96 \lambda _{2,2}}{12 \left(\Delta _2
		\left(\Delta _2 \left(\left(\Delta _2-8\right) \Delta _2+2\right)+56\right)+96\right)},\nonumber\\
	\lambda_{6,2} &={\left(\Delta _2-4\right) \Delta _2 \over 128 \left(\Delta _2 \left(\Delta _2 \left(\left(\Delta _2-8\right) \Delta _2+2\right)+56\right)+96\right)}\Big[((256\nonumber\\
	&-(\Delta _2-4) \Delta _2 ((\Delta _2-4) \Delta _2 (3 (\Delta _2-4)
	\Delta _2-56)+688)) \lambda _{1,2}\label{eq:consSpin2}\\
	&-4 ((\Delta _2-4) \Delta _2 (5 (\Delta _2-4) \Delta _2-52)+416) \lambda
	_{2,2})\Big]\nonumber\\
	&-\frac{48 \lambda _{2,2}}{\Delta _2 (\Delta _2 ((\Delta _2-8) \Delta
		_2+2)+56)+96},\nonumber\\
	\lambda_{7,2} &={\left(\Delta _2+2\right) \left(\Delta _2+4\right) \over 12 \left(\Delta _2-2\right) \Delta _2 \left(\Delta _2 \left(\Delta _2 \left(\left(\Delta _2-8\right) \Delta
		_2+2\right)+56\right)+96\right)}\Big[((\Delta _2-4) \Delta _2 \times\nonumber\\
	&\times((3 (\Delta _2-4) \Delta _2-44) \lambda _{1,2}+4
	\lambda _{2,2})-96 \lambda _{2,2})\Big],\nonumber\\
	\lambda_{8,2} &=-{\left(3 \left(\Delta _2-4\right) \Delta _2+16\right) \over 48 \left(\Delta _2 \left(\Delta _2 \left(\left(\Delta _2-8\right) \Delta _2+2\right)+56\right)+96\right)}\Big[((\Delta _2-4) \Delta _2 ((3 (\Delta _2-4) \Delta _2\nonumber\\
	&-44) \lambda
	_{1,2}+4 \lambda _{2,2})-96 \lambda _{2,2})\Big].\nonumber
\end{align}

The corresponding contribution to the correlator $\hat{G}(P_i,Z_i)$ is given by 
\begin{equation}\label{eq:SpinningBlocksSpin0secondformula}
	\begin{aligned}
		\hat{G}(P_i,Z_i)|&_{[T^2]_2} =\sum_{i=1}^8\rho_{i,2}D_i W_{[T^2]_2},
	\end{aligned}
\end{equation}
where the conformal partial wave $W_{[T^2]_2}$ is given by \eqref{eq:PartialWave}. Again, the coefficients $\rho_{i,2}$ are related to $\lambda_{i,2}$ by an overall factor of the one-point function in the scalar state. They therefore, satisfy the same conservation condition as the $\lambda$'s in \eqref{eq:consSpin2}. The projection to the physical space and the relevant kinematics are described in the first part of this appendix.

\section{Spin-\texorpdfstring{$4$}{4} double-stress tensor block}
For the spin-$4$ double-stress tensor operator $[T^2]_{J=4}$ there are a priori $10$ structures labelled by $\lambda_{i,4}\equiv \lambda^{(i)}_{TT[T^2]_{J=4}}$ with $i=1,2,\ldots 10$ in  \eqref{eq:SpinningThreePt}. Conservation reduces the number of structures to $3$ as follows:
{
	\allowdisplaybreaks
	\begin{align}
		\lambda_{4,4} &= {1\over96 ((\Delta _4-4) \Delta _4
			((\Delta _4-4) \Delta _4-44)+192)}\Big[(-((\Delta _4-4) \Delta _4 ((\Delta _4-4)\times\nonumber\\
		&\times \Delta _4 ((\Delta _4-4) \Delta _4 (3 (\Delta _4-4)
		\Delta _4-200)+5712)-92032))-485376) \lambda _{3,4}\nonumber\\
		&-2 (\Delta _4-6) (\Delta _4+4) (((\Delta _4-4)
		\Delta _4 ((\Delta _4-4) \Delta _4 (3 (\Delta _4-4) \Delta _4-68)-1024)\nonumber\\
		&+13056) \lambda _{1,4}+2 ((\Delta
		_4-4) \Delta _4 (3 (\Delta _4-4) \Delta _4-116)+768) \lambda _{2,4})\Big],\nonumber\\
		\lambda_{5,4} &= \frac{1}{8} (2 ((\Delta _4-4) \Delta _4+16) \lambda _{1,4}+(\Delta _4-8) (\Delta _4+2) \lambda _{3,4}+4 \lambda _{2,4}),\nonumber\\
		\lambda_{6,4} &={1\over256 (\Delta _4-6) (\Delta _4+4)
			((\Delta _4-4) \Delta _4 ((\Delta _4-4) \Delta _4-44)+192)}\Big[2 (\Delta _4-6)\times\nonumber\\
		&\times (\Delta _4+4) (((\Delta _4-4) \Delta _4 ((\Delta _4-4) \Delta _4 ((\Delta
		_4-4) \Delta _4 ((\Delta _4-4) \Delta _4 ((\Delta _4-4) \Delta _4\nonumber\\
		&-64)+1040)+11392)-262144)+2162688)
		\lambda _{1,4}+2 ((\Delta _4-4) \Delta _4\times\nonumber\\
		&\times((\Delta _4-4) \Delta _4 ((\Delta _4-4) \Delta _4 ((\Delta _4-4)
		\Delta _4-88)+2448)-17408)\nonumber\\
		&+86016) \lambda _{2,4})+((\Delta _4-4) \Delta _4 ((\Delta _4-4) \Delta _4
		((\Delta _4-4) \Delta _4\times\nonumber\\
		&\times ((\Delta _4-4) \Delta _4 ((\Delta _4-4) \Delta _4 ((\Delta _4-4) \Delta
		_4-108)+5104)-131904)\nonumber\\
		&+2009088)-18300928)+81788928) \lambda _{3,4}\Big],\nonumber\\
		\lambda_{7,4} &= {1\over 24
			((\Delta _4-4) \Delta _4 ((\Delta _4-4) \Delta _4-44)+192)}\Big[(\Delta _4-4) (\Delta _4+6) \times\nonumber\\
		&\times(2 (\Delta _4-6) (\Delta _4+4) (((\Delta _4-4) \Delta _4+20)
		\lambda _{1,4}+2 \lambda _{2,4})+((\Delta _4-4) \Delta _4\times\label{eq:consSpin4}\\
		&\times ((\Delta _4-4) \Delta _4-36)+704) \lambda _{3,4})\Big],\nonumber\\
		\lambda_{8,4} &= {1\over 32 (\Delta _4-6) (\Delta _4+4)
			((\Delta _4-4) \Delta _4 ((\Delta _4-4) \Delta _4-44)+192)}\Big[2 (\Delta _4-6)\times\nonumber\\
		&\times (\Delta _4+4) (((\Delta _4-4) \Delta _4 ((\Delta _4-4) \Delta _4 ((\Delta
		_4-4) \Delta _4-20) ((\Delta _4-4) \Delta _4\nonumber\\
		&+24)+4736)+135168) \lambda _{1,4}+2 ((\Delta _4-4) \Delta _4
		((\Delta _4-6) (\Delta _4-4) \Delta _4 (\Delta _4+2)\nonumber\\
		&-320)+7680) \lambda _{2,4})+(\Delta _4 (\Delta _4
		(\Delta _4 (\Delta _4 (\Delta _4 (((\Delta _4-20) \Delta _4+120) \Delta _4^3-1968 \Delta
		_4\nonumber\\
		&+2112)+23296)-78848)-327680)+1638400)+5111808) \lambda _{3,4}\Big],\nonumber\\
		\lambda_{9,4} &={1\over 3 (\Delta _4-6) (\Delta _4+4) ((\Delta
			_4-4) \Delta _4 ((\Delta _4-4) \Delta _4-44)+192)}\Big[(\Delta _4-6)\times\nonumber\\
		&\times (\Delta _4+4)\times (((\Delta _4-4) \Delta _4 (9 (\Delta _4-4) \Delta _4+68)-768)
		\lambda _{1,4}\nonumber\\
		&+16 ((\Delta _4-4) \Delta _4-6) \lambda _{2,4})+((\Delta _4-4) \Delta _4 ((\Delta _4-4) \Delta _4\times\nonumber\\
		&\times(3 (\Delta _4-4) \Delta _4-116)+3136)-15360) \lambda _{3,4}\Big],\nonumber\\
		\lambda_{10,4} &={1\over 12 ((\Delta _4-4) \Delta _4 ((\Delta _4-4) \Delta _4-44)+192)}\Big[((\Delta _4-4) \Delta _4+12)\times\nonumber\\
		&\times(-2 (\Delta _4-6)\times (\Delta _4+4) (((\Delta _4-4) \Delta
		_4+20) \lambda _{1,4}+2 \lambda _{2,4})\nonumber\\
		&-((\Delta _4-4) \Delta _4 ((\Delta _4-4) \Delta _4-36)+704) \lambda
		_{3,4})\Big].\nonumber
	\end{align}
}

The corresponding contribution to the correlator $\hat{G}(P_i,Z_i)$ is given by 
\begin{equation}\label{eq:SpinningBlocksSpin4}
	\begin{aligned}
		\hat{G}(P_i,Z_i)|&_{[T^2]_4} =\sum_{i=1}^{10}\rho_{i,4}D_i W_{[T^2]_4},
	\end{aligned}
\end{equation}
where the conformal partial wave $W_{[T^2]_4}$ is given by \eqref{eq:PartialWave}. The coefficients $\rho_{i,4}$ are related to $\lambda_{i,4}$ by an overall factor of the one-point function in the scalar state. They therefore, satisfy the same conservation condition as the $\lambda$'s in \eqref{eq:consSpin4}. The projection to the physical space and the relevant kinematics are described in the first part of this appendix.

\section{Integrated double stress tensor contribution}\label{as.labelik}
In this section we list the explicit expression for the integrated $\OO(C_T\mu^2)$ part of the conformal block expansion of $G_{xy,xy}$, $G_{tx,tx}$ and $G_{tz,tz}$ obtained using the procedure described above. We regulate divergences by including a factor of $|x|^{-\epsilon}$, which produces simple poles as $\epsilon\to0$.  For $G_{xy,xy}$ we find as $\epsilon\to0$:
\begin{equation}\label{eq:xyxyCFT}
	G_{xy,xy}|_{\mu^2 C_T}  = p^{(0)}_{xy,xy}(t,z)+p^{(1)}_{xy,xy}(t,z)\log(t^2+z^2)+{c_1 t^2+c_2 z^2\over \epsilon}
\end{equation}
where $c_1,c_2$ are some constants depending on the CFT data and
\begin{equation}
	p^{(0)}_{xy,xy}(t,z) = \frac{\pi ^5 \mu ^2 C_T}{1693440000(t^2+z^2)}\sum_{j=0}^2 p^{(0,2j)}_{xy,xy}t^{4-2j}z^{2j}
\end{equation}
with 
\begin{equation}
	\begin{aligned}
		p^{(0,0)}_{xy,xy}&=-8 (22050 \rho _{1,0}^{(1)}-162243 \rho _{1,2}^{(1)}-11683490 \rho _{1,4}^{(1)}+129168 \rho _{2,2}^{(1)}+4702775 \rho _{2,4}^{(1)}\\
		&+6991740 \rho _{3,4}^{(1)})+4410 \gamma^{(1)}
		_0+304479 \gamma^{(1)} _2-3577875 \gamma^{(1)} _4,\\
		p^{(0,2)}_{xy,xy}&=2\Big(-8(22050 \rho _{1,0}^{(1)}-89343 \rho _{1,2}^{(1)}-3641540 \rho _{1,4}^{(1)}+56268 \rho _{2,2}^{(1)}+1646645 \rho _{2,4}^{(1)}\\
		&+2005920 \rho _{3,4}^{(1)})+4410 \gamma^{(1)}
		_0+14364 \gamma^{(1)} _2-964005 \gamma^{(1)} _4\Big),\\
		p^{(0,4)}_{xy,xy}&=7\Big(8 (-3150 \rho _{1,0}^{(1)}+2349 \rho _{1,2}^{(1)}-215350 \rho _{1,4}^{(1)}+2376 \rho _{2,2}^{(1)}+90475 \rho _{2,4}^{(1)}\\
		&+123300 \rho _{3,4}^{(1)})+630 \gamma^{(1)} _0-39393 \gamma^{(1)}
		_2+74415 \gamma^{(1)} _4\Big),\\
	\end{aligned}
\end{equation}
and 
\begin{equation}
	p^{(1)}_{xy,xy}(t,z) = -\frac{\pi ^5 \mu ^2 C_T}{15680000}\sum_{j=0}^1 p^{(1,2j)}_{xy,xy}t^{2-2j}z^{2j}
\end{equation}
with
\begin{equation}
	\begin{aligned}
		p^{(1,0)}_{xy,xy}&=3\Big(16 (-702 \rho _{1,2}^{(1)}-19565 \rho _{1,4}^{(1)}+702 \rho _{2,2}^{(1)}+8085 \rho _{2,4}^{(1)}\\
		&+11480 \rho _{3,4}^{(1)})+490 \gamma^{(1)} _0-5607 \gamma^{(1)} _2+12040 \gamma^{(1)}
		_4\Big),\\
		p^{(1,2)}_{xy,xy}&=\Big(16 (486 \rho _{1,2}^{(1)}-6055 \rho _{1,4}^{(1)}-486 \rho _{2,2}^{(1)}\\
		&+2695 \rho _{2,4}^{(1)}+3360 \rho _{3,4}^{(1)}+280 \gamma^{(1)} _4)+1470 \gamma^{(1)} _0-189 \gamma^{(1)} _2\Big).
	\end{aligned}
\end{equation}

Next, for $(\partial_t^2+\partial_z^2)G_{tx,tx}$ we find
\begin{equation}\label{eq:txtxCFT}
	(\partial_t^2+\partial_z^2)G_{tx,tx}|_{\mu^2 C_T}  = p^{(0)}_{tx,tx}(t,z)+p^{(1)}_{tx,tx}\log(t^2+z^2)+{c_3\over \epsilon}
\end{equation}
for some constant $c_3$ where
\begin{equation}
	p^{(0)}_{tx,tx}(t,z) = -\frac{\pi ^5 \mu ^2 C_T}{423360000 \left(t^2+z^2\right)^3}\sum_{j=0}^3 p^{(0,2j)}_{tx,tx}t^{6-2j}z^{2j}
\end{equation}
with 
\begin{equation}
	\begin{aligned}
		p^{(0,0)}_{tx,tx}&=176400 \rho _{1,0}^{(1)}-265032 \rho _{1,2}^{(1)}+30139760 \rho _{1,4}^{(1)}+529632 \rho _{2,2}^{(1)}-12698840 \rho _{2,4}^{(1)}\\
		&-17881920 \rho _{3,4}^{(1)}+97020 \gamma^{(1)} _0-345492 \gamma^{(1)}
		_2-792435 \gamma^{(1)} _4,\\
		p^{(0,2)}_{tx,tx}&=\frac{1}{48} (25401600 \rho _{1,0}^{(1)}+203700096 \rho _{1,2}^{(1)}-9623496960 \rho _{1,4}^{(1)}-165597696 \rho _{2,2}^{(1)}\\
		&+3983253120 \rho _{2,4}^{(1)}+5576739840 \rho
		_{3,4}^{(1)}+21591360 \gamma^{(1)} _0+81539136 \gamma^{(1)} _2\\
		&+356907600 \gamma^{(1)} _4),\\
		p^{(0,4)}_{tx,tx}&=\frac{1}{48} (25401600 \rho _{1,0}^{(1)}+270884736 \rho _{1,2}^{(1)}+270063360 \rho _{1,4}^{(1)}-232782336 \rho _{2,2}^{(1)}\\
		&-40360320 \rho _{2,4}^{(1)}-293207040 \rho _{3,4}^{(1)}+29211840
		\gamma^{(1)} _0+125629056 \gamma^{(1)} _2\\
		&-45889200 \gamma^{(1)} _4),\\
		p^{(0,6)}_{tx,tx}&=176400 \rho _{1,0}^{(1)}+1134648 \rho _{1,2}^{(1)}-6309520 \rho _{1,4}^{(1)}-870048 \rho _{2,2}^{(1)}+2824360 \rho _{2,4}^{(1)}\\
		&+3044160 \rho _{3,4}^{(1)}+255780 \gamma^{(1)} _0+573048 \gamma^{(1)} _2-71715
		\gamma^{(1)} _4,
	\end{aligned}
\end{equation}
and 
\begin{equation}\begin{aligned}
		p^{(1)}_{tx,tx}&=-\frac{3 \pi ^5 \mu ^2 C_T }{15680000}\Big[-8 \left(4 \left(81 \rho _{1,2}^{(1)}+35 \rho _{1,4}^{(1)}-81 \rho _{2,2}^{(1)}-35 \rho _{3,4}^{(1)}\right)+315 \gamma^{(1)} _4\right)\\
		&+980 \gamma^{(1)} _0+63 \gamma^{(1)}
		_2\Big].
\end{aligned}\end{equation}

Lastly, for  $(\partial_t^2+\partial_z^2)^2G_{tz,tz}$ we find
\begin{equation}\label{eq:tztzCFT}
	(\partial_t^2+\partial_z^2)^2G_{tz,tz}|_{\mu^2 C_T}  = \frac{\pi ^5 \mu ^2 C_T}{2940000 \left(t^2+z^2\right)^5}\sum_{j=0}^4 p^{(0,2j)}_{tz,tz}t^{8-2j}z^{2j},
\end{equation}
where
\begin{equation}
	\begin{aligned}
		p^{(0,0)}_{tz,tz}&=-7776 \rho _{1,2}^{(1)}+197120 \rho _{1,4}^{(1)}+7776 \rho _{2,2}^{(1)}-86240 \rho _{2,4}^{(1)}-110880 \rho _{3,4}^{(1)}\\
		&+1470 \gamma^{(1)} _0+189 \gamma^{(1)} _2-1400 \gamma^{(1)} _4,\\
		p^{(0,2)}_{tz,tz}&=-248832 \rho _{1,2}^{(1)}+19983040 \rho _{1,4}^{(1)}+248832 \rho _{2,2}^{(1)}-8451520 \rho _{2,4}^{(1)}\\
		&-11531520 \rho _{3,4}^{(1)}-5880 \gamma^{(1)} _0-152712 \gamma^{(1)}_2-845320 \gamma^{(1)} _4,\\
		p^{(0,4)}_{tz,tz}&=233280 \rho _{1,2}^{(1)}-82577600 \rho _{1,4}^{(1)}-233280 \rho _{2,2}^{(1)}+34496000 \rho _{2,4}^{(1)}\\
		&+48081600 \rho _{3,4}^{(1)}-14700 \gamma^{(1)} _0+82530 \gamma^{(1)} _2+3193400 \gamma^{(1)} _4,\\
		p^{(0,6)}_{tz,tz}&=435456 \rho _{1,2}^{(1)}+29986880 \rho _{1,4}^{(1)}-435456 \rho _{2,2}^{(1)}-12246080 \rho _{2,4}^{(1)}\\
		&-17740800 \rho _{3,4}^{(1)}-5880 \gamma^{(1)} _0+218736 \gamma^{(1)} _2-1147160 \gamma^{(1)}_4,\\
		p^{(0,8)}_{tz,tz}&=-38880 \rho _{1,2}^{(1)}-257600 \rho _{1,4}^{(1)}+38880 \rho _{2,2}^{(1)}+86240 \rho _{2,4}^{(1)}+171360 \rho _{3,4}^{(1)}\\
		&+1470 \gamma^{(1)} _0-16695 \gamma^{(1)} _2+12320 \gamma^{(1)} _4.
	\end{aligned}
\end{equation}

\section{Comparison with the bulk calculations}
Solving \eqref{eq:matching} we find the anomalous dimensions \eqref{eq:solAnomDim}, the relations \eqref{eq:solCoeff} and the following bulk coefficients $(a^{(xy)}_{8,1}(6,0),a^{(xy)}_{8,1}(8,0),a^{(tx)}_{8,1}(8,0))$:
\begin{equation}\label{eq:solACoeff}
	\begin{aligned}
		a^{8,1(xy)}_{6,0} &= \frac{\pi ^4 \mu ^2 \left(2 \rho _{1,0}^{(1)}-3 \rho _{1,2}^{(1)}+\rho _{1,4}^{(1)}\right)}{1440}-\frac{3150449 \mu ^2}{47628000}+\frac{1441 \mu ^2}{37800 \epsilon },\\
		a^{8,1(xy)}_{8,0} &=-\frac{\pi ^4 \mu ^2 \left(2 \rho _{1,0}^{(1)}-3 \rho _{1,2}^{(1)}+\rho _{1,4}^{(1)}\right)}{1920}+\frac{1820863 \mu ^2}{127008000}-\frac{1801 \mu ^2}{50400 \epsilon },\\
		a^{8,1(tx)}_{8,0} &=\frac{\pi ^4 \mu ^2 \left(2 \rho _{1,0}^{(1)}+3 \rho _{1,2}^{(1)}-5 \rho _{1,4}^{(1)}\right)}{2880} -\frac{132403 \mu ^2}{1411200}-\frac{47 \mu ^2}{45360 \epsilon},
	\end{aligned}
\end{equation}
which are divergent as $\epsilon\to0$. Note that by also studying the $G_{xx,xx}$ polarization we get one more linearly independent equation:
\begin{equation}\label{eq:Axxxx}
	a^{8,1(xx)}_{0,0}=\frac{\pi ^3 \mu ^2 \left(\rho _{1,0}^{(1)}+2 \rho _{1,4}^{(1)}\right)}{80}-\frac{6713281 \mu ^2}{5292000 \pi }+\frac{11741 \mu ^2}{6300 \pi  \epsilon }.
\end{equation}

%%%%%%%%%%%
%%%%%%%%%%%
%%%%%%%%%%%
%%%%%%%%%%%
%%%%%%%%%%% p2 appendices:
%%%%%%%%%%%
%%%%%%%%%%%
%%%%%%%%%%%
%%%%%%%%%%%

\chapter{More Shear-Channel Results in Gauss-Bonnet Gravity}\label{Appendixxzxz}
    
When the source $\hat{H}_{xz}$ is turned on,  using the method discussed in \ref{shearHoloss},  we find
{
	\allowdisplaybreaks
	\begin{align}
	%G_{xz,xz}^{(bulk)}\Big|_{\tilde{\mu}^0}=&\frac{1}{\partial_t^2+\partial_z^2}\frac{3 \pi  t C_T \left(7 t^2-z^2\right)}{5 \sqrt{t^2} \left(t^2+z^2\right)^5}\ , \\
G_{xz,xz}^{(bulk)}\Big|_{\tilde{\mu}^0}=&\frac{1}{\partial_t^2+\partial_z^2}\frac{3 \pi   C_T \left(7 t^2-z^2\right)}{5 \left(t^2+z^2\right)^5}\ , \\
	%G_{xz,xz}^{(bulk)}\Big|_{\tilde{\mu}^1}=&(\kappa -2) \frac{1}{\partial_t^2+\partial_z^2}\frac{3 \pi   t \tilde{\mu } C_T \left(t^4-6 t^2 z^2+z^4\right)}{100 \kappa ^2 (\kappa +1) L^8 \sqrt{t^2} \left(t^2+z^2\right)^4}\ , \\
G_{xz,xz}^{(bulk)}\Big|_{\tilde{\mu}}=&(\kappa -2) \frac{1}{\partial_t^2+\partial_z^2}\frac{3 \pi    \tilde{\mu } C_T \left(t^4-6 t^2 z^2+z^4\right)}{100 \kappa ^2 (\kappa +1) L^8  \left(t^2+z^2\right)^4}\ , \\
	G_{xz,xz}^{(bulk)}\Big|_{\tilde{\mu}^2}=&-\frac{1}{\partial_t^2+\partial_z^2}\Bigg[\frac{\pi  \tilde{\mu }^2 C_T}{2100 \kappa ^4 (\kappa +1)^2 L^{16} \left(t^2+z^2\right)^3} \Big(6 (\kappa  (277 \kappa -700)\nonumber\\
	&+388) t^6+24 (\kappa  (161 \kappa -418)+230) t^4 z^2+6 (\kappa  (311 \kappa -836)\nonumber\\
    &+524) t^2 z^4+3 (\kappa  (277 \kappa -700)+388) \left(t^2+z^2\right)^3 \log
   \left(t^2+z^2\right)\nonumber\\
   &-16 (\kappa  (38 \kappa -119)+71) z^6\Big)+\frac{3}{5} \pi C_T  a^{8,1(xz)}_{8,0}\Bigg].\label{xzxzmu2hr}
\end{align} 
}In the Einstein-gravity limit, $\kappa\rightarrow1$, these results agree with those in Chapter \ref{ch4}. The $\tilde{\mu}$ contribution vanishes when $\kappa=2$, the critical value of the GB coupling for this channel.

Next consider the $\tilde{\mu}^2$ contribution. In the lightcone limit, we find 
    \begin{equation}
        \begin{split}
            G_{tx,tx}^{(bulk)}(x^+, x^-)\Big|_{\tilde{\mu}^2} \underset{x^- \to 0}{=}
&\,\,-\frac{1}{\partial_+ \partial_-}\bigg(\frac{ (\kappa -2)^2 17 \pi  C_T (x^+)^3 \tilde{\mu }^2}{33600 \kappa ^4 (\kappa +1)^2 L^{16} (x^-)^3}\\
            &-\frac{\pi  C_T \big(\kappa  (107 \kappa -340)+212\big) (x^+)^2 \tilde{\mu }^2}{11200 \kappa ^4 (\kappa +1)^2 L^{16} (x^-)^2}+{\cal O}\big({1\over x^{-}}\big)\bigg) \ .
        \end{split}
    \end{equation}
The leading-lightcone contribution vanishes at the critical $\kappa=2$.

\vspace{5mm}

\noindent{\it Reduced equation of motion:} 

With $\hat{H}_{xz}$ turned on, the corresponding reduced equation of motion is  
\begin{align}
\mu^2_{\rm{eff}({\rm shear})}\Theta_{2({\rm shear})} + \mu_{\rm{eff}({\rm shear})}\Theta_{1({\rm shear})}+ \Theta_{0({\rm shear})}= 0  \ , ~~~
\mu_{\rm{eff}({\rm shear})}=  \big(\kappa- 2 \big) \tilde\mu
\end{align}  
where $\Theta_{2({\rm shear})}$ is the same as the $\hat{H}_{tx}$ result:
{\footnotesize{\begin{align}
{ \Theta_{2({\rm shear})}\over 4  v^4  } =& 
\Big[w^2 \left(w^2 \del^4_w-38 w \del^3_w+591 \del^2_w\right)-4431 w \del_w +13440  \Big]Q_{\text{tot}}\ .
\end{align}}}
In this case, $\Theta_{1({\rm shear})}$ and $\Theta_{0({\rm shear})}$ are given by 
{\footnotesize{\begin{align}
{\Theta_{1({\rm shear})}\over 2  \kappa ^2 (\kappa +1) L^8 v^2}
=&\Big[ 2 \left(w^2-1\right) w^5   \del^3_w  -2 v w^6 \del_v \del^2_w     +  6  \left(9-5 w^2\right) w^4 \del^2_w + 34 v w^5 \del_v \del_w \nn\\
&~~~~~ +  6  \left(21 w^2-89\right) w^3 \del_w  -160  w^4 v \del_v +1920 w^2 \Big]\bar Q\nn
\end{align}}}
\vspace{-8mm}
{\footnotesize{\begin{align}
 &+\Big[ \left(w^4-1\right) w^4 \del^4_w    -2 \left(w^2-3\right) v w^5   \del_v \del^3_w + v^2 w^6  \del^2_v \del^2_w  -2  \left(6 w^4+5 w^2-19\right) w^3 \del^3_w \nn\\
& ~~~~~-17 v^2 w^5 \del^2_v \del_w  
+(31 w^2 -162 )  v w^4  \del_v \del^2_w
+  (-591 + 270 w^2 + 17 w^4) w^2 \del^2_w \nn\\
&~~~~~+ 80 v^2 w^4 \del^2_v 
+(1602 - 143 w^2)  v w^3  \del_v \del_w  
+ (4431 - 2670 w^2 + 335 w^4) w \del_w \nn\\
&~~~~~+ 80  (w^2 -72 ) v w^2 \del_v  - 640 (2 w^4- 15 w^2+21)\Big] Q_{\text{tot}}\ , \\
{\Theta_{0({\rm shear})}\over \kappa ^4 (\kappa +1)^2 L^{16} w^2}
=& \Big[ 2  ( w^2-1)^2 w^3 \del^3_w +  2  (w^2-1) v w^4\del_v \del^2_w -4 v^2 w^5 \del_w \del^2_v  \nn\\
& -(  26 w^4- 80 w^2+54 )  w^2 \del^2_w + 32 v^2 w^4 \del^2_v  - 2 (  7 w^2-17)v w^3  \del_v \del_w \nn\\
& +2  (61 w^4- 296 w^2  +267  )  w \del_w-160 v w^2 \del_v -64 (3 w^4 - 25 w^2+30  )\Big]\bar Q\nn
\end{align}
\vspace{-10mm}
\begin{align}
&+ \Big[ ( w^2-1)^2 w^4\del^4_w +  2 v^3 w^5 \del_w \del^3_v + 4  (w^2-1) v w^3 \del^3_w \del_v - ( 3 w^2-7)  v^2 w^4 \del^2_w\del^2_v \nn\\
&~~~  -12 (  w^4 - 3 w^2+2) w^3  \del^3_w  -16  w^4  v^3 \del^3_v  + ( 5 w^4 - 89 w^2 +108) v w^2  \del^2_w \del_v  +  ( 29 w^2 -119 )v^2 w^3 \del_w \del^2_v \nn\\
&~~~ + 3  ( 19 w^4- 82 w^2 +71)  w^2 \del^2_w -16 (4 w^2-35 )   w^2 v^2 \del^2_v - (55 w^4- 745 w^2+1068  ) v w \del_w \del_v \nn\\
&~~~ -21  ( 5 w^4-30 w^2
+33) w\del_w + 80 (2 w^4- 29 w^2+48  ) v \del_v  \Big] Q_{\text{tot}} \ . 
\end{align}}} 
\vspace{-8mm}

\chapter{Coefficients of Stress-Tensor Three-Point Function}\label{App:CFT}

The stress-tensor three-point function is parameterized by three coefficients $(\hat{a},\hat{b},\hat{c})$ \cite{Osborn:1993cr}. 
An alternative basis uses $(t_2,t_4,C_T)$, which can be related to the previous basis in the following way \cite{Hofman:2008ar}: 
\begin{equation}
	\begin{aligned}
		t_2 = \frac{30(13\hat{a}+4\hat{b}-3\hat{c})}{14\hat{a}-2\hat{b}-5\hat{c}} \ , ~~~~ t_4 = -\frac{15(81\hat{a}+32\hat{b}-20\hat{c})}{2(14\hat{a}-2\hat{b}-5\hat{c})},
	\end{aligned}
\end{equation}
and \cite{Osborn:1993cr}
\begin{equation}
	C_T = 4S_d \frac{(d-2)(d+3)\hat{a}-2\hat{b}-(d+1)\hat{c}}{d(d+2)}.
\end{equation}  Focus now on $d=4$. The stress-tensor three-point function was studied in the context of  holographic Gauss-Bonnet gravity in \cite{Buchel:2009sk}, which found $(t_{2,GB},t_{4,GB},C_T)$. 
Setting $t_2=t_{2,GB}$ given in \eqref{eq:t2GB} and $t_{4,GB}=0$, one finds\footnote{One can obtain the stress-tensor contribution to the thermal $TT$ correlators using \eqref{eq:defMuTilde}, together with Eqs.\ \eqref{eq:TxyxyInt}, \eqref{e.p1c28} and \eqref{e.p1c30}.} 
\begin{equation}\label{eq:abcGB}
	\begin{aligned}
		\hat{a} = \frac{8 C_T \left(-6+\frac{5}{\kappa}\right)}{45 \pi ^2} \ , ~~~~
		\hat{b} = \frac{C_T \left(33-\frac{50}{\kappa}\right)}{90 \pi ^2} \ ,  ~~~~ 
            \hat{c} = \frac{2 C_T \left(-84+\frac{61}{\kappa}\right)}{45 \pi ^2}\ .
	\end{aligned}
\end{equation}

%%%%%%%%%%%
%%%%%%%%%%%
%%%%%%%%%%%
%%%%%%%%%%%
%%%%%%%%%%% p3 appendices:
%%%%%%%%%%%
%%%%%%%%%%%
%%%%%%%%%%%
%%%%%%%%%%%

\chapter{Position-Space Analysis: Reduced Equations of Motion}\label{ap.pertpos}

To have a consistency check on the momentum-space results in Chapter \ref{ch6}, here we discuss perturbative solutions of the reduced equations of motion in position space, based on the  computation performed in Chapters \ref{ch4} and \ref{ch5}. In position space, we can systematically calculate the 
near-lightcone correlators 
order by order in a $\mu$ expansion where $\mu=(\pi/\beta)^4$.  

    \section{OPE in position space from holography}

Let the five-dimensional bulk coordinates\footnote{In this position-space calculation we adopt the Euclidean signature and for simplicity denote the Euclidean time by $t$, as in the Chapters \ref{ch4} and \ref{ch5}.} be $(r,t,x,y,z)$ and assume the metric fluctuations do not depend $x$ and $y$.
We consider a near-boundary regime:
    \begin{equation}\label{eq.opel}
        r\rightarrow\infty\quad\text{with}\quad rt,\,rz\,\,\text{fixed} \ .
    \end{equation}
 Defining $v=\frac{z}{r}$ and $w^2=1+r^2t^2+r^2z^2$, the bulk limit that isolates the near-lightcone correlator contribution is \cite{Fitzpatrick:2019zqz}
    \begin{equation}\label{eq.lcl}
        r\rightarrow\infty\quad\text{with}\quad v\,\,\text{fixed} \ .
    \end{equation}
Performing this limit on the equations of motion, one gets the reduced equations of motion for the bulk-to-boundary propagators $\mathcal{Z}$ 
which can be solved by the ansatz:
    \begin{align}
        \mathcal{Z}&=\mathcal{Z}^{AdS}\left(Q+\bar{Q}(\log r-\log w)\right),\label{eq.ourversionQtot}\\
    Q=\sum_{n=0}^\infty\sum_{m=-n}^n&a_{nm}v^{2n}w^{2m}\quad\text{and}\quad
    \bar{Q}= \sum_{n=0}^\infty\sum_{m=3}^nb_{nm}v^{2n}w^{2m}\label{Qdef}
    \end{align}
where $\mathcal{Z}^{AdS}$ is the bulk-to-boundary propagator in pure AdS, which in the scalar channel is $\frac{2r^2}{\pi w^6}$.\footnote{The full solution $Z$ is connected to $\mathcal{Z}$ by $Z(t,z,r)=\int\dd t^{\prime}\dd z'\mathcal{Z}(t-t^{\prime},z-z')\hat{Z}(t^{\prime},z')$, where $\hat{Z}$ is the boundary value of the invariant $Z$, see Chapters \ref{ch4} and \ref{ch5}.}

Using the above ansatz and the scalar-channel reduced equation of motion (in the position space) obtained in Chapter \ref{ch5}, one finds that the coefficients $a_{n,3}$ (for all $n\geq3$)  in \eqref{Qdef} are undetermined. 
This reflects the fact that, by performing the limit \eqref{eq.opel} one loses the information deep in the bulk. 
However, one can check that these undetermined coefficients are always suppressed in the lightcone limit.

We next compute the holographic stress-tensor correlators perturbatively in the $\mu$ expansion. 
In doing so, we note that in the lightcone limit ($i.e.$, $\xm\rightarrow0$ where $x^{\pm}= -it\pm z$) the correlator is fully determined by the coefficients $a_{nn}$.
The first few terms of the near-lightcone correlator $G_{\rm scalar}$ in position space are 
{
\allowdisplaybreaks
    \begin{equation}\label{eq.G3pertPosSpRES}
    \begin{aligned}
      \lim_{\xm \to 0}  G_{\rm scalar}=~ - {\pi C_T \over \xm^2} \Big(\frac{1}{5\xp^3\xm}
        &+\frac{1}{100} \mu\\
        &+\frac{\xm \xp^3}{1200} \mu^2\\
        &-\frac{21 \xm^2 \xp^6\log(-\xp\xm)}{286000} \mu^3\\
        &+\frac{71\xm^3 \xp^9\log(-\xp\xm)}{9792000} \mu^4\\
        &-\frac{2303\xm^4\xp^{12}\log(-\xp\xm)}{5684800000} \mu^5 \Big)+{\cal O} (\mu^6) \ .
    \end{aligned}
    \end{equation}}Using the same method, one can generalize the position-space computation to other two channels. However, due to the computational complexity in position space, we analyze the correlators in other channels in momentum space.

The correlator \eqref{eq.G3pertPosSpRES}  depends on the combination $\xm\xpp^3$, consistent with the Fourier transformed results using the variable $\alpha$ which we discuss next.

\section{Fourier transform to momentum space}

We here transform the position-space correlator \eqref{eq.G3pertPosSpRES}  to momentum space, where the conjugate variables to $(x^+,x^-)$ are $(q_+,q_-)=-\frac12(q^-,q^+)$. 
Fourier transform of the zeroth-order contribution diverges and thus needs to be regularized. 
We use the dimensional regularization, where instead of  $(\xm\xp)^{-3}$ we consider $(\xm\xp)^{-3-\epsilon}$ and then take $\epsilon \to 0$.  The result is 
    {\small\begin{equation}
        -\frac{\pi ^2 C_T (q^+q^-)^2}{320\,\epsilon}-\frac{1}{320} \pi ^2 C_T (q^+q^-)^2 \left(\log \left(-q^+q^-\right)+2 \gamma -3-\log (4)\right)+\mathcal{O}\left(\epsilon\right) \ ,
    \end{equation}}where $\gamma$ is the Euler's constant. Terms $\mathcal{O}(\epsilon)$ can be neglected, while the pole can be eliminated by counterterms. 
The regulator-independent (physical) $\log (-q^+q^-) $ term is
   {\small \begin{equation}
    -\frac{\pi C_T}{5\xm^3\xp^3}\longrightarrow-\frac{1}{320} \pi ^2 C_T (q^+q^-)^2\log (-q^+q^-).
    \end{equation}}We have verified that this result exactly matches the leading term in \eqref{eq.0corrG3} computed in momentum space.

The $\OO(\mu^1)$ and $\OO(\mu^2)$ terms can be directly Fourier transformed: 
    {\small\begin{align}
        -\frac{\pi C_T\mu}{100\xm^2}\longrightarrow-\frac{\pi ^2 C_T \mu  (q^++q^-)}{100 q^-}\approx-\frac{\pi ^2 C_T \mu  q^+}{100 q^-}\ , ~~
        -\frac{\pi C_T\mu^2\xp^3}{1200\xm}\longrightarrow\frac{2\pi^2 C_T \mu ^2}{25 (q^-)^4} \ .
    \end{align}}
\begin{center}
\begin{figure}[ht]
\hspace*{-1.5cm}
	\includegraphics[scale=0.72]{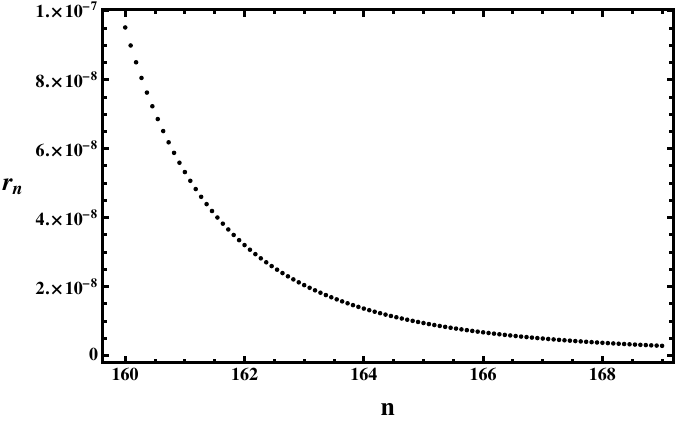}
    \centering
    \caption{Estimation of the radius of convergence in the scalar channel}
    \label{fig.rc}
\end{figure}
\end{center}
For the terms $\OO(\mu^n)$ with $n\geq3$, we use
    {\small\begin{equation}
    \log(-\xp\xm)\longrightarrow\frac{4\pi}{q^+q^-} \ .
    \end{equation}}Applying derivatives with respect to $\xp$ and $\xm$  gives   
    {\small\begin{equation}\label{eq.alllogft}
        \xpp^{-6}(\xp^3\xm)^n\log(-\xp\xm)\longrightarrow-\frac{3 \pi  16^{n+2} \Gamma (n+3) \Gamma (3 n+6) }{(n+1) q^+(q^-)^7\left(q^+(q^-)^3\right)^{n}} \ ,
    \end{equation}}which is valid for any $n\geq0$. Using this result, we can Fourier transform the log-terms appearing in \eqref{eq.G3pertPosSpRES}.
We obtain
{\small\begin{align}
G=G_0\Big( -\frac{1}{2}\log(-q^+q^-)  + \sum_{k= 1} c_{k}\alpha^{-k} \Big) \ , ~~~\alpha=-\mathfrak{q}^+(\mathfrak{q}^-)^3 
\end{align}}where $G_0$ and the first  several coefficients $c_{k}$ are
{\small\begin{align}
G^{\rm{scalar}}_0&= \frac{\pi^2C_T(q^+q^-)^2}{160} \ ,\,\,\,\,\, c^{\rm{scalar}}_{1}= \frac{1}{10} \ ,\,\,\,\,\,\, c^{\rm{scalar}}_{2}=  \frac{1}{20} \ , \,\,\,\,\,\,\,\, c^{\rm{scalar}}_{3}= \frac{378}{715}\ ,\ldots \label{scalarforri}\\ 
G^{\rm{shear}}_0&= - \frac{\pi^2C_T(q^+q^-)^2}{160} \ ,  \,c^{\rm{shear}}_{1}=- \frac{1}{40} \ , ~ c^{\rm{shear}}_{2}= -\frac{17}{560} \ , ~  c^{\rm{shear}}_{3}=-\frac{2241}{5720}\ ,\ldots \\
G^{\rm{sound}}_0&= - \frac{\pi^2C_T(q^+q^-)^2}{160} \ , \,c^{\rm{sound}}_{1}= \frac{1}{60} \ , \,\,\,\,\, c^{\rm{sound}}_{2}=\frac{11}{420} \ , \,\,\,\,\,\, c^{\rm{sound}}_{3}= \frac{2297}{6435}\ ,\ldots 
\end{align}}We further list the fourth-order correction in the scalar channel for comparison in Appendix \ref{apB}:
{\small\begin{align} \label{eq:scalarTQuar}
 c^{\rm{scalar}}_{4}=\frac{4473}{170} \ . 
\end{align}}Focusing on the scalar channel, in Fig.\ \ref{fig.rc} we also estimate the radius of convergence (see \eqref{rok})  by computing more higher-order terms in the expansion \eqref{scalarforri}. We find that $\lim_{n\to \infty} r_n \approx 0$.

\chapter{Momentum-Space Thermal Conformal Blocks} \label{apB}

In this appendix, we use the momentum-space conformal blocks to examine the scalar channel in Einstein gravity, whose equations of motion and action are equivalent to the ones for a massless scalar field. Note, however, that this equivalence do not persist when considering higher-derivative terms, such as Gauss-Bonnet gravity.\footnote{More precisely, while the OPE coefficients for minimal-twist stress tensors in the minimally coupled scalar case do not depend on higher-derivative terms such as the Gauss-Bonnet coupling \cite{Fitzpatrick:2019zqz} except through the temperature, the scalar wave equation in Gauss-Bonnet gravity is different from the scalar-channel EoM of metric perturbations.}  

The thermal conformal blocks in momentum space were computed in \cite{Manenti:2019wxs}. 
Expanded in thermal conformal blocks, the scalar correlator can be written as
{\small\begin{align}
		\langle \cal{O}\cal{O}\rangle_\beta(\omega_n,|\mathbf{q}|=q) = \sum_{\cal{O}_{\Delta,J}} a_{\cal{O}_{\Delta,J}}G_{\Delta,J}^{\Delta_\cal{O}}(\omega_n,q) \ , ~~
		a_{\cal{O}_{\Delta,J}} = \frac{\lambda_{\cal{O}\cal{O}\cal{O}_{\Delta,J}}b_{\cal{O}_{\Delta,J}}}{c_{\cal{O}_{\Delta,J}}\beta^\Delta}\frac{J!}{2^J(\frac{d-2}{2})_J} 
\end{align}}where $\omega_n=\frac{2\pi n}{\beta}$ is the Matsubara frequency and $a_{\cal{O}_{\Delta,J}}$ are thermal coefficients. (The explicit form of  $G_{\Delta,J}^{\Delta_\cal{O}}(\omega_n,q)$ is given by Eq.\ (2.11) in \cite{Manenti:2019wxs}.)  Minimal-twist operators are the ones that dominate 
in the near-lightcone limit that we are interested in.

The relevant coefficients for the near-lightcone correlator up to $\cal{O}(\beta^{-16})$, where the contributing operators are $[T^k]_{J}$ with dimension $4k$ and spin $J=2k$ with $k=1,\ldots 4$, are 
{\allowdisplaybreaks
	{\small\begin{align}\label{eq:AT}
			a_1 &= \frac{3C_T}{10} \ , \\
			a_{T} &= \frac{3C_T}{10}\frac{\Delta_{\cal O}}{120}\left(\frac{\pi}{\beta}\right)^4 \ ,\\
			a_{[T^2]_{4}} &= \frac{3C_T}{10}\frac{\Delta_{\cal O}  \left(7 \Delta_{\cal O} ^2+6 \Delta_{\cal O} +4\right) }{201600 (\Delta_{\cal O} -2)}\left(\frac{\pi}{\beta}\right)^8 \ ,\\
			a_{[T^3]_{6}} &=  \frac{3C_T}{10}\frac{\Delta_{\cal O}  \left(1001 \Delta_{\cal O} ^4+3575 \Delta_{\cal O} ^3+7310 \Delta_{\cal O} ^2+7500 \Delta_{\cal O} +3024\right)}{10378368000 (\Delta_{\cal O} -3) (\Delta_{\cal O} -2)}\left(\frac{\pi}{\beta}\right)^{12}\ , \\
			a_{[T^4]_{8}}&=  \frac{3C_T}{10}\frac{\Delta_{\cal O}\left(119119 \Delta_{\cal O}  ^6+969969 \Delta_{\cal O}  ^5+4184550 \Delta_{\cal O}  ^4 \right) }{592812380160000
				(\Delta_{\cal O}  -4) (\Delta_{\cal O}  -3) (\Delta_{\cal O}  -2)}\left(\frac{\pi}{\beta}\right)^{16}\nonumber\\ \label{eq:AT4}
			&~~+\frac{3C_T}{10}\frac{\Delta_{\cal O}\left(10867340 \Delta_{\cal O}^3+16958856 \Delta_{\cal O} ^2+14428176 \Delta_{\cal O}+5009760   \right)}{592812380160000
				(\Delta_{\cal O}  -4) (\Delta_{\cal O}  -3) (\Delta_{\cal O}  -2)}\left(\frac{\pi}{\beta}\right)^{16}.
\end{align}}}We have used $\mu 
=(\pi /\beta)^4$ and normalized the correlator to agree with the stress-tensor correlator in the scalar channel, $\langle T_{xy}T_{xy}\rangle$.  
The stress-tensor coefficient is fixed by Ward identities and the stress-tensor one-point function, while the $[T^2]_4$ and $[T^3]_6$ were computed in \cite{Fitzpatrick:2019zqz,Karlsson:2019dbd}. Here we have further obtained the coefficient for the $[T^4]_8$ operator with dimension $16$ and spin $8$ based on the method of \cite{Karlsson:2019dbd}.  

Let us first discuss operators $[T^k]_{J=2k}$ with $k=0,1,2,3$.
For the identity contribution, the block has a simple pole at $\Delta_\cal{O}=4$ with a residue that is purely a contact term. 
Removing the contact term gives
\be 
\langle \cal{O}\cal{O}\rangle|_{1} = -\frac{\pi ^2C_T}{640} \left(q^2+\omega_n^2\right)^2 \log \left(q^2+\omega_n^2\right) \ .
\ee 
After Wick-rotating $\omega_n\to -i\omega$ and taking the lightcone limit we reproduce \eqref{eq.0corrG3}. 
Likewise, for the stress-tensor exchange we find 
\be 
\langle \cal{O}\cal{O}\rangle|_{T} = -\frac{C_T \pi^2}{200}\frac{q^-}{q^+}\left(\frac{\pi}{\beta}\right)^4 \ ,
\ee 
which is in agreement with \eqref{eq:TExch}, where we remind the reader that $G_{xy,xy}=\frac{1}{2}G_{\rm scalar}$.  Moreover, we have verified that the $[T^2]_4$ and $[T^3]_6$ contributions reproduce the coefficients listed in \eqref{scalarforri}.  

Something interesting happens when we consider $[T^k]_{2k}$ with $k\geq 4$. First notice that, as seen in \eqref{eq:AT4}, there is a pole at $\Delta_\cal{O}=4$.
Interpreted as a correlator of a scalar with dimension four, this is related to the operator mixing between the $[T^4]_8$ operator with dimension $16$ and spin $8$ and a double-trace operator of the schematic form $\cal{O}\partial_{(\mu}\partial_\nu\partial_\rho\partial_{\sigma)}\cal{O}$ with the same quantum numbers. However, the block with $\Delta=16$, $J=8$ has a zero at $\Delta_\cal{O}\to 4$:
{\small\begin{align}
		G_{16,8}^{\Delta_\cal{O}}(\omega_n,q)\propto\frac{(\Delta_{\cal O}-4) \left(q^8-36 q^6 \omega_n^2+126 q^4 \omega_n^4-84 q^2 \omega_n^6+9 \omega_n^8\right)}{\left(q^2+\omega_n^2\right)^{10}}+\cal{O}\big((\Delta_\cal{O}-4)^2\big) \ .
\end{align}}Thus, the pole in $a_{[T^4]_{8}}$ cancels with the zero of the blocks. By Wick-rotating and taking the lightcone limit, we obtain
\be 
\langle \cal{O}\cal{O}\rangle|_{[T^4]_{8}}=\frac{2290176 \pi ^2 C_T}{425 (q^-)^2(q^+)^{10}}\left(\frac{\pi}{\beta}\right)^{16} \ ,
\ee which is in agreement with \eqref{eq:scalarTQuar}.

%%%%%%%%%%%%%%%%%
%%%%%%%%%%%%%%%%%

\chapter{Equations of Motion for Gauss-Bonnet Gravity}\label{ap.eomsGB} 

In Section \ref{sec:Momentum}, we showed that the equations of motion of metric fluctuations in Gauss-Bonnet gravity in the limit \eqref{LCbulklimit} reduce to the equations of motion in Einstein gravity. Here we list the coefficients $A, B$ in the Einstein-Gauss-Bonnet  equations of motion using the notation adopted in this thesis.\footnote{In Gauss-Bonnet gravity, the equations of motion in terms of gauge-invariant variables were first derived in \cite{Buchel:2009sk}. Some simplified expressions can be found in,  $e.g.$, Appendix D of \cite{Grozdanov:2016fkt}.} 
\\\\
\noindent {\bf Scalar Channel:}
{\allowdisplaybreaks
{\small
\begin{align}
\label{sc1}
A_{\rm{scalar}}&=\frac{u}{u^2-1}  \left(\frac{1}{\left(\kappa ^2-1\right) \left(1-u^2\right)+1}+\frac{1}{U(u)}\right)-\frac{1}{u} \ , \\
\label{sc2}
B_{\rm{scalar}}&= \frac{1}{4 u (U(u)-1)} \Big(\frac{(\kappa -1) (\kappa +1)^2 \left(3 \left(\kappa ^2-1\right) u^2-\kappa ^2\right)}{U(u)^2}\qfr^2 +\frac{\left(\kappa ^2-1\right)^2 }{U(u)-1}\wfr^2 \Big)
\end{align}}}
where $U(u)= \sqrt{\kappa ^2-\kappa ^2 u^2+u^2}$. 

\vspace{0.5cm}

\noindent{\bf Shear Channel:}
{\allowdisplaybreaks
{\small
\begin{align}
\label{sh1}
A_{\rm{shear}}&=  {1\over {u (1-U(u)) U(u)^3 \big(\kappa ^2 (\kappa +1) q^2 (U(u)-1)-\left(\kappa ^2-1\right) \omega ^2 U(u)^2\big)}}\\
&~~~ \times \Big[2 \kappa ^4 (\kappa +1) \left(\frac{1}{2} \left(1-\kappa ^2\right) \left(u^2-1\right) (U(u)-2)+U(u)-1\right)\qfr^2  \nn\\
&~~~~ +\left(1-\kappa ^2\right)  \left(\kappa ^4+\left(1-\kappa ^2\right)^2 u^4-2 \left(1-\kappa ^2\right) u^2 \left(U(u)-\kappa ^2\right)-\kappa ^2 U(u)\right)  U(u)^2\wfr ^2 \Big]
\ , \nn\\
\label{sh2}
B_{\rm{shear}}&=  \frac{\kappa^2 (\kappa +1) (U(u)+1)}{4 u \left(u^2-1\right) U(u)^2} \qfr^2  +\frac{ U(u)^2+2 U(u)+1}{4 u \left(u^2-1\right)^2} \wfr^2
\end{align}}} 

\noindent{\bf Sound Channel:}
{\allowdisplaybreaks
{\small
\begin{align}
\label{so1}
A_{\rm{sound}}&={1\over {2 u D_1(u) (U(u)-1) U(u)^2}}\Big[3 D_1(u) U(u)^2 (U(u)-1)  + \Big(\left(\kappa ^2-1\right)^2 u^4 \left(-3 \kappa ^2+5 U(u)-7\right)\nn\\
&~~~~~~~~~~~ +\kappa ^2 \left(\kappa ^2-1\right) u^2 \left(18 \kappa ^2-13 U(u)+10\right)-15 \kappa ^4 \left(\kappa ^2-2 U(u)+1\right)\Big)\qfr^2\nn\\
&~~~~~~~~~~~ -3 (1-\kappa ) \left(\kappa ^2-\left(\kappa ^2-1\right) u^2\right) \big(5 \kappa ^2 (U(u)-1)-\left(\kappa ^2-1\right) u^2 (5 U(u)-7)\big) \wfr ^2 \Big]  \ ,\\ 
\label{so2}
B_{\rm{sound}}&=  \frac{\left(\kappa ^2-1\right)^2}{D_0(u)} \Big[
\left(\kappa ^2-1\right)^3 \qfr^2 u^6 \left(3 (\kappa -1) \wfr^2+\qfr^2\right)+12 (\kappa -1)^2 \kappa ^2 (\kappa +1) \qfr^2 u^5\nn\\
&~~~~~~-4 (\kappa -1) \kappa ^2 \qfr^2 u^3 \left(3 \kappa ^2-7 U(u)+4\right)+\left(\kappa ^2-1\right)^2 u^4 \Big(\qfr^4 \left(3 \kappa ^2 (U(u)-2)+U(u)\right)\nn\\
&~~~~~~+2 (\kappa -1) \qfr^2 \wfr^2 U(u)-3 (\kappa -1)^2 \wfr^4 U(u)\Big)-\kappa ^2 \left(\kappa ^2-1\right) u^2 \big(\qfr^4 \left(\kappa ^2+2 U(u)\right) \nn\\
&~~~~~~+(\kappa -1) \qfr^2 \wfr^2 \left(9 \kappa ^2-4 U(u)\right)-6 (\kappa -1)^2 \wfr^4 U(u)\big)-3 \kappa ^4 \big(\qfr^4 \left(\kappa ^2 (U(u)-2)+U(u)\right)\nn\\
&~~~~~~+2 (\kappa -1) \qfr^2 \wfr^2 \left(U(u)-\kappa ^2\right)+(\kappa -1)^2 \wfr^4 U(u)\big) \Big]
\end{align}}}
where
{\allowdisplaybreaks
{\small 
\begin{align}
D_0(u)&= 4 (\kappa -1) u (U(u)-1)^2 U(u)^3 D_1(u) \ , \\
D_1(u)& =  (\kappa ^2-1) u^2 \left(\qfr^2+3 (\kappa -1) \wfr^2\right)+3 \kappa ^2 \left( (U(u)-1)\qfr^2-(\kappa -1) \wfr ^2\right)\ .
\end{align}}}